%% file: tese_arxiv.tex
\documentclass{iagtese_moser}	
                        
\usepackage[usenames,dvipsnames,svgnames,table]{xcolor}
\newcommand{\sori}{$\sigma$\,Ori\,E}

\usepackage{amsmath}

\def\mean#1{\left< #1 \right>}

\usepackage[flushleft]{threeparttable}
\usepackage[makeroom]{cancel}
\usepackage{graphicx}
\usepackage{bm}
\usepackage{placeins}
\usepackage{mathrsfs}
\usepackage{multirow}
\usepackage{booktabs}
\usepackage{lipsum}
\usepackage{pdfpages}
\usepackage{titlesec}
\usepackage{ulem}
\usepackage{wrapfig}
\usepackage{csquotes}
\usepackage{decorule}

\hypercolor				
\begin{document}			
						%
\input{tex/capa.tex}			
\pagestyle{empty}			
						%
\maketitle					%
\Dedicatoria				%
\input{tex/dedicatoria.tex}		%
\Agradecimentos			%
\input{tex/agradecimentos.tex}	
						%
\Epigrafe					%
\input{tex/epigrafe.tex}		%
						%
\Abstract					%
\input{tex/abstract.tex}		%
\Resume
\input{tex/resume.tex}	
\Resumo					%
\input{tex/resumo.tex}		
						%
\tableofcontents 			
\listoffigures 				
\listoftables 				
\Prologue
\input{tex/prologue.tex}

\cleardoublepage	
\pagestyle{fancy}			
                                %

\input{chap/intro.tex}
\input{chap/tools.tex}

\input{chap/magstars_arxiv.tex}
\input{chap/bestars_arxiv.tex}
\input{chap/phots_arxiv.tex}
\input{chap/aeri.tex}
   						%
						%
						%
\bibliography{/data/Dropbox/bib}	
						%
\begin{apendice}			
\input{tex/ap-equations.tex}		
\input{tex/ap-polobs.tex}		
\input{tex/ap-beatphot.tex}
\input{tex/ap-other.tex}

\input{tex/ap-activelog.tex}
\end{apendice}				
						%
\end{document}

%% file: tex/capa.tex
\institution{University of São Paulo (USP) \\
Institute of Astronomy, Geophysics and Atmospheric Sciences \\
Department of Astronomy \\
\& \\
University Nice-Sophia Antipolis (UNS) \\
UFR Sciences \\
Doctoral School in Fundamental and Applied Sciences}

\title{An interferometric view of hot star disks}

\translator{A thesis presented to The Department of Astronomy, Institute of
Astronomy, Geophysics and Atmospheric Sciences, University of São Paulo (USP)
 and to The Doctoral School in Fundamental and Applied Sciences, University of
Nice-Sophia Antipolis (UNS), in partial fulfillment of the requirements for the
double-degree of Doctor in Sciences in the subject of Astronomy.\\ \\
Under the supervision of:\\
Dr. Alex Cavaliéri Carciofi (USP)\\
\hspace{1 cm} \& \\
Dr. Armando Domiciano de Souza (UNS)
\newline}

\author{Daniel Moser Faes}

\date{São Paulo \ano}

%% file: tex/dedicatoria.tex
\hfill
\vfill
\hfill{\textit{To my ancestors, whom left Trento }

\hfill{\textit{in search of a new life in} Terra Incognita Brasil}
\vspace{0.5cm}
\decorule
\vspace{0.5cm}
\hfill{\textit{And to the king whose reign I found }}

\hfill{\textit{going through the} virage Fairmont \textit{and the} virage du Portier...}
\vspace{2cm}

%% file: tex/agradecimentos.tex

First of all, I thank my advisors Alex C. Carciofi \& Armando Domiciano de Souza. In addition to science, they shared with me moments of joy and pain in my developing process. I also acknowledge other great researchers I met and with whom I could learn or be inspired by: J. Bjorkman, Th. Rivinius and D. Baade. Also important, K. Bjorkman, A. Granada; in Nice, P. Bendjoya, P. Stee, F. Millour, F. Farrock and T. Lanz; in Brazil, R. Costa, W. Maciel and A. Damineli. A special thanks to G. Meynet, who kindly helped me in Switzerland after been stolen! 

I acknowledge the support from FAPESP (grants 2010/19060-5 and 2012/04916-7). This research made use of \textsc{simbad} and \textsc{vizier} databases (CDS, Strasbourg), as well as NASA Astrophysics Data System. Also, the Jean-Marie Mariotti Center (JMMC) Softwares. This work was possible by the use of the computing facilities of the Laboratory of Astroinformatics (IAG/USP, NAT/Unicsul), whose purchase was done by the Brazilian agency FAPESP (grant 2009/54006-4) and the INCT-A.

\decorule
\vspace{0.25cm}

En français, je remercie tous mes amies que j'ai rencontrais pendant mon séjour en France: Felipe \& Virginie; Marcia, Ana Paula, Jacques \& Mar\'ilia; Ricardo, Let\'icia \& Ricardo. Tous les bénévoles du Fourneau Économique, en particulier 
la souer Anne-Marie et Galina. Aussi, l'atelier de chant choral de l'UNS. Les étudiants et post-docs du labo Lagrange et de l'OCA, en special Paul \& Lina; Samir \& Zeinab; Gustavo, Andr\'e \& Suelen.
Comment me manque le beau temps à Nice, du pain, du vin et des fromages! Mais surtout les tours de vélo... Cyril \& Laura, je ne vous ai pas oublié!

\decorule
\vspace{0.25cm}

Em portugu\^es, agradeço aos meus pais, Marlize e Severino. Ao meu pedaço, Thamirys. A todos os amigos do BeACoN: Rodrigo, Cyril, Despo; Leandro, Bruno, Mohammad, Daiane; Bednarski, André Luiz e Artur. E os agregados, Robert, Fellipy e Marcelo. Aos meus \textit{homemates} Daniel e Arianna. Aos amigos brasileiros que visitei na Europa: Aiara (Heidelberg), Michael \& Sonia (Heilbronn); Andressa (em Munique, e sem passaporte!) e Melina (Berlim e Leipzig); Tha\'ise (em P\'adua) e Marcus (Paris). A todos que, de alguma forma, me ajudaram nesta conquista, os meus sinceros agradecimentos!

\vfill

\begin{flushleft}
\rule{6cm}{0.5pt}\\
{\footnotesize{This thesis was written in \LaTeX{} making use of the
\textsc{iagtese} class developed by students of IAG/USP.}}
\end{flushleft}

%% file: tex/epigrafe.tex

\vfill

\begin{flushright}
``\textit{It is only by going through a volume of work that
your work will be as good as your ambitions. And I took
longer to figure out how to do this than anyone I've ever met.}''\\

\vspace{0.4cm}
Ira Grass

\end{flushright}
\vspace{0.5cm}

\begin{flushright}



``\textit{Soit dit en passant, c'est une chose assez hideuse que le succès.\\
Sa fausse ressemblance avec le mérite trompe les hommes.}''\\

\vspace{0.4cm}
Victor Hugo

\end{flushright}
\vspace{0.5cm}

\begin{flushright}
``\textit{Foi vendo as coisas que eu vi, foi procurando entendê-las,\\
foi contemplando as estrelas, que eu aprendi a rezar.}''\\

\vspace{0.4cm}
José Fernandes de Oliveira

\end{flushright}
\vspace{1cm}

%% file: tex/abstract.tex
Optical long baseline interferometry was recently established as a technique capable of resolving stars and their circumstellar environments at the milliarcsecond (mas) resolution level. This high-resolution opens an entire new window to the study of astrophysical systems, providing information inaccessible by other techniques.

Astrophysical disks are observed in a wide variety of systems, from galaxies up to planetary rings, commonly sharing similar physical processes. Two particular disk like systems are studied in the thesis: (i) B He-rich stars that exhibits magnetic fields in order of kG and that trap their winds in structures called magnetospheres; and (ii) Be stars, fast rotating stars that create circumstellar viscous disks.

This study uses the interferometric technique to investigate both the photosphere proper and the circumstellar environment of these stars. The objective is to combine interferometry with other observational techniques (such as spectroscopy and polarimetry) to perform a complete and well-constrained physical description of these systems. This description is accompanied by radiative transfer models performed by the {\sc hdust} code.

The first firm detection of a hot star magnetosphere in continuum linear polarization is reported, as a result of the monitoring campaign of \sori{} at the Pico dos Dias Observatory (OPD/LNA). The polarimetric data was modeled by a single-scattering ``dumbbell+disk'' model, that describes the magnetosphere as constituted of two blob-like structures (at the intersection to the rotation and magnetic equators) and a circumstellar disk at the magnetic equator. The polarimetric modeling predicts a lower blob to disk mass ratio than the expect from the RRM (Rigid Rotating Magnetosphere) model, that provides a good spectroscopic description of these magnetospheres.
\\

In addition to \sori{}, we report the first polarimetric detections of the magnetospheres of HR\,7355 and HR\,5907. Our analysis indicate that these structures share similar properties, such as polarimetric modulation amplitude and mass distribution. 
In the case of the HR\,5907, we also report the first interferometric detection of a magnetosphere with differential phase amplitude of about 3 degrees with AMBER-VLTI/ESO. 


A new interferometric phenomenon, called CQE-PS (Central Quasi-Emission Phase Signature), was described and identified as a useful tool for the study of Be circumstellar disks, in particular for shell stars. Departures from a S-shaped differential phases occur when the disk obscures part of the stellar photosphere. This disk absorption alters the phase signal mainly near the rest wavelengths and can introduce a central reversal in the phase profile. This phenomenon can be used to probe disk size, density and radial slope. It may even provide an estimate for the (maximum) stellar angular size.

The photosphere of the Be star Achernar was studied in detail using high-precision visibilities and closure phase information from PIONIER-VLTI/ESO. This is the first precise photospheric characterization of a Be star and is, up to date, the characterization of the star with the highest mass and rotation rate. The variable line profile of Achernar and high-precision photometric frequencies are analyzed in the light of these new parameters. These results have great significance for stellar models and for the Be stars, such as the determination of the gravitational darkening coefficient.

We present the \textsc{BeAtlas} project as part of BeACoN group from IAG-USP. It consists of a systematic grid of Be stars models generated by the \textsc{hdust} code aiming at a comprehensive investigation of Be stars and their state-of-art modeling, the VDD (Viscous Decretion Disk). As first project application, we argue that Achernar is not a typical main-sequence star, displaying characteristics of a star leaving the main sequence: it exhibits an big size and a luminosity higher than the expected for its mass.

The recent outburst of Achernar was investigated and modeled. For the analysis of the activity, a large set of observational data was obtained, such as AMBER spectro-interferometry, FEROS spectroscopy (ESO) and optical polarimetry (OPD/LNA). These observations contain the first spectro-interferometric study of a Be star disk evolution, angularly resolving the growing disk.

The secular evolution of the newly formed disk is characterized and the VDD modeling prescription was employed. The AMBER interferometric analysis shows no evidence of a polar wind in this active phase. The H$\alpha$ spectroscopy exhibits a slow and gradual evolution, and reaches a near stationary regime just after $\sim$1.6~years. This feature will allow to estimate the disk viscous diffusion coefficient with dynamic VDD modeling.
\vspace{30pt}


%% file: tex/resume.tex

Interférométrie optique/IR à longue base a été récemment mis en place comme une technique capable de résoudre spatialement les étoiles et leurs environnements circumstellaires au niveau de la milliseconde d'angle (\textit{mas}). Cette haute résolution ouvre toute une nouvelle fenêtre pour l'étude de les systèmes astrophysiques, fournissant des informations inaccessibles par d'autres techniques. 

Disques astrophysiques sont observées dans une grande variété de systèmes, de galaxies jusqu'à anneaux planétaires, partageant communément processus physiques similaires. Deux disques particuliers sont étudiés dans la thèse: (i) les étoiles B He-riches qui présente des champs magnétiques de l'ordre de kG et que confine leurs vents dans des structures appelées magnétosphères; et (ii) les étoiles Be, rotateurs rapides qui présentent des disques circumstellaires épisodiques.

Cette étude utilise la technique interférométrique pour étudier à la fois la photosphère et l'environnement circumstellaire de ces étoiles. L'objectif est de combiner l'interférométrie avec d'autres techniques d'observation (telles que la spectroscopie et la polarimétrie) pour effectuer une description physique complète et bien contraindre ces systèmes. Cette description est acquise par l'interprétation des l'ensemble des observations par des modèles de transfert radiatif effectués par le code de \textsc{hdust}. 

La première détection d'une magnétosphère d'une étoile chaude en polarisation linéaire du continuum est rapporté, à la suite de la campagne de surveillance de \sori{} au Observatoire Pico dos Dias (OPD/\-LNA). Les données polarimétrique ont été interprétées par un modèle simple de diffusion composé d'un ``haltère + disque'', qui décrit la magnétosphère comme constitué de deux structures comme 
sphéroïdes (à l'intersection de les équateurs rotationnel et magnétiques) et un disque circumstellaire à l'équateur magnétique. La modélisation polarimétrique prédit une masse pour le 
sphéroïde au rapport du disque inférieure à celle attendue par la modélisation RRM (\textit{Rigid Rotating Magnetosphere}), qui fournit une bonne description spectroscopique de ces magnétosphères. 

En plus de \sori{}, nous présentons les premières détections polarimétriques pour les magnéto\-sphères de HR\,7355 et HR\,5907. Notre analyses indique que ces structures partagent des propriétés similaires, comme l'amplitude de la modulation polarimétrique et la distribution de masse. Dans le cas de la HR\,5907, nous présentons également la première détection interférométrique d'une magnétosphère avec une amplitude de phase différentiel d'environ 3 degrés avec AMBER-VLTI/ESO. 

Une nouvelle signature interférométrique, appelé CQE-PS (\textit{Central Quasi-Emission Phase Signature}), a été décrite et identifiée comme un outil efficace pour étudier des disques circumstellaires Be, en particulier pour les étoiles \textit{shell} (étoiles avec disques vues par l'équateur). Départs de phases différentielles en forme de ``S'' se produisent lorsque le disque obstrue une partie de la photosphère stellaire. Cette absorption pour le disque modifie le signal de phase essentiellement à proximité des longueurs d'onde de repos et peut introduire une inversion de phase central dans le profil des phases. Ce phénomène peut être utilisé pour sonder la taille du disque, et son profil radial de densité. Il peut également fournir une estimation de la taille (diamètre) angulaire maximale de l'étoile.

La photosphère de l'étoile Be Achernar a été étudiée en détail en utilisant des visibilités de haute précision et de l'information de clôture de phase de PIONIER-VLTI/ESO. Ceci est la première caractérisation précise de la photosphère une étoile Be et est, à ce jour, la caractérisation de l'étoile avec le taux de masse et de rotation le plus élevé. Le profil de raie variable de Achernar et fréquences photométriques de haute précision sont analysés à la lumière de ces nouveaux paramètres. Ces résultats ont une grande importance pour les modèles stellaires et pour les étoiles Be, telles que la détermination du coefficient de assombrissement gravitationnel.

Nous présentons le projet \textsc{BeAtlas} dans le cadre du groupe Beacon de l'IAG-USP. Il se compose d'une grille systématique de modèles Be étoiles générés par le code \textsc{hdust} visant à une recherche approfondie des étoiles Be et de leur modélisation ``état de l'art'', le VDD (disque de décrétion visqueux). Comme première application du projet, nous soutenons que Achernar est pas une étoile de la séquence principale typique, présentant des caractéristiques d'une étoile laissant la séquence principale: elle présente une grande taille et une luminosité plus élevée que le prévu pour sa masse.

La récente activité de Achernar a été étudié et modélisé. Pour l'analyse de l'activité récente du disque, un grand ensemble de données d'observation a été obtenu, comme spectro-interférométrie AMBER, spectroscopie FEROS (ESO) et polarimétrie optique (OPD/LNA). Ces observations permettent la première étude spectro-interférométrique de l'évolution temporelle d'un disque de étoile Be (croissance du disque résolue angulairement).

L'évolution séculaire du disque juste formé est caractérisé. Visant à obtenir un modèle physique réaliste de l'environnement circumstellaire de Achernar, le code \textsc{hdust} a été utilisé avec un modèle de prescription VDD. L'analyse interférométrique AMBER ne montre aucun signe d'un vent polaire dans cette phase active. Le spectroscopie H$\alpha$ présente une évolution lente et progressive, et atteint un régime stationnaire après seulement $\sim1,6$~années. Cette fonctionnalité permettra d'estimer le coefficient de diffusion visqueuse du disque avec la modélisation de la dynamique VDD.
\vspace{30pt}

Mots-clés: étoiles: individuelle (Achernar) - étoiles: paramètres fondamentales - techniques: interférométrique - matière circumstellaire - étoiles: raie d'émission, Be - étoiles: champ magnétique

%% file: tex/resumo.tex
Interferometria óptica de longa linha de base recentemente estabeleceu-se como uma técnica capaz de resolver estrelas e seus ambientes circunstelares no nível de mili segundos de arcos (\textit{mas}). Esta alta resolução abre uma janela inteiramente nova para o estudo de sistemas astrofísicos, fornecendo informações inacessíveis por outras técnicas.

Discos astrofísicos são observados numa ampla variedade de sistemas, de galáxias à discos planetários, em geral compartilhando de processos físicos similares. Dois sistemas de discos foram estudados nesta tese: (i) o estrelas B ricas em He e que possuem campos magnéticos da ordem de kG e que confinam seus ventos em estruturas chamadas magnetosferas; e (ii) estrelas Be, estrelas de rotação rápida que criam um disco circumstelar viscoso. 

Este estudo usa a técnica interferométrica para investigar ambas a própria fotosfera e o ambiente circunstelar destas estrelas. O objetivo é combinar a interferometria com outras técnicas observacionais (tal como espectroscopia e polarimetria) para realizar uma descrição física completa e precisa destes sistemas. Esta descrição é acompanhada por modelos de transferência radiativa executados pelo código \textsc{hdust}.

A primeira deteção segura de uma magnestosfera de estrela quente em polarização no contínuo é relatada, como um resultado da campanha de monitoramento de \sori{} no Observatório Pico dos Dias (OPD/LNA). Os dados polarimétricos foram modelados por um modelo ``alteres+disco'' de espalhamento simples, que descreve as magnetosferas como constituídas de duas estruturas esferoidais (na intersecção dos equadores de rotação e o magnético) e um disco circunstelar no equador magnético. A modelagem polarimétrica prevê uma menor razão entre a massa do blob e o disco que o previsto pelo modelo RRM (\textit{Rigid Rotating Magnetosphere}), que provê uma boa descrição espectroscópica destas magnetosferas.

Além de \sori{}, nós apresentamos as primeiras detecções polarimétricas das magnetosferas de HR\,7355 e HR\,5907. Nossa análise indica que estas estruturas compartilham propriedades similas, como amplitude da modulação polarimétrica e distribuição de massa. No caso de HR\,5907, nós também apresentamos a primeira detecção interferométrica de uma magnetosfera com amplitude de fase diferencial de certa de 3 graus com o AMBER-VLTI/ESO. 

Um novo fenômeno interferométrico, chamado de CQE-PS (\textit{Central Quasi-Emission Phase Signature}), foi descrito e identificado como um ferramenta útil para o estudo de discos circunstelares de Be, em particular para estrela do tipo \textit{shell}. Devido um formato de 'S' nas fases diferenciais ocorre quando o disco obscurece parte da fotosfera estelar. Esta absorção do disco altera o sinal da fase principalmente próximo aos comprimentos de onda do repouso e pode introduzir um reverso central no perfil das fases. Este fenômeno pode ser usado para sondar o tamanho do disco, densidade e inclinação radial. Ele pode até fornecer uma estimativa para o (máximo) tamanho angular estelar. 

A fotosfera da estrela Be Achernar foi estudada em detalhes usando visibilidades de alta precisam e informação de fechamento de fase do PIONIER-VLTI/ESO. Esta é a primeira caracterização fotosférica de precisão de uma estrela Be e é, até o momento, a caracterização da estrela de maior massa de taxa de rotação. O perfil de linha variável de Achernar e frequências fotométricas de alta-precisão foram analizadas à luz destes novos parâmetros. Estes resultados tem grande significância para modelos estelares and para as estrelas Be, tal como a determinação do coeficiente de escurecimento gravitacional.

Nós apresentamos o projeto \textsc{BeAtlas} como parte do grupo BeACoN do IAG-USP. Ele consiste de uma grade sistemática de modelos de estrelas Be gerados pelo código \textsc{hdust} com o objetivo de uma investigação abrangente de estrelas Be e seu ``estado da arte'' em modelagem, o VDD (\textit{Viscous Decretion Disk}). Como primeira aplicação do projeto, nós argumentamos que Achernar não é uma estrela típica de sequência principal, mostrando características de uma estrela deixando a sequência principal: ela exibe um grande tamanho e uma luminosidade mais alta que o esperado para sua massa. 

A recente ejeção por Achernar foi investigada e modelada. Para a análise da atividade, a amplo conjunto de dados observacionais foi obtido, tal com espectrointerferometria AMBER, espectroscopia FEROS (ESO) e polarimetria óptica (OPD/LNA). Estas observações contém o primeiro estudo espectro-interferométrico da evolução de um disco Be, angularmente resolvendo o disco em crescimento.

A evolução secular do recém formado disco é caracterizada e a prescrição de modelagem VDD foi empregada. A análise interferométrica AMBER não mostra evidência de um vento polar nesta fase ativa. A espectrocopia H$\alpha$ exibe uma lenta e gradual evolução, e alcança um regime quase estacionário somente após $\sim1,6$~anos. Esta característica permitirá estimar o coeficiente de difusão viscosa através da modelagem dinâmica VDD. 
\vspace{30pt}

Palavras-chave: estrelas: individual (Achernar) - estrelas: parâmetros fundamentais - técnicas: interferométrica - material circunstelar - estrelas: linha de emissão, Be - estrelas: campo magnético

%% file: tex/prologue.tex

\footnote{Adapted from the Scott Adams' book ``The Dilbert Principle''.}\textit{As some of you may know, my main goal is to become a scientist. It's a challenge for a student to write a whole thesis. Scientists are trained to be brief. Everything I have learned in my entire academic life can be boiled down to a dozen bullet points, several of which I already forgot.}

\textit{You would feel kind of perturbed if you took a big thick thesis and all it had in it was a dozen bullet points, particularly if several of them seemed to be ``filler''. So, my ``plan for excellence'' is to repeat myself and cite previously published works to take up some page space. In marketing terms, this is called ``adding value''. In science, it is called ``methodological rule''. And for your reading pleasure I will include many colorful but unnecessary figures.} 

Bonne lecture!

%% file: chap/intro.tex
\chapter{Introduction}
\setcounter{page}{37}

\section{Massive stars}
Stars are not equal. This statement is clear when looking at the night sky. Stars have different brightness and different colors. While the observed brightness is not an absolute quantity as it is influenced by distance, the visuals colors are directly related to the stellar temperatures.

The term \textit{hot stars} refers to the stars with surface temperatures above 10,000 Kelvin (K), appearing as blue-white stars in the night sky with temperatures much higher than our Sun. According to the usual spectral classification, these stars correspond to the O and B types. This classification was defined in Harvard, USA, in late nineteenth century when spectroscopy established as a regular observational technique. 

The luminosity of the stars was then compared with their colors in the so-called Hertzsprung-Russell (HR) diagram, in reference to the pioneer works of Ejnar Hertzsprung and Henry Norris Russell (independently done in 1911 and 1913, respectively). It was found that these quantities are not randomly distributed as most stars appear to be ordered in a path, called  ``the main sequence" (MS), where the bluest stars were also the more luminous.

OB main sequence stars are only approximately 0.1\% of the stars in the Solar neighborhood but are the most common type in the night sky \citep{led01a}. The reason is they are the brightest stars (i.e., they can be easily seen even distant) and they are rare in being formed. The empirical function that describes the distribution of initial masses for a population of stars is known in astronomy as Initial Mass Function (IMF). For example, \citet{sal55a} found that the number of stars  decreases rapidly with increasing mass. Defining $\Phi(m)dm$ as the fraction of stars with a mass between $m-dm/2$ and $m+dm/2$, Salpeter parametrized the IMF by a power-law on the form
\begin{equation}
\Phi(m)\,dm \propto m^{-\alpha}\,dm \,,
\end{equation}
with $\alpha=2.35$. 

Studies of main sequence binary stars (e.g., \citealp{har88a}) that allowed the determination of their respective masses,  established the relation between their temperatures and masses. The conclusion is temperature, luminosity and mass are mutually dependent quantities for these stars.


As examples of empirical mass-radius and mass-luminosity relations for MS stars we reproduce the expressions from \citet{mac99a}
\begin{equation}
\log \left(\frac{L}{L_\odot} \right) \sim 3.8\log \left(\frac{M}{M_\odot}\right)+0.08\,;
\end{equation}
\begin{equation}
\log \left(\frac{R}{R_\odot} \right) \sim 0.64\log \left(\frac{M}{M_\odot}\right)+0.011.
\end{equation}
They are valid for $M>0.2M_\odot$ and for $1.3<M/M_\odot<20$, respectively.

As the knowledge about the stellar evolution consolidated, it became clear that the main sequence is the longest period of evolution of a non-degenerated star and when its properties are more stable. The main feature of this evolutionary phase is hydrogen-burning as the stellar power source. 
 
The Vogt-Russel theorem, named after the astronomers Heinrich Vogt and Henry Norris Russell, states that the structure and observed properties of a star, in equilibrium, is uniquely determined by its mass and chemical composition. This implies that the main parameter that distinguishes the majority of stars is the mass. 

The term \textit{massive stars} is assigned to stars with masses higher than 8 Solar masses ($M_\odot$). This mass limit constraints massive stars to the spectral O and B types, reinforcing that massive stars are hot stars.

\subsubsection*{O and B types overview}
The standard spectral classification was based on relative strengths of absorption lines. All of the spectral types were subdivided with a numerical suffix from 0 to 9. A star with line strengths midway between A0 and B0 types is classified as A5 type star. 

The stellar spectra is also classified into classes. These classes distinguished basically the luminosity of the objects. This is known as the Yerkes spectral classification, or Morgan-Keenan classification (lead by the astronomers William Wilson Morgan and Philip C. Keenan) done in middle of the tweentieth century as an extension to the Harvard classification. Denoted by Roman letters I to V, the brightness of stars falls along the sequence from I luminous supergiants, III normal Giants up to V main sequence stars (dwarfs), the main objects of this study.

An O-type main sequence star (O\,V) is a very hot, blue-white star. They have temperatures higher than 30,000 K and so appear on the left of the HR diagram. Their large luminosity exceeds the Solar one ($L_\odot$) more than 30,000 times and their masses are higher than 16 $M_\odot$. Due to their high photospheric temperature, O type stars can ionize Helium (He) atoms and their spectra show prominent He\,{\sc ii} absorption lines, as well as strong lines of other ions with high energy ionization potential. Hydrogen (H) and neutral He lines are weak since these elements are almost completely ionized.

Due to high temperature and luminosity, O-type stars end their lives rather quickly (few dozen million years) in violent supernova explosions, resulting in black holes or neutron stars. Most O-type stars are young, massive, main sequence stars, but the central stars of planetary nebulae, old low-mass stars near the end of their lives, may have O spectra as well as white dwarfs stars.
    
B\,V type stars have masses roughly from 3 to 16 $M_\odot$ and surface temperatures between 10,000 and 30,000 K. Their luminosity range between 200 and 30,000 $L_\odot$. Their spectra have neutral He, and moderate H lines, since a substantial part of the hydrogen is ionized. 

Many B-type stars (and some O-type) exhibit peculiar features in their spectra. A characteristic that is considerably common in B-type stars is their high rotation rates. 


\subsection{Astrophysical context}
Massive stars play a crucial role in astrophysics. Due to they large luminosity, they contribute in an important way to the integrated spectrum of galaxies. With their strong winds, they feed the interstellar medium with momentum and kinectic energy, impacting on the star formation rate. Massive stars are also the main producers of cosmic heavy elements, including those necessary for life, and drive galatic chemical evolution. At the end of their short lifetimes they originate very energic events as supernovae and probably are the progenitors of gamma-ray bursts. See \citet{lei92a} and \citet{pul08a} for an overview.

\subsubsection*{ISM and star formation}
Spiral arms are high density regions of galaxy disks that are rich in gas and dust. The densest regions, in the form of molecular clouds, lead to the formation of stars, some of which are O- and B-type stars and that are easily identifiable due to their high brightness. Also, as these stars have short lifetimes, they cannot move great distances before their death and so they stay in or relatively near the place where they originally formed. 

Massive stars are then typically located in regions of active star formation, such as the spiral arms of spiral galaxies. These stars illuminate any surrounding material and are largely responsible for the distinct colors of galaxy's arms. 

As massive star dies, it may explode as a supernova. As the core collapses, it releases vast amounts of particles and energy that blow the star apart as they blast through space. The massive explosion produces shock waves that compress the ISM material surrounding the dying star. This compression leads to a new round of star birth.
These characteristic make massive stars tracers of ISM and star-formation in galaxies.

\subsubsection*{OB star associations}
The chief distinguishing feature of the members of a stellar association is that the most stars have similar chemical composition and ages, since they are gravitationally connected since their formation. In young stellar associations, most of the light comes from O- and B-type stars, so such associations are called OB associations.

OB associations are sparsely populated groupings of stars, typically between a few tens and a few hundreds of light-years across. They consist mainly of very young stars that have formed in the relatively recent past (a few million or tens of millions of years ago) from the same large interstellar cloud. Their study allow evaluating the initial mass function of star formation and chemical enrichment in galaxies in time and space.

\subsection{Peculiar hot stars \label{sec:pecstars}}
The spectral study of stars revealed that a significant part of the hot stars had peculiar features, even among the ones that appeared to share most of their features with main sequence stars. This study focuses on the study of B-type peculiar stars with circumstellar disks. Here is made a short presentation of the peculiar characteristics and their occurrence in hot stars. The detailed presentation of the studied subtypes are done in their respective chapters.

Peculiar hot stars as sometimes called \textit{active OB stars}. Spectroscopic peculiarities are very common in OB-type stars, many of them displaying emission from their circumstellar environments. An interesting fact that should deserve greater research in the future is the role of binarity in this peculiars stars, since at least 44\% of hot stars are in binary or multiple systems \citep{mas09a, san11a}. The binarity should also play an important role in the star rotation rate, since it allows angular momentum exchanges. The review by \citet{riv13a} includes a good summary of peculiar hot stars:
\begin{itemize}
    \item \textbf{SPB}: Slow Pulsating early-type B stars with periods of a few hours;
    \item \textbf{$\bm{\beta}$\,Cep}: B type stars with photometric variability period of a few days;
    \item \textbf{Herbig Ae/Be}: pre main sequence stars with accretion disks;
    \item \textbf{OB Supergiant}: post main sequence stars including, \textbf{LBV} (Luminous Blue Variable), \textbf{WR} (Wolf-Rayet), \textbf{P Cyg}, and \textbf{VV Cep} subtypes stars.
    \item \textbf{Oe}: O-type stars with strong winds and that exhibit emission lines;
    \item \textbf{He-rich}: early B-type stars with unusual strong lines of non-ionized He. They are often associated with strong magnetic fields in their photospheres. 
    \item \textbf{Bn}: B-type stars displaying broad line profiles but no emission signal, usually interpreted as very fast edge-on rotators. 
    \item \textbf{Be}: B-type main sequence stars whose spectra have shown or show emission lines. At least 20\% of B-type stars are Be stars in our Galaxy with higher proportion in lower metallicity environments. 
\end{itemize}

\begin{table}
\centering
\caption[Selected spectral properties of peculiar hot stars]{Selected spectral properties of peculiar hot stars and their observed presence in different classes of (non-supergiant) B-type stars. Circumstellar line emission formed in: (1a)~equatorial decretion disk, (1b)~accretion disk, (1c)~corotating clouds. Other properties include (2)~low-order line profile variations, (3)~radial and/or short-period pulsation, (4)~rapid rotation, (5)~large-scale magnetic field, (6)~surface abundance anomalies. Adapted from \citet{por03a}, originally on \citet{baa03a}.}
\begin{threeparttable}
\begin{tabular}[]{lcccccccc}
\toprule
    & \multicolumn{7}{c}{Observed general property} \\
    Star group & 1a & 1b & 1c & 2 & 3 & 4 & 5 & 6  \\
\midrule
    Classical Be  &        $\surd$ &         - &         - &   $\surd$ &         - &   $\surd$ &         - &         - \\
    Herbig Ae/Be  &              - &   $\surd$ &         - &         - &         - &         - &         - &         - \\
    He-rich  &                   - &         - &   $\surd$ &         - &         - &   $\surd$ &   $\surd$ &   $\surd$ \\
    SPB  &                       - &         - &         - &   $\surd$ &         - &         - &         - &         - \\
    $\beta$\,Cep  &              - &         - &         - &         - &   $\surd$ &         - &         - &         - \\
    Bn  &                        - &         - &         - &         - &         - &   $\surd$ &         - &         - \\
    He-abnormal and Bp &         - &         - &         - &         - &         - &         - &   $\surd$ &   $\surd$ \\
\bottomrule
\end{tabular}
    \begin{tablenotes}
        \item[]{\footnotesize \emph{Note}: The presence (or not) of each property should not be regarded as \textit{sine qua non}, but rather expresses a statistically expected property.}
    \end{tablenotes}
 \end{threeparttable}
\label{tab:pecs}
\end{table}
Table~\ref{tab:pecs} contains the main spectral properties that characterize the active OB stars.

It is noteworthy that the term \textit{peculiar} is sometimes used to describe specifically the chemical composition of the stellar photospheres, with a proper phenomenology associated. One example is Sirius, with weak Ca\,{\sc ii} K-line and
strong metallic lines, classified as an Am star.

\FloatBarrier
\section{Radiative transfer}
Radiative transfer is the branch of physics that describes the propagation of electromagnetic radiation (light) through a material medium. The interaction of the light with matter can be described in terms of absorption, emission, and scattering processes. Analytic solutions to the radiative transfer can be applied for simple cases but for more realistic media, with complex multiple effects, numerical methods are required.

Consider the case of traveling particles over a medium consisted of isolated particles. The numerical density of the particles is $n$ and their section collision cross section $\sigma$. The quantity
\begin{equation}
l = \frac{1}{n\sigma}\,,
\end{equation}
is defined as the \textit{mean free path} of the traversing particles. This scenario can be applied to a incident light beam of intensity $I_\lambda$
\begin{equation}
dI_\lambda=-n\sigma_\lambda I_\lambda\,dz=-\rho\kappa_\lambda I_\lambda\,dz\,,
\label{eq:dabs}
\end{equation}
where $dI_\lambda$ is the change in the intensity, $dz$ is the distance traveled, $\rho$ the density of the particles and $\kappa_\lambda$ is called \textit{absorption coefficient} or \textit{opacity}. The opacity is thus the cross section for absorbing particles (photons, in our case) per mass of the medium and usually is a function of wavelength, density and temperature.

A useful definition is the \textit{optical depth}
\begin{equation}
\tau_\lambda=-\int_0^Z\kappa_\lambda\rho\,dz.
\end{equation}
This allows describing the light absorption as
\begin{equation}
I_\lambda = I_{\lambda0}\exp^{-\tau_\lambda}\,,
\end{equation}
solution of Eq.~\ref{eq:dabs}. Beyond absorption, the (pure) emission is described as 
\begin{equation}
dI_\lambda=j_\lambda\rho\,dz\,,
\end{equation}
where $j_\lambda$ is the \textit{emission coefficient} of the medium. Combining the absorption and the emission terms
\begin{equation*}
dI_\lambda=-\kappa_\lambda\rho I_\lambda\,dz+j_\lambda\rho\,dz
\end{equation*}
\begin{equation}
-\frac{1}{\kappa_\lambda\rho}\frac{dI_\lambda}{dz}=I_\lambda-\frac{j_\lambda}{\kappa_\lambda}.
\end{equation}
The ratio of the emission coefficient to the absorption coefficient is called the \textit{source function} $S_\lambda\equiv j_\lambda/\kappa_\lambda$. Applying this and the optical depth definitions on above equation yields the \textit{radiative transfer equation} (RTE)
\begin{equation}
\frac{dI_\lambda}{d\tau_\lambda}=I_\lambda-S_\lambda.
\end{equation}


The physical conditions of a material medium can be described by the thermodynamic quantities. The optical properties of the medium, necessary for the radiative transfer problem solution (as $S_\lambda$ and $\tau_\lambda$), depends on the states of the medium. 

A common assumption made when solving the thermodynamic statistical equilibrium is the \textit{local thermodynamic equilibrium} (LTE). The reason for adopting LTE is that it enormously simplifies the calculation of states which matter (electron, atoms, ions, molecules) can be configured. It assumes that locally all thermodynamical states (as velocity distribution, Saha-Boltzmann distribution over degrees of excitation and ionisation, etc.) are described by a single temperature value. This is particularly important for describing gaseous media, and allows the optical properties be easily calculated to the radiative transfer solution.
If LTE is assumed, the mean intensity $J_\lambda$ of the system is described by the Planck function $B_\lambda$. The Planck function $B_\lambda$ is defined as 
\begin{equation}
B_\lambda = \frac{2\pi hc^2}{\lambda}\frac{1}{\exp\frac{hc}{\lambda kT}-1}\,,
\end{equation}
where $\lambda$ is the electromagnetic wavelength, $h$ is the Planck constant, $c$ the speed of light, $k$ the Boltzmann constant and $T$ the temperature of the body. 

In low-density environments, however, the particle collision rates are low and energy is not equally distributed. The states assume a complex configuration and can no longer be associated with a single temperature value, condition which is defined as non-LTE equilibrium ($J_\lambda\neq B_\lambda$). There are many astrophysical environments where departure from LTE are important. For example, the escaping photons of a star destroy the LTE equilibrium at its atmosphere \citep{mih73a}. The non-LTE effects are specially important when analyzing spectral lines, where the particle level populations determines its optical properties. 


\FloatBarrier
\section{Concepts of stellar astrophysics \label{sec:stconcep}}
In this section we briefly present the concepts of stellar astrophysics that are particularly important to this study. A key component that profoundly changes the stellar structure and evolution is the stellar rotation. The effects originated by rotation include geometrical deformation and photospheric gravity darkening, which can be observed only by high angular resolution (HRA) techniques, as well as evolutionary effects such as rotational chemical composition mixing. The proper consideration of these and other photospheric characteristics are fundamental when comparing models and observations since the modern HRA techniques are very sensible to brightness distribution of the targets (i.e., optical interferometry). 

\subsubsection*{Stellar atmospheres and limb darkening}
Stellar atmospheres are the connecting links between the observations and stellar astrophysics. In particular, the emergent spectrum of the \textit{photosphere} (the atmospheric layer from which light is radiated) provide the way to estimated the internal stellar parameters of interest.

Limb darkening (LD) is an optical effect that occurs in the photospheres of stars, including our Sun, where the center of the stellar surface appears brighter than the edges or \textit{limb}. LD is important in several areas of stellar physics, including the study of transiting extrasolar planets, eclipsing binary stars, interferometry and some gravitational microlenses.


The radiation reaching and observer from a star can be described as the emission along his line of sight up to that point where it becomes opaque. When looking the edge of a star, the emission reaching the observer do not come from the same physical depth as when considering the stellar center because the line of sight travels over an oblique angle through the stellar atmosphere. In other words, the stellar radius at which one see opaque (optical depth $\tau\approx 1$) increases as the line of sight moves towards the edges (an analogy can be made with the air mass of a star at the observer's zenith and another near the horizon).

The effective temperature of the stellar photospheric region decreases for an increasing distance from the star. And the radiation emitted from a high-density gas is a strong function of temperature. So when one look at a larger radius at the limb regions, one is actually seeing a cooler region when compared to the center of the stellar surface.

The optical depth along non-vertical ray (as along a stellar atmosphere) is given by
\begin{equation}
\tau_\lambda=\tau_\lambda'\sec\theta\,,
\end{equation}
where $\theta$ is the angle to vertical. So, the RTE becomes
\begin{equation}
\mu\frac{dI_\lambda}{d\tau_\lambda'}=I_\lambda-S_\lambda\,,
\end{equation}
where $\mu\equiv\cos \theta$. The solution of the RTE presumes that we know the source function. For the top of the stellar atmosphere, where $\tau'\sim0$, we expand $S_\lambda$ using Taylor series 
\begin{equation}
S_\lambda(\tau')\approx S_\lambda(0)+\dot{S}_\lambda(0)\tau_\lambda'=a+b\tau_\lambda'.
\end{equation}
Using it at the RTE yields
\begin{equation*}
I_\lambda(\tau_\lambda'=0,\mu)=S_\lambda(\tau_\lambda'=\mu)
\end{equation*}
\begin{equation}
I_\lambda(0,\mu)=S_\lambda(\tau_\lambda=1).
\label{eq:edd-bar}
\end{equation}
Eq.~\ref{eq:edd-bar} is known as \textit{Eddington-Barbier relation}, stating that for a source function of $S_\lambda=a+b\tau'$, the emergent specific intensity along a given ray is just equal to the source function at optical depth unity ($\tau\approx1$) along that ray. It offers an interesting insight to understand the limb darkening effect. LD expresses how the specific intensity $I_\lambda$ varies as a function of angle to the the normal direction $\mu$. An example of a LD law is \citep{cla00a}
\begin{equation}
\frac{I_\lambda(\mu)}{I_\lambda(1)}=1-\sum_{k=1}^{n}a_k(1-\mu^{k/2})\,,
\end{equation}
where the terms $a_k$ are the \textit{limb darkening coefficients}.

\subsubsection*{Stellar wind}
Stellar wind is a branch of contemporary astrophysics with wide ramifications. It is a common stellar phenomenon, even in low mass stars, such as the Sun. In the case of the Sun, the mass loss is negligible, with a small impact in its evolution. Nevertheless its effects are easily seen as the wind influences our planetary magnetosphere and the heliosphere has a rich phenomenology when interacting with the local ISM \citep{fri11a}.

Stellar winds constitute a hydrodynamic phenomenon of circumstellar gas flow. In the case of hot or red giants stars, the observed mass loss rates are significant, affecting drastically the evolution of these stars. The main characteristic of hot stars winds is the dominant role that radiation plays in the energy and momentum balance of the circumstellar plasma. Thus, massive stars mass loss rates are determined from the calculation of the radiative acceleration in their winds. 

The fundamental theoretical works on stellar winds are the ones by \citet{luc70a}, who identified line scattering as the mechanism that could drive stellar winds. However these authors predicted mass loss rates that were too low compared to the observations, as they assumed that only a few optically thick lines were present. The situation improved due to the landmark paper by \citet{cas75a} (hereafter CAK), who included an extensive line list that yielded a significant larger value for mass loss rate. 

The mass loss rate $\dot{M}$ is related to the density ($\rho$) profile and velocity ($v$) profiles. From the mass conversation relation:
\begin{equation}
\dot{M}=4\pi r^2\rho(r)v(r)\,,
\end{equation}
for a spherically symmetric wind in the steady-state regime. The velocity law usually assumed in the CAK theory, known as \textit{beta law}:
\begin{equation}
v(r)\simeq v_0+(v_\infty-v_0)\left(1-\frac{R_*}{r}\right)^\beta\,,
\end{equation}
where $v_0$ is the wind initial speed and $v_\infty$ is called the terminal speed. See \citealp{lam99a} for an overview.

As example of mass loss rates, \citet{kud00a} show the following relation to the mass loss of O-type stars of the solar neighborhood (III, V classes)
\begin{equation}
\log \left[\dot{M}v_\infty \left(\frac{R_*}{R_\odot}\right)^{1/2}\right]=19.87+1.57\log \left(\frac{L_*}{L_\odot}\right)\,,
\end{equation}
with $\dot{M}$ is given in $M_\odot$\,yr$^{-1}$ and $v_\infty$ in km s$^{-1}$. 

\subsection{Stellar Rotation \label{sec:strot}}
Stars can rotate up to their critical rotational limit, i.e., a rotation that produces a centrifugal force that can completely cancel the gravitational force at the stellar surface. Thus, rotational effects should be much more pronounced closer to the equator then the poles.

The rotational effects can profoundly transform the internal stellar configuration (for example, a core 
rotating faster than the outer envelope), changing the equilibrium configuration and driving internal instabilities which transport both chemical elements and angular momentum. The book from \citet{mae09a} is an excellent reference about this subject.

Rotational effects have usually been studied spectroscopically via the broadening of photospheric absorption lines. Recently, rotational effects were also detected interferometrically. \citet{van01a} measured the oblateness of a star, Altair, for the first time. \citet{dom03a} found the first indication of strong stellar oblateness for Achernar with VLTI-VINCI interferometer; and \citet{mon07a} did the first directly detection of the effect on a main sequence star via interferometric imaging of Altair using the CHARA interferometric array. 

Up to now, seven rapid rotating stars have their rotational oblateness measured by interferometry \citep{van12a}. A summary of these measurements is at Table~\ref{tab:introW}.

\begin{table}
\centering
\begin{threeparttable}
\caption{List of rapid rotating stars observed stars interferometrically to date.}
\begin{tabular}[]{lclcc}
\toprule
Proper & \multicolumn{1}{l}{Bayer} & Spectral & Rotation & Ref. \\ 
name & designation & type & rate (W) &  \\ 
\midrule
Achernar & $\alpha$\,Eri & B3Vpe & 0.838(10) & 1 \\ 
Alderamin & $\alpha$\,Cep & A7IV-V & 0.730(39) & 2 \\ 
Altair & $\alpha$\,Aql & A7IV-V & 0.696(12) & 3 \\ 
Caph & $\beta$\,Cep & F2III-IV & 0.688(66) & 4 \\ 
Rasalhague & $\alpha$\,Oph & A5IV & 0.634(17) & 2 \\ 
Regulus & $\alpha$\,Leo & B8IVn & 0.783(38) & 4 \\ 
Vega & $\alpha$\,Lyr & A0V & 0.701(28) & 5 \\ 
\bottomrule
\end{tabular}
\label{tab:introW}
    \begin{tablenotes}
        \footnotesize
        \item
        (1)~\citet{dom14a}, (2)~\citet{zha09a}, (3)~\citet{mon07a}, (4)~\citet{che11a},
        (5)~\citet{pet06a}.
    \end{tablenotes}
 \end{threeparttable}
\end{table}

\subsubsection*{Rotation rate notation}
Rotational velocities can be expressed in many ways, both in angular ($\Omega$) and linear ($v$) velocities. Stellar rotational rates are usually defined as fractions of their critical values
\begin{equation}
v_{\rm crit}=\sqrt{ \frac{2}{3} \frac{GM_*}{R_p} }; ~ \Omega_{\rm crit}=\sqrt{ \frac{8}{27} \frac{GM_*}{R_p^3} }.
\end{equation}
The factor $3/2$ comes from the oblateness $R_{\rm eq}=3/2R_p$ for critical solid body rotation, known as Roche approximation. The Roche model describe equipotential surfaces considering centrifugal forces due to rotation in addition to gravity. As pointed out by \citet{riv13a}, the description of the rotational rate in terms of this quantities are only meaningful in the Roche approximation. The authors suggest then the use the quantity $W$, defined in terms of the equatorial orbital speeds
\begin{equation}
v_{\rm orb}=\sqrt{ \frac{GM_*}{R_{\rm eq}} }; ~ \Omega_{\rm orb}=\sqrt{ \frac{GM_*}{R_{\rm eq}^3} }\,,
\end{equation}
\begin{equation}
W=\frac{v_{\rm rot}}{v_{\rm orb}}.
\end{equation}
Equatorial orbital quantities (and so $W$) are independent on the details of how the star rotates (e.g., rigid vs.\ differential rotation as scenarios usually explored in the literature) and $W$ directly defines what velocity boost is required for a given star to launch material into the closest possible orbit. In this study we endorse the use of this quantity.

Useful relations between rotational rates and rigid rotators are
\begin{equation}
W = \sqrt{2 \left(\frac{R_{\rm eq}}{R_p} -1\right)}\,,
\label{eq:Wrr}
\end{equation}
\begin{equation}
\frac{\Omega}{\Omega_{\rm crit}} = \sqrt{\frac{27}{8} \frac{W^2}{\left(1+0.5W^2\right)^{3}}}.
\end{equation}

\subsubsection*{Geometrical oblateness}
Here we present the structural geometrical effect of the stellar rotation under the assumption of a Roche's equipotential surface. In this approximation, rotating stars are assumed to be rigid rotators, i.e., the angular velocity $\Omega$ is a constant with respect to stellar latitude.

The stellar surface constant equipotential $\phi$ can be written as
\begin{equation}
\phi(\theta,r)=-\frac{GM}{r(\theta)}-\frac{1}{2}\Omega^2r^2(\theta)\sin^2\theta\,,
\end{equation}
and then compared to the one at polar direction ($\theta=0$)
\begin{equation}
\frac{GM}{r(\theta)}+\frac{1}{2}\Omega^2r^2(\theta)\sin^2\theta=\frac{GM}{R_{\rm p}}.
\label{eq:geff}
\end{equation}

This can be related to the local effective gravity ($\vec{g}_{\rm eff}$)
\begin{equation}
\vec{g}_{\rm eff}=-\nabla\phi\,,
\end{equation}
\begin{equation}
\vec{g}_{\rm eff}=\left[-\frac{GM}{r^2(\theta)}+\Omega^2r(\theta)\sin^2\theta\right]\hat{r}+
[\Omega^2r(\theta)\sin\theta\cos\theta]\,\hat{\theta}\,,
\end{equation}
The surface effective gravity modulus is then
\begin{equation}
g_{\rm eff}(\theta)=\left[\left(-\frac{GM}{r^2(\theta)}+\Omega^2r(\theta)^2\sin^2\theta\right)^2+\Omega^4r^2(\theta)\sin^2\theta\cos^2\theta\right]^{1/2}.
\end{equation}
Equation~\ref{eq:geff} can be rearranged trigonometrically to determine $r(\theta)/R_p$ for a given $W$ (or $\Omega$, from \citealp{col66a})
\begin{equation}
\frac{r(\theta)}{R_p}=\cos \left(\frac{\arccos (\Omega/\Omega_{\rm crit}\sin\theta)+4\pi}{3}\right) \times \left(\frac{-3}{\Omega/\Omega_{\rm crit}\sin\theta}\right)\,,
\label{eq:radrat}
\end{equation}
with $0<\theta\leq\pi/2$, i.e., from pole ($\theta=0$) to the equator ($\theta=\pi/2$).

\subsubsection*{Polar radius and luminosity}
Polar radii and luminosity values can only be determined by model of internal structure that considers rotation. In a first approximation, however, one may consider these quantities as independent of rotation. 

Indeed, the polar radius has a weak dependence on the rotation rate, which implies small changes of internal structure brought by centrifugal force. While the equatorial radius strongly inflates with rotation, the polar radius decreases by a few percent in general ($\lesssim2\%$ at critical rotation for typical B and late type O stars, between 3 and 20 M$_\odot$; \citealp{eks08a}).

The luminosity can also be considered as constant in the range of masses corresponding to B type stars. As shown for instance by \citet{mae09a}, a simple relation can be obtained,
\begin{equation}
\frac{L(\Omega/\Omega_{\rm crit})}{L(0)}=1-b\left(\frac{\Omega}{\Omega_{\rm crit}}\right)^2\,,
\end{equation}
 with $b$=0.07, 0.065 and 0.06 for a 3, 9 and 20 M$_\odot$ respectively. The above ratios reaches at most 7\% for critically rotating stars. For stars rotating below $\Omega/\Omega_{\rm crit} = 0.5$ ($W\sim0.3$), the difference between luminosity of a rotating star and a non-rotating counterpart is less than 1\%.

\subsubsection*{Gravity darkening}
While rotation increases the total stellar surface area, it also decreases the average radiative flux (i.e., temperature) per unit area to keep the (near) constant luminosity. The decrease of the surface effective gravity implies in differences in the emergent flux and the effective temperature according to the latitude, $T_{\rm eff}(\theta)$. Thus, the emission of a rotating star is the integration of spectra of different gravities and temperatures, and is dependent on the viewing angle.

This decreasing flux as function of co-latitude in rotating stars is called \textit{Gravity darkening} or \textit{von Zeipel effect}, after the pioneer work by \citet{von24a}. According to von Zeipel, the local emergent flux should be proportional to the local effective gravity.
To a pure radiative envelope (i.e, a black body), this implies that $T_{\rm eff}\propto g_{\rm eff}^{1/4}$, where the equatorial regions of the star are fainter and cooler that the polar ones. 

\begin{figure}
    \centering
    \includegraphics[width=.5\linewidth]{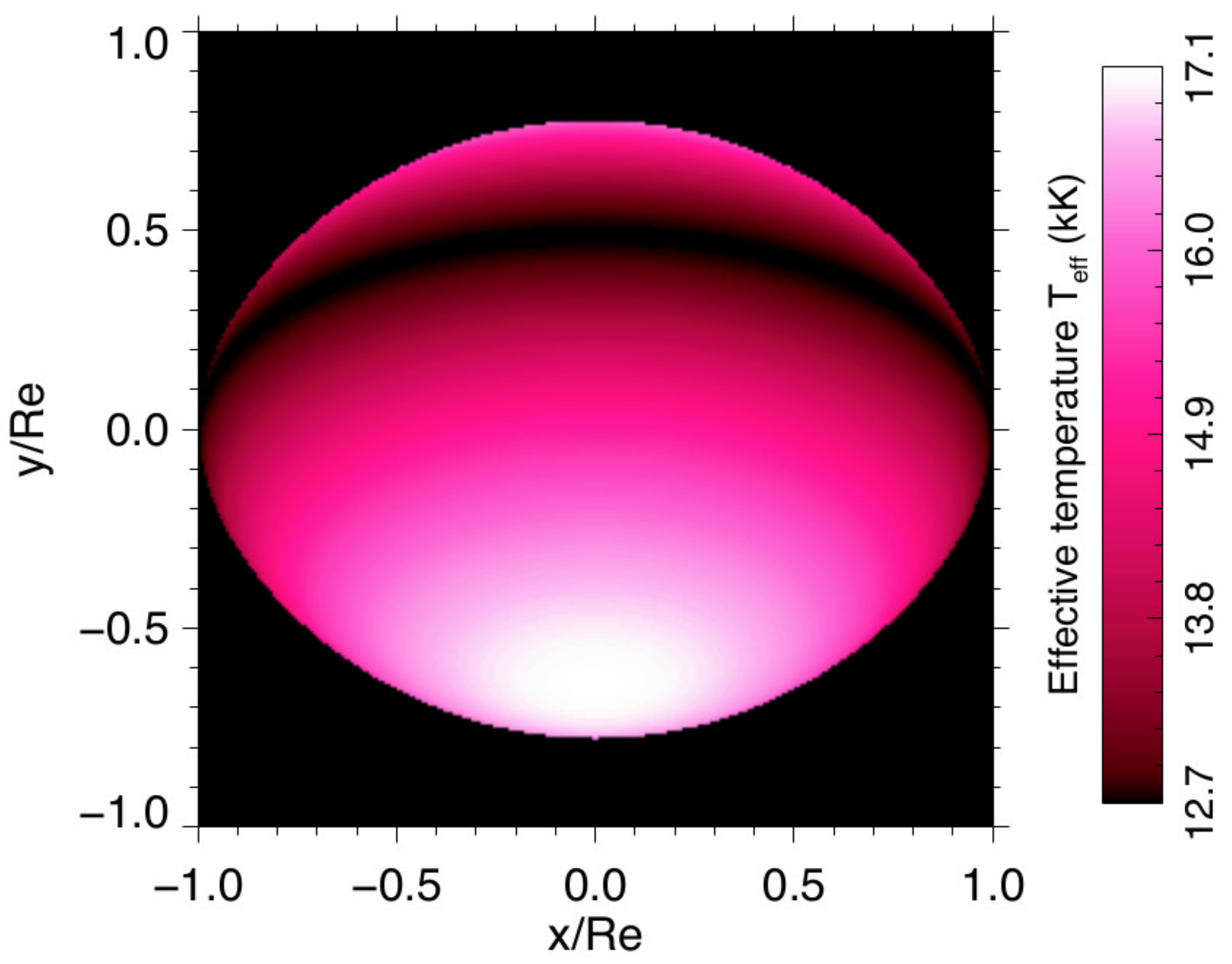}
    \caption[The temperature map of the fast-rotating star Achernar ($W\simeq0.84$) seen at $i\simeq60^\circ$]{The temperature map of the fast-rotating star Achernar ($W\simeq0.84$) seen at $i\simeq60^\circ$. The effects of the geometrical oblateness ($R_{\rm eq}/R_p=1.352$) and the gravity darkening ($\beta=0.166$) are easily identified \citep{dom14a}.}
    \label{fig:gdob}
\end{figure}

The von Zeipel relation can be generalized using a power law in the form of $T_{\rm eff}\propto g_{\rm eff}^\beta$. Even so, \citet{luc67a} applied the von Zeipel principle not to a black body emission but to a (pure) convective envelope. The author found the value of $\beta=0.08$, indicating that the traditional von Zeipel exponent could overestimate the temperature differences between stellar pole and equator. Indeed, interferometric results showed $0.08<\beta<0.25$ for the photosphere of rotating stars \citep{van12a}. This was interpreted as the existence of a thin convective layer at the surface of these stars, but alternative scenarios exist. The effects of the geometrical oblateness and the gravity darkening can be seen in Fig.~\ref{fig:gdob} applied to the Be star Achernar. More details are in Chapter~\ref{chap:phots}.

\section{The Be stars\label{sec:}}

Be stars are main sequence B type stars that show, or has shown in the past, Balmer lines in emission \citep{jas81a}, attributed to a circumstellar (CS) disk. Be stars are fast rotators \citep{riv06a} and this fact brings several consequences to its structure and evolution. These stars are the fastest rotating stars in the main sequence, making them key targets to study stellar evolution at high rotation rates.

In the Galaxy, about 15-20\% of all field B-type stars are Be stars, but with the Be phenomenon as such extending from late O- to early A-type stars. The proportion of Be stars increases in the Large Magellanic Cloud (LMC) and especially in the Small Magellanic Cloud (SMC), and can be high as one-half in the B0-B4 subtypes range \citep{mar06a, mar07a}. 
It is agreed that the relative frequency of Be stars are anti-correlated with metallicity. No evidence of magnetic field in Be stars exists, although recent magnetism surveys point to a rate of $\sim$10\% over all the B-type range (e.g., \citealp{wad14a}). 

Be stars are known to be rapid rotators and this property is believed to be fundamentally linked to the existence of the disk. Another mechanism(s) must act together with rotation to explain the phenomenon, where the material ejected by the star ought to reach Keplerian rotation in progressively higher energy orbits. Although great strides to assess the photosphere-disk interface has been made, 
the additional mechanism(s) has not yet been conclusively identified.

\subsection{Be stars evolution and rotation rates}
\begin{figure}
    \centering
    \includegraphics[width=0.6\linewidth]{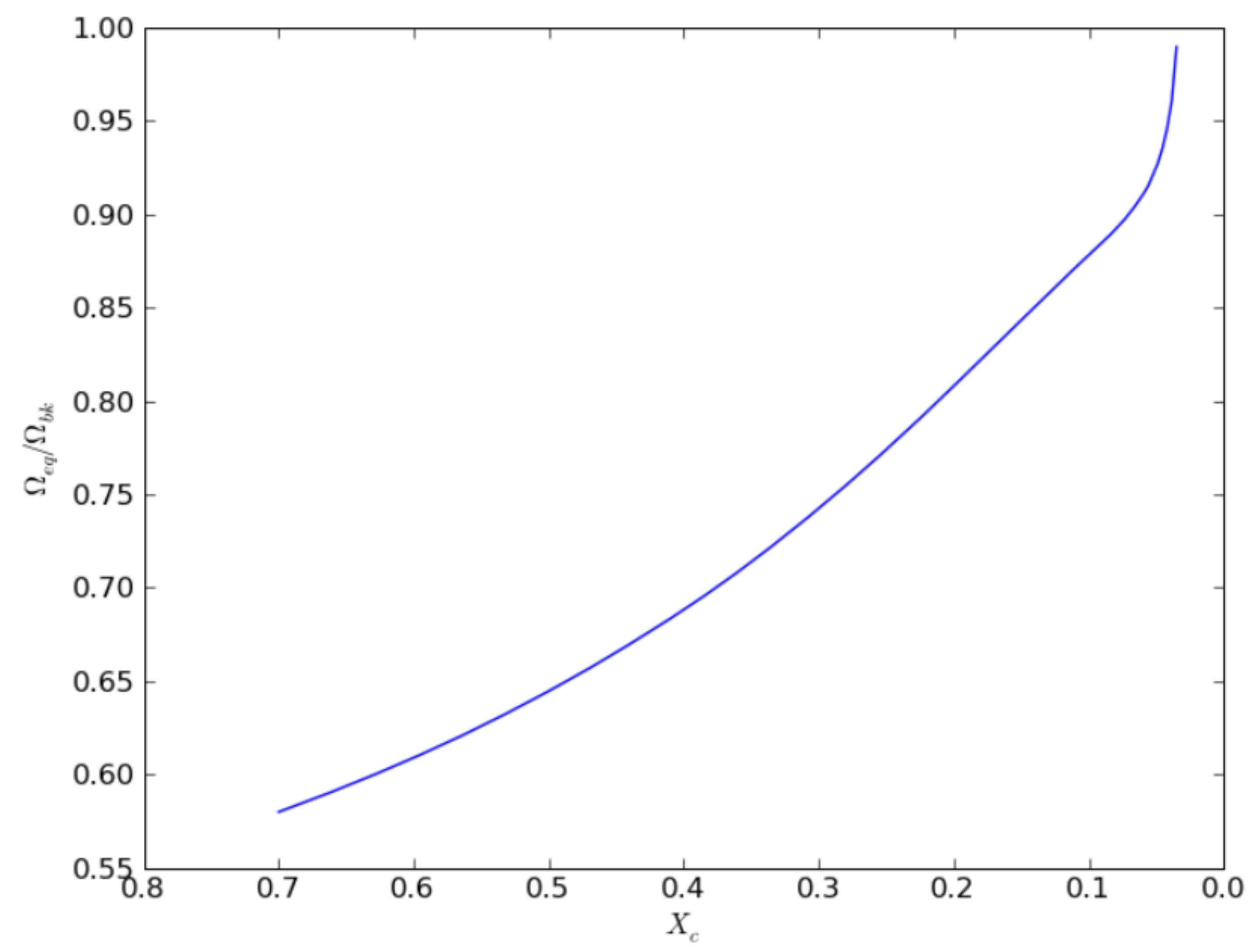}
    \caption[Equatorial angular velocity divided by the Keplerian angular velocity at equator as a function of the Hydrogen mass fraction of the star for a constant total angular momentum]{Equatorial angular velocity divided by the Keplerian angular velocity at equator as a function of the Hydrogen mass fraction of the star ($X_c$) for a constant total angular momentum ($M=7M_\odot$ and $Z\sim0.014$). The x-axis is inverted because the value of $X_c$ decreases due to H fusion by the star along the main sequence. \citep{rie13a}.}
    \label{fig:introMSrot}
\end{figure}

It is known that rotating stars transport angular momentum from their interior towards the surface during their evolution. The classical picture is that single B-type stars that have a sufficiently large rotational rate ($\Omega/\Omega_{\rm crit}>0.8$, or $W>0.53$) at the zero age main sequence (ZAMS) will approach their critical limit towards the end of the main sequence (e.g. \citealp{eks08a} and \citealp{geo13a}). This effect can be seen in Fig.~\ref{fig:introMSrot}, there the expected rotation rate calculated with the ESTER code \citep{esp07a} is related to the Hydrogen fraction present in the stellar core ($X_c \gtrsim 0.1$, the main sequence). 

The increasing the rotation rate with evolution does not imply the absence of rapid rotators early in the MS. It indicates that evolution will make stars that were moderately rotating stars to rotate faster towards the end of the MS, a phase at which fast rotators would likely to exist. Observations of stars in very young open clusters show that there exists a mass-dependent distribution of rotational velocities with age \citep{hua10a}, and despite being rare, it is possible to have very fast rotators early in the MS.

Traditional stellar evolution models do not contradict this picture, and early evolutionary models (for example, \citeauthor[op. cit.]{eks08a}) did not contain predictions of stars rotating close to the critical limit during the whole MS. For example, all the evolutionary tracks presented by \citet{geo13a} assume that the star rotates as a solid body at the ZAMS. Just after the ZAMS, because of meridional circulation, the internal rotational velocity distribution is readjusted, reducing the rotational velocity rate at the surface, in particular for the most rapid rotators.

This constraint was recently overcome by making a better treatment of pre-MS phases. A more realistic internal structure of the star at the ZAMS was obtained by \citet{hae13a} and applied by \citet{gra14a}, so their models can describe fast rotators all over the MS phase. The evolutionary phase where the Be phenomenon occurs is a topic that undoubtedly deserves a detailed examination.

\FloatBarrier
\section{Astrophysical disks}
Astrophysical disks are the (near-)continuous media in orbit around a massive central body. They are observed in a wide variety of astrophysical systems, and they are created in both inflows and outflows whenever the rotation speed of the orbiting material greatly exceeds the radial flow speed. Examples of disk systems formed during infall include young stellar objects, mass transfer binaries, and active galactic nuclei (AGN). Similarly, outflowing disks also form around  luminous stars, such as B[e] stars, and possibly luminous blue variable and asymptotic giant branch (AGB) stars.

The presence and structure of astrophysical disks are driven by angular momentum and mass transfer with the central body, where viscosity is often the key element (see \citealp{pri81a} for a review). Disks are long-lived structures (i.e., are stable for many orbits), and may exhibit waves and perturbations. These instabilities are related to important but not yet fully understood processes of star and planet formation, as well as formation of galactic structure and intracluster medium structure (from AGN feedbacks). For review on instabilities and waves in astrophysical disks, see \citet{bal03a}.

\subsection{Thin disks \label{sec:}}
In a thin disk (i.e., $z\ll r$), the surface density profile $\Sigma(r)$ can be obtained from the Navier-Stokes equations in cylindrical symmetry, integrated over disk height. The mass conservation equation is
\begin{equation}
\frac{\partial\Sigma}{\partial t}+\frac{1}{r}\frac{\partial }{\partial r}\left(r\Sigma v_r\right)=0\,,
\end{equation}
and the angular momentum conservation is
\begin{equation}
\frac{\partial}{\partial t}\left(\Sigma r^2 \Omega\right)+\frac{1}{r}\frac{\partial}{\partial r}\left(r\cdot\Sigma r^2 \Omega\cdot v_r\right)=\frac{1}{r}\frac{\partial}{\partial r}\left(r\cdot\Sigma r\cdot\nu r\frac{\partial \Omega}{\partial r}\right)\,,
\end{equation}
where $v_r$ is the radial velocity, $\Omega$ is the angular velocity, $r\partial\Omega/\partial r$ the velocity shear, and $\nu$ is the kinematic viscosity. 

The above equations can be combined to a diffusion equation for the surface density
\begin{equation}
\frac{\partial\Sigma}{\partial t}=\frac{3}{r}\frac{\partial}{\partial r}\left[r^{1/2}\frac{\partial}{\partial r}\left(\nu\Sigma r^{1/2}\right) \right]\,,
\label{eq:difsurf}
\end{equation}
for a point-mass potential $\Omega = \left(GM/R^3\right)^{1/2}$. The vertical thickness $H$ follows from hydrostatic equilibrium
\begin{equation}
\frac{1}{\rho}\frac{\partial P}{\partial z}=\frac{\partial}{\partial z}\left[ \frac{GM}{\left(r^2+z^2\right)^{1/2}}\right]\approx-\frac{GMz}{r^3}\,,
\end{equation}
and the scale height $H(r)$ can be determined as
\begin{equation}
-\frac{1}{\rho}\frac{c_s^2\rho}{H}\simeq-\frac{GMH}{r^3}\,,
\end{equation}
with $c_s=\sqrt{k_{\rm B}T/\mu m_H}$ the disk sound speed. Therefore
\begin{equation}
H\simeq c_s \left(\frac{GM}{r^3}\right)^{-1/2}\simeq \frac{c_s}{\Omega} \simeq \frac{c_s}{v_\phi}r.
\label{eq:H}
\end{equation}

\subsubsection*{Viscosity}
The timescale for viscous disk evolution is 
\begin{equation}
\tau\sim \dfrac{r^2}{\nu}\,,
\end{equation}
estimated from Eq.~\ref{eq:difsurf} by taking $\dfrac{\partial^2}{\partial r^2}\rightarrow \dfrac{1}{r^2}$ and $\tau^{-1}=\dfrac{1}{\Sigma}\dfrac{\partial \Sigma}{\partial t}$. 
The kinematic viscosity $\nu$ is sometimes described as ``molecular viscosity'' 
\begin{equation}
\nu\simeq v_T\lambda \,,
\end{equation}
where $v_T$ is the thermal velocity and $\lambda$ is the mean free path. 
More often, specially in the Be stars context, the kinematic viscosity is described in terms of ``$\alpha$-disk'' parametrization \citep{sha73a}
\begin{equation}
\nu=\alpha c_s H\,,
\end{equation} 
where $\alpha$ is a dimensionless constant. $\alpha$ can be interpreted as the efficiency on which viscosity arising from collisions in a turbulent medium moving in eddies of length-scale~$\sim H$ at velocity $v\sim c_s$. It is an often \textit{ad-hoc} assumption $\alpha$ constant with time or with radius.

The main physical explanation of the viscosity mechanism in astrophysical disks is the magneto-rotational instability (MRI; e.g., \citealp{bal02a}). Roughly speaking, MRI is originated by magneto-hydrodynamical instabilities in a differentially rotating and magnetized disk. This would drive turbulence, producing viscosity at tapping the rotational energy. 

\subsection{Circumstellar disks}
The disks around hot stars are the result of the hydrodynamical transport of the material and its radiative heating. From the point of view of their opacity, the disks of hot stars are relatively simple when compared to their cool stars counterparts. In the case of the last, the dominant opacity sources and dust and molecular gas, which depend on the rather complicated chemistry within the protostellar disk. In the case of the former, however, the disk is comprised mostly of ionized atoms and free electrons. This makes hot star disks ideal laboratories for studying the physics of outflowing disks.

As discussed by \citet{bjo05a}, the mechanism by which the stellar outflow is produced have important consequences on the disk structure. The azimuthal velocity $v_\phi$ of the main circumstellar environments are 
\begin{equation}
v_\phi = 
    \begin{cases}
    v_{\rm rot}(r/R_*)        & \text{(magnetically dominated; } r<R_{\rm A})\,, \\
    v_{\rm rot}(r/R_*)^{-1}   & \text{(radiatively driven),} \\
    v_{\rm rot}(r/R_*)^{-1/2} & \text{(viscously driven),} \\
    \end{cases}
\end{equation}
where $v_{\rm rot}$ is the disk azimuthal velocity at the stellar equator, and $R_{\rm A}$ is the Alfv\'en radius, the maximum extent of closed magnetic loops (see e.g., \citealp{ud-08a}).
The velocities reflect different interactions of the circumstellar material with their central star. For example, a azimuthal velocity proportional to the radius set a centrifugal driven envelope, as one trapped by a magnetic field. If the envelope has a Keplerian rotation profile this is equivalent to angular momentum exchanges to balance the gravitational force at different distances. And, if the azimuthal velocity decreased linearly with radius, the envelope angular momentum is conserved, setting a radiatively driven wind. Specifically for this work are evaluated disks dominated by magnetism (B-type He-rich stars; Chap.~\ref{chap:mag}) and by viscosity (Be stars). 

\subsection{Be star disks \label{sec:inkep}}
The investigation of the photosphere of Be stars, and of their circumstellar (CS) disks, has entered a new era during last years, both observationally and theoretically. Optical and infrared interferometry played a key role in the recent development of the Be field. It is a technique capable of bringing qualitatively new information by resolving the stars and their disks at the milliarcsecond (mas) level.  Due to the relatively strict magnitude limits of current interferometers, bright nearby Be stars are among the most popular and  most frequently observed targets.

From both observational and theoretical studies it is known that a considerable part of circumstellar disks, notably those of Be stars, are geometrically thin and rotate in a Keplerian fashion. Although spectroscopic evidences already suggested that Be stars disk rotated in a Keplerian way (e.g., the CQE line profiles in shell stars; \citealp{han95a}), interferometric observations provided a more precise and reliable evidence of this rotational configuration (\citealp{mei07a} observed the first interferometric evidence of Keplerian profile in a Be star, followed by \citealp{kra12a,whe12a}). This allowed identifying kinematic viscosity as the mechanism that makes these disks grow: viscous torques in the Keplerian disk transfer angular momentum from the base of the disk outwards, thus allowing the gas particles to reach progressively wider orbits. 

The viscous decretion disk (VDD) model \citep{lee91a, bjo97a, oka01a} is now regarded by many as the paradigm that offers the needed conceptual framework to explain the observed properties of the disks. Be star disks are highly-ionized environments, radially driven by viscosity and where non-LTE effects are also important. The vertical direction is pressure-supported, confirming that the complete solution of their structure is also determined by the radiative equilibrium - and then dependent of the radiative transfer solution. The VDD model, solved by radiative transfer, have proved to be fully consistent with multiple disk observations. See \citet{riv13a} for a dedicated review. 

The emergence of the VDD model to explain the Be star circumstellar disks ensures their similarity with (viscous) accretion disks that are present in various astrophysical scales, from galactic discs, AGNs, to X-ray binary stars. In most of these systems - including Be stars - little is known about mechanisms that originate the viscosity, that drives angular momentum and mass transfer and that fine-tune their structures.

The CS disk surface density, $\Sigma(r,t)$, as function of time is given by a simplified form of the Navier-Stokes equations \citep{oka07a}
\begin{equation}
\frac{\partial\Sigma(r,t) }{\partial t}=\frac{1}{\tau_{\rm vis}}\left( \frac{3}{r}
\frac{\partial}{\partial r}\left[ r^{1/2} \frac{\partial}{\partial r} 
(r^2\Sigma(r)) \right] + \Sigma_{\rm in}(t)\frac{r_{\rm in}^2}{r_{\rm in}^{1/2}-1}
\frac{1}{r}\delta(r-r_{\rm in}) \right)\,,
\label{eq:vdd}
\end{equation}
where $r_{\rm in}$ is the disk injection radius, $\Sigma_{\rm in}$ the injected surface density, and 
\begin{equation}
\tau_{\rm vis}=\frac{\left(GMR_\mathrm{eq}\right)^\frac{1}{2}}{\alpha c_{s}^2}\,,
\end{equation}
controls the radial flow ($\alpha$ is the Shakura-Sunyaev's viscous parameter and $c_s$ the disk sound speed).

The steady-state solution of Eq.~\ref{eq:vdd} (i.e., $\partial\Sigma/\partial t=0$ with $\Sigma_{\rm in}\neq0$) yields
\begin{equation}
\Sigma(r)=\frac{\dot{M}v_{\rm orb}R_{\rm eq}^{1/2}}{3\pi\alpha c_s^2r^{3/2}}\left[ \left(\frac{R_0}{R_{\rm eq}}\right)^{1/2}-1\right]\,,
\label{eq:introMdot}
\end{equation}
where $\dot{M}$ is the disk mass injection, and $R_0$ is a constant from the torque momentum equation integration, sometimes called as \textit{disk truncation}.

The disk material is often parametrized in the literature as the density $\rho$ (mass per volume), related to the surface density by
\begin{equation}
\Sigma(r)=\int_{-\infty}^{\infty}\rho(r,z)dz\,,
\end{equation}
where the density can be written in the parametric form 
\begin{equation}
\rho(r,z)=\rho_0 \left(\frac{R_{\rm eq}}{r}\right)^m\exp \left[ -\frac{1}{2} \left(\frac{z}{H}\right)^2 \right]\,,
\label{eq:rho}
\end{equation}
with $H$ as the disk scale height. The steady-state solution of the surface density leads to $m=3.5$. This and further details of the VDD models are presented in Appendix~\ref{ap:kepeq}. Second order effects, as non-isothermal disk scale height, can be investigated in steady-state disks \citep{car08a}. 

\citet{hau12a} studied the viscous dynamical evolution of Be stars, solved by radiative transfer. Their study included the temporal evolution of the disk density for different scenarios, including disk build-up and dissipation processes. They showed that the main parameters that controls the Be stars light curves at different wavelengths are mass disk injection history, inclination angle $i$ and the $\alpha$ viscosity parameter. 

The temporal variation of some fundamental quantities in Be stars are related to their disk surface density profiles. In particular, we highlight that the density power-law index $m>3.5$ mimics a dynamical disk at growing process and $m<3.5$ the opposite, namely a disk in dissipation. 
Other important results are that long wavelength light curves (as \textit{mm}) are useful for disk size determination. Physical parameters from these systems can be inferred from their light curves, however a degenerescence exists in the parameters and the sets of decretion scenarios for a given (short-term) observed light curve. This degenerescence can be avoided with more observables and a longer time coverage of the observations.

\FloatBarrier
\section{The optical long baseline interferometry (OLBI) technique \label{sec:optinterf}}
The investigation of the photospheric and the circumstellar emission of hot stars entered a new phase in the first decade of the twentieth first century. Optical and infrared interferometry were established as a novel technique capable of bringing qualitatively new information by resolving stars and their surroundings at the milliarcsecond (mas) level. Due to the relatively bright magnitude limits of interferometers, nearby hot stars (notably Be stars) are among the most popular and most frequently observed targets.

\subsubsection*{Principles of interference}
In common cases, light can be treated as composed of particles (photons). We can think about photometry as a photons count process and a small telescope image as a photon count with spatial information. However, interesting phenomena are associated with light when we consider its wave properties. One example is the light polarization.
Other one is the interference: waves can combine themselves in a constructive and destructive way depending on their frequencies and phases (position). 

The motivation behind the interference is to be able to increase of spatial resolution when looking at a celestial object and it is discussed below. Radio astronomy took to advantage of this principle since at least the 1960's. Optical interferometry needed to wait until the twentieth first century to systematically make use of it. This time difference is explained by the technological boundaries needed by the two different wavelength domains. While in radio frequencies we can easily record the actual wave format information (and perform the interference when appropriated at a later time), in optics this is impossible since the involved frequencies are too high to be processed by any electronic device. In optical frequencies the interference must be done in real time in the optical system. Further more, the optical length in which the interference occurs is crucial to result. The required precision is of the order of the wavelength to be combined. This means that only optical paths determined in fractions of $\mu$m are capable to perform the desired interference. And this was an insuperable technological barrier until a decade ago.

\subsubsection*{The double-slit experiment}
In 1801 Thomas Young demonstrated the light has wave properties trough the double-slit experiment (also referred as Young's experiment)\footnote{We recommend watching the video ``The Original Double Slit Experiment'' by Veritassium (\url{https://www.youtube.com/watch?v=Iuv6hY6zsd0}) about their research on Young's original work.}.

Consider the light as and electromagnetic wave $\vec{E}$, linearly polarized in $\hat{x}$ and propagating along $\hat{z}$ 
\begin{equation}
\vec{E} = \textrm{Re}\left\{a\exp\left[2\pi i\left(vt-\frac{z}{\lambda}\right)\right]\right\} \hat{x} = \textrm{Re}\{a\exp(-i\phi)\exp(2\pi i vt)\} \hat{x}.
\label{eq:ReE}
\end{equation}
With $z$ fixed and $\phi=2\pi z/\lambda$, we can write
\begin{equation}
\vec{E} = \tilde{E}\exp(2\pi i vt) \hat{x}\,,
\label{eq:E}
\end{equation}
where $\tilde{E}=a\exp(-i\phi)$. 

In the double slit experiment, a light source (as the Sun) illuminates a plate pierced by two parallel slits, and the light passing through the slits is observed on a screen behind the plate. At any given point on the image plane, the observed intensity $I$ will be given by the modulus squared of the summation of the electric field $\vec{E}$ arriving from the two slits. If we call these $\vec{E_1}$  and $\vec{E_2}$, we can write the detected intensity as
\begin{align}
I &= \mean{(\tilde{E_1}+\tilde{E_2})^*\times (\tilde{E_1}+\tilde{E_2})} \nonumber \\
  &= \mean{\tilde{E_1}^2}+\mean{\tilde{E_2}^2}+\mean{\tilde{E_1}\tilde{E_2}^*}+\mean{\tilde{E_1}^*\tilde{E_2}} \nonumber \\
  &= \mean{\tilde{E_1}^2}+\mean{\tilde{E_2}^2}+\mean{2\|\tilde{E_1}\|\|\tilde{E_1}\|\cos(\phi)}\,,
\end{align}
where $\phi$ is the phase difference between the electric field components $\vec{E_1}$ and $\vec{E_2}$ and the angle brackets refer to the time average. Here, as above, $\phi$ is phase or spatial positioning of the wave. The first two terms of this equations refer to the mean intensity seen in the double-slit fringe pattern, while the third term, which is associated with the modulation of the fringes from light to dark, clearly encodes the values of the complex products $\mean{\tilde{E_1}\tilde{E_2}^*}$ and $\mean{\tilde{E_1}^*\tilde{E_2}}$, i.e., the visibility function and its complex conjugate. If we assume $\|\tilde{E_1}\|=\|\tilde{E_2}\|=E$, the modulation is $0<I<4E^2$.


\subsubsection*{Angular resolution}
Diffraction refers to the phenomena which occur when a wave encounters an obstacle. The spatial resolution of a lens (like a telescope) is set by diffraction limit of the incident light. The diffraction pattern is the Fourier transform\footnote{A brief definition of Fourier transform is done at Appendix~\ref{ap:eqinterf}.} of the aperture intensity distribution. The pattern resulting from a uniformly-illuminated circular aperture has a bright region in the center and is known as Airy disk (named after George Biddell Airy). 

The angle at which the first Airy lobe goes to zero, measured from the direction of incoming light, is given by the approximate formula:
\begin{equation}
    \sin \theta \approx \theta \approx 1.22 \frac{\lambda}{D} \,,
\label{eq:diflim}
\end{equation}
where $\theta$ is in radians, $\lambda$ is the wavelength of the light and $D$ is the diameter of the aperture. In case of telescope, $D$ is the telescope diameter and $\theta$ is the angular size of a emitter that the telescope can isolate (or \textit{resolve}). 

In the double slit experiment - and in the two telescopes configuration - it is possible to demonstrate that the first Airy disk minimum, used as criterion for the optical system resolution and defined as \textit{diffraction limit}, will be
\begin{equation}
    \theta \approx 1.22 \frac{\lambda}{B} \,,
\label{eq:resint}
\end{equation}
where $B$ is the separation of the incoming light wave (in the case of celestial observations, it is the separation of the telescopes projected in the direction of the target, referred as $\vec{B}_{\rm proj}$; see below).


Atmospheric seeing sets the maximum effective resolution of an image at optical wavelengths to $\approx0.5$~arcsec or worse, depending on the meteorological conditions. The diffraction limit of a large telescope is much smaller, being $\approx20$~miliarcsec for a 4~m telescope, and can be pursued by modern active-optics techniques. However, if we combine the light from two far apart telescopes ($B=300$~meters), the angular resolution is $\approx0.3$~miliarcsec. Of course, this is a challenging value since the influence of the atmosphere and instrument must be taken into account.

\subsection{Visibilites and phases}
The quantity registered in interferometry is the normalized complex visibility $\tilde{V}$. The visibility amplitude $\|\tilde{V}\|$, in terms of the registered fringes intensities $I$, is
\begin{equation}
\|\tilde{V}\|= \frac{I_{\rm max}-I_{\rm min}}{I_{\rm max}+I_{\rm min}}\,,
\end{equation}
ranging from 0 to 1. When $\|\tilde{V}\|\approx0$, the source is considered as an extended object, which is called \textit{fully resolved} target. When $\|\tilde{V}\|\approx1$, a source is considered as a point-like source and is called \textit{unresolved} target. Since visibility $\tilde{V}$ is a complex quantity, its amplitude is often expressed as $V^2$. 

The complex visibility $\tilde{V}(\vec{u})$ is function of the vector $\vec{u}$, the spatial frequency of the projected baseline $\vec{B}_\textrm{proj}, \vec{u}=\vec{B}_\textrm{proj}/\lambda$. $\vec{B}_\textrm{proj}$ is not just the separation of the telescopes ($B$), but this separation projected in the object direction. Typically, $\|\vec{B}_\textrm{proj}\|\propto B\cos\zeta$, where $\zeta$ is the zenithal angle of the object.

The van Cittert-Zernike theorem \citep{van34a, zer38a} relates the complex visibility $\tilde{V}(\vec{u})$ to the Fourier transform of the brightness distribution $\mathscr{I}(\vec{r}, \lambda)$ of an extended source on the plane of sky
\begin{equation}
\tilde{V}(\vec{u})=\frac{\iint \mathscr{I}(\vec{r}, \lambda) \exp{(-2\pi i \, \vec{u}\cdot\vec{r})} \,d^2\vec{r}}{\iint \mathscr{I}(\vec{r}, \lambda) \,d^2\vec{r}}=\|\tilde{V}\|\exp{(i\phi)} \,,
\label{eq:compvis}
\end{equation}
where $\vec{r}$ is the on-sky plane angular position and $\phi$ is the visibility \textit{phase}. A common quantity used in high spectral resolution interferometry are the
differential phases $\phi_\textrm{diff}$ defined as
\begin{equation}
\phi_\textrm{diff}(\lambda,\lambda_\textrm{r})=\phi(\lambda)-\phi(\lambda_\textrm{r}) \,,
\label{eq:diffph}
\end{equation}
where $\lambda_\textrm{r}$ is a wavelength of reference simultaneously observed. 

\subsubsection*{The atmospheric influence and closure phases}
So far, we have neglected the influence of the atmosphere over the aperture of the telescope. Indeed, the amplitude of the visibilites can obtained with relative high accuracy and the vast majority of optical interferometry results before the 2000's are from the visibility amplitudes alone. Unfortunately, turbulent atmosphere causes a big displacement (or decomposition) of the incoming wavefront, destroying the fringe phase information. This effect is sometimes called \textit{piston effect}. 

One condition that is crucial to successfully perform the full interference is to cancel the optical path difference (OPD) between the beams for the different telescopes. Even though adaptive optics makes possible to trace the fringe pattern and correct wavefront delays arriving each telescope, the absolute phase information is lost (e.g., \citealp{mar94a}). However, with the advent of telescope arrays (three or more telescopes), the \textit{closure phase} can be used to overcome this difficulty. 

In a three-telescope system, each one receives the wavefront with phase $\psi_n$ and delay $e_n$. Summing the phase information of telescopes combined two by two, the piston effect can be canceled to reveal fundamentally new information about the sources under study, not contained in the visibility amplitudes. Indexing the phases according to the telescopes combined
\begin{align}\begin{split}
    \psi_{12} &= \phi_{12}+e_{2}-e_{1}\,, \\
    \psi_{23} &= \phi_{23}+e_{3}-e_{2}\,, \\
    \psi_{31} &= \phi_{31}+e_{1}-e_{3}\,,
\label{eq:phases}
\end{split}\end{align}
the closure phase $O_{123}$ for the three antennas can then be obtained as
\begin{align}\begin{split}
    O_{123} &= \psi_{12}+\psi_{23}+\psi_{31} \\
            &= \phi_{12}+\cancel{e_2}-\bcancel{e_1} + \phi_{23}+\xcancel{e_3}-\cancel{e_2} + \phi_{31}+\bcancel{e_1}-\xcancel{e_3} \\
            &= \phi_{12}+\phi_{23}+\phi_{31}.
\end{split}\end{align}   
Or, in equivalent way, by the \textit{bispectrum} $\tilde{B}_{123}= \tilde{V}_{12}\tilde{V}_{23}\tilde{V}_{31}$, the complex product of the visibilities from the array ($\tilde{B}_{123}=\|\tilde{B}_{123}\|\exp iO_{123}$). Similarly, multiple telescopes beams can be rearranged three by three to provide closures phases independently in a multiple array (for example, a four telescopes system provide 6 simultaneous closure phase measurements). 

Closure phases are often interpreted as the amount of asymmetry in the target's brightness distribution. It has a crucial role in different interferometric analysis, specially for the image reconstruction techniques. For an overview on the role of closure phases and the information it contains, see \citet{mon03a}. The main point is that if the Fourier space of the image (called $uv$ plan) is properly sampled with closure phases, we can apply an inverse Fourier transform on the data and rebuild an image that may have as much information as a regular image, depending on the extension of $uv$ coverage. See \citet{thi13a} for a recent review on interferometric image reconstruction techniques and the mathematical problems they contain.

\subsection{The VLTI AMBER and PIONIER interferometeric beam-combiners}
This study contains an extensive amount of data from the Very Large Telescope Interferometer (VLTI) facility of ESO (European Southern Observatory) in Paranal, Chile \citep{hag12a}. VLTI interferometri telecopes array can be set with utility telescopes (UTs, 8.2~m), or with the auxiliary telescopes (ATs, 1.8~m). The baseline lengths ATs can vary between 8 to 200 meters, depending on the observational period availability. With UTs, the baselines ranges between 46 to 130~m. 

AMBER is an near-infrared spectro-interferometer that operates in the $J$, $H$, and $K$ bands (i.e., from 1.0 to 2.4 $\mu$m), described by \citet{pet07a}. The instrument was designed to be used with two or three beams, allowing to measure both differential and closure phases. AMBER can be used with an external fringe tracker, the FINITO. On the UTs it is possible to reach $H=7$ and on the ATs  the limiting magnitude is $H=5$. There are three possible spectral resolution ($R=\Delta\lambda/\lambda$) modes: (i) high resolution (HR) $R=12000$; (ii) medium resolution (MR) $R=1500$; and (iii) low resolution (LR) $R=35$. 

PIONIER is a $H$-band four-telescope combiner with high precision visibilities measurements and multiple (four) closure phase information and is described by \citet{leb11a}. PIONIER tracks its own fringes. The magnitude limit is $H=8$. This value is independent of whether ATs or UTs are used, and UTs should only be used when other issues (such as availability of guide stars or need for small interferometric field of view) prohibit the use of ATs. It has only two spectral capabilities: $R=5$ or $R=30$ across the $H$ band.

\subsection{Remarks on interferometry \label{sec:interfrem}}
We described hitherto the main concepts of optical interferometry that are used for this study. The only missing topic is the \textit{spectro-interferometry}, or the interferometry with high spectral resolution. In spectro-interferometry it is sometimes useful to describe the interferometric quantities in terms of differential quantities (notably the differential phases). The idea behind this is that within a small wavelength range, all the atmospheric effects are roughly the same. So, instead of losing all, for example, phase information, we can measure its difference with respect to a reference wavelength measurement observed simultaneously. This is particularly suitable for line profiles, where the kinematics of the target can be addressed and the continuum region is the natural choice as reference.

We emphasize that the end product of an interferometric observation is not a regular image but the Fourier transform of the brightness distribution on the field-of-view of all the telescopes. This introduces new features when analyzing interferometric data. For example, in Sect.~\ref{sec:diffph} we discuss how emissions can be measured with higher resolution than the nominal resolution of the interferometer (Eq.~\ref{eq:resint}) using phase information. 

Another important feature is that the interpretation of the data in physical terms can be very dependent on the model used (i.e., the assumptions made), which makes it crucial to have a realistic brightness distribution of the target - something only achieved by proper solution of radiative transfer. The discussion of the inference of physical parameters from interferometric data will permeate this study.

\FloatBarrier
\section{Other observational techniques}
Important techniques used to investigate the properties of hot stars include asteroseismology, spectroscopy, polarimetry (their combination, spectropolarimetry) and interferometry. The combination of these multiple techniques, which mostly are probes different physical processes, offers new and fundamental investigation tools for these stars. Behind this idea was organized the IAU Symposium 307 ``New Windows on Massive Stars'' (Geneva, Switzerland, on June 2014) and is the context where this study is inserted.

\subsection{Spectroscopy}
Spectroscopy in astrophysical context is the technique that investigates the intensity of a light beam as a function of wavelength or frequency. This technique is present in the study of hot stars since at least the observations of Pietro Angelo Secchi of the Be star $\gamma$\,Cas in 1866, whose H$\beta$ line featured an intense emission characteristic\footnote{About this discovery, we reproduce here the comments from \citet{riv13a}: \textit{``In the heyday of nationalism, this [observation] was communicated in French language to a German Journal by an Italian astronomer, working at \emph{the} international organization of the time, the Vatican''}.}.

Spectrographic instruments that analyze the light along visible range are usually single or cross-dispersed grating spectrographs. (Single) Diffraction gratings are the most useful optical dispersing element due to its high luminous efficiency when compared to other elements, as prisms. The linear dispersion can be easily adjusted and the plane grating can be used over a wide range of angles of incidence, simply rotating the grating. 

This simplicity makes spectrographs very popular instruments among amateur astronomers. Amateur astronomers significantly contribute, either joining databases with professional astronomers (e.g., \href{http://basebe.obspm.fr/basebe/}{BeSS database}\footnote{\url{http://basebe.obspm.fr/basebe/}}), or in the form of individual observations. In particular, we mention the discovery of the active phase of Achernar in 2013 by the amateur astronomers Tasso Napole\~ao and Marcon (Brazil), and their contribution with subsequent observations, as other obtained from BeSS. 

The resolution of the spectrographs can be substantially increased by the use of cross-dispersed diffraction gratings, which become common after the 1980's. This instruments are known as \textit{echelle} spectrographs. In this study we employed observations from different echelle spectrographs. The spectroscopic observations used are described in Sect.~\ref{sec:aerispec}.


\subsection{Polarimetry \label{sec:pol}}
As mentioned previously, the polarization of light results from its (vectorial) wave properties. 
From Eqs.~\ref{eq:ReE} and \ref{eq:E}, the more general case of any light wave propagating in the direction $\hat{z}$ 
\begin{equation}
\vec{E}=(\tilde{E_1}\hat{x}+\tilde{E_2}\hat{y})\exp(2\pi ivt) \,,
\label{eq:elip0}
\end{equation}
\begin{align}\begin{split}
    \tilde{E_1}&=a_0\exp(-i\phi) \,,\\
    \tilde{E_2}&=a_0\exp(-i\phi+i\gamma).
\label{eq:elip}
\end{split}\end{align}

Equations~\ref{eq:elip0} and \ref{eq:elip} represents the electromagnetic wave position over time. They describe an ellipse, or an elliptically polarized light.

There are two particular cases of the elliptical polarization:
\begin{itemize}
    \item $\bm{\gamma=\pm\pi/4}$, where the electromagnetic vectors describe a circle (or \textit{circular polarization}); 
    \item $\bm{\gamma = 0}$ or $\bm{\pm\pi/2}$, where the electromagnetic vectors describe a single line (or \textit{linear polarization}.
\end{itemize}

\subsubsection*{The Stokes parameters}
Using an equation equivalent to the \ref{eq:elip0}, George Gabriel Stokes defined in 1852 a set of parameters as follows and that are known as \textit{Stokes parameters}
\begin{subequations}
\begin{align}
    I&\equiv \|\tilde{E_1}\|^2+\|\tilde{E_2}\|^2, \\
    Q&\equiv \|\tilde{E_1}\|^2-\|\tilde{E_2}\|^2, \\
    U&\equiv 2{\rm Re}(\tilde{E_1}\tilde{E_2}^*), \\
    V&\equiv -2{\rm Im}(\tilde{E_1}\tilde{E_2}^*).
\end{align}
\end{subequations}

The parameter $I$ represents the light total intensity, the parameters $Q$ and $U$ the linear polarization and $V$ the circular polarization. Stokes parameters allow to relate the light polarization characteristics on a convenient way
\begin{equation}
    Q^2+U^2+V^2 \leq I^2.
\end{equation}
In particular, the degree of polarization $P$ and its orientation $\theta$ are defined as
\begin{equation}
    P = \frac{\sqrt{Q^2+U^2+V^2}}{I} \,,
\end{equation}
\begin{equation}
    \theta = \frac{1}{2}\arctan\frac{U}{Q} \,,
\end{equation}
where $P$ is the polarized light fraction, $0\leq P\leq 1$, 100\% polarized. For a given observational error on the linear polarization measurement ($\sigma_P \equiv \sigma_Q \equiv \sigma_U$), it is possible to show that the orientation angle error $\theta$ is
\begin{equation}
\sigma_\theta = 28.6\sigma_P \,,
\end{equation}
with $\theta$ in degrees, neglecting the standard stars orientation determination errors used to put the observation in the celestial frame.

\FloatBarrier
\section{Polarization in selected contexts \label{sec:}}
\subsection{Linear polarization of circumstellar disks \label{sec:polsaw}}
Hot star disks are usually composed of highly ionized and low density gas. This condition favors the Thomson's scattering by free electrons whose scattering opacity is independent of the wavelength ($\lambda$). In this process, the electrons oscillate in the same frequency of the incident light beam, which leads to the reemission in the same frequency of incidence. Protons, that are much heavier than electrons, are little affected by the incoming light. As the polarization of the re-emitted beam depend on its angle of incidence, the light is linearly polarized in this process.

The degree of polarization by scattering depends upon the density of the medium and its geometry with respect to the incoming light. Given its simple geometry, the polarization of a circumstellar disk indicates precisely its orientation on the sky plane. But more complex configurations can be probed by linear polarization without angularly resolving it (e.g., \citealp{bro77a}).

Linear polarization still offers additional diagnosis of the circumstellar medium. The region of the disk where the scattering occurs depends on the geometric shape of the disk. Defining 
\begin{equation}
\zeta = \frac{d}{dr}\left(\frac{H(r)}{r}\right)\,,
\label{eq:zeta}
\end{equation}
$\zeta$ can used to classify disks in there main shapes: (i) $\zeta > 1$ for \textit{flaring} disks; (ii)  $\zeta<1$ for \textit{self-shadowed} disks; and the special case (iii) $\zeta=0$ for \textit{flat} disks (``flat'' here means that $H(r)\propto r$). From the observational point of view, $\zeta$ should not be determined as function of the scale height itself, but as function of the height above the midplane where the disk becomes transparent, sometimes called \textit{surface height}. In Eq.~\ref{eq:zeta} we assumed their equality, what would be just an approximation.

Early interferometric measurements constrained the Be stars disk thickness within the 3$^\circ$ to 20$^\circ$ range \citep{woo97a}, indicating these disks might be flared. As the stellar flux reaching the disk is much more intense in its inner regions and the flaring can block direct irradiation of their outermost regions, light polarization come from the closest region to the star.
This is in agreement with the theoretical predictions of a thin disk ($H\ll R_{*}$), pressure-supported scale height (Eq.~\ref{eq:H}). Thus, the polarization level is not set by disk radius, but almost exclusively by the gas density (except for very small disks, i.e., $R_{d} \sim R_{*}$). 

\begin{figure}
    \centering
    \includegraphics[width=0.6\linewidth]{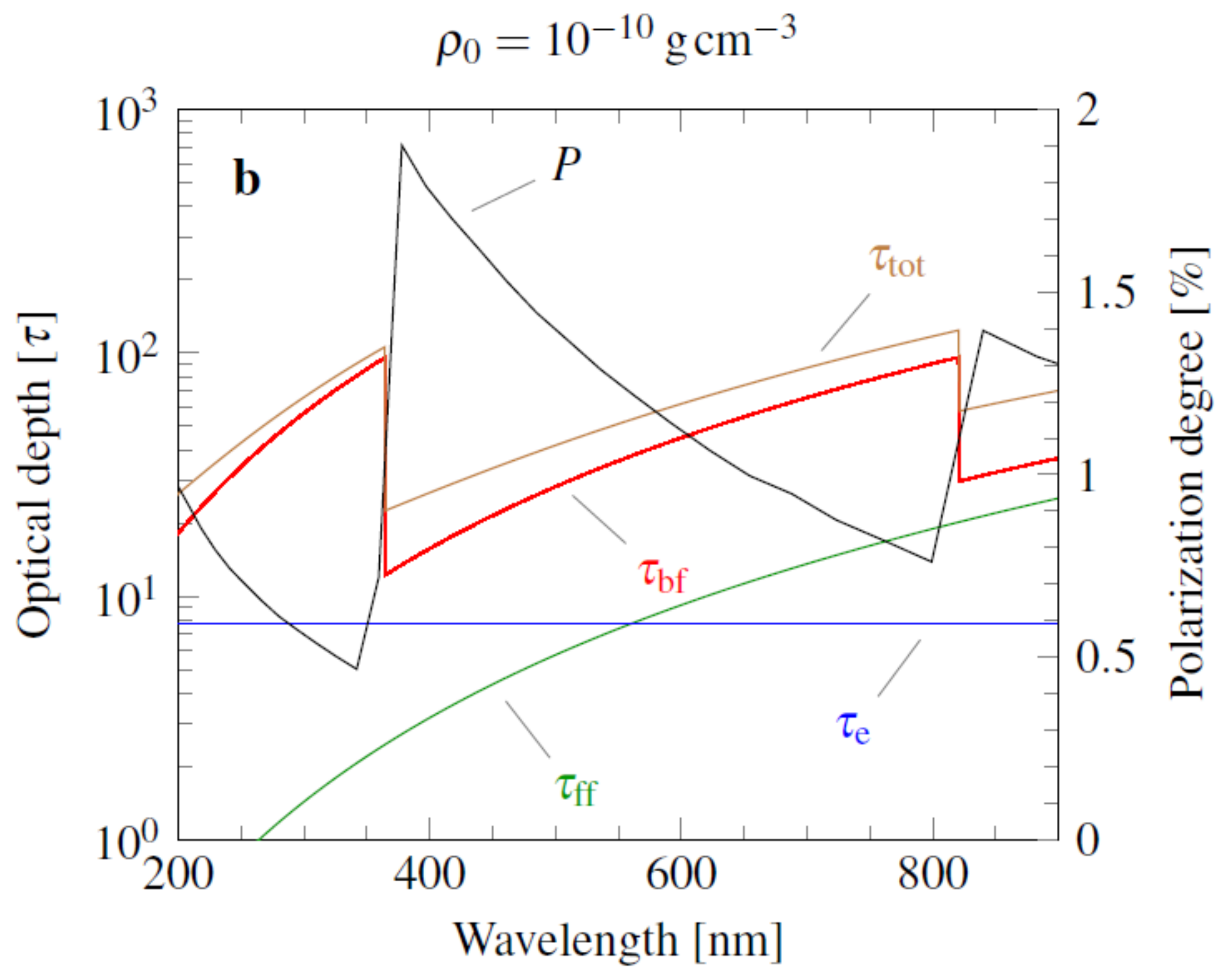}
    \caption[Polarized spectrum $P$ and radial optical depth contributions along the midplane of a typical Be disk]{Polarized spectrum $P$ and radial optical depth contributions along the midplane of a typical Be disk. The total optical depth $\tau_{\rm tot}$ is the sum of the optical depth for each continuum opacity source (free-free $\tau_{\rm ff}$, bound-free $\tau_{\rm bf}$, and Thomson scattering $\tau_{e}$). \citep{riv13a}.} 
    \label{fig:intro-saw}
\end{figure}

The polarized spectra of Be stars often displays a sawtooth shape (Fig.~\ref{fig:intro-saw}, black line), with abrupt changes very close at Hydrogen photoionization thresholds. The sawtooth polarization can be understood as roughly the combination of electronic scattering ($\tau_{e}$) in addition to the pre-scattering H\,{\sc i} absorption ($\tau_{\rm bf}+\tau_{\rm ff}$). The opacity dependence of the gas with $\lambda$ roughly results in 
$\kappa_{\rm HI}\rho \propto \lambda^{3}$ (e.g., \citealp{bjo94a}, Eq.~30) within the thresholds.

Even if spectropolarimetric information is not available, broadband polarimetric imaging can be useful in determining the density of these environments. \citet{hau14a} show examples of how broad-band polarization can be employed to probe Be star disks growth and destruction. 

\subsection{Interstellar polarization}
The interstellar medium (ISM) is the matter that fills the space between stars in a galaxy. This matter is not homogeneous and is composed by energetic particles (cosmic rays), gas and dust. This material interacts with the background emitted light on their way towards Earth. In the context of this study, this interaction mainly bring two results: (i) a selective absorption, stronger in smaller wavelengths (extinction), and (ii) changes in the linear polarization.

The mechanism by which ISM changes the light polarization was originally proposed by \citet{hal49a} and \citet{hil49a}, and consists of the alignment of non-spherical dust grains over the light path generating a selective polarized absorption. The first explanation of this phenomenon was done by \citet{dav51a}, accounting that in the presence of an external magnetic field, paramagnetic dust grains have an induced magnetic moment and tend to align to the field. This was further developed by \citet{dol76a} who established the current view of the alignment process where the radiative torque has an important contribution to dust grains alignment.  

Given the wave nature of light, where the electromagnetic fields oscillate perpendicular to the propagation direction, the ISM light polarization due to the magnetic grains alignment only reveals the $\vec{B}$ perpendicular to the line-of-sight, i.e., $B_\perp$. The $B_\parallel$ is hard to be measured and cannot be traced by linear polarization.

The photospheric emission of main sequence stars is (mostly) unpolarized. If a given star has a substantial amount of circumstellar material (e.g., a Be star with its disk), its emergent spectrum will become linearly polarized. So, when observing such a star, the observed polarization $P_{\rm obs}$ will be the result of the ISM and the intrinsic ($_*$) circumstellar components
\begin{align}
P_{\rm obs}^2 &= \frac{(Q_{\rm obs})^2 + (U_{\rm obs})^2}{I^2} \,, \nonumber \\
P_{\rm obs}   &= \frac{\sqrt{(Q_{\rm ISM}+Q_*)^2 + (U_{\rm ISM}+U_*)^2}}{I} . 
\end{align}
As light absorption, the ISM polarization depends on the wavelength. The detailed spectral characteristics depend on the ISM conditions, primarily the grains shape and composition. \citet{ser75a} establish an empirical relation $P_{\rm ISM}(\lambda)/P_{\rm max}$, where $P_{\rm max}$ is loosely related to the extinction coefficient $E(B-V)$ 
\begin{equation}
3.0 \leq \frac{P_{\rm max}}{E(B-V)} \leq 9.0,
\end{equation}
with units $\%$ and mag, respectively. The value $9.0$ represents the maximum ISM efficiency to polarize the light with the increasing $E(B-V)$. The wavelength dependence of polarization ($P_{\rm ISM}$) is described by
\begin{equation}
P_{\rm ISM}(\lambda)=P_{\rm max} \exp \left[-K\ln^2 \left(\frac{\lambda_{\rm max}}{\lambda}\right)\right]\,,
\label{eq:serk}
\end{equation}
where $K$ variable was introduced by \citet{wil80a}. $K=1.15$ corresponds to the original \citeauthor{ser75a} formulation. 

\subsubsection*{Methods for determining the ISM polarization}
In the context of circumstellar disks, there are four methods that can be applied to determine the ISM polarization components \citep{qui97a}: (i) Pure photospheric emission, if the CS disk is dissipated; (ii) $Q/U$ components and $\lambda_{\rm max}$ are the same for field stars; (iii) Hydrogen line profile depolarization; (iv) $QU$ diagram analysis;

Method (i) is suitable to the case of variable stars. If at a given time there is no evidence of circumstellar emission (for example, pure photospheric line profiles), the observed polarization will be a direct measure on the ISM one. See Chap.~\ref{chap:phots} for an example to the Be star Achernar. 

Field stars are stars angularly close to the target star (typically, within one degree). They may be selected by having a very low or zero intrinsic polarization. If nearby stars show a regular pattern in the $Q/U$ ratio and $\lambda_{\rm max}$, their light should be traveling over the same ISM component(s) and then the field stars polarization can be used to infer the ISM polarization of the target (method ii). It is usually not enough that the probed stars are angularly nearby the target: they should also be in a similar distance to the target, when this information is available.

The third method (iii) is the line profile depolarization \citep{mcl79a}. The principle of the method is that Hydrogen recombination transitions are, in the absence of magnetic fields, unpolarized. But it has some drawbacks. For this method, high spectral resolution observations are required and H line profiles wavelengths do not contain exclusively the (unpolarized) line emission. An underlying polarized emission may be also present. 
The ratio of line ($P_*(\lambda)$) to continuum polarization ($P_*(\lambda_C)$) is given by the empirical relation
\begin{equation}
\frac{P_*(\lambda)}{P_*(\lambda_C)}=\frac{1}{1+\chi(\lambda)}\,,
\end{equation}
where $\chi(\lambda)$ is the ratio of the intensity of the unpolarized emission to the polarized one in the underlying $\lambda$. Thus, $\chi\rightarrow 0$ means that there is no significant unpolarized line-emission [$P_*(\lambda)=P_*(\lambda_C)$]; and $\chi\rightarrow\infty$ means that to no significant polarized continuum emission exists [$P_*(\lambda)\rightarrow0$]. An adequate estimate of $\chi(\lambda)$ is necessary for the precise application of this method.



\subsection{Circular polarization and the Zeeman effect \label{sec:zeeman}}
In 1896, Pieter Zeeman discovered that, when light was in the presence of magnetic fields, single spectral lines were splitted into two or more lines. This phenomenon is due to the interaction between the magnetic field and the inherent magnetic moment of the microscopic particle (neutral atom, ion, or molecule). 
As an example, a transition that have total spin of zero between its states has the magnetic moment determined by the orbital motion of the electrons. The line transition have energy $E$ and in the presence of a magnetic field its energy can be change by the among of $\pm\Delta E$
\begin{equation}
\Delta E = \mu_B \|\vec{B}\| \,,
\end{equation}
where $\mu_B = \dfrac{eh}{4\pi m_e}\simeq 9.274\times10^{-21}$~erg\,G$^{-1}$ is the Bohr magneton, with $e$ the electron charge, $h$ the the Planck constant and $m_e$ the electron mass. The energy difference changes the line wavelength $\lambda$ according to the relation
\begin{equation}
\Delta \lambda = \frac{hc}{\Delta E}\,,
\end{equation}
where $c$ is the light speed. A more detailed description of the Zeeman effect can be found in quantum physics textbooks. 

The Zeeman components of a spectral line, radiated along the magnetic field direction, are circulary polarized. So, when observing a stellar atmosphere with the presence of a large-scale magnetic field with a non-zero component perpendicularly to the observers line-of-sight ($\|B_\perp\|>0$), the circular polarization within the spectral lines will be circularly polarized (i.e., Stokes $V\neq0$).

Polarization studies inside spectral lines can be performed with spectropolarimetry, that is the combination of the polarimetry and spectroscopy observational techniques. In the weak field approximation ($\|\vec{B}\|\lesssim30$~kG), valid for most of stars, the circular polarization will be proportional to the strength of the (dipolar) magnetic field and the intensity variation ($dI$) over the wavelength step ($d\lambda$; \citealp{lan73a})
\begin{equation}
V(\lambda)\propto\frac{dI}{d\lambda} \|B_\perp\|.
\end{equation}

Zeeman signatures in stellar spectra are generally extremely small, typical polarization amplitudes of 0.1\%. Thus, detecting them requires measurements of polarization with signal-to-noise level of about 10$^{4}$. A multi-line approach for increasing the signal-to-noise of the measured polarization was proposed by \citet{sem89a}. This approach, known as the LSD method (Least Squares Deconvolution), was successfully used for detecting stellar circular polarization as an indication of magnetism by \citet{don97a}.

%% file: chap/tools.tex
\chapter{Methods and tools \label{chap:tools}}
In this chapter we contextualize the work from its host institution, the Institute of Astronomy, Geophysics and Atmospheric Sciences at the University of S\~ao Paulo (IAG-USP). It made use of several computing resources and scientific softwares. We present here the main employed tools, including some developments made during the PhD.

\section{The BeACoN group}
In 2009, Prof. Alex C. Carciofi started at IAG-USP the BeACoN group, the \textit{Be Aficionados Collaborative Network}. The group has a number of important collaborators, including Dr.~Armando Domiciano de Souza and other astronomers of the interferometric group of the Observatoire de la Côte d'Azur, which made this study possible. The aim of the group is to investigate the Be phenomenon and to obtain a realistic physical model of the circumstellar environment of the Be in interaction with the photosphere. 

The main modeling tool of the group is the radiative transfer code \textsc{hdust}, described below. This is the first PhD thesis of the group, as part of the continuous application of \textsc{hdust} models to reproduce different observational quantities since 2006. It is also, to our knowledge, the first thesis in a Brazilian institution based on optical/IR long baseline interferometry (OLBI). 

\subsection{The IRAF reduction package \textsc{beacon} \label{sec:iraf}}
The BeACoN group is conducting a spectroscopic and polarimeter survey of Be stars, hot magnetic and other peculiar stars the OPD (Observat\'orio Pico dos Dias, LNA-MCT)\footnote{More info available at \url{http://www.lna.br/opd/opd.html}}. The survey already contains more than 3000 individual observations since 2006 of about 50 stars. A complementary survey is also in execution to determine the interstellar polarization of these stars through field observations of stars.
The data reduction is done via the \textsc{beacon} package\footnote{Available at \url{https://github.com/danmoser/beacon}.} based on \textsc{iraf} (Image Reduction and Analysis Facility; \citealp{tod93a}). 

\subsubsection*{Polarimetry}
The polarimetric data shown in this thesis are results of the survey. The survey uses a CCD camera with a polarimetric module described by \citet{mag96a}, consisting of a rotating half-wave plate and a calcite prism placed in the telescope beam. A typical observation consists of 8 or 16 consecutive wave plate positions separated by 22.5$^\circ$. In each observing run at least one polarized standard star is observed in order to calibrate the observed position angle. The polarimetric reduction algorithm is described by \citet{mag84a}. It is an adaptation of the \textsc{pccdpack} package \citep{per00a} for the particular observing mode we use in the survey. Basically, since many targets have quite small polarization values (as is the case of Achernar, see Chapters~\ref{chap:phots} and \ref{chap:aeri}), a high S/N ratio is required to reach the typical target polarization accuracy aimed in the survey (0.01\%, for which a S/R~$\sim10^5$ is required). This high S/N is reached by taking a large number of frames for each waveplate position.

\subsubsection*{Spectroscopy}
The spectroscopic component of the survey are based on two main instruments: (i) the ECass, a long slit Cassegrain spectrograph and (ii) the MUSICOS (Multi-SIte Continuous Spectroscopy). The reduction of long slit spectra was incorporated to the \textsc{beacon} package using the algorithm from \citet{fae11a}. After a French donation to the Brazil, the MUSICOS spectrograph become available at the OPD in 2012. The donation was reported by \citet{fra10a}, and its availability by \citet{pra10a}. 

MUSICOS is a bench spectrograph fed by a fiber optic at the telescope focus \citep{bau92a}. It covers a wide spectral range in two non-simultaneous modes: ``blue'' (3800-5400~\AA), and ``red'' ($\sim$5400-8800~\AA). It has a resolving power of $R\approx30000$. Given its simultaneous high resolution and wide spectral coverage, MUSICOS replaced the Ecass in the survey.

Unfortunately, no support concerning the reduction of the data was supplied. So I developed a reduction algorithm for the instrument in the \textsc{beacon} package, adapted to the OPD detectors. A reduction manual is under development, and will be available in the BeACoN website for the community.

\subsection{Radiative transfer code \textsc{hdust}}
\textsc{hdust} is one of the most performing Monte Carlo radiative transfer codes to date \citep{car06a,car08a}. It is a fully three-dimensional (3D), non-local thermodynamic equilibrium (NLTE) code designed to solve the coupled problems of radiative transfer, radiative equilibrium and statistical equilibrium for arbitrary gas density and velocity distributions.

\begin{figure}
    \centering
    \includegraphics[width=.5\textwidth]{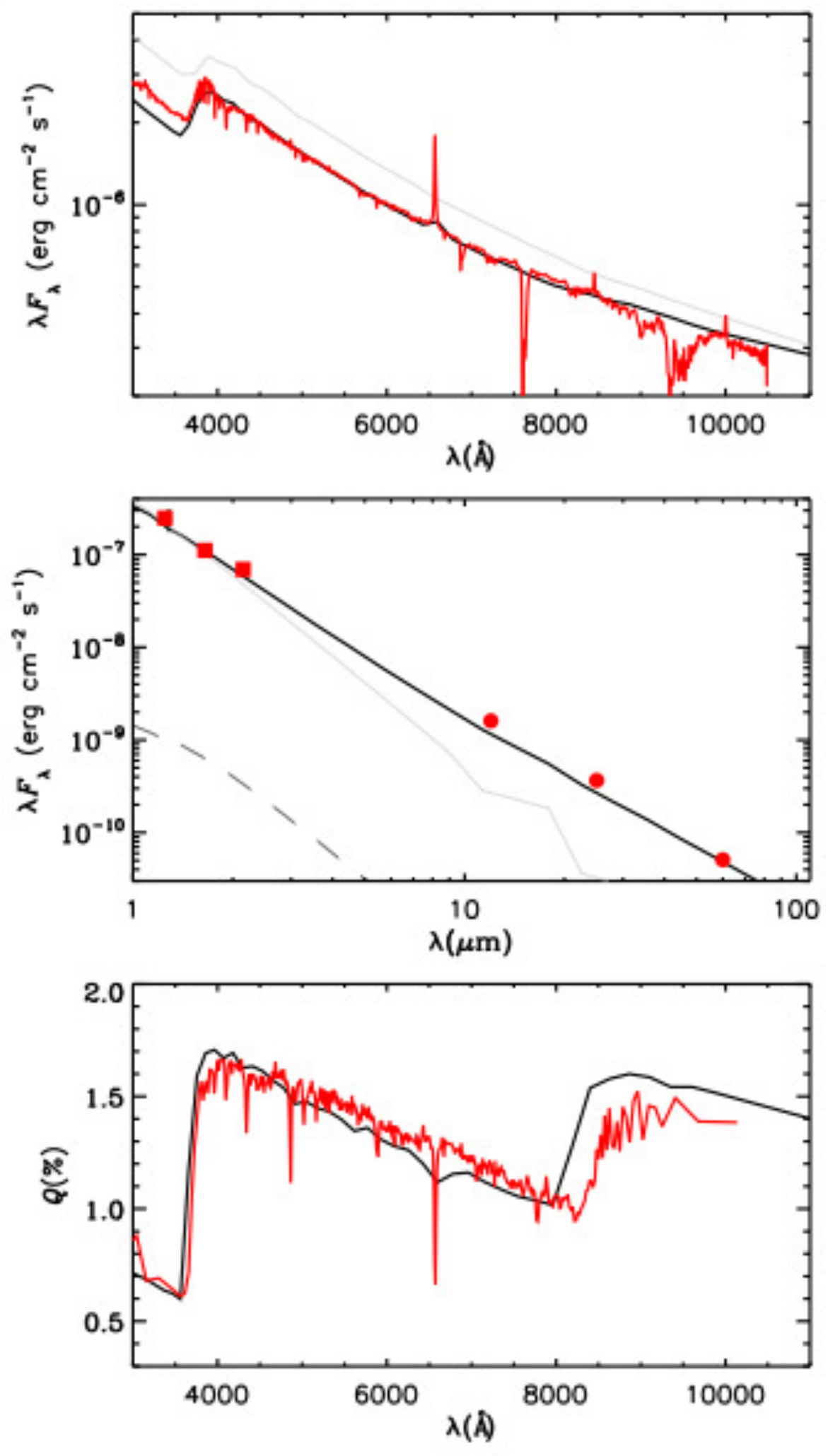}
    \caption[\textsc{hdust} fit of the emergent spectrum and polarization of $\zeta$\,Tau]{\textsc{hdust} fit of the emergent spectrum (visible \textit{top}; IR \textit{middle}) and polarization (bottom) of $\zeta$\,Tau \citep{car09a}. The observations are in red and model in black. }
    \label{fig:0}
\end{figure}

The \textsc{hdust} code has already been successfully applied to the analysis of a number of Be and B[e] stars and is used to analyze and interpret the observational dataset present in this study. Since its first publication in 2004, \textsc{hdust} has contributed to 22 refereed publications. One example is shown in Fig.~\ref{fig:0}, where multiple observables of the star $\zeta$\,Tau are reproduced.

\subsubsection*{Spectral lines synthesis of rotating stars}
Until early 2013, \textsc{hdust} could provide accurate line profiles only for non-rotating stars. This is because in its default mode, the code used photospheric models of low resolution from \citet{kur79a} for greater computational efficiency. In order to compute the photospheric line profile, a file with the (non-rotating) photospheric model needed to be provided by the user. As a result, the photospheric line profile did not include the effects of gravity darkening.

Prof.~Carciofi and I upgraded the \textsc{hdust} code to read accurate photospheric line profiles for different effective temperatures and effective surface gravity. The profiles were computed by \textsc{synspec}, a general spectrum synthesis program. \textsc{synspec} is an auxiliary routine, based on IDL, of the stellar atmosphere models TLUSTY \citep{hub88a}. The line grid is encoded in the External Data Representation (XDR) binary format.

The new version of the code upgrade was validated by comparisons with the \textsc{charron} code (Code for High Angular Resolution of Rotating Objects in Nature; \citealp{dom02a,dom12b}). 

\subsection{\textsc{BeAtlas} project \label{sec:beatlas}}
Aiming at a comprehensive investigation of Be stars and VDD scenario, the project \textsc{BeAtlas} came about. The project consists of a systematic grid of models generated by the \textsc{hdust} code. The grid was designed to cover the whole known range of parameters of these stars and has the broad goals of (i) testing the universality of the VDD by applying it to large samples of Be stars; (ii) serving as a tool for planning interferometric observations of Be stars, and (iii) serving as a starting point for fitting data of individual stars. 

The Be emission strongly depends on the stellar photospheric configuration, which changes along the stellar evolution. The stellar evolution models used to derive the photospheric parameters of the central stars are from the group at the University of Geneva \citep{geo13a,gra13a}. The first phase of the project, we selected twelve stellar masses from 3.4 up to 14.6\,$M_{\odot}$ (corresponding to the spectral types B9 and B0.5, respectively) and five stellar rotation rates to keep the central star parameters spacing representative. 
The stars are assumed to be rigid rotators in the Roche approximation, so that there is a well know relation between $W$ and the stellar oblateness ($R_{\rm eq} / R_{\rm pole} = 1.1$ to 1.45, corresponding to $W\sim0.45$ to 0.95; Eq.~\ref{eq:Wrr}). The gravity darkening $\beta$ exponent was assumed to be a function of $W$, according to the relation of \citet[more details on this subject are given in Chapter~\ref{chap:phots}]{esp11a}.


In addition to mass and $W$, we also modeled stars with different ages in the main sequence. Following the natural parameterization of stellar evolution models, we chose $X_c$, the fraction of H in the core, as an age indicator. The values simulated are listed in Table~\ref{tab:beat}. Note that in the first phase of the grid, all purely photospheric models were computed. However, disk models were computed inicially for $X_c = 0.3$ only.

The disk were parameterized according to the VDD model. We chose seven different base surface density ($\Sigma_0 = $ 0.05 to 2.0 g\,cm$^{- 2}$) corresponding to different stellar mass-loss rates and four exponents for the volume density (Eq.~\ref{eq:rho}): $m=3.5$, that corresponds to a steady-state isothermal disk, a smaller value $m=3.0$ that can be associated to either a dissipating disk \citep{hau12a} or a disk truncated by a secondary star \citep{oka02a}, and two larger values ($m=4.0$ and 4.5) which can be associated to young disks \citep{hau12a}. In addition, we computed models for which the radial density structure were given by the non-isothermal solution of the radial viscous diffusion problem \citep{car08a}. 
The simulation disk radius is $50R_{\rm eq}$ for each star. One can define \textit{pseudo-photosphere} as the region of the disk that is radially opaque to continuum or line radiation and therefore dominates the disk emission. The radius of this pseudo-photosphere\footnote{The term pseudo-photosphere can have other definitions. For example, in \citet{har94a} it denotes the region within which the disk viewed face-on is optically thick and so looks like an extension of the stellar photosphere.} depends on the wavelength considered. For instance, \citet{hau12a} points that, for a steady-state disk, the pseudo-photosphere extends to $\sim2 R_{\rm eq}$ at the $V$ band. For line transitions, such as H$\alpha$, this region can be as large as $10-20 R_{\rm eq}$. This means that models with disks larger than the pseudo-photosphere are essentially identical. Thus, BeAtlas models are suitable for isolated Be stars, or those with long period binary companions, until the $mm$ wavelengths. The effects of truncated disks, resulted from short-period binary interactions, are not covered in the first phase of the project, but are planned for the subsequent phase. Finally, the observables are generated to ten different observer's line-of-sight equally spaced in the cosine range ($\cos i=0$ to 1). Table~\ref{tab:beat} contains the list of the main simulated parameters of the project.

\begin{table}
\centering
\caption{List of parameters of Be stars of the \textsc{BeAtlas} project.}
\begin{threeparttable}
\begin{tabular}[]{ll}
\toprule
    Parameter & List of values \\
\midrule
    Spec. Type & B0.5, B1, B1.5, B2, B2.5, B3, B4, B5, B6, B7, B8, B9 \\
    Mass ($M_\odot$) & $14.6, 12.5, 10.8, 9.6, 8.6, 7.7, 6.4, 5.5, 4.8, 4.2, 3.8, 3.4$ \\
    $X_{\rm c}$ (Fraction of H in the core)\tnote{a} & $0.08, 0.30, 0.42, 0.54, 0.64, 0.77$ \\
    Metalicity Z & 0.014 \\
    Oblateness ($R_{\rm eq}/Rp$) & $1.1, 1.2, 1.3, 1.4, 1.45$ \\
    Rotation rate $W$ & $0.447,0.633, 0.775, 0.894, 0.949$ \\
    $\Sigma_0$ (g\,cm$^{-2}$) & $0.02, 0.05, 0.12, 0.28, 0.68, 1.65, 4.00$ \\
    Mass density radial exponent ($m$)\tnote{b} & $3.0, 3.5, 4.0, 4.5$ + non-isothermal steady-state \\
    Disk radius ($R_{\rm eq}$) & 50 \\
    Inclination angle ($i$; deg.) & 0.0, 27.3, 38.9, 48.2, 56.3, 63.6, 70.5, 77.2, 83.6, 90.0 \\
\bottomrule
\end{tabular}
    \begin{tablenotes}
        \footnotesize
        \item[a] The list corresponds to the photospheric models. $X_{\rm c}$=0.30 is the only value for models with disk currently available.
        \item[b] Based on the parametric prescription of $\rho_0 (R_{\rm eq}/r)^m$.
    \end{tablenotes}
 \end{threeparttable}
\label{tab:beat}
\end{table}


This grid corresponds to approximately 25,000 models, for which high resolution spectroscopy, polarimetry and imaging (mainly targeting the interferometric quantities) were computed. The calculations required an extensive use of computational resources executed at the AstroInformatics Laboratory at IAG-USP (Brazil)\footnote{More information at \url{http://lai.iag.usp.br}.}. The time estimated for computing were nine months at the regular queue schedule.

I was involved in all steps of the \textsc{BeAtlas} project so far. I participated in the project design discussions (i.e., the quantities to be simulated and the relevant parameters needed) and I also developed the tools to handle the project data, since code execution, validation tests and analysis. These tools are part of the \textsc{pyhdust} library, described below. The project also has the participation of the post-doc Rodrigo~Vieira, the PhD students Leandro~R. R\'imulo and Bruno~C. Mota and the undergraduated student Andr\'e~L. Figueiredo, from IAG-USP.

The first phase of \textsc{BeAtlas} models was finished in middle of this year (2015), and a few applications are currently being developed: (i) Bruno Mota, PhD student of Prof.~Alex Carciofi at USP (Brazil), is leading the application of the models to constrain the stellar parameters of rotating stars by comparison with ultra-violet data from IUE (International Ultraviolet Explorer) satellite; (ii) Tiago Souza, MSc student of Prof.~Marcelo Borges at ON (Observat\'orio Nacional, Brazil), is fitting the $H$ band features of Be stars seen by APOGEE survey; and (iii) Andr\'e Figueiredo, undergrate student of Prof.~A. Carciofi, is assessing how the slope of the polarization spectrum can be used to constrain the central star rotational rate after the discovery by \citet{kle15a}. Many other applications are expected in the near future.

\subsection{\textsc{pyhdust} library \label{sec:pyhdust}}
\textsc{pyhdust} is an open-source Python language-based library to manipulate observations and models of astrophysical objects and was entirely developed during this thesis work. It is available at \url{https://github.com/danmoser/pyhdust}. Its main modules are:
\begin{itemize}
    \item \textbf{pyhdust}: functions to manipulate \textsc{hdust} output files;
    \item \textbf{input}: functions to generate \textsc{hdust} input files;
    \item \textbf{beatlas}: functions related to the \textsc{beatlas} project;
    \item \textbf{phc}: \textit{physical constants} and general use functions module;
    \item \textbf{interftools}: functions to manipulate and plot interferometric data;
    \item \textbf{poltools}: the same as above, for polarimetric data;
    \item \textbf{fieldstars}: functions to analyze and plot polarization data of field stars;
    \item \textbf{spectools}: functions to manipulate and plot spectroscopic data;
    \item \textbf{singscat}: single scaterring modeling tools, presented in Chapter~\ref{chap:mag}.
\end{itemize}

Most of the analysis and graphs presented in this work were performed using \textsc{pyhdust}, as well all the \textsc{beatlas} project management so far. It interfaces other Python libraries, such as \textsc{pyfits}, used for reading \textit{fits} format files, and the \textsc{emcee} minimization library, among others. A more detailed documentation is available at \url{http://astroweb.iag.usp.br/~moser/doc/}.
\section{Monte Carlo Markov chain code \textsc{emcee}}
The Monte Carlo Markov Chain method (MCMC) is a class of algorithms that can be used for fitting arbitrary multi-parametric functions to data. This work makes use of the \textsc{emcee} code, a Python language implementation of the MCMC method whose algorithm is described by \citet{for13a}. The code has been used in a growing number of astrophysics papers\footnote{A list of \textsc{emcee} application in astrophysical literature is available at \url{http://dan.iel.fm/emcee/current/testimonials}} and it is applied in different minimization process in this work.

Briefly, from a set of parameters and a given likelihood function, MCMC method computes the probability density function (PDF) by sampling each of the model parameters, with resolution set by the number of \textit{walkers}. The likelihood function do not need to be normalized. The convergence of the sample around the maximum likelihood distribution will be satisfied if a sufficient number of \textit{iterations} are performed. The best-fitting parameters values must be chosen from the PDFs by the user, depending on the nature of the minimization.

In our analyzes, we established the likelihood function as inversely proportional to the $\chi_{\rm red}^2$ value, commonly defined to measure a goodness of data fit
\begin{equation}
   \chi_\mathrm{red}^2 = \frac{\chi^2}{\nu} = \frac{1}{\nu} \sum {\frac{(\mathrm{model} - \mathrm{data})^2}{\sigma^2}}\,,
\end{equation}
where $\nu$ is the number of degrees of freedom and $\sigma$ the error associated to the data. The elected criterion to determine the best-fitting values was to choose the median value of the sampled PDFs. The errors were estimated from their (non-symmetric) PDF histograms area, containing approximately 34.1\% of the occurrences from the median (i.e., equivalent to one $\sigma$ if the PDF approach a normal distribution). The PDF can also be jointed in frequency maps, in order to verify the correlation between the multiple parameters.

%% file: chap/magstars_arxiv.tex
\chapter{Magnetospheres of hot stars \label{chap:mag}}
Magnetism is a fundamental property of the stars and play an important role in stellar evolution. For low-mass stars, the magnetic field is believed to be generated through dynamo processes within their convective envelopes as it occurs in our Sun. Indeed, most cool stars exhibit a large number of solar-like activity phenomena, i.e., a considerable magnetic variability. Unlike their low-mass counterparts, magnetic fields of massive stars are very stable on timescales of months and even years. This stability, the simple topographical features, and their no scalability with rotation, are evidences that magnetic fields of hot stars were generated in early stages of stellar evolution. This scenario is known as \textit{fossil field}, although the role of binarity is not well determined. See \citealp{don09a} for a review.

Recent surveys using Zeeman diagnostics (e.g., MiMeS, Magnetism in Massive Stars, \citealp{wad11a}) show that approximately 10\% of hot stars possess strong (dipolar) magnetic fields ($\sim$1-10 kG). The magnetic field presence, however, differ for different sub-types of hot stars. Important in the context of this study is that not a single magnetic Be star was detected. And among the Bp stars, the incidence fraction of magnetic fields is high, especially in the ones with chemical peculiarities. See \citet{shu14a} and references therein. 

The term \textit{magnetosphere} refers to the channeled stellar wind into a circumstellar structure that results from the dynamical interaction of the stellar magnetic field with stellar rotation and mass loss. \citet{pet13a} analyzed a comprehensive number of magnetic hot stars and proposed a classification of the magnetospheres based on the radial structure of the circumstellar environment. The magnetospheres can be described in terms of the Alfv\'en radius $R_{\rm A}$ and  (Keplerian) co-rotation radius $R_{\rm K}$.

$R_{\rm A}$ can be estimated as \citep{ud-08a}
\begin{equation}
R_{\rm A}=R_{\rm eq}[0.3+(\eta_*+0.25)^{1/4}]\,,
\end{equation}
where $\eta_*$ is the \textit{wind magnetic confinement parameter}, and $R_{\rm K}$ is
\begin{equation}
    R_{\rm K}=\frac{3}{2} R_p \left(\frac{\Omega}{\Omega_{\rm crit}}\right)^{-2/3}=R_{\rm eq}W^{-2/3}.
\label{eq:corot}
\end{equation}  
If $R_{\rm A}<R_{\rm K}$, the star is classified as having a \textit{dynamical magnetosphere}. Otherwise, $R_{\rm A}>R_{\rm K}$, the star possess a \textit{centrifugal magnetosphere}. Thus, magnetic stars with high rotational rates tend to generate centrifugal magnetospheres.

\begin{figure}
    \centering
    \includegraphics[width=.6\linewidth]{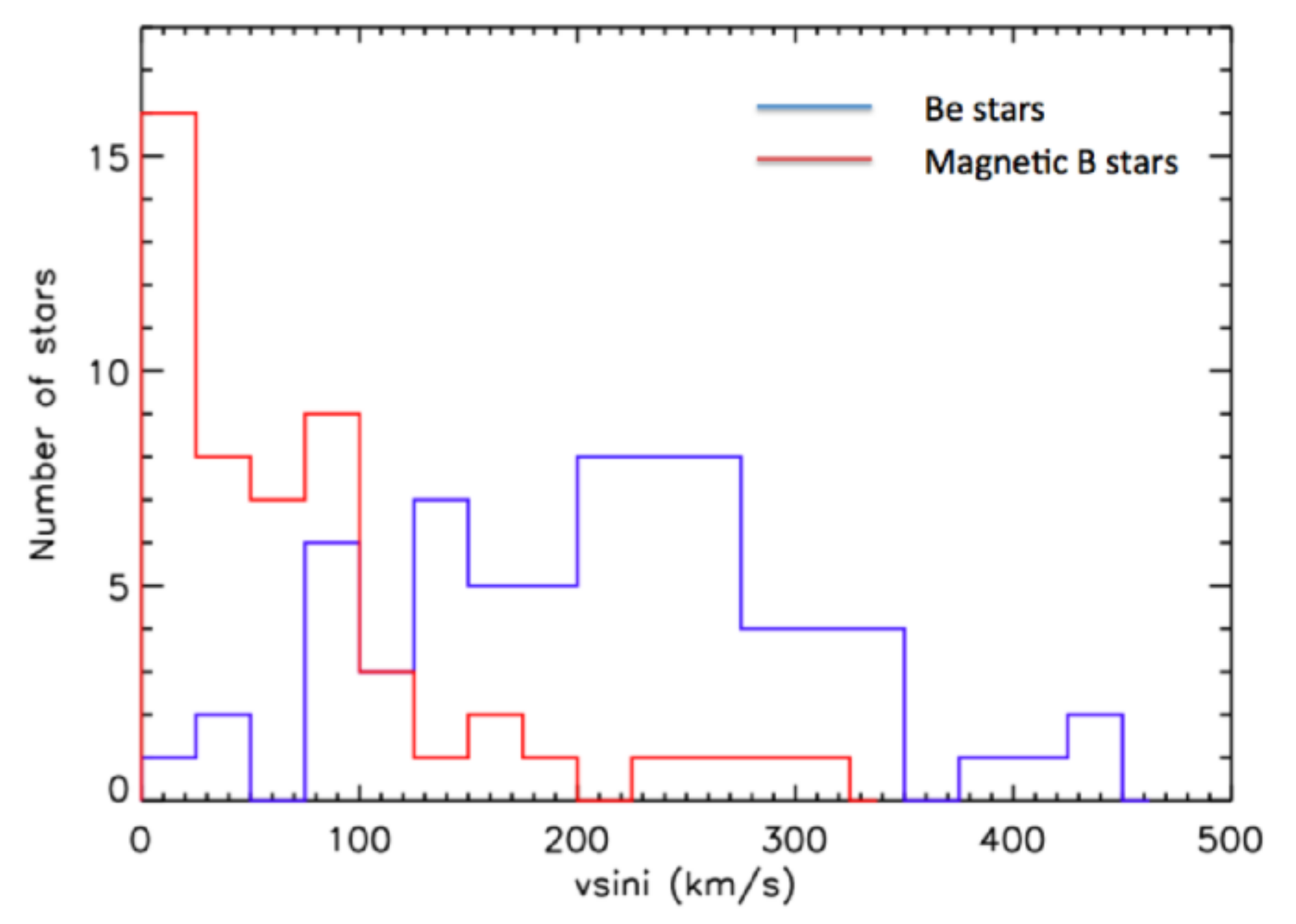}
    \caption[The projected rotational velocities of the MiMeS Be stars sample compared to those of the known magnetic B stars]{The projected rotational velocities of the MiMeS Be stars sample compared to those of the known magnetic B stars (\citealp{wad14a}).}
    \label{fig:magvsini}
\end{figure}

Unlike the Be phenomenon, the magnetism in fast rotating stars appears to be rare. This is illustrated in Fig.~\ref{fig:magvsini}, where the frequent low $v\sin i$ values of B-type magnetic stars contrasts with the broad distribution of Be stars in the MiMeS survey.
Indeed, until the discoveries of HR\,7355 (\citealp{riv10a} and \citealp{oks10a}, $P=0.52$\,d, $v\sin i=310$\,km s$^{-1}$) and HR\,5907 (\citealp{gru12a}, $P=0.508$\,d, $v\sin i=280$\,km s$^{-1}$), no star with magnetosphere had been found with a rotation rate that exceeds the moderate equatorial velocity of the archetype magnetic star \sori{} with $v\sin i = 160$\,km s$^{-1}$. 

\FloatBarrier
\subsubsection*{The centrifugal magnetospheres}
The centrifugal magnetospheres are a characteristic of He-rich stars, which have the strongest magnetic fields detected in non-degenerate stars since the 1970's observations (\citealp{lan78a,bor79a}). The strong magnetic fields have a number of consequences for photospheric and circumstellar structures, with rich phenomenology synchronized to the stellar rotational period. For example, a $\sim 10\,{\rm kG}$ dipole magnectic field detected in this kind of stars lead to not yet fully understood chemical stratification processes. At the magnetic poles, there are Helium-rich spots, which, as the star rotates, lead to a periodic modulation of the Helium (and other) lines \citep{tow05b}. 

These magnetospheres are observed via Balmer emission line variations and gyrosynchrotron emission in the radio waveband, as well as ultraviolet resonant lines. State-of-the-art models go a long way in quantitatively explaining the rich series of high-resolution spectra of the magnetospheres (e.g., \citealp{gru12a, riv13c}). However, only interferometry and polarimetry observations can directly assess the geometrical configuration of the circumstellar plasma (the latter without actually resolving it).

\begin{figure}
    \centering
    \includegraphics[width=.8\linewidth]{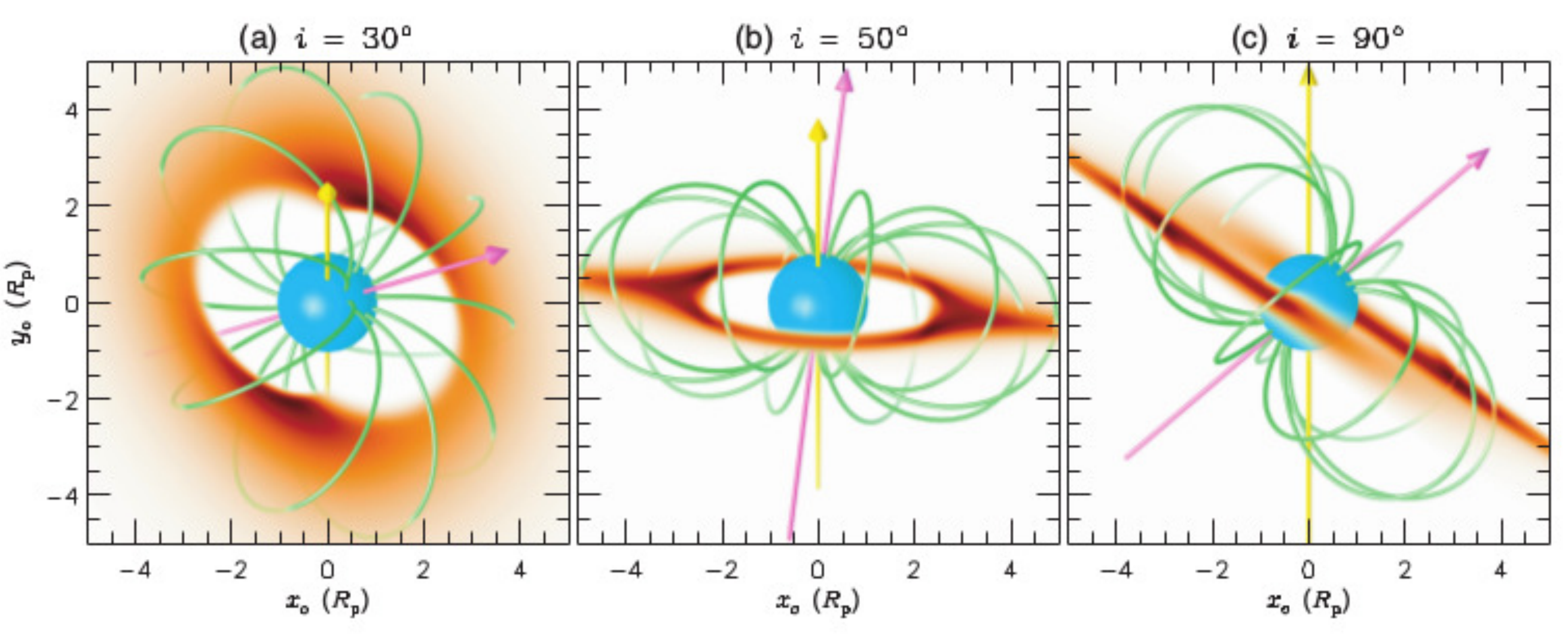}
    \caption[Maps of the column density to observers for different RRM configurations]{Maps of the column density to observers for three RRM configurations: (a) \{$i, \beta$\} = {30$^\circ$, 50$^\circ$} at a rotational phase $\phi$ = 81$^\circ$; (b) \{$i, \beta$\} = {50$^\circ$, 50$^\circ$} at $\phi$ = 171$^\circ$; (c) \{$i, \beta$\} = {90$^\circ$, 50$^\circ$} at $\phi$ = 81$^\circ$. In all three cases, $W\approx0.3$. Black/opaque corresponds to the highest column density, and white/transparent to the lowest, with intermediate levels shown in orange. The axes are drawn as arrows: yellow/upright for the rotation axis and magenta/oblique for the magnetic axis. Magnetic field lines with a summit radius of 5 polar radius (i.e., stellar radius at the rotational poles) are indicated in green (\citealp{tow08a}).}
    \label{fig:magshape}
\end{figure}

The magnetospheres can be modeled via the Rigidly Rotating Magnetosphere model (RRM; \citealp{tow05b}), a time-independent, semi-analytic formalism which is able to treat arbitrary magnetic topologies. RRM solves the density along the surface defined by the intersection of the magnetic field with the minima of the stellar rotating gravitational potential. Hydrostatic equilibrium is assumed along the magnetic field lines, with the wind material settling on this disk-like structure. The typical configuration is the stellar rotation not aligned to the magnetic field, configuring a oblique rotator. In this case, where the dipolar magnetic field direction does not coincide with the rotational axis, RRM predicts a warped disk with the densest regions corresponding to the intersections of the magnetic and rotational equators (here called ``blobs''). Fig.~\ref{fig:magshape} shows the predicted column density of the magnetospheres for three different configurations of the inclination angle $i$ and the angle $\beta$ between the magnetic axis and the rotational axis. When first applied to \sori{} \citep{tow05a} the RRM model was able to reproduce the shape of emission lines as well as the photometric depths of eclipses by the plasma clouds of magnetospheres. 

\FloatBarrier
\section{Polarimetry of magnetospheres and \sori{} \label{sec:sori}}
The $\sigma$\,Ori system is a multiple system containing five stars. $\sigma$\,Ori\,A is an O-type star, while the  B, D and E are B-type stars and C is an A-type, all of them dwarfs. The closest stars are A and B, separated by $\sim250$~mas, while \sori{} is the more distant, apart $\sim42.5$~arcseconds from AB \citep{hof95a}. The first linear polarimetric observations of the $\sigma$\,Ori system was carried out by \citet{kem77a}, who marginally detected a periodic modularization of \sori{}. The measured modulation amplitude was of the order of 0.1\%, with accuracy of $\sim0.05\%$.

Polarization is a very useful technique that allows one to probe the geometry of the circumstellar scattering material. One consequence of the material accumulation in hot star magnetospheres is its high temperature and high ionization. The stellar light is scattered by the circumstellar free electrons and produces a linearly polarized signal. So, the magnetospheres are excellent targets for the BeACoN group tools, namely our high-precision polarimetry and the radiative transfer modeling - complemented by optical interferometry. Our first application to \sori{} resulted in the paper by \citet{car13a}, who reported the first firm detection of a hot star magnetosphere in continuum linear polarization.

\begin{figure}
    \centering
    \includegraphics[width=0.8\linewidth]{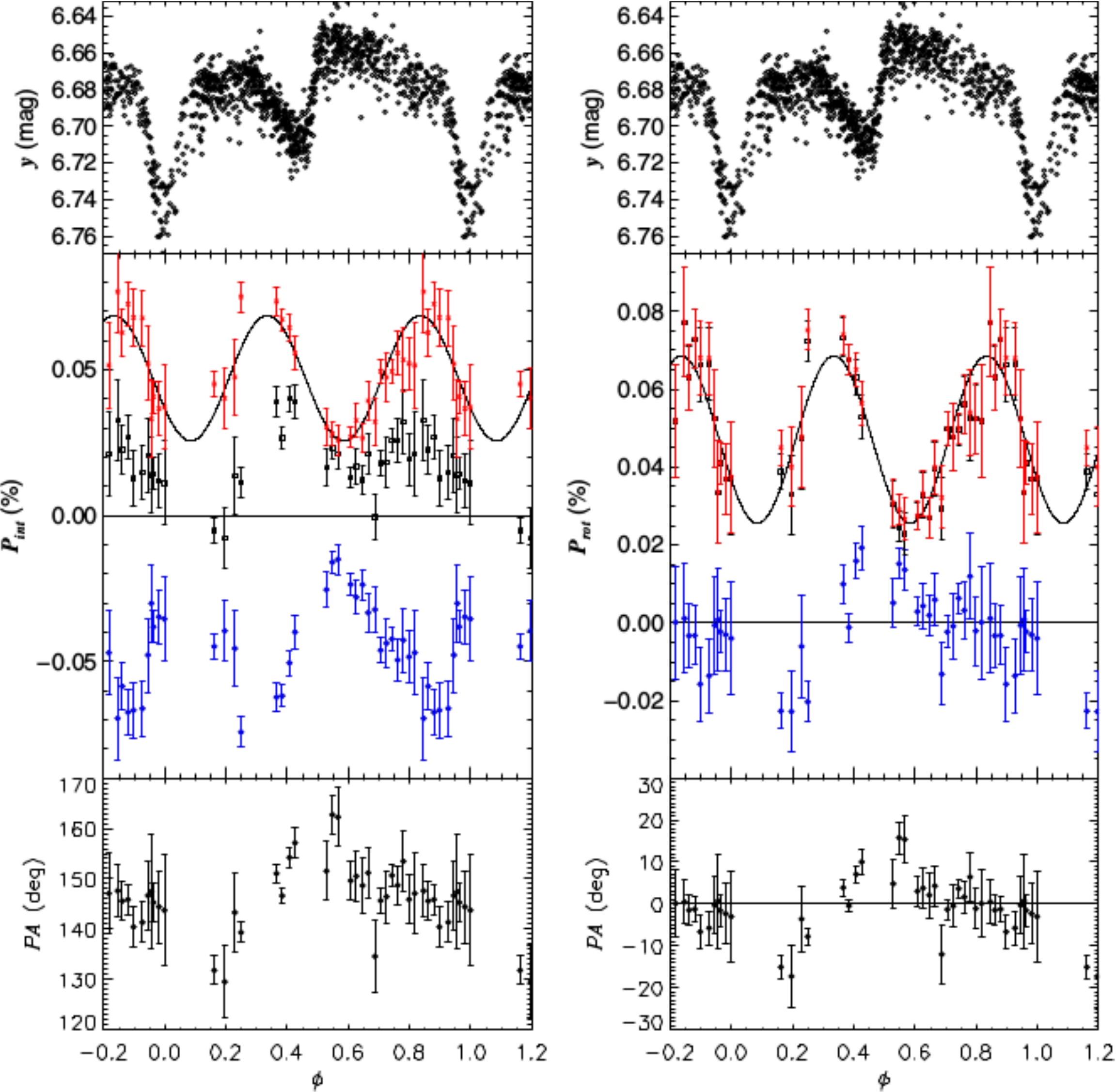}
    \caption[Variation of the intrinsic polarization of \sori{} compared with photometric variations]{Variation of the intrinsic polarization of \sori{} compared with photometric variations ($y$ filter). \textit{Top panels}: $y$ filter photometric data from \citet{hes77a}, folded in phase. \textit{Middle}: intrinsic $P$ (red), $Q$ (black), and $U$ (blue). \textit{Bottom}: intrinsic position angle. \textit{Left}: the polarization data are displayed in the equatorial reference frame. The polarization of $\sigma$\,Ori\, AB was used as a measure of the IS polarization. \textit{Right}: same as the left plot, rotated such that $\mean{U_{\rm int}} = 0$. The solid lines are the result of a cosine function fitting (Eq.~\ref{eq:polcos}). \citep{car13a}.}
    \label{fig:mags-sori1a}
\end{figure}

In Fig.~\ref{fig:mags-sori1a}, we present the results of the observational campaign of \sori{} until 2013. 
We started studying the polarization signal by a model-independent analysis. Here we summarize the steps:
\begin{itemize}
    \item The raw data, determined in the instrumental frame, is rotated to the equatorial reference frame using calibration data from standard polarized stars. The calibrated data is shown in Fig.~\ref{fig:mags-sori1a}, left panel.
    \item The ISM polarization components $Q_{\rm ISM}$ and $U_{\rm ISM}$ are removed from the data. The ISM polarization towards \sori{} was estimated using the measured polarization of $\sigma$\,Ori\,AB.
    \item The data is again rotated, in this step to align the system to the celestial North. This was done by forcing the average value of the $U$ parameter across the rotation period to be zero. The intrinsic polarization is shown in Fig.~\ref{fig:mags-sori1a}, right panel.
    \item The intrinsic polarization maximum and minimum can then be evaluated and, given the double polarimetric oscillation within one photometric period, the following equation was fitted to the data 
    \begin{equation}
    P(\phi) = P_0 + A\cos[4\pi(\phi - \delta)]\,,
    \label{eq:polcos}
    \end{equation}
    making use of the Levenberg-Marquardt algorithm \citep{mar63a}.
    \item This polarization curve, model-independent, can used to determine the intrinsic polarization characteristics, such as symmetry and delay relative to photometric curve ($\delta$, measured as phase).
\end{itemize}
The results of this analysis are summarized in Table~\ref{tab:pol1}. The considered ISM polarization values are in Table~\ref{tab:ISM}. We highlight the nearly symmetric modulation ($\chi^2_{\rm red}=0.97$) at twice the frequency of the photometric cycle with amplitude $A\sim0.02\%$, and the presence of a constant polarization component $P_0\sim0.047\%$, i.e., the polarization degree is never zero. This means that the magnetosphere is always asymmetric, irrespective of viewing angle. 

Eq.~\ref{eq:polcos} is a cosine function, defined in terms of the photometric curve. The minimum of polarization and photometry coincide when $\delta=\pm0.25$ (the value of $\delta=0$ would be equivalent to the polarization maximum at the photometric minimum). The derived value of $\delta\sim-0.17$ indicates that the maximum of photospheric absorption does not occur when the main component generating the polarization is aligned with the photosphere. This is the result of an asymmetric circumstellar cloud, and is discussed in more detail below. 

The derived fitting of Eq.~\ref{eq:polcos} can be used as reference for a more precise model fitting, physically based, and where the ISM components and the on-sky orientation are simultaneously determined.

\begin{table}
\centering
\caption[Polarimetry best-fitting parameters of a cosine function with double the rotational frequency for \sori{}, HR\,5907 and HR\,7355]{Polarimetry best-fitting parameters of a cosine function with double the rotational frequency (Eq.~\ref{eq:polcos}) for \sori{}, HR\,5907 and HR\,7355. The considered ISM components are in Table~\ref{tab:ISM}.}
 \begin{threeparttable}
\begin{tabular}[]{ccccccc}
\toprule
 & $\mean{P_{\rm raw}}$ & $P_0$ & $A$ & $\delta$ & $\Delta$$\theta$$\rightarrow$$\mean{U}$=0 & $\chi^2_{\rm red}$ \\ 
 & \% & \% & \% & phase & deg. & - \\    
\midrule
\sori{}\tnote{a} & - & 0.0471(9) & 0.021(1) & -0.17(1) & 150.0(-) & 0.67 \\ 
\sori{} & 0.328(3) & 0.0484(11) & 0.0205(17) & -0.167(7) & 148.1(1) & 0.97 \\ 
HR\,5907 & 0.619(5) & 0.0410(60) & 0.0184(87) & -0.180(37) & 77.4(2) & 2.80 \\ 
HR\,7355 & 0.029(8) & 0.1457(31) & 0.0199(42) & 0.005(16) & 86.4(2) & 0.67 \\ 
\bottomrule
\end{tabular}
    \begin{tablenotes}
        \item[a] From \citet{car13a}.
    \end{tablenotes}
 \end{threeparttable}
\label{tab:pol1}
\end{table}

\begin{table}
\centering
\caption[]{Interstellar polarimetric components at $V$-band of \sori{}, HR\,5907 and HR\,7355, obtained by the methods indicated.}
 \begin{threeparttable}
\begin{tabular}[]{cccccc}
\toprule
Star & Method & ${P_{\rm IS}}$ (\%) & $Q_{\rm IS}$ (\%) & $U_{\rm IS}$ (\%) & $\theta_{\rm IS}$ (deg) \\ 
\midrule
\sori{}\tnote{a} & Field stars & 0.350(15) & -0.348(15) & 0.040(15) & 86.7(0.4) \\ 
\sori{}\tnote{a} & Grid minimization & 0.351(10) & -0.350(10) & 0.025(10) & 88.0(0.3) \\ 
\sori{} & MCMC minimization & 0.314(11) & -0.314(12) & 0.015(9) & 88.6(0.3) \\ \hline
HR\,5907 & Field stars & 0.602(45) & -0.275(36) & 0.535(61) & 58.6(3.0) \\ 
HR\,5907 & MCMC minimization & 0.606(27) & -0.318(27) & 0.516(23) & 60.8(0.8) \\ \hline
HR\,7355 & Field stars & 0.142(37) & 0.141(37) & -0.017(37) & 176.6(1.1) \\ 
HR\,7355 & MCMC minimization & 0.011(24) & 0.006(16) & 0.009(24) & 28.2(0.7) \\ 
\bottomrule
\end{tabular}
    \begin{tablenotes}
        \item[a] From \citet{car13a}.
    \end{tablenotes}
 \end{threeparttable}
\label{tab:ISM}
\end{table}

\subsubsection*{The radiative transfer solution of the RRM model}
In an attempt to reproduce the observed polarization modulation of \sori{}, we fed the predicted density distribution of the RRM model of \citet{tow05a} to \textsc{hdust}. All stellar and geometrical parameters were kept fixed to the values determined by \citet{tow05a}, and the only free parameter in the model was the density scale of the magnetosphere.

\begin{figure}
    \centering
    \includegraphics[width=0.6\linewidth]{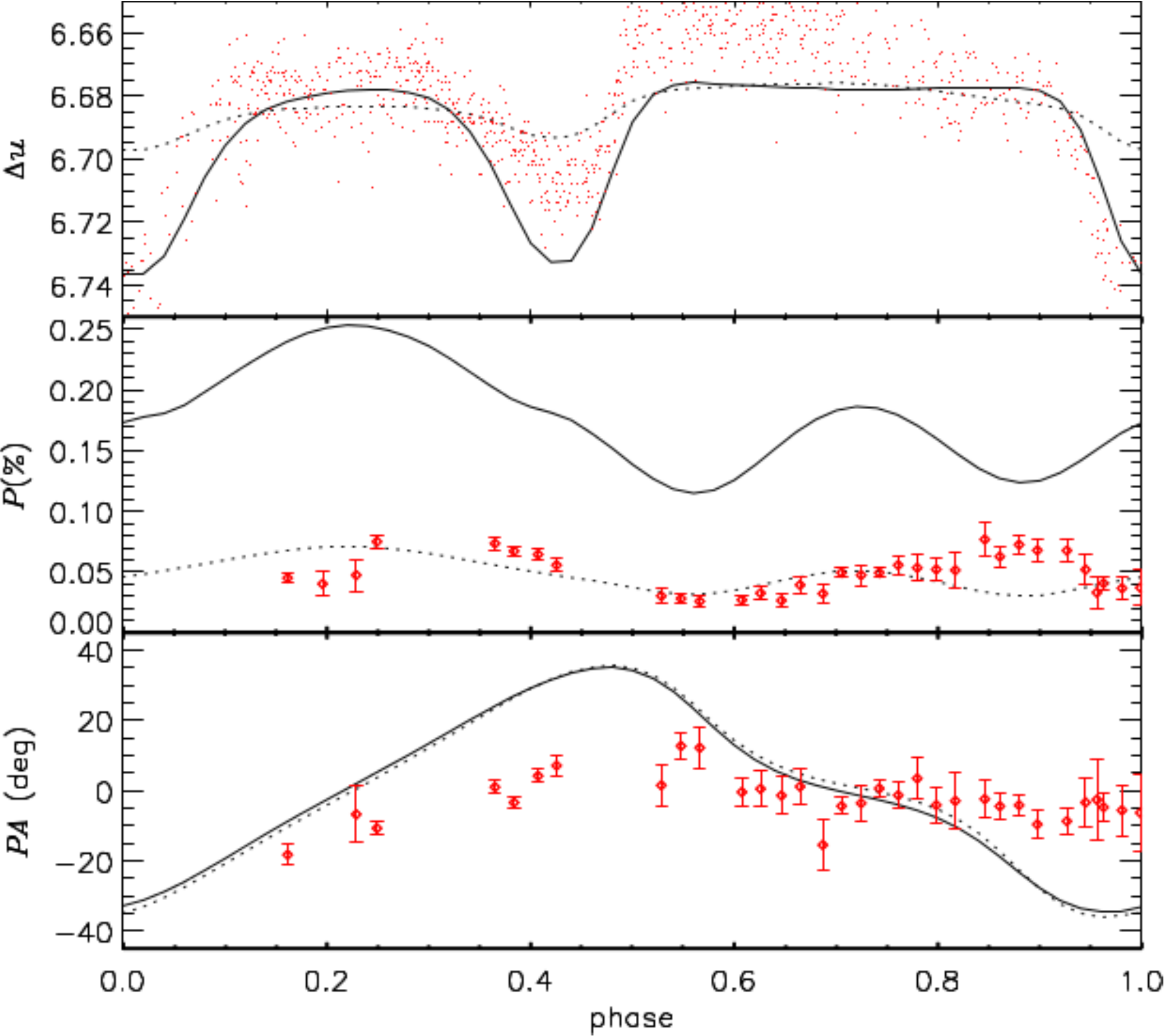}
    \caption[Modeling of the intrinsic polarization of \sori{} using the RRM model]{Modeling of the intrinsic polarization of \sori{} using the RRM model (observations are in red). The only free parameter is the maximum number density in the magnetosphere, which was set to $10^{12}$~cm$^{-3}$ (solid lines) to reproduce the depth of the eclipses and $2.5\times10^{11}$ cm$^{-3}$ (dotted lines) to reproduce the amplitude of the linear polarization \citep{car13a}.}
    \label{fig:mags-soriRT}
\end{figure}

The results of this modeling are shown in Fig.~\ref{fig:mags-soriRT}. We could not find a model that simultaneously matched both the photometry and the polarimetry. A higher density model (solid line) that matches the depth of the photometric eclipses predicts a polarization amplitude that is three times larger than what is observed. Conversely, a lower density model that reproduces the amplitude of the polarization fails to reproduce the photometric amplitude. 

Importantly, the detailed shape of the polarization curve was not well matched, in particular the position angle variation. Since the polarization position angle is sensitive to the geometry of the scattering material, this discrepancy indicated that the primary difficulty with the basic RMM model is the shape of its density distribution predicted by the model. Based on these results, we decided to build an ad-hoc, albeit physically motivated, model for the magnetosphere.

\FloatBarrier
\subsection{Single scattering model \label{sec:blobmod}}
The RRM predicts the existence of diametrically opposed plasma clouds confined in the rotational equator (and a diffuse plasma disk close to the magnetic equator). Based on this idea, and after the suggestion from S. Owocki (priv. comm.), we developed a simple ad-hoc magnetosphere model that consists of two spherical blobs situated in the equatorial plane at the intersection between this plane and the magnetic equator, where most of the circumstellar material is located. This configuration mimics a \textit{dumbbell} a shape. The plane where the ``dumbbdell'' rotates is tilted by an angle $\beta$ from the rotational equator, equivalent to its magnetic equatorial plane.

The $Q$ and $U$ Stokes parameters can be described as a function of orbital phase for a given inclination angle $i$ following a simple single-scattering approach (valid only in the optically thin limit, or $\tau\ll 1$). The scattered flux is first determined in the frame of the star using the formalism described in \citet[Eq.~6]{bjo94a} and then rotated to the frame of the observer \citep[Eq.~20]{bjo94a}. 

For the geometrical configuration of the magnetosphere, we adopted a few fixed parameters whose values were based on the model of \citet{tow05a}. We set the blobs as spheres, with radius $R_{\rm blob}$. We emphasize that the value of $R_{\rm blob}$ is somewhat arbitrary, since in the limit of $R_{\rm blob}\ll R_*$ and in the single-scattering approximation, a large, tenuous blob is roughly equivalent to a dense, small blob. In other words, what really controls the polarization level of each component is their total scattering mass. For the inclination angle we adopt two possible values: i = 70$^\circ$, corresponding to a projected counterclockwise rotation of the blobs on the sky and $180^\circ-70^\circ=110^\circ$, corresponding to a clockwise rotation. 

\begin{figure}
    \centering
    \includegraphics[width=0.42\linewidth]{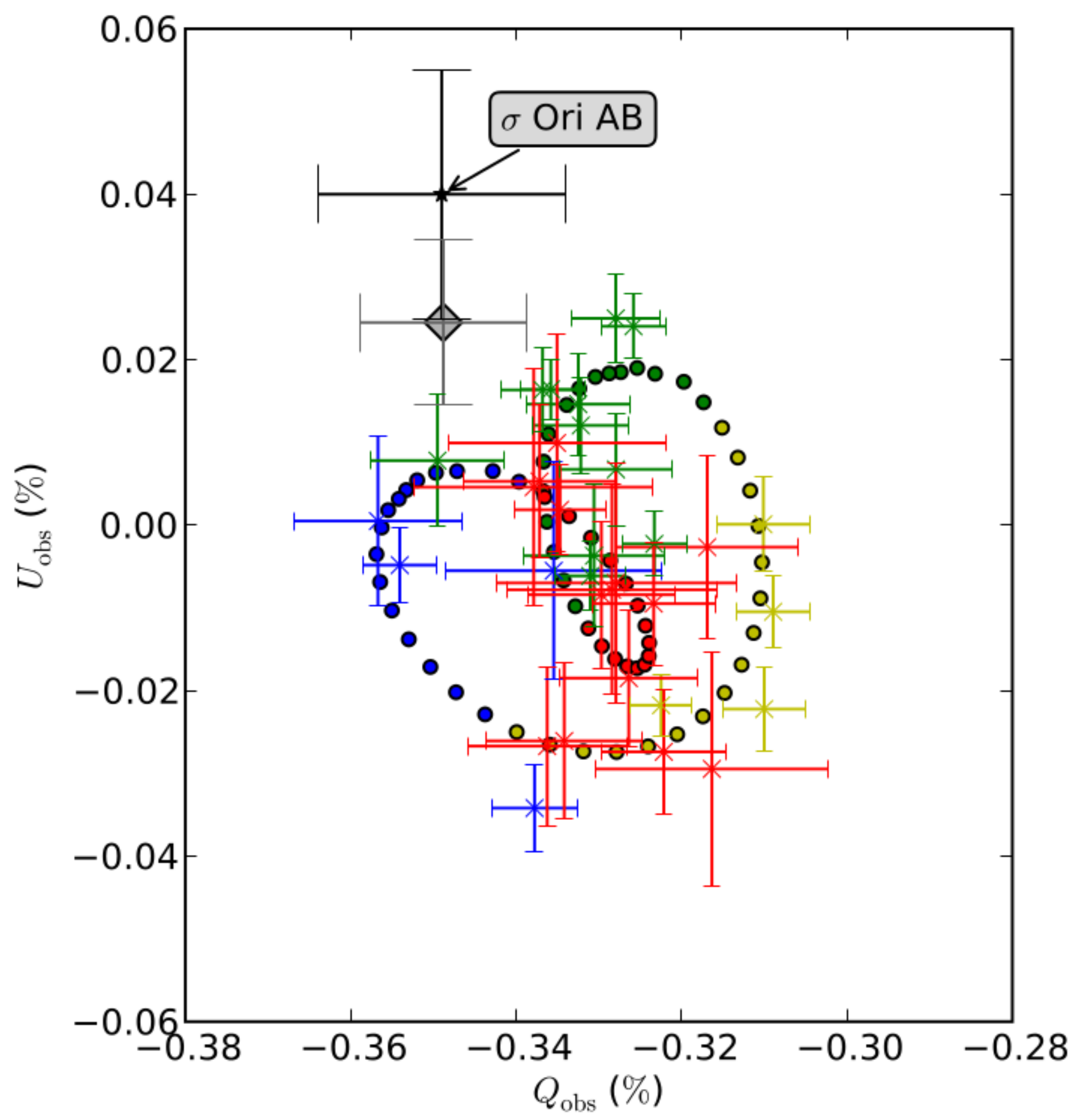}
    \includegraphics[width=0.56\linewidth]{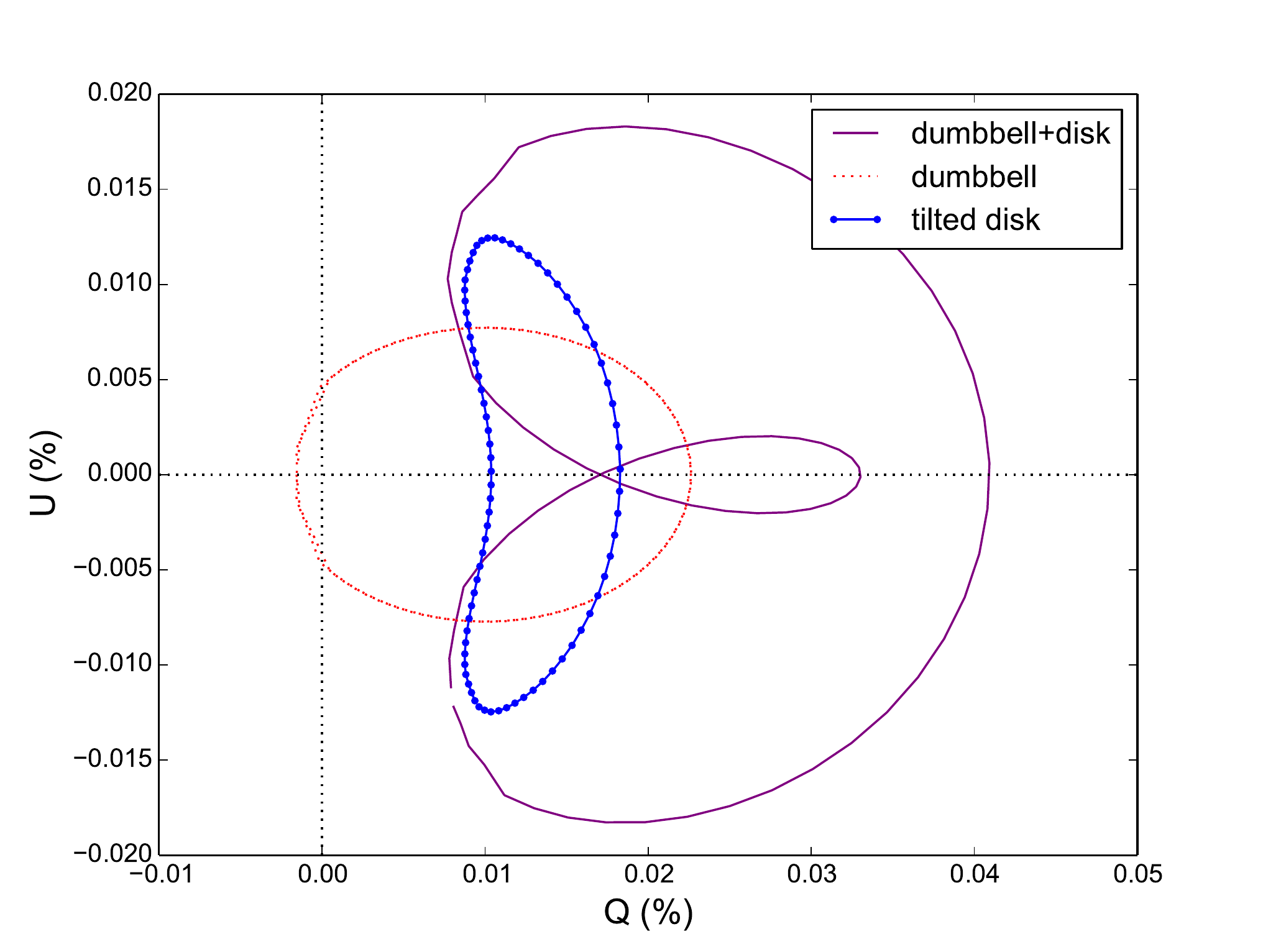}
    \caption[$QU$ plot of the observed data of \sori{} and the intrinsic polarization of the ``dumbbell+disk'' model compared to a pure dumbbell and pure tilted disk]{\textit{Left:} $QU$ plot of the observed data of \sori{} (points with error bars). The orbital phase correspondence of the points is as follows: blue, $\delta$-($\delta$+0.25), orange, ($\delta$ +0.25)-($\delta$+0.50), green, ($\delta$+0.50)-($\delta$+0.75), and red, ($\delta$+0.75)-$\delta$. The filled circles show the best-fitting model, including the IS component as gray diamond. The IS polarization of $\sigma$\,Ori\,AB is indicated \citep{car13a}. \textit{Right:} The intrinsic polarization of the ``dumbbell+disk'' model with the typical double-loop heart shape (purple line). The curves of the isolated components are also shown: the dumbbell component (red dashed-line) and the tilted disk (blue line with dots). Due to their symmetry, the property $\mean{U}=0$ is always valid (considering a integer number of cycles). Also, the disk component makes that always $U>0$.}
    \label{fig:QUmods}
\end{figure}

We started the scattering modeling of \sori{} neglecting the disk component of the magnetosphere, as predicted by the RRM model (Fig.~\ref{fig:magshape}). However, our initial results showed that only two blobs could not explain the observed polarization incursions in the $QU$ plan. As shown in Fig.~\ref{fig:QUmods}, right panel, the result of two blobs rotation over the $QU$ plane is an double coincident ellipse. If only a tilted disk is present, a similar result is found: a double ``banana''-shaped track in the $QU$ plane. Only the combination of these two structures can reproduce the measured $QU$ track, which led to inclusion of the disk component in the model. 

\begin{figure}
    \centering
    \includegraphics[width=0.6\linewidth]{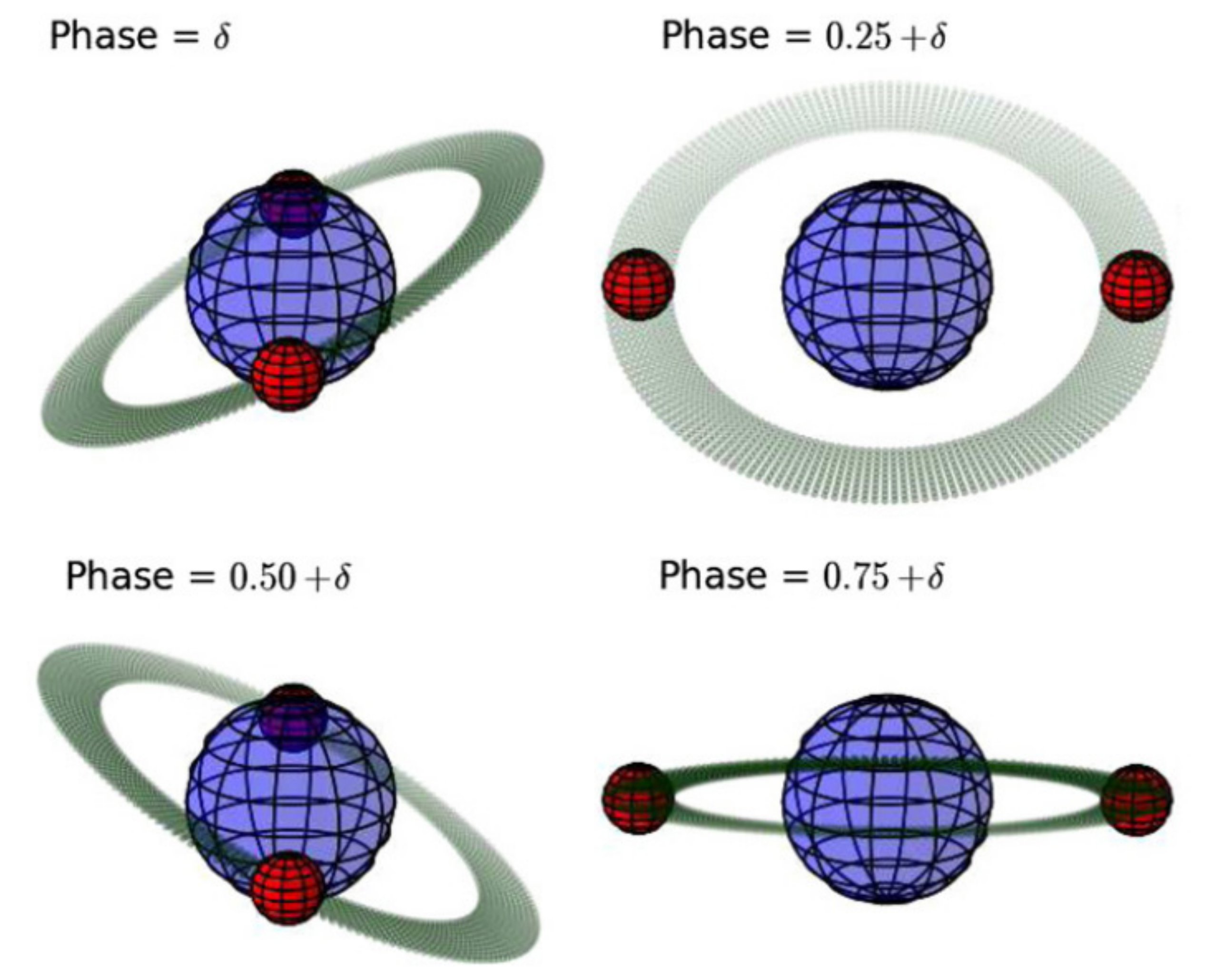}
    \caption[Geometric conception of the ``dumbbell + disk'' model to scale ($i=70^\circ$)]{Geometric conception of the ``dumbbell + disk'' model to scale ($i=70^\circ$). The corresponding phase is indicated, where $\delta$ is the phase shift between photometric and polarimetric minima \citep{car13a}.}
    \label{fig:shape}
\end{figure}

\begin{figure}
    \centering
    \includegraphics[width=0.5\linewidth]{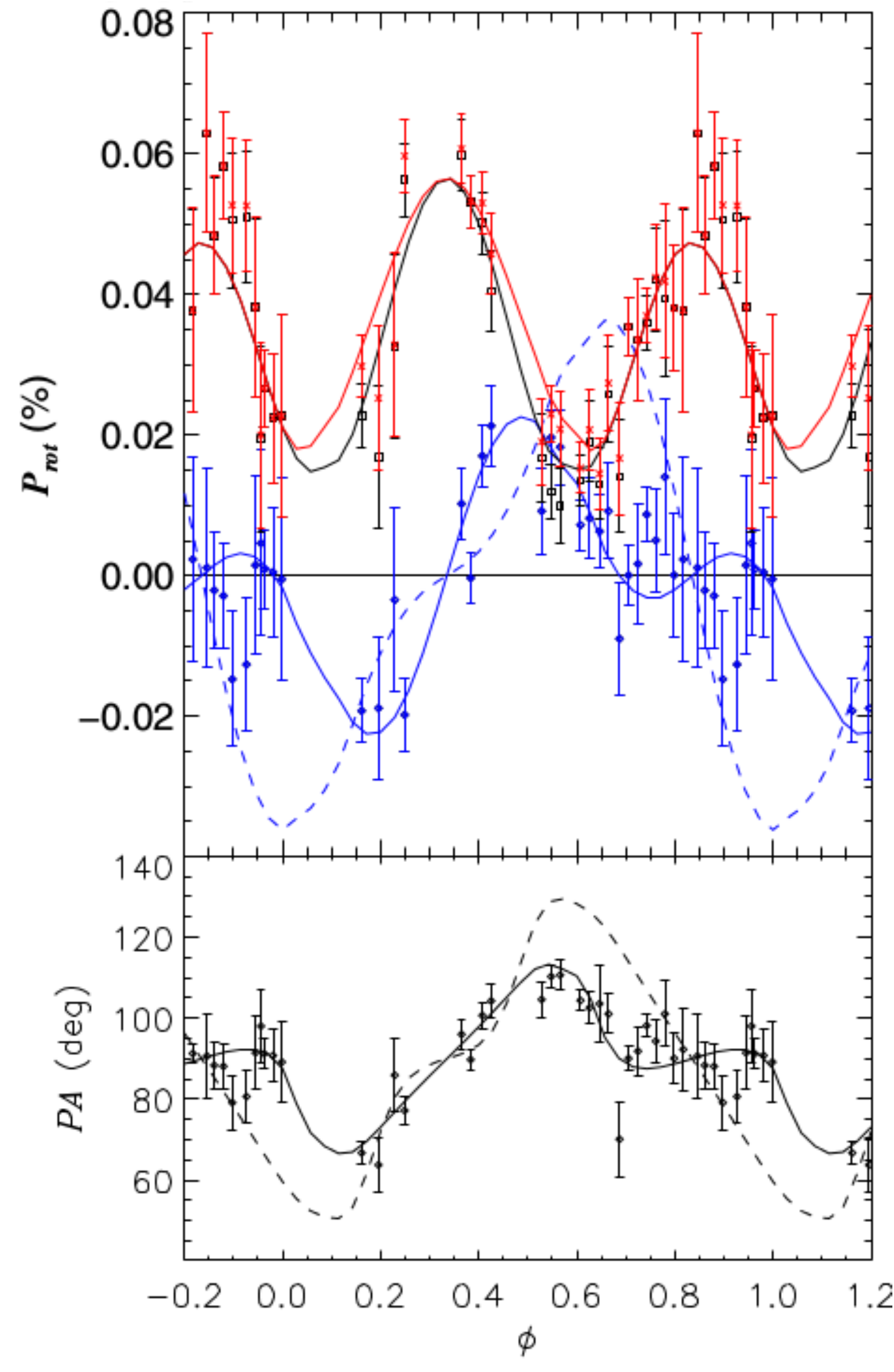}
    \caption[Polarization of \sori{} compared to the single-scattering model]{Polarization of \sori{} compared to the single-scattering model. The solid line is for our best-fitting model with $i = 70^\circ$ and the dashed line is for $i = 110^\circ$ \citep{car13a}.}
    \label{fig:sorifit}
\end{figure}

As discussed for the blob size, the disk geometrical thickness is also somewhat arbitrary: if the single-scattering approximation holds, a thinner, denser disk is equivalent to a thicker, less dense disk. Below, we summarize the criteria used to define the geometrical configuration of the ``dumbbell+disk'' model:
\begin{itemize}
    \item The stellar radius adopted is the equatorial one ($R_*=R_{\rm eq}$). The current version of the code does {not} take into account the star oblateness neither the gravity darkening effect. These effects are important for near-critical rotating stars, and are planned for the near future.
    \item The center of the blob is at $R_{\rm K}$, the co-rotation (or ``Keplerian'') radius (Eq.~\ref{eq:corot}).
    For slow and moderate rotators (e.g., \sori{}), the blob position is $\gtrsim2.5R_{\rm eq}$. For fast rotators, the blob may be very close to the stellar photosphere due to small value of the co-rotating radius.
    \item The blob diameter is fixed as $2/3R_*$.  
    \item The polarimetric period is the same from photometry (and spectroscopy), but an arbitrary phase shift can be applied.
    \item The disk component (rather, a ring) has the same radial position and size of the blob, with fixed scale height of $H=0.01R_*$.
\end{itemize}
This configuration is illustrated in Fig.~\ref{fig:shape} for different orbital phases and at the viewing angle $i=70^\circ$. The resulted modeling with dumbbell+disk components is shown in Fig.~\ref{fig:sorifit}. The here described scattering code is available at the \textsc{pyhdust} library. 

The main results of this study are: (i) it confirms the RRM prediction of two main circumstellar structures for the magnetosphere, namely a thin disk and two over-density regions at the intersection of the rotation and magnetic planes; (ii) the RRM model prediction for the total scattering mass of each of these components do not agree with the polarization data. In particular, the derived mass for the ``disk'' component is much larger than the expected. This challenges RRM description for this star, as pointed out by \citep{tow13a}.

\subsection{New minimization process of \sori{} polarization light-curve}
\begin{figure}
    \centering
    \includegraphics[width=\linewidth]{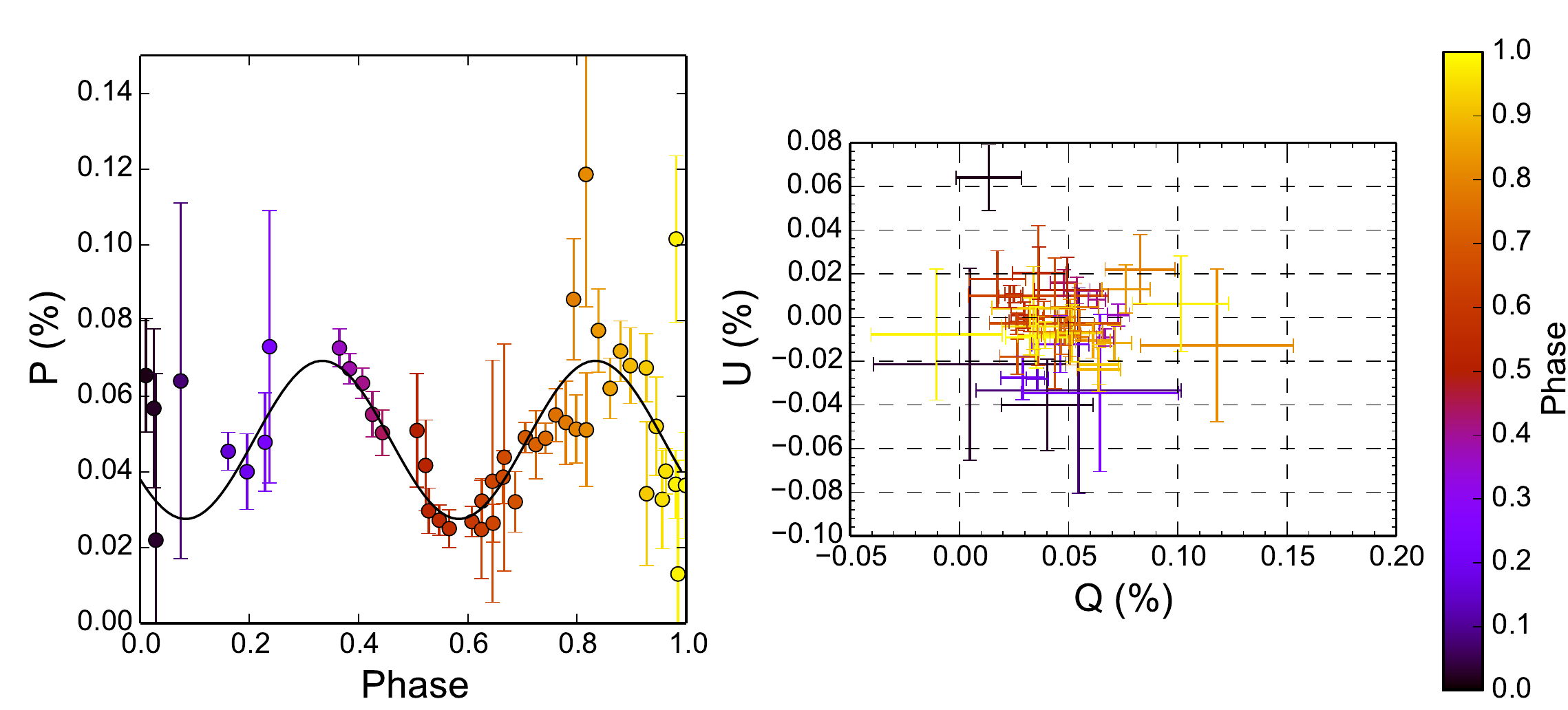}
    \includegraphics[width=\linewidth]{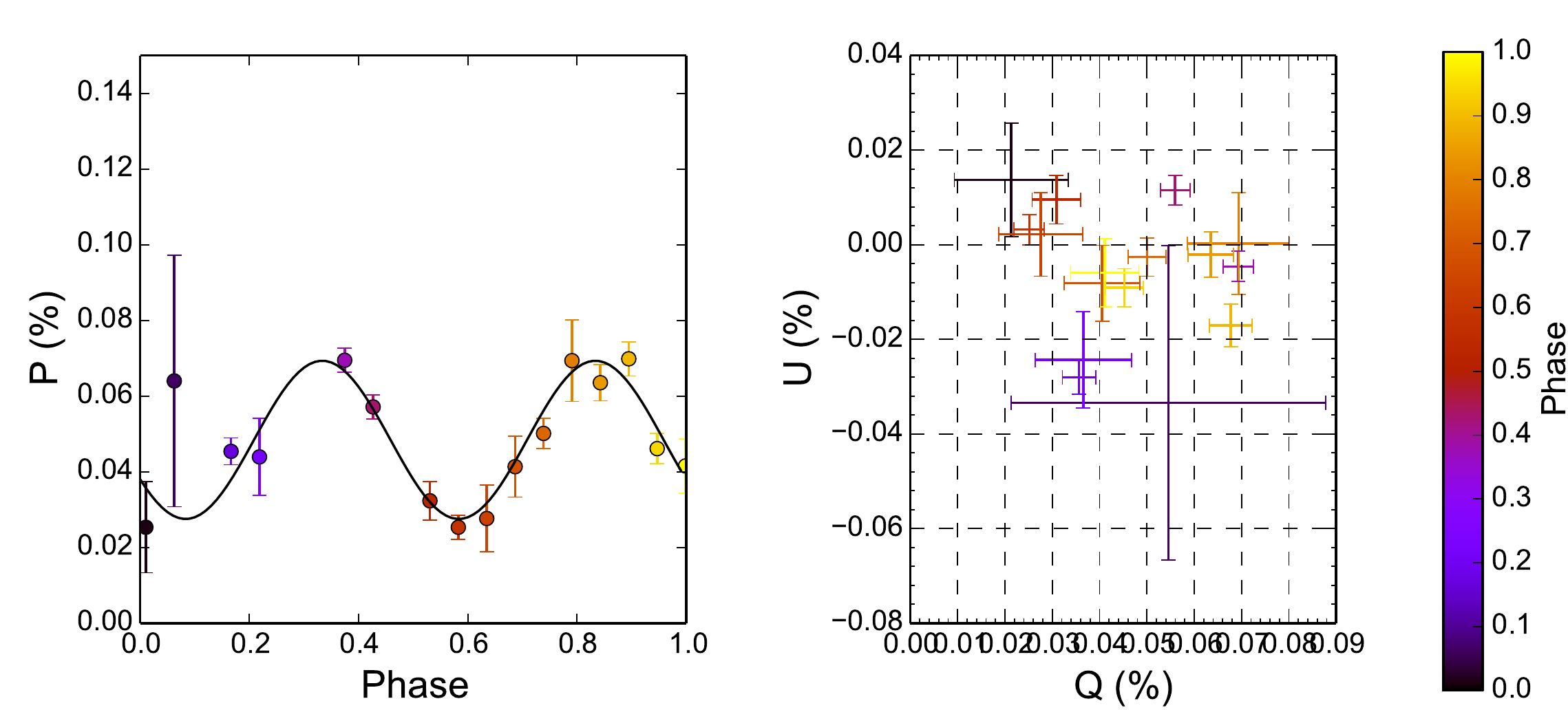}
    \caption[$V$-band intrinsic linear polarization measurements of \sori{} and $QU$ diagram on the celestial frame]{$V$-band linear polarization measurements of \sori{}. \textit{Left:} subtracted interstellar components (Table~\ref{tab:pol1}). The line corresponds to the best fit of Eq.~\ref{eq:polcos}. \textit{Right:} $QU$ diagram of the observed polarization on the celestial frame. \textit{Top}, all observational points. \textit{Bottom}, the data points binned in phase. Their values in the celestial frame are in Table~\ref{tab:apsori}.}
    \label{fig:soricos}
\end{figure}

The polarization survey of sigma Ori E continued after the publication of the first results. Fig.~\ref{fig:soricos} shows the fit for all intrinsic data (top panel), as well the observed $QU$ points. We also present the data binned in phase (bottom panel). The procedure adopted is to divide the $[0, 1]$ phase interval in 20 bins of $\Delta \text{phase}=0.05$. The polarization value for each bin in then the weighed average of all points belonging to that bin. 

The previous and new best-fitting values of Eq.~\ref{eq:polcos} for \sori{} are presented in Table~\ref{tab:pol1}. The addition of a few more observational points did not changes the derived values of polarization bias level, amplitude of modulation and phase shift in respect to the photometric curve.

The fit of the ``dumbell+disk'' model shown in \citet{car13a} was performed manually, by changing the parameters around a $\chi^2$ minimum that was found by computing a model grid. Here we present new results for \sori{}, applying the Monte Carlo Markov chain (MCMC) method for our most recent data set. The minimization simultaneously determines all the selected physical parameters involved in the modeling. This method allows to evaluate the correlation between the parameters and the uncertainty in their determination, in a self-consistent analysis.

To standardize the \sori{} analysis with the other smaller stars, which have lower signal-to-noise (S/R) ratio, the first step is to group the polarization data into regular phase intervals, i.e., the $Q$ and $U$ Stokes components. In this way, the data points have a higher signal and also a uniform phase sampling. Then, for each \textit{model} (or set of parameters randomly generated in the MCMC simulation), the following steps are performed:
\begin{enumerate}
    \item The intrinsic polarization model is generated in the stellar reference system (i.e., for which North is aligned to the stellar spin axis) for all rotational phases.
    \item The model is rotated to the equatorial system. Here, it is important to distinguish between counter- and clockwise rotation. This is done by allowing $i$ to vary.
    \item Interstellar polarization components $Q_{\rm ISM}$ and $U_{\rm ISM}$ are randomly sampled within a pre-determined interval (by default, determined by field stars analysis).
    \item The $\chi^2$ value is computed to determine the likelihood associated with model parameters, which is higher the lower the $\chi^2$ value, defined as 
    \begin{equation}
        \chi^2=\sum_{i}^N\left[  \frac{Q_{\rm obs}(\phi_i)-Q_{\rm mod}(\phi_i)}{\sigma_{Q(\phi_i)}} \right]^2 + \sum_{i}^N\left[  \frac{U_{\rm obs}(\phi_i)-U_{\rm mod}(\phi_i)}{\sigma_{U(\phi_i)}} \right]^2 \,,
    \label{eq:chi2pol}
    \end{equation}
    where $N$ is number of phase bins and $\phi_i$ their values. 
    \item Based on the likelihood of the current parameters, a new set is generated until reaching the number of iterations defined by the user. 
\end{enumerate}

\begin{figure}
    \centering
    \includegraphics[width=\linewidth]{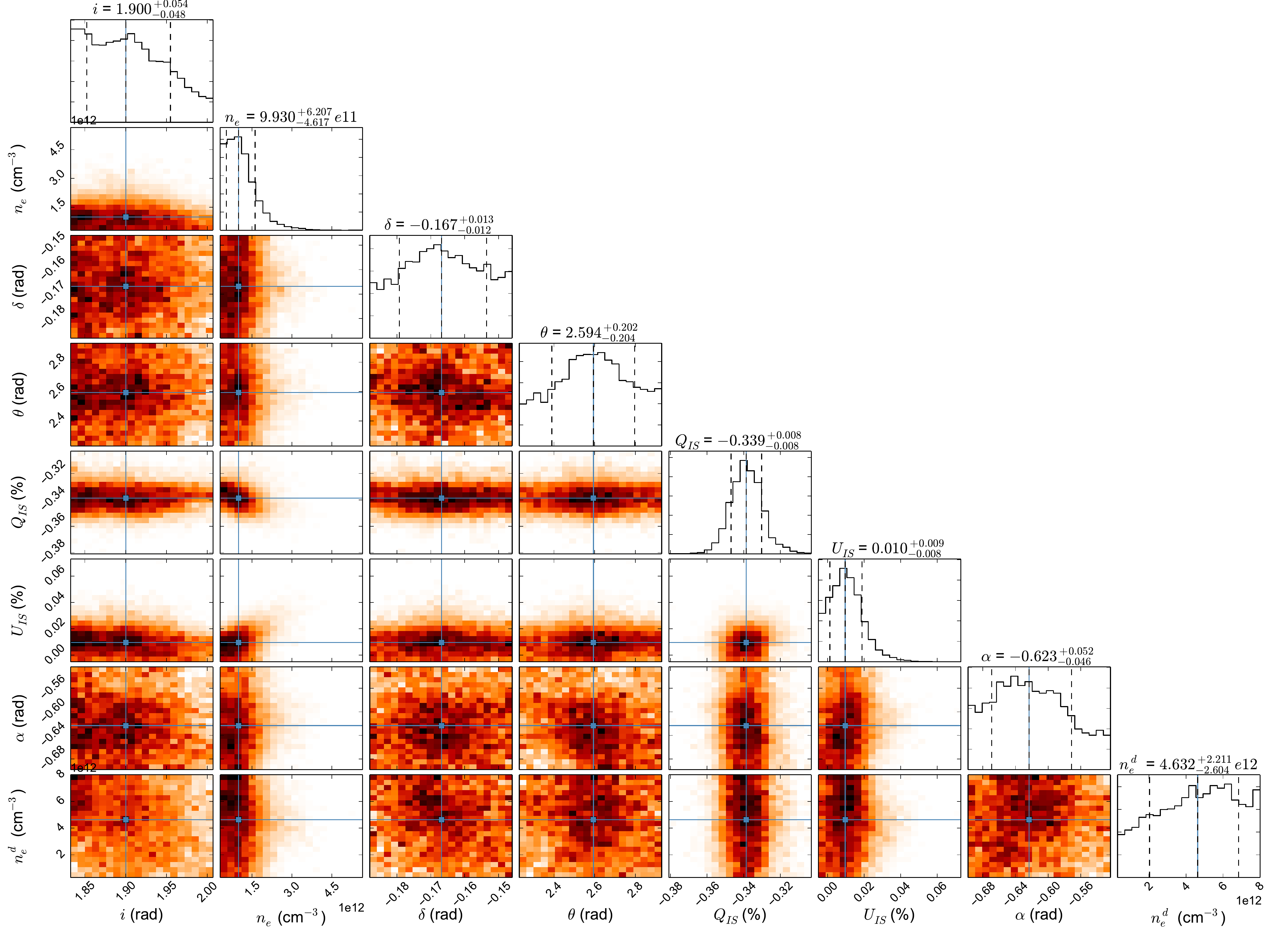}
    \caption[The MCMC probability density distributions for the ``dumbbell+disk'' polarization modeling to \sori{} at $i\sim70^\circ$, and the corresponding correlation maps]{The MCMC probability density distributions for the ``dumbbell+disk'' polarization modeling to \sori{} at $i\sim70^\circ$, and the corresponding correlation maps. Here the inclination angle $i$ and the obliquity $\alpha=90^\circ-\beta$ are left free, but the results show that they are degenerate within the selected intervals. The heat-like color map indicate the highest probabilities (black) to the lowest (white). \textit{From left to right (or top-bottom):} inclination angle (rad), blob electronic density (cm$^{-3}$), phase shift, on-sky orientation angle ($\theta$), the interstellar $Q$ component (\%),  interstellar $U$ component (\%), the complementary angle between the rotation and magnetic axes (rad) and disk electronic density (cm$^{-3}$). The dashed vertical lines in the histograms indicate the percentile of 16\% ($-1\sigma$), the median and the percentile of 84\% ($1\sigma$) of each distribution.}
    \label{fig:sorifullemcee}
\end{figure}

\begin{figure}
    \centering
    \includegraphics[width=\linewidth]{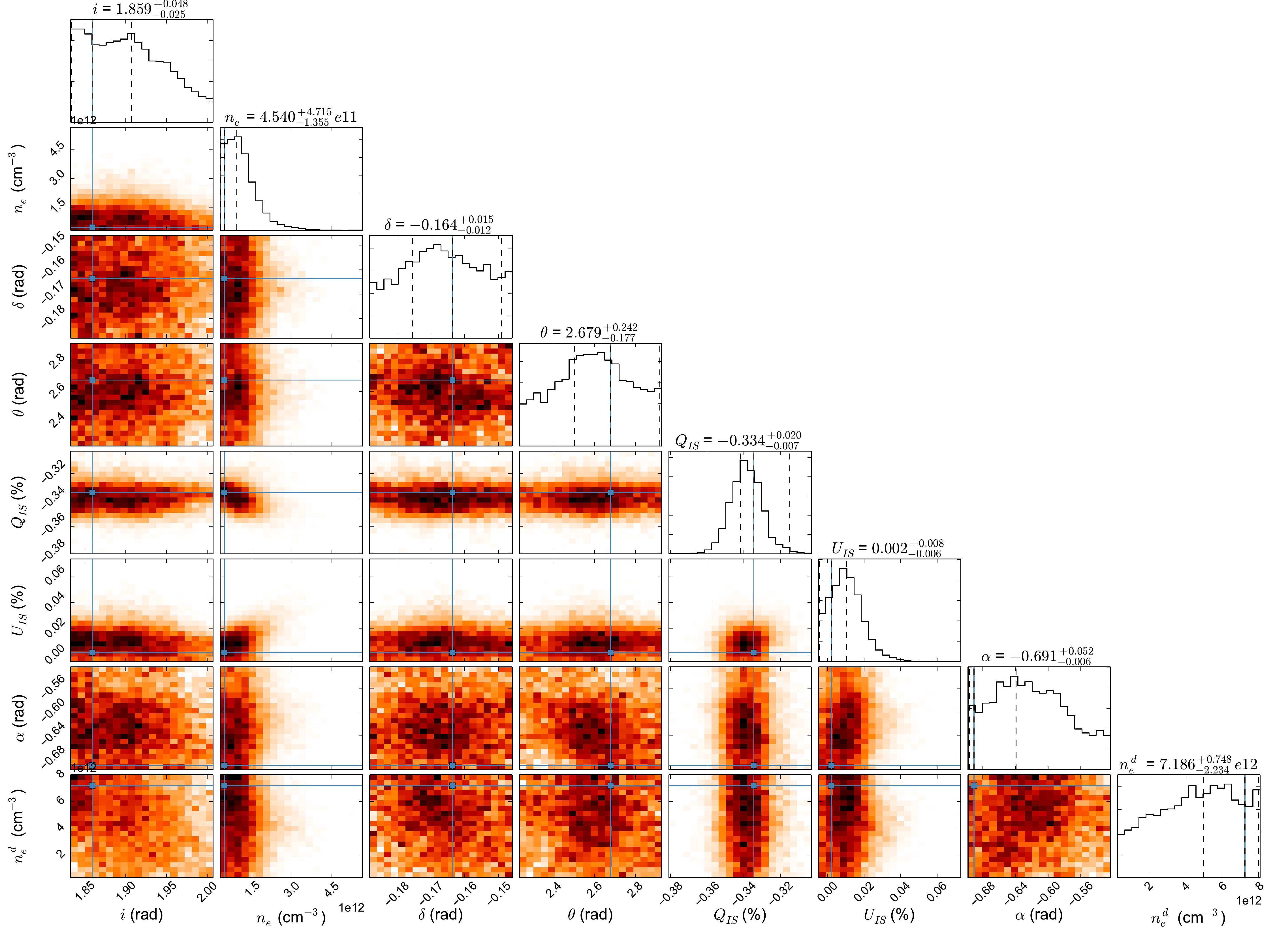}
    \caption{Same as previous figure, but for $i\sim110^\circ$.}
    \label{fig:sorifullemcee2}
\end{figure}

\begin{figure}
    \centering
    \includegraphics[width=\linewidth]{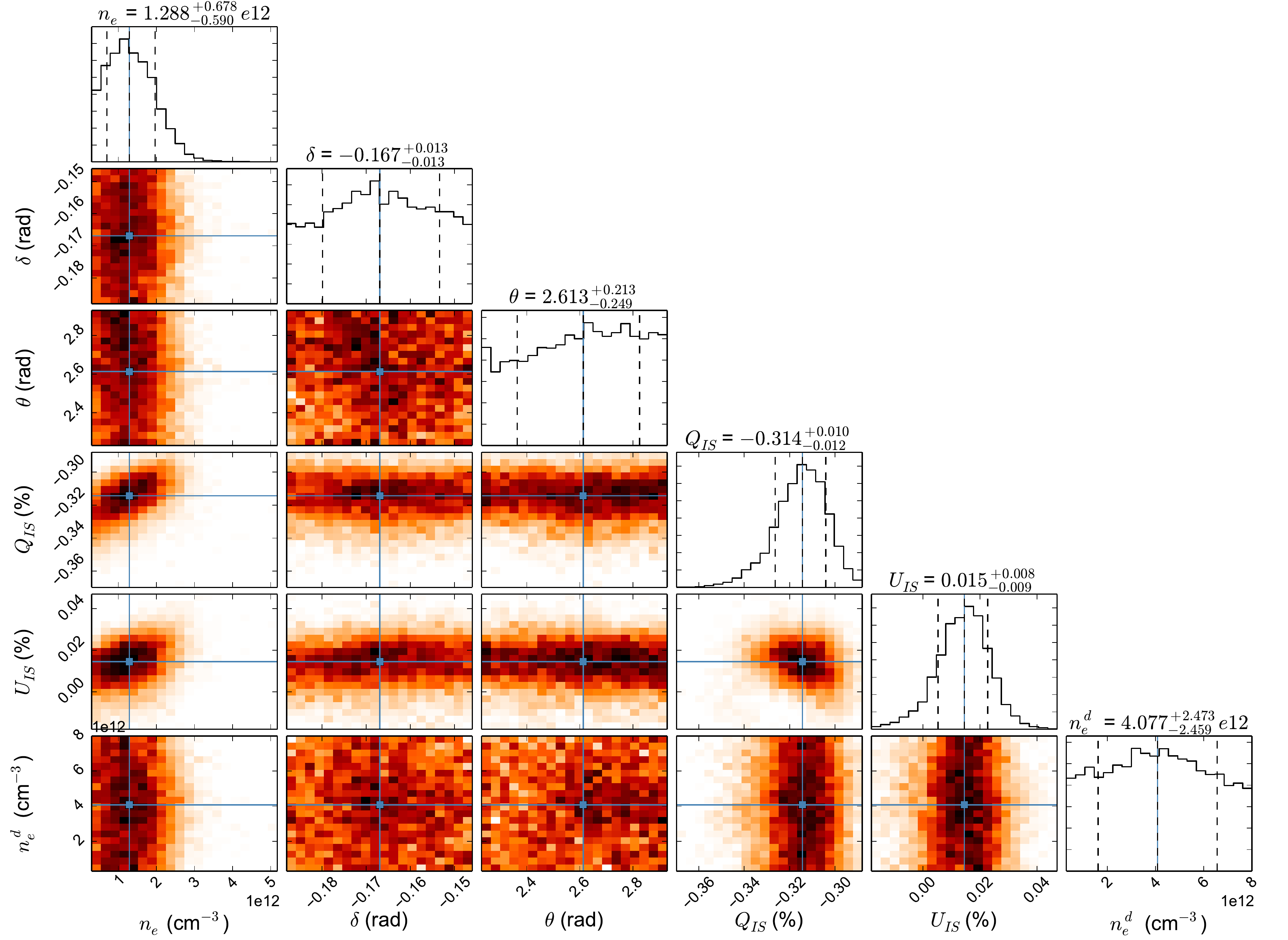}
    \caption{Same as previous figure, but for $i=70^\circ$ and $\alpha=90^\circ-\beta$ ($\beta=55^\circ$), kept fixed.}
    \label{fig:sori-fin}
\end{figure}

Fig.~\ref{fig:sorifullemcee} and \ref{fig:sorifullemcee2} show the MCMC PDFs and their correlations, minimizing simultaneously seven model parameters (the five parameters fitted in Table~\ref{tab:magsspars}, in addition to $i$ and $\beta$). They confirm the rotational direction of the star as counter-clockwise, since $i\sim70^\circ$ (median\,$\chi^2_{\rm red}=4.74$, in contrast to the median\,$\chi^2_{\rm red}=9.65$ at $i\sim110^\circ$). They also show that the inclination angle $i$ and the magnetic obliquity $\beta$ can not be constrained within the selected intervals. In our code, the obliquity is set by the complementary angle $\alpha=90^\circ-\beta$ shown in the figures.
We fixed parameters $i$ and $\beta$ to the values adopted by \citet{tow13a} to determine the other main polarimetric physical parameters (Fig.~\ref{fig:sori-fin}). The previous and newly derived parameters are listed in Table~\ref{tab:magsspars}.
\begin{table}
\begin{threeparttable}
\centering
\caption[Best-fitting parameters of single scattering ``dumbbell+disk'' model for \sori{}, HR\,5907 and HR\,7355 magnetospheres with MCMC minimization]{Best-fitting parameters of single scattering ``dumbbell+disk'' model for \sori{}, HR\,5907 and HR\,7355 magnetospheres with MCMC minimization. The fitted and fixed parameters are indicated. The parameters marked as ``ref.'' would not change the polarization curve in the current version of the modeling.}
\begin{tabular}{cccccc}
\toprule
Star & \sori{}\tnote{a} & \sori{}\tnote{b} & HR\,5907\tnote{c} & HR\,7355\tnote{d} & Params.  \\ 
\midrule
Hipparcos parallax (mas) & - & -  & 7.64(0.37) & 3.66(0.38) & ref. \\ 
$B_p$ (kG; dipole) & 9.6 & 9.6 & 10 & 11 & ref. \\ 
$\Omega/\Omega_{\rm crit}$ & 0.5 & 0.454 & 0.80 & 0.89 & ref. \\ 
W & 0.29 & 0.26 & 0.53 & 0.63 & ref. \\ 
Zero phase (MJD) & 42778.329 & 42778.329 & 47913.19 & 54940.33 & fixed \\ 
Rotational period $P$ (days) & 1.1908229 & 1.1908229 & 0.508276 & 0.5214404 & fixed \\ 
$i$ (deg) & 70/$\cancel{110}$ & 70 & ${70}$/110\tnote{e} & 60\tnote{e}/${120}$ & fixed \\ 
$\beta$ (deg) & 62 & 55 & 7 & 75 & fixed \\ 
$\alpha$ (deg) & 28 & 35 & 83 & 15 & fixed \\ 
$R_{\rm eq}$ ($R_\odot$) & 4.28 & 3.77 & 3.1 & 3.69 & fixed \\ 
$R_{p}$ ($R_\odot$) & 4.11 & 3.65 & 2.72 & 3.06 & ref. \\ 
Spec. CS emission ($R_{\rm eq}$) & $\gtrsim 2$ & $\gtrsim 2$ & 1.2-4.4 & 2-4 & ref. \\ 
$R_{\rm K}$ ($R_{\rm eq}$) & 2.38 & 2.54 & 1.74 & 1.63 & ref. \\ 
Blob mean dist. ($R_{\rm eq}$) & 2.38 & 2.54 & 1.90\tnote{f} & 2.54\tnote{f} & fixed \\ 
min. blob dist. ($R_{\rm eq}$) & 2.05 & 2.21 & 1.57 & 2.21 & fixed \\ \vspace{5pt}
max. blob dist. ($R_{\rm eq}$) & 2.71 & 2.87 & 2.23 & 2.87 & fixed \\ \hline \vspace{5pt}
$n_e$ (10$^{12}$ cm$^{-3}$) & 1.0$^{+0.6}_{-0.9}$ & 1.29$^{+0.68}_{-0.59}$ & 0.72$^{+0.55}_{-0.32}$ & 1.53$^{+1.17}_{-0.86}$ & fitted \\ \vspace{5pt}
$\delta$ (phase) & -0.17$^{+0.02}_{-0.02}$ & -0.167$^{+0.013}_{-0.013}$ & 0.08$^{+0.16}_{-0.17}$ & -0.02$^{+0.18}_{-0.16}$ & fitted \\ \vspace{5pt}
$\theta$ (deg) & 150.0$^{+7.0}_{-7.0}$ & 149.7$^{+12.2}_{-14.3}$ & 78$^{+12}_{-13}$ & 88$^{+12}_{-14}$ & fitted \\ \vspace{5pt}
$n_e^d$ (10$^{12}$ cm$^{-3}$) & 2.7$^{+1.0}_{-1.0}$ & 4.1$^{+2.5}_{-2.5}$ & 3.8$^{+2.5}_{-2.4}$ & 4.4$^{+2.3}_{-2.6}$ & fitted \\ \hline 
$M_{\rm blob}$ (10$^{-12}$ $M_\odot$) & 4.00 & 3.47 & 1.08 & 3.86 & ref. \\ 
$M_{\rm disk}$ (10$^{-12}$ $M_\odot$) & 6.50 & 7.07 & 2.66 & 7.11 & ref. \\ 
$2M_{\rm blob}/M_{disk}$ & 1.23 & 0.98 & 0.81 & 1.09 & ref. \\ 
\bottomrule
\end{tabular}
\begin{tablenotes} \footnotesize
    \item[a] From \citet{car13a}. Their analysis exclude the inclination angle of $i\sim110^\circ$. 
    \item[b] Reference values from \citet{tow13a}.
    \item[c] Reference values from \citet{gru12a}.
    \item[d] Reference values from \citet{riv13c}.
    \item[e] The accuracy of the data is unable to clearly define the direction of rotation. The indicated values are the most likely.
    \item[f] Based on velocity measurements of spectroscopic lines. 
\end{tablenotes}
\label{tab:magsspars}
\end{threeparttable}
\end{table}

When comparing the results of MCMC minimization with results from \citet{car13a}, it is important to be aware that better stellar parameters (from \citealp{tow13a}) were used. The stellar equatorial radius decreased from 4.28 to 3.77~$R_\odot$, modifying the parameterized volume of the magnetosphere, and the rotational rate changed from $W=0.29$ to $0.26$, modifying the co-rotational radius. Thus, the new findings indicate: (i) 30\% denser blob, but with a smaller total mass; (ii) a density 50\% higher to the disk, and that led to a slightly higher mass. In the end, the ratio $2M_{\rm blob} /M_{\rm d} \simeq 1.0$, a value $\sim20\%$ lower than previously determined.

\FloatBarrier
\section{Polarimetry of magnetospheres and HR\,7355 \label{sec:hr7355}}
HR\,7355 is a hot magnetic star with very similar properties to \sori{} \citep{riv13c}. The main difference is that HR\,7355 possesses the highest rotational rate known among stars harboring centrifugal magnetospheres ($W=0.63$), closely accompanied by HR\,5907. This has important implications to the RRM model: the rotation near break-up of fast rotating stars poses a puzzle since magnetic fields should brake the stellar rotation.

Unlike \sori{}, HR\,7355 does not belong to a multiple stellar system and exhibits a low raw polarization level, what makes it hard to determine the interstellar polarization. The IS component of the polarization was estimated by the selected field stars listed in Table~\ref
{tab:polishr7355}. The selection criteria was (i) stars angularly close to target, (ii) stars with compatible Hipparcos distance, and (iii) stars with no known peculiarity, preferably at main sequence.

\begin{figure}
    \centering
    \includegraphics[width=\linewidth]{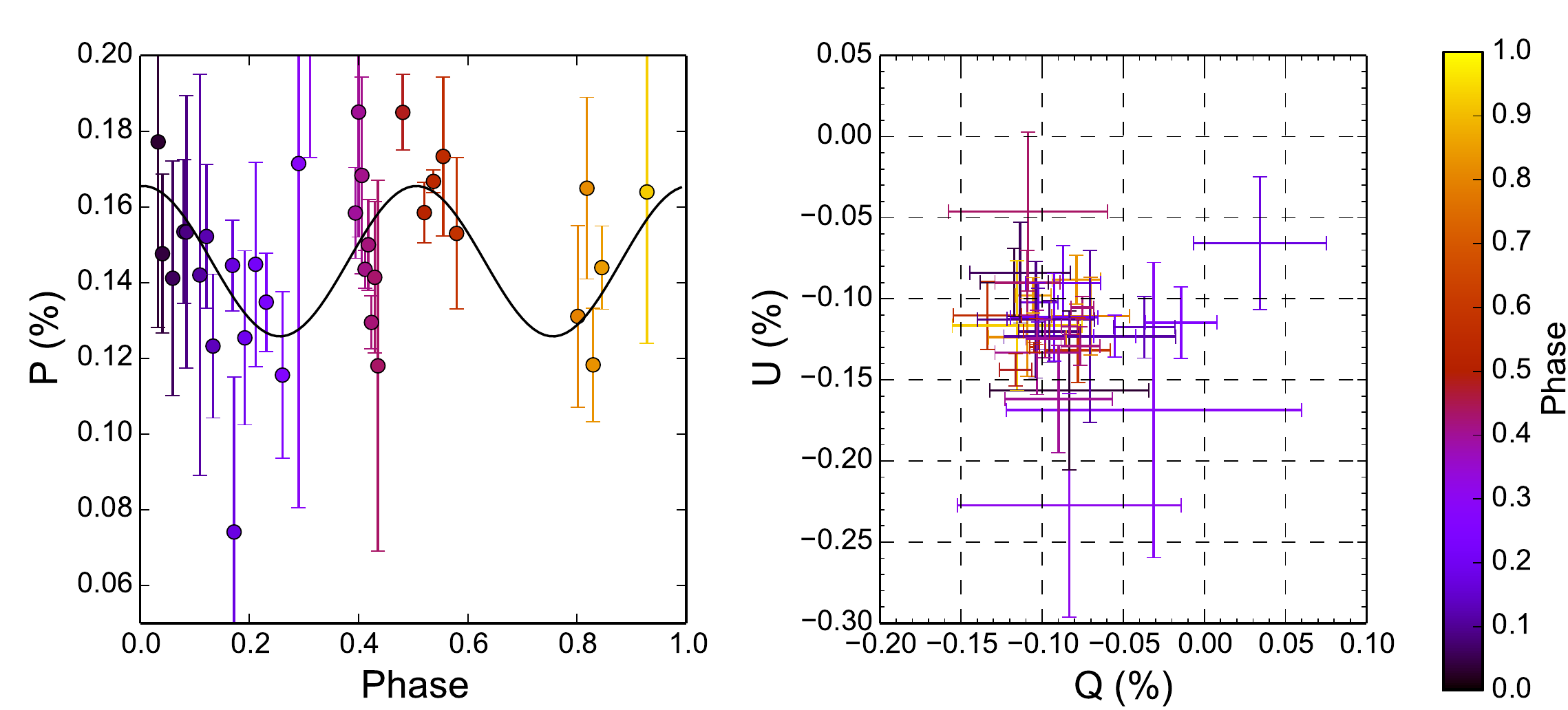}
    \includegraphics[width=\linewidth]{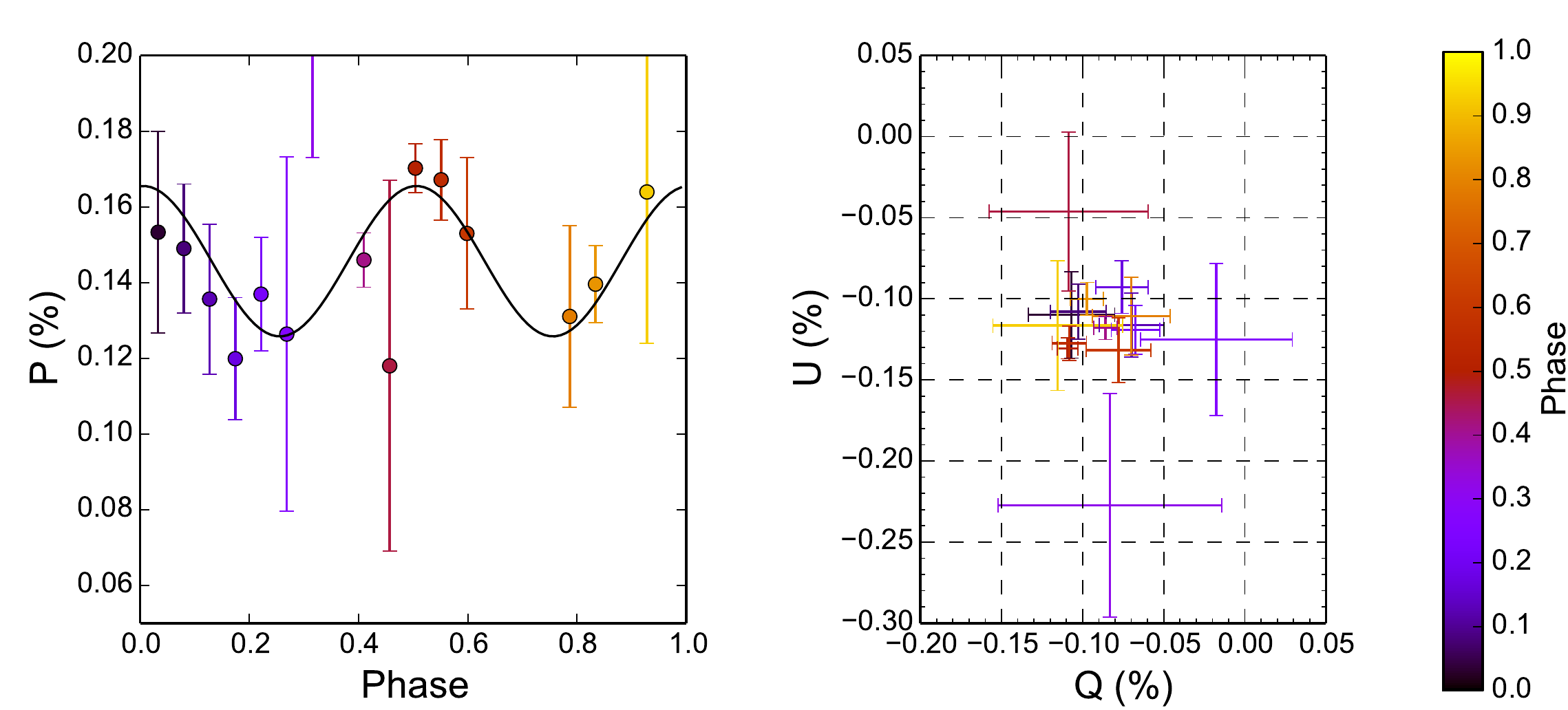}
    \caption[$V$-band intrinsic linear polarization measurements of HR\,7355 and $QU$ diagram on the celestial frame]{$V$-band linear polarization measurements of HR\,5907. \textit{Left:} subtracted interstellar components (Table~\ref{tab:pol1}). The line corresponds to the best fit of Eq.~\ref{eq:polcos}. \textit{Right:} $QU$ diagram of the observed polarization on the celestial frame. \textit{Top}, all observational points. \textit{Bottom}, the data points binned in phase. Their values in the celestial frame are in Table~\ref{tab:aphr7355}.}
    \label{fig:hr7355}
\end{figure}

Our results for the intrinsic polarization of HR\,7355 show a much higher average polarization than \sori{} (0.146\% vs. 0.047\%). However, the fit of the polarization with Eq.~\ref{eq:polcos} reveals a similar polarimetric amplitude, i.e., $\sim0.02\%$ (Fig.~\ref{fig:hr7355}). The derived position angle PA is $\sim86.4^\circ$. Before applying the ``dumbbell+disk'' for this star, we investigate the effects that the high rotation rate could cause in the polarimetric signal of this star.

\begin{table}
\centering
\caption{$V$-band polarization measurements of HR\,7355 field stars.}
\begin{tabular}{cccccccccccc}
\toprule
Star & Date & $Q$ (\%) & $U$ (\%) & $P$ (\%) & $\sigma$ (\%) & $\theta$ (deg) & $\sigma_\theta$ (deg) \\
\midrule
HIP95027 & 14mai20 & 0.1742 & -0.2521 & 0.3064 & 0.0082 & 152.3 & 0.8 \\ \hline
\multirow{ 3}{*}{HIP95604} & 14jul31 & 0.0950 & -0.0902 & 0.1310 & 0.0132 & 158.3 & 2.9 \\ 
 & 14mai20 & 0.0517 & -0.0152 & 0.0539 & 0.0073 & 171.8 & 3.9 \\ 
 & 11set29 & 0.0639 & -0.0162 & 0.0659 & 0.0112 & 172.9 & 4.9 \\ \hline
\multirow{ 2}{*}{HIP95386} & 11jun13 & 0.2612 & 0.0783 & 0.2727 & 0.0147 & 8.3 & 1.5 \\ 
 & 11jun13 & 0.2337 & 0.0679 & 0.2434 & 0.0118 & 8.1 & 1.4 \\ \hline
\multirow{ 2}{*}{HIP95412} & 11jun15 & 0.2429 & -0.0321 & 0.2450 & 0.0111 & 176.2 & 1.3 \\ 
 & 11jun16 & 0.2350 & -0.0712 & 0.2456 & 0.0189 & 171.6 & 2.2 \\ \hline
HIP95782 & 14set24 & -0.0867 & 0.1814 & 0.2010 & 0.0240 & 57.8 & 3.4 \\ 
\bottomrule
\end{tabular}
\label{tab:polishr7355}
\end{table}

\subsection{Effects of high rotation in the polarization}
We know that high rotational rates profoundly changes the characteristics of the stellar surface. This, in turn, must be reflected in the circumstellar disk polarization signal once the light flux and its incidence angle on the disk are modified. \citet{bjo94a} studied this problem in the context of polarization of Be stars disks.

In disk-like circumstellar environments, such as the magnetospheres, the scattered radiation is dominated by light from the stellar equator. On the one hand, these equatorial regions have a lower effective temperature - and lower flux - than the higher latitudes of the star due to gravity darkening. With less flux been polarized by the disk and more unpolarized stellar light reaching the observer, the net result should be a smaller observed polarization level. This depolarization would be larger the faster the stellar rotation rate. On the other hand, the bigger equatorial radius yields a bigger stellar surface area and a disk material reconfiguration due to a different potential. The consequence of these changes in polarization are difficult to anticipate.

\begin{figure}
    \centering
    \includegraphics[width=.75\linewidth]{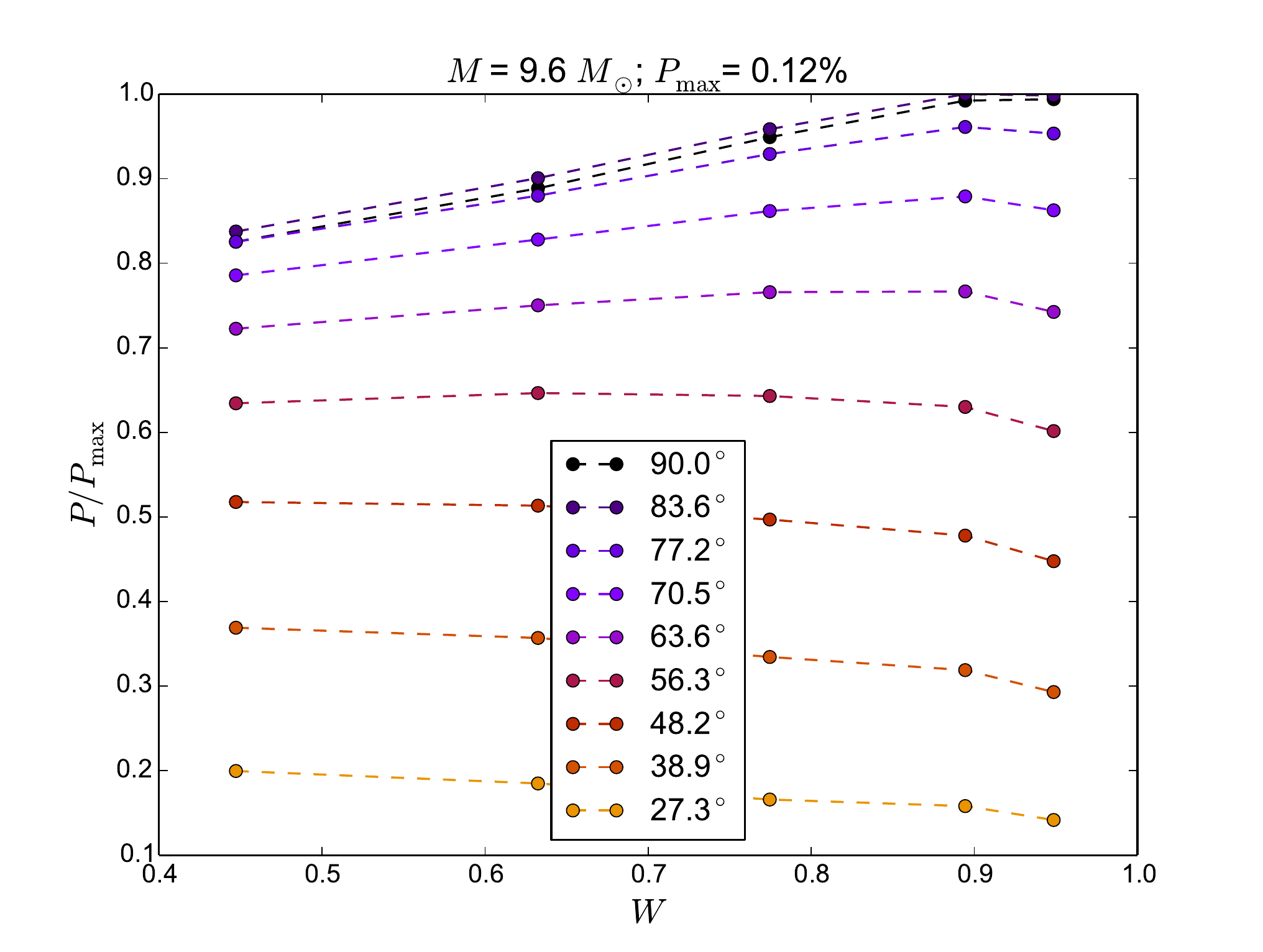}  
    \caption[Relative $V$-band polarization level of a Be star as function of the stellar rotation rate seen at different inclination angles]{Relative $V$-band polarization level of a Be star as function of the stellar rotation rate seen at different inclination angles. The considered star has mass $M=9.6M_\odot$ with a steady-state viscous decretion disk with base superficial density of $\Sigma_0=0.02$~g\,cm$^{-3}$.}
    \label{fig:mag_polM9p6}
\end{figure}

In Fig.~\ref{fig:mag_polM9p6} we show the $V$-band polarization of a Be star with a mass and disk density compatible with the ones expected to centrifugal magnetospheres (i.e., B2V star with $n\sim10^{12}$ particles per cm$^3$) as function of the stellar rotation rate and seen at different inclination angles. The Be models employed are from the \textsc{BeAtlas} project, presented in Chapter~\ref{chap:tools}. It is known that the polarization of ionized circumstellar disks increases with the inclination angle $i$ due to increased projected asymmetry (a face-on disk has a null polarization). The maximum level occurs at $i\sim70^\circ$, angle at which effects of photospheric and disk self absorption become important.
In the figure, we see that up to angles $i\sim60^\circ$ the polarization level decreases with rotation angle, but for angles near edge-on viewing, the reverse is observed.

The effect of high stellar rotational rate could be important for the analysis of HR\,7355, and other fast rotating magnetosphere stars, such as HR\,5907 (next section). Indeed, the polarization level can decrease up to $\sim50\%$ with rotation, for small inclination angles ($i\lesssim30^\circ$). However, for the inclination angle derived by spectroscopic modeling of HR\,7355 ($i\sim60^\circ$) and its rotation rate ($W=0.63$), the expected effect is of the order of $\Delta P=\|P(W=0.63)/P(0)\|\lesssim6\%$, and was neglected in this work. 

The absence of material near the stellar equator in the centrifugal magnetospheres could lead to different conclusions than those of Be disks. To verify this, we executed radiative transfer simulations for which the Be disk was truncated to a disk between 2 to 3 $R_{\rm eq}$ and with uniform density as function of the radius. A very similar result was found than the one shown in Fig.~\ref{fig:mag_polM9p6}. The inclusion of a fast rotating star is planned for the ``dumbbell+disk'' modeling tool in the near-future, providing a better physical description of the system, particularly for those seen at a small inclination angles.

\subsection{The minimization of the HR\,7355 polarization curve}
The higher rotational rate of HR\,7355 situates the circumstellar co-rotational radius $R_{\rm K}$ closer to the star. The expected value for HR\,7355 is $R_{\rm K}=1.63$~$R_{\rm eq}$. According to the ``dumbbell+disk'' polarization modeling, the blobs are centered at this distance, with radius of 1/3 $R_{\rm eq}$. This means that to the the circumstellar material almost touches the stellar surface. 

To explore the effect of the different distances of the blobs to the star, we considered two distinct mean distances: (i) $R_{\rm blob} = R_{\rm K}=1.63$~$R_{\rm eq}$ and (ii) $R_{\rm blob}=2.54$~$R_{\rm eq}$, the same value as \sori{}. This second value is supported by the emitting spectroscopic lobes characteristics of $R=2$-$4$~$R_{\rm eq}$ from \citet{riv13c}. A smaller CS distance in the ``dumbbell+disk'' model reflects in a smaller density needed to reach the same polarization level, but it also changes the detailed shape of the polarimetric light-curve components $Q$ and $U$.

\begin{figure}
    \centering
    \includegraphics[width=\linewidth]{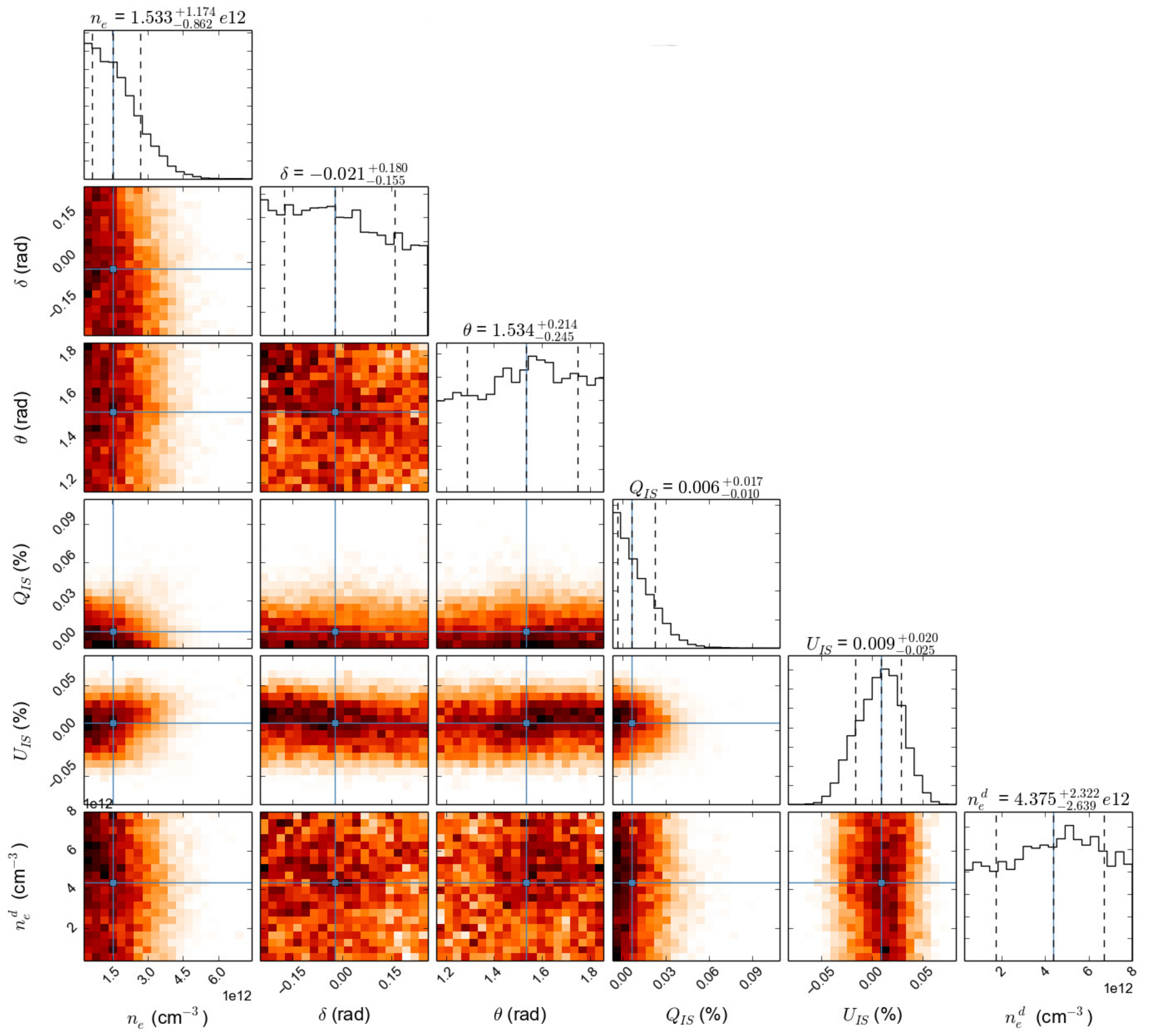}
    \caption[The MCMC probability density distributions for the ``dumbbell+disk'' polarization modeling to HR\,7355 at $i\sim60^\circ$, and the corresponding correlation maps]{The MCMC probability density distributions for the ``dumbbell+disk'' polarization modeling to HR\,7355 at $i=60^\circ$, and the corresponding correlation maps.  The heat-like color map indicate the highest probabilities (black) to the lowest (white). \textit{From left to right (or top-bottom):} inclination angle (rad), blob electronic density (cm$^{-3}$), phase shift, on-sky orientation angle ($\theta$), the interstellar $Q$ component (\%),  interstellar $U$ component (\%), the complementary angle between the rotation and magnetic axes (rad) and disk electronic density (cm$^{-3}$). The dashed vertical lines in the histograms indicate the percentile of 16\% ($-1\sigma$), the median and the percentile of 84\% ($1\sigma$) of each distribution.}
    \label{fig:mc73a}
\end{figure}

\begin{figure}
    \centering
    \includegraphics[width=\linewidth]{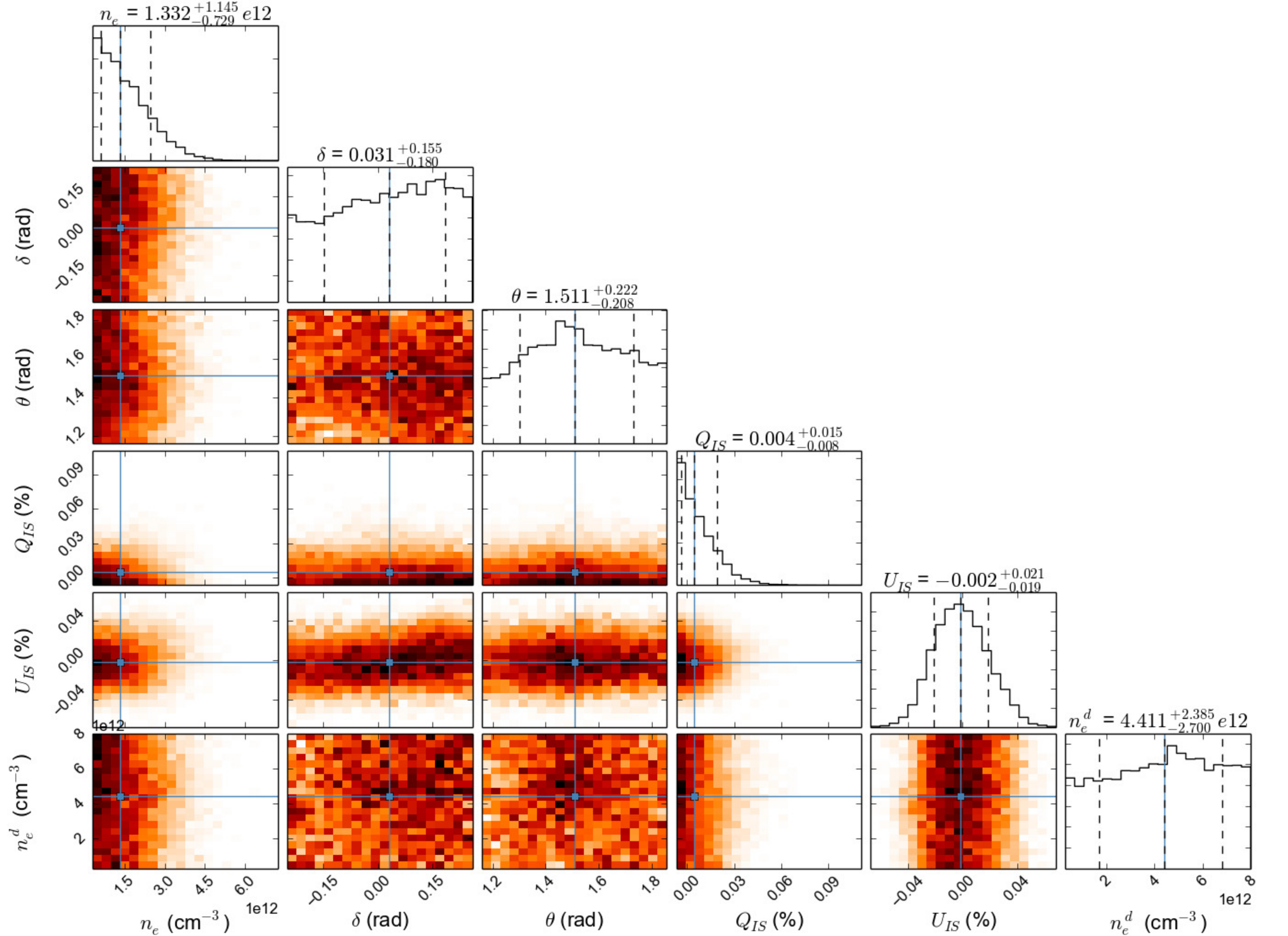}
    \caption{The same as previous figure, but for $i=120^\circ$.}
    \label{fig:mc73b}
\end{figure}

As first result, we found that assuming the blobs at the the co-rotational radius yields a much worse overall fit (median $\chi^2_{\rm red}=4.75$ for $R_{\rm blob}=1.63R_{\rm eq}$) than assuming the blob is located at large distances ($\chi^2_{\rm red}=2.75$ for $R_{\rm blob}=2.54R_{\rm eq}$).
Secondly, as occurred to \sori{}, the minimization was unable to constraint the inclination angle ($i$) and the obliqueness of the magnetic field with the rotation ($\beta$).
Also, we found a sightly better result for a inclination angle of $i\sim60^\circ$ than for $i\sim120^\circ$, but the differences in fit quality are small ($\chi^2_{\rm red}=2.53$ vs$.$ 3.07). The MCMC minimizations for $i\sim60^\circ$ and $i\sim120^\circ$ are shown in Fig.~\ref{fig:mc73a} and Fig.~\ref{fig:mc73b}, respectively. The best-fitting values are listed in Table~\ref{tab:magsspars}. The derived densities (and masses) for the blob and disk components are very similar to the ones determined to \sori{}. Only the blobs appear do to $\sim10\%$ denser, leading to $2M_{\rm blob} /M_{\rm d} \simeq1.1$.

A higher accuracy in the parameters is expected as we gather new data from our running survey. The spectropolarimetric analyzes of field stars with the Wilking-Serkowski fitting might help solve the discrepancy between the IS value derived from them ($P\sim0.142\%$) in comparison the one fitted by the MCMC method, compatible with null value ($P\sim0.006\%$). 
In addition, we will carry on a more in-depth investigation of an interesting feature: these results, both from the model-independent and from the MCMC minimization ($\delta=0.005(16)$ and 0.08(17), respectively), indicate that the minimum of the photometry is anti-correlated with polarization (i.e., the maximum of polarization occurs at photometry minimum, defined as the phase $\phi=0$, as seen in Fig.~\ref{fig:hr7355}). This would imply in a complex configuration for the circumstellar environment, with the blobs not being responsible for the region simultaneously dominating the polarization and photosphere absorption in the light-curve. In this case, the presence of photosphere spots may be important.

\FloatBarrier
\section{Magnetospheres seen by interferometry and HR\,5907 \label{sec:hr5907}}
\begin{figure}
    \centering
    \includegraphics[width=\linewidth]{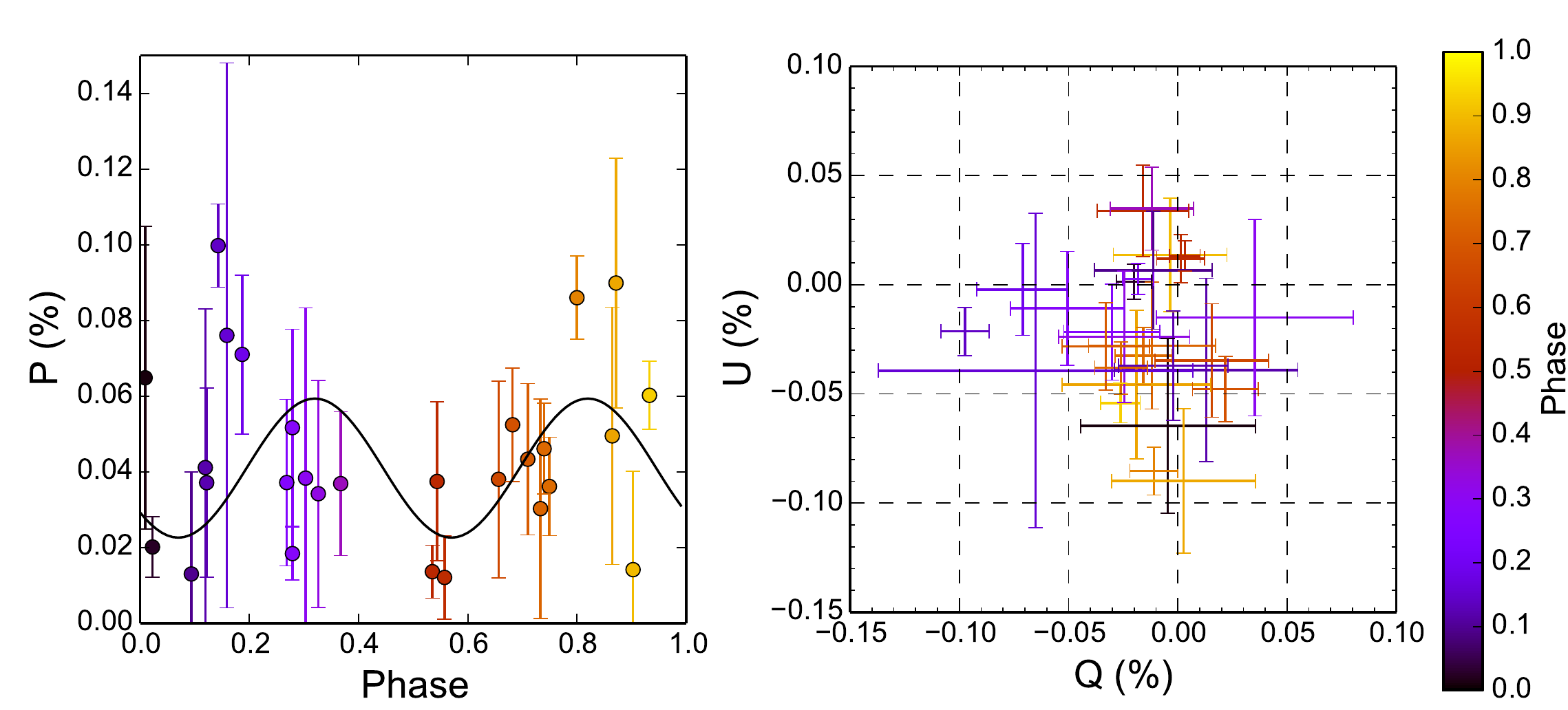}
    \includegraphics[width=\linewidth]{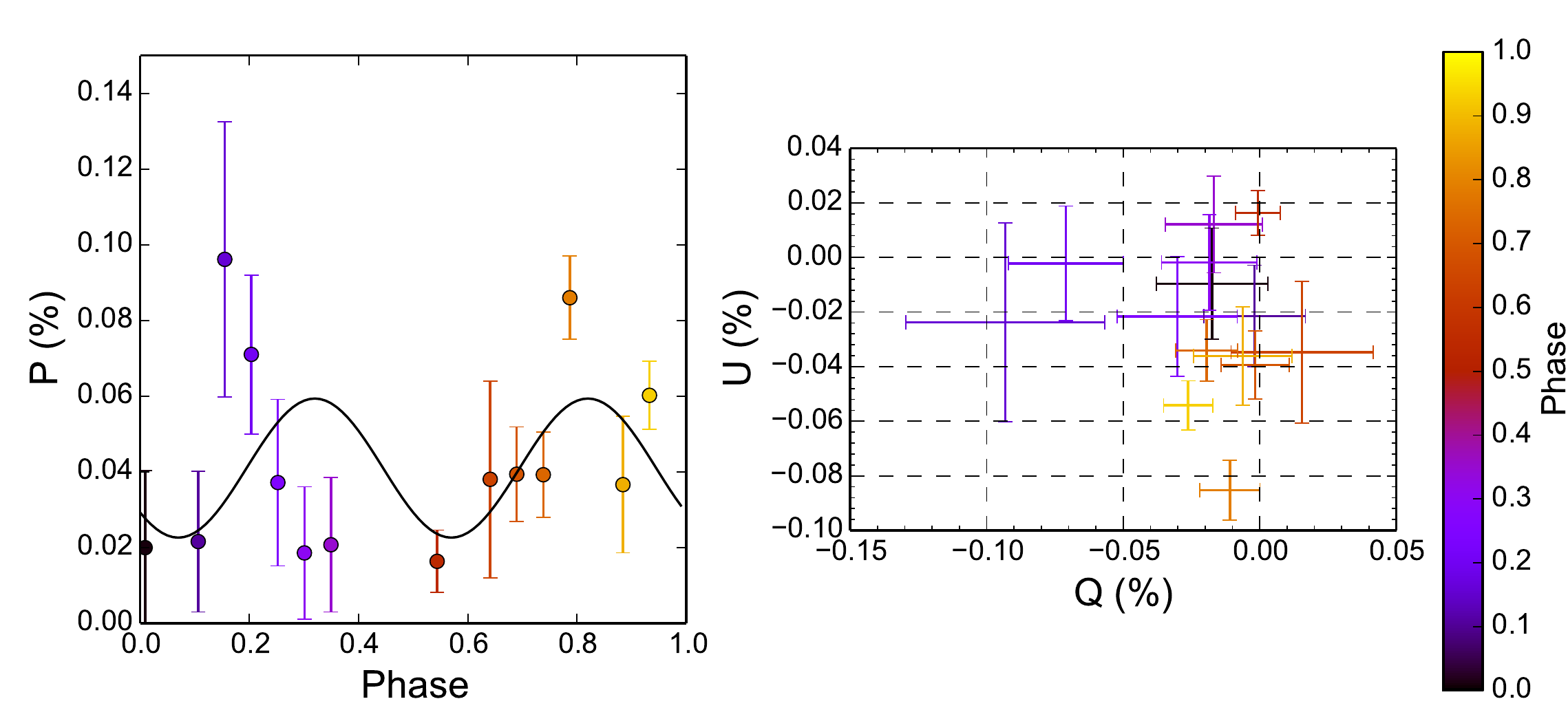}
    \caption[$V$-band intrinsic linear polarization measurements of HR\,5907 and $QU$ diagram on the celestial frame]{$V$-band linear polarization measurements of HR\,5907. \textit{Left:} subtracted interstellar components (Table~\ref{tab:pol1}). The line corresponds to the best fit of Eq.~\ref{eq:polcos}. \textit{Right:} $QU$ diagram of the observed polarization on the celestial frame. \textit{Top}, all observational points. \textit{Bottom}, the data points binned in phase. Their values in the celestial frame are in Table~\ref{tab:aphr5907}.}
    \label{fig:hr5907}
\end{figure}

The bright B2V star HR\,5907 ($V_{\rm mag}=5.4$) hosts a confirmed dense, magnetically bound circumstellar material, displaying a high rotation rate for a magnetic star ($W=0.53$; \citealp{gru12a}). 
Among the three magnetic stars analyzed in this work, HR\,5907 is the one with the worst S/N in the polarimetric data. The raw polarization data of HR\,5907 shows a marginal signal of modularization as compared to \sori{} (Fig.~\ref{fig:hr5907}). Also differently from that star, HR\,5907 do not belong to a multiple stellar system what make much harder estimate the ISM polarization contribution. The field stars used to estimate the interstellar components of HR\,5907 are in Table~\ref{tab:fs5907}. The criteria employed in their selection were the same as used to HR\,7355. The resulting values for the interstellar polarization components are in Table~\ref{tab:pol1}.

\begin{table}
\centering
\caption{$V$-band polarization measurements of HR\,5907 field stars.}
\begin{tabular}{cccccccccccc}
\toprule
Star & Date & $Q$ (\%) & $U$ (\%) & $P$ (\%) & $\sigma$ (\%) & $\theta$ (deg) & $\sigma_\theta$ (deg) \\
\midrule
\multirow{3}{*}{hip78099} & 11mai11 & -0.1763 & 0.4408 & 0.4747 & 0.0097 & 55.9 & 0.6 \\ 
 & 11mai11 & -0.2392 & 0.4100 & 0.4747 & 0.0097 & 60.1 & 0.6 \\ 
 & 11set29 & -0.2438 & 0.4246 & 0.4896 & 0.0141 & 59.9 & 0.8 \\ \hline
hip77858 & 11jun29 & -0.5358 & 0.3780 & 0.6557 & 0.0123 & 72.4 & 0.5 \\ \hline
hip77960 & 11set28 & -0.1772 & 1.0196 & 1.0349 & 0.0117 & 49.9 & 0.3 \\ 
\bottomrule
\end{tabular}
\label{tab:fs5907}
\end{table}

The intrinsic polarization HR\,5907 is shown in Fig.~\ref{fig:hr5907}. Despite the fewer polarimetric observations among the magnestospheres analyzed here, and the corresponding low signal-to-noise ratio the model-independent fitting (Eq~\ref{eq:polcos}; $\chi^2_{\rm red}=2.80$) resulted in very close values of \sori{} magnetosphere, with a oscillation amplitude  $A\sim 0.02\%$ and a base level of $P_0\sim 0.04\%$. The derived position angle was $77.4^\circ$, value which is useful when interpreting the available interferometry observations of the star.

\subsection{The spectrointerferometry of HR\,5907}
The proximity of HR\,5907 (parallax of 7.64(37) mas) and its short rotation period ($P = 0.508$~d) sets a good opportunity to characterize its magnetosphere using interferometry. Thus, AMBER observations were proposed aiming at determining the detailed spacial configuration of the co-rotating magnetosphere. Spectrointerferometry offers a unique way to investigate the stellar brightness distribution at multiple spectral channels, proving both spatial and kinematical information. 

\begin{figure}
    \centering
    \includegraphics[width=.55\linewidth]{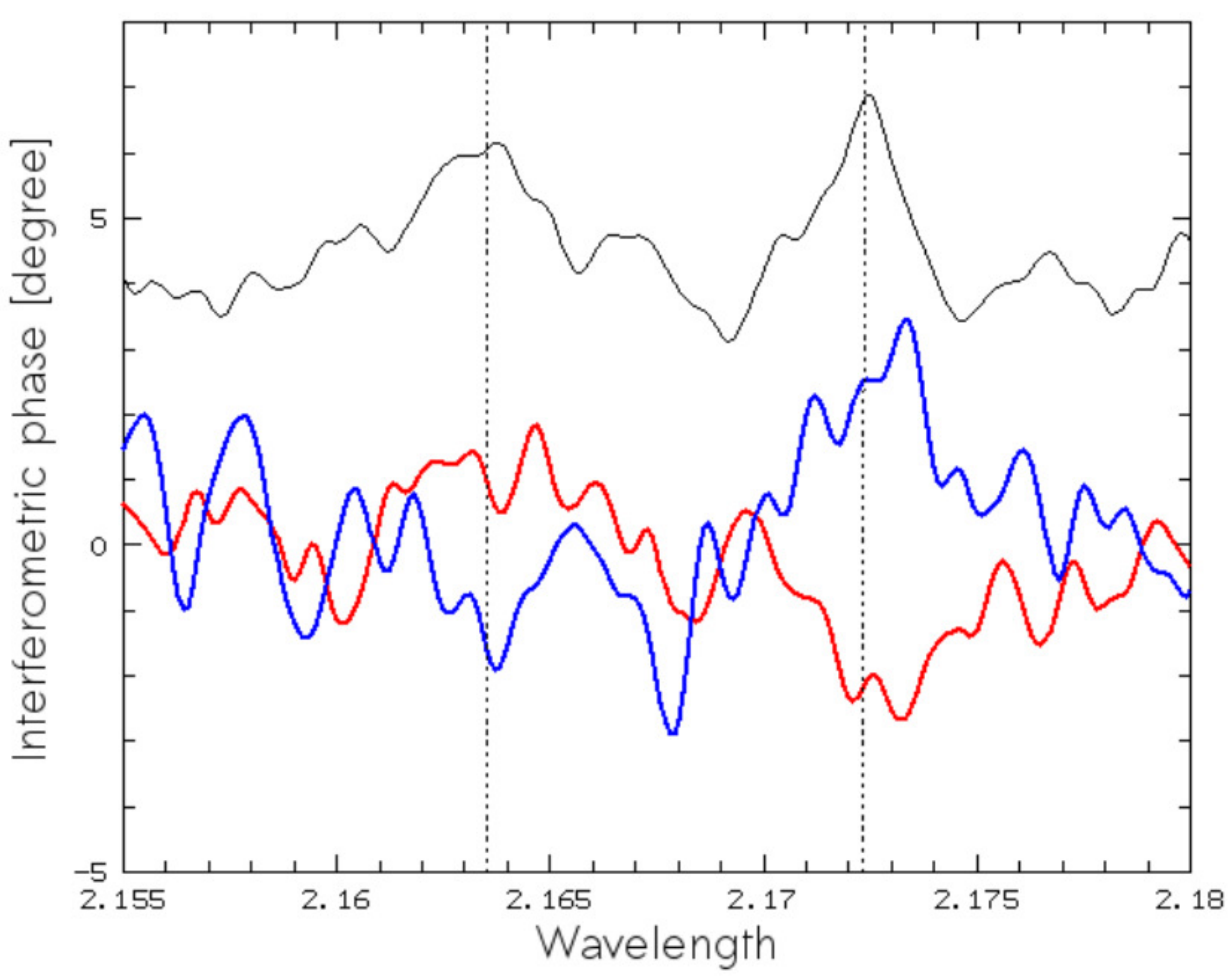}
    \caption[Interferometric phases and flux of Br$\gamma$ line for HR\,5907 from AMBER/VLTI]{\textit{Upper profile}: flux observed by AMBER across the Br$\gamma$ line (scaled and filtered with a 5 pixel Gaussian for display). The emission peaks correspond to those seen in H$\alpha$ at that rotational phase. \textit{Lower profiles}: Filtered interferometric phases for baselines K0$\rightarrow$A1 and A1$\rightarrow$I1. An interferometric phase signal of the order of about 3$^\circ$ is seen at the position of the red emission peak, one of about 1.5$^\circ$ going into the other direction for the blue peak \citep{riv12a}.}
    \label{fig:sihr}
\end{figure}

Visual spectroscopic and circular polarimetric data was analyzed by \citet{gru12a} and predicted a magnetosphere detectable by interferometry across the Br$\gamma$ line. This scenario was tested by observations, aiming at deriving data for a quantitative analysis of the magnetosphere. The data resulted in the first interferometric detection of magnetosphere, but in a somewhat smaller signature of $\sim$3$^\circ$ phase difference (Fig.~\ref{fig:sihr}). The details of observations and the results obtained are given in the proceedings paper \citet{riv12a}, attached to the end of this chapter. Even though we can surely say that the magnetosphere was detected in phase, the data quality, due to the really bad observational conditions, prevents the use of the data for a more qualitative characterization of the magnetosphere.

About the interferometric observations, if the star is aligned at PA\,$=77.4^\circ$ as the of linear polarization indicates, the disk is oriented at 167.4$^\circ$. For the interferometer configuration described in Table~\ref{tab:ambhr5907} (telescopes K0, A1 and I1), the 100~m of effective distance can be applied: A1$\rightarrow$I1, $B_{\rm proj}=105.1$~m, $\Delta\theta_{\rm disk} = 11^\circ$ and K0$\rightarrow$A1, $B_{\rm proj}=127.4$~m, $\Delta\theta_{\rm disk} = 34.6^\circ$. The third baseline, besides having a smaller size ($B_{\rm proj}=46.6$~m), was almost aligned to the stellar pole, where no phase shift is expected ($\Delta\theta_{\rm pole} = 3.8^\circ$).

The observed phase shift can be estimated from the angular displacement of photocenter (Eq.~\ref{eq:photc}, discussed in Chapter~\ref{chap:bes}). \citet{gru12a} show that the circumstellar emission of HR\,5907 can be up to 15\% brighter than the star at the H$\alpha$ line profile. Assuming this brightness to Bracket $\gamma$ Hydrogen line (Br$\gamma$, $\lambda_{\rm vac}=21661.178$~\AA) and that the emission is concentrated in the co-rotation radius ($R\sim2R_{\rm eq}$), this results in a photocenter shift of $\Delta p=0.26$ stellar radius $\left( \Delta p=\frac{1\times0+0.15\times2}{1+0.15} R_{*}\right)$. This shift, applied to the values of HR\,5907 (Table~\ref{tab:magsspars}), results in a angular shift of $0.03$~mas. This phase shift, seen by an interferometer of 100~m of projected baseline, yield a phase of $\sim$2.5$^\circ$, what is roughly the observed value.

\begin{table}
\centering
\caption{Projected baselines for AMBER observations of HR\,5907 \citep{riv12a}.}
\begin{tabular}[]{ccc}
\toprule
Stations & \multicolumn{1}{l}{$\|\vec{B}_{\rm proj}\|$ (m)} & \multicolumn{1}{l}{PA (deg)} \\     
\midrule
K0$\rightarrow$A1 & 127.4 & -158.0 \\ 
A1$\rightarrow$I1 & 105.1 & 1.6 \\ 
I1$\rightarrow$K0 & 46.6 & 106.4 \\     
\bottomrule
\end{tabular}
\label{tab:ambhr5907}
\end{table}

\subsection{The minimization of the HR\,5907 polarization curve}
\begin{figure}
    \centering
    \includegraphics[width=\linewidth]{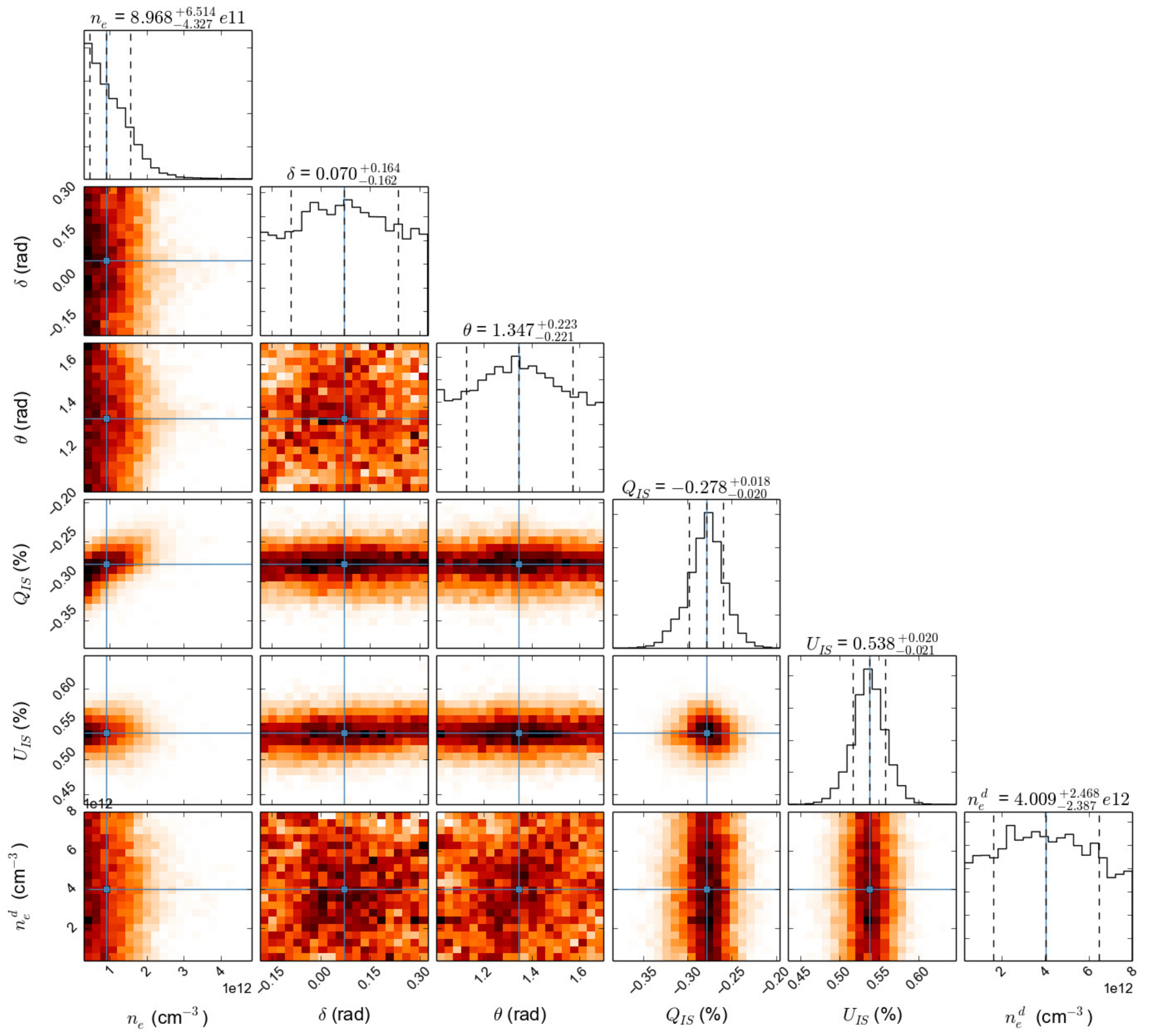}
    \caption[The MCMC probability density distributions for the ``dumbbell+disk'' polarization modeling to HR\,5907 at $i\sim70^\circ$, and the corresponding correlation maps]{The MCMC probability density distributions for the ``dumbbell+disk'' polarization modeling to HR\,5907 at $i=70^\circ$, and the corresponding correlation maps.  The heat-like color map indicate the highest probabilities (black) to the lowest (white). \textit{From left to right (or top-bottom):} inclination angle (rad), blob electronic density (cm$^{-3}$), phase shift, on-sky orientation angle ($\theta$), the interstellar $Q$ component (\%),  interstellar $U$ component (\%), the complementary angle between the rotation and magnetic axes (rad) and disk electronic density (cm$^{-3}$). The dashed vertical lines in the histograms indicate the percentile of 16\% ($-1\sigma$), the median and the percentile of 84\% ($1\sigma$) of each distribution.}
    \label{fig:mc5907_1}
\end{figure}

\begin{figure}
    \centering
    \includegraphics[width=\linewidth]{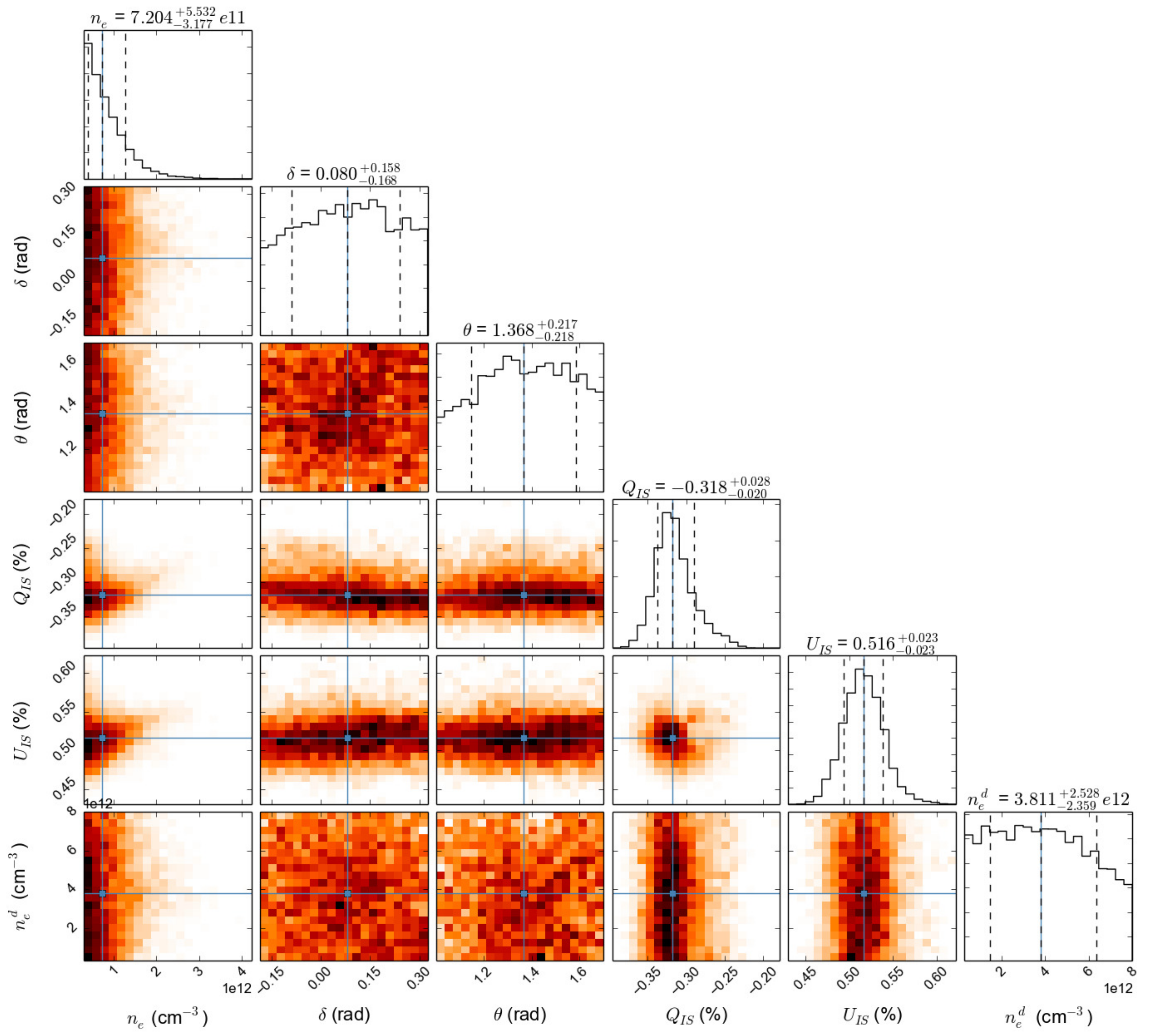}
    \caption{Same as previous figure, but for $i=110^\circ$.}
    \label{fig:mc5907_2}
\end{figure}

The MCMC minimization PDFs for HR\,5907 polarimetric curve using the ``dumbbell+disk'' model are in Fig.~\ref{fig:mc5907_1} for $i\sim60^\circ$ and Fig.~\ref{fig:mc5907_2} for $i\sim120^\circ$. The mean $\chi^2_{\rm red}$ are 7.95 and 6.42, respectively. The best-fitting parameters are listed in Table~\ref{tab:magsspars}. As discussed for HR\,7355, we neglected the effects of stellar rotation on the modeling of HR\,5907 due to its inclination angle ($i\sim70^\circ$). 

Further polarimetric HR\,5907 observations are been carried out at OPD/LNA in order to complete phase coverage and increase precision of the polarimetric signal. The derived polarimetric phase shift obtained $\delta=0.08$ is very similar to the value found for \sori{}, indicating an asymmetric magnetosphere. The low values found for circumstellar environment ($n_e \sim 1-3\times10^{12} M_\odot$) are likely the result of the irregular modulation seen in the data. Still, with the data available the MCMC constrain of the ISM polarization was compatible with field stars observations. The current minimization points at a lower density of magnetosphere by a factor of $1/2.5$. The region of intersection of the magnetic field and rotational planes (the blobs) has a smaller factor, yielding a $2M_{\rm  blob}/M_{\rm disk}\sim0.8$, i.e., $\sim20\%$ lower than for \sori{}.



\FloatBarrier
\section{Chapter summary}
High-precision linear polarization measurements were carried out at the OPD/LNA (Brazil) for three hot magnetic stars that exhibit centrifugal magnetospheres: \sori{}, HR\,7355 and HR\,5907. To \sori{}, this resulted in the first firm detection of a magnetosphere in polarimetry, reported by \citet{car13a}. Subsequently, this structure was also detected for the other two stars, a yet unpublished result and that we expect to publish soon.

Although RRM model is able to reproduce the shape of emission lines as well as the photometric depths of eclipses by the plasma clouds of magnetospheres, it fails at describing the detected polarization curves. For this reason, and the fact the densities of magnetospheres are low, we made a single scattering ad-hoc model based on the two main geometrical components predicted: two diametrically opposed blobs connected by a tilted disk.  

Our dubbed ``dumbbell+disk'' model appeared as a useful tool to interpreted the polarization signal of the magnetospheres and determine their intrinsic physical parameters. In particular, we could determine the orientation of the \sori{} system and the mass ratio of the blob vs$.$ disk components $2M_{\rm blob}/M_{disk}\sim1.0$, a value slight lower than previously determined by \citet{car13a} but that keeps the model interpretation, i.e., the inability of the RRM modeling simultaneously fit the photometric and the polarimetric light curves.

We show that two other magnetospheres, namely HR\,5907 and HR\,7355, have a polarimetric oscillation amplitude very similar to \sori{}, although having a lower signal-to-noise lightcurve (Table~\ref{tab:pol1}). The absence of fast-rotating stellar effects in the polarization modeling appears to do not alter considerably simulations for $i\sim70^\circ$, the case of the analyzed stars. The low polarimetric signal prevented a precise determination of the magnetosphere parameters with MCMC minimization, such as the direction of rotation. Even so, an interesting result was obtained: HR\,7355 displays $2M_{\rm blob}/M_{disk}\sim1.1$, similar to \sori{}, while HR\,5907 has an lower $2M_{\rm blob}/M_{disk}\sim0.8$ and overall scattering mass. This can be the result of the low S/N light curve or due to the obliqueness of the magnetic field with rotation, $\beta$: while \sori{} and HR\,7355 have $\beta\gtrsim55^\circ$, the rotation and magnetic axis of HR\,5907 are almost coincident, with $\beta\sim7^\circ$. 

The future perspectives are: (i) extend our polarimetric observation dataset of magnetic stars observable in the southern hemisphere (Table~\ref{tab:magsurvey}), having a comprehensive phase coverage for them; (ii) improve the scattering modeling to include stellar rotational effects, and improve HR\,5907 and HR\,7355 polarimetric model fitting; (iii) create an interface to feed the ``dumbdell+disk'' structure into \textsc{hdust}, which does not employ the single scattering approximation and treats continuum opacity in addition to electron scattering. This allows to investigate the photometric light curve shape and its relation to the polarimetry.
\begin{table}
\centering
\caption{List of hot magnetic stars observed in linear polarization in the survey at OPD/LNA.}
\begin{tabular}[]{lcclcclcc}
\toprule
    \multicolumn{1}{c}{Star} & $V_{\rm mag}$ & & \multicolumn{1}{c}{Star} & $V_{\rm mag}$ & & \multicolumn{1}{c}{Star} & $V_{\rm mag}$\\
\midrule
$\delta$ Ori C & 6.85 & & HD\,64740 & 4.61 & & $\tau$ Sco & 2.81 \\
$\theta^1$ Ori C & 5.13 & & HD\,96446 & 6.68 & & HR\,7355 & 6.02 \\
\sori{} & 6.66 & & $\sigma$ Lupi & 4.42 & & HD\,191612 & 7.84 \\ 
HD\,37776 & 6.98 & & HR\,5907 & 5.40 & & &  \\ 
HD\,57682 & 6.40 & & HD\,148937 & 6.77 & & &  \\ 
\bottomrule
\end{tabular}
\label{tab:magsurvey}
\end{table}

The spatial information of the resolved magnestopheres provides constraints to analysis of magnetic field and the resulted trapped material around the rotating star. The magnestosphere geometry is directly related to fundamental stellar parameters, as field strength and configuration; angles between rotation and magnetic dipole axis; the role of the rotation velocity to shape the magnetosphere and to govern the mass flow through it; among many others. The relevance of here presented studies is to help constraining some of the parameters that drive this interaction between wind, field and rotation.

\FloatBarrier
\section{Publication: Polarimetric Observations of \sori{} \label{sec:pubsori}}
\href{http://adsabs.harvard.edu/abs/2013ApJ...766L...9C}{ADS Page: http://adsabs.harvard.edu/abs/2013ApJ...766L...9C}

\FloatBarrier
\section{Publication: The interferometric signature of the rapidly corotating magnetosphere of HR 5907 \label{pub:hr5907}}
\href{http://adsabs.harvard.edu/abs/2012AIPC.1429..102R}{ADS Page: http://adsabs.harvard.edu/abs/2012AIPC.1429..102R}

%% file: chap/bestars_arxiv.tex
\chapter{Spectrointerferometry applied to the Be star disks \label{chap:bes}}

Spectrointerferometry is sensitive to the brightness distribution in narrow spectral channels of a given target. This can be used to relate the velocity with the angular position of the emission, and may map the velocity profile of a rotation system. Indeed, the detailed measurements that have demonstrated that Be disks rotate in a Keplerian fashion made use of this technique \citep{mei07a,kra12a,whe12a}. 

An important feature of spectrointerferometry that needed to be investigated was departures from the expected differential phases (DP) of Be stars, which are discussed below. As and example, \citet{kra12a} reported that a phase reversal was registered at the DP signal of the Be star $\beta$\,CMi only for the longest baseline available, indicating that this could be an over-resolution effect. \citet{mei12a} and \citet{ste14a} also report these complex shape phases. 

In \citet{fae13a} we presented a systematic study of how the DPs are affected by how resolved is the target by the interferometer. In addition to differentiate the resolved vs$.$ non-resolved effects, we discovered a new phenomenon, which we dubbed CQE-PS, that can profoundly affect the phase profile, and provide useful new quantitative information about the disk and the star. This paper is as appendix, and our main findings and conclusions are summarized in this chapter.

\subsubsection*{The Central-Quasi Emission (CQE)}
Before the spectro-interferometric era, one of the first strong indications that Be disks are Keplerian came from high-resolution spectroscopy ($R>30000$) of shell stars, which are stars with strongly rotationally-broadened photospheric lines and additional narrow absorption lines. They are understood as ordinary Be stars seen edge-on, so that the line-of-sight towards the star probes the CS equatorial disk.
\citet{han95a} and \citet{riv99a} studied the so-called \textit{Central Quasi-Emission} (hereafter, CQE) peaks, where the disk, under certain circumstances, causes a cusp in the deepest region of the line profile of shell stars. The existence of the CQE implies slow radial motions of the gas (much smaller than the sound speed), which means the disk is supported by rotation.

The same mechanism that produces the CQE observed in shell star line profiles can cause important changes in the monochromatic intensity maps of the Be star plus disk system, with observable effects on the interferometric quantities. This effect was dubbed \textit{CQE Phase Signature} \citep{fae12a}. Similar to the spectroscopic CQEs, the CQE Phase Signature (hereafter, CQE-PS) can only be studied by high {spectral} resolution observations ({$R>2500$}), such as can be obtained {by the new generation of interferometers (AMBER/VLTI at near infrared, or VEGA/CHARA, \citealp{mou09a}, at visible range)}. 

The initial studies made on the CQE profiles \citep{han95a,riv99a} demonstrated its strong dependence on the distribution and velocity field of the CS material. Likewise, the CQE-PS has a high sensitivity to structural components of the CS disk - and even to the stellar size. 

\section{A Be star reference model \label{refcase}}
To study the DPs of Be stars at different angular resolutions and to characterize the CQE-PS with our modeling tool, \textsc{hdust}, a realistic model for the Be + disk system needed to be defined. 
For the central star we adopt a rotationally deformed and gravity darkened star whose parameters are typical of a B1\,Ve star. The stellar geometry is described by {an oblate} spheroid with stellar flux determined by traditional {von Zeipel} effect, where the local effective temperature is proportional to the local surface gravity as $T_\textrm{eff}\propto {g_\textrm{eff}}^{0.25}$ (at that time, the variations in von Zeipel coefficient $\beta$ with the rotation rate of the star were not yet well established, see next chapter). For the CS disk description, we adopted the VDD model. For the case of isothermal viscous diffusion in the steady-state regime, the disk volume density has the particularly simple form $\rho(r) = \rho_0 (R_{\rm eq}/r)^{-m}$, where $m=3.5$.

The interferometric quantities depend on the spatial resolution with which the object is seen. To explore different configurations we define the quantity $\nu_{\rm obs}$, the ratio between the baseline length of the interferometer {and the distance to the star}. The unit used is m\,pc$^{-1}$. The fundamental quantities for the model are listed in Table~\ref{tab:bemod}.

\citet{riv13a} considered this a very representative model for a Be star and chose it to perform many of the analyzes on their review paper, with simulations conducted by me.

\begin{table}
\centering
\caption{Reference Be model parameters \citep{fae13a}.}
\begin{tabular}[]{ccc}
\toprule
    Parameter & Symbol & Ref. Case \\
\midrule
    Spectral type & - & B1\,V \\
    Mass & $M$ & 11.0 $M_{\odot}$\\
    Polar radius & $R_{\rm pole}$ & 4.9 $R_{\odot}$\\
    Pole temperature & $T_{\rm pole}$ &  27440 K  \\
    Luminosity & $L_{\star}$ & 10160 $L_{\odot}$ \\
    Critical velocity & $v_{\rm crit}$ & 534.4 km/s \\
    Rotation rate & $W$; $\Omega/\Omega_{\rm crit}$ & 0.53; 0.80 \\
    Oblateness & $R_{\rm eq}/R_{\rm pole}$ & 1.14 \\
    Gravity darkening & $T_{\rm pole}/T_{\rm eq}$ & 1.16  \\ \hline
    Disk radius & $R_{\rm disk}$ & 10 $R_{\star}$  \\
    Disk density scale & $n_0$ & 10$^{13}\,\rm cm^{-3}$ \\
    Density exponent & $m$ & 3.5 \\ \hline
    Inclination angle & $i$ & 45$^\circ$  \\
    Spectral resolving power & $R$ & 12\,000  \\
    Baseline/distance & $\nu_{\rm obs}$ & 1 m pc$^{-1}$ \\
\bottomrule
\end{tabular}
\label{tab:bemod}
\end{table}

\section{Differencial phases in the resolved and unresolved regimes \label{sec:diffph}}

One important characteristic of interferometric differential phases arises when the target is \textit{marginally-resolved}, i.e., $\vec{u}\cdot\vec{r} \ll 1$. In this case, interferometric differential phases (hereafter DP or just phases) can directly map the target's photocenter (e.g., \citealp{dom04a}, Eq.~4):
\begin{equation}
\phi_\textrm{diff}(\lambda, \lambda_r)=-2\pi\vec{u}\cdot[\vec{\epsilon}(\lambda)-\vec{\epsilon}(\lambda_r)] \,,
\label{eq:photc}
\end{equation}
where $\vec{\epsilon}(\lambda)$ and $\vec{\epsilon}(\lambda_r)$ vectors are respectively the \textit{photocenters} for $\lambda$ and $\lambda_r$, a reference wavelength. The usual procedure is to take $\lambda_r$ on the adjacent continuum of the spectral line, where $\vec{\epsilon}(\lambda_r)=0$. These photocenter measurements can be made at very high angular resolution without resolving the target, with astrometric determinations with higher precision than the nominal interferometer resolution. One example is reported by \citet{car09a}, where the photocenter displacements were measured with precision smaller than 0.1 miliarcsecond on the Be star $\zeta$\,Tauri.

\begin{figure}
\centering
\includegraphics[width=.31\linewidth]{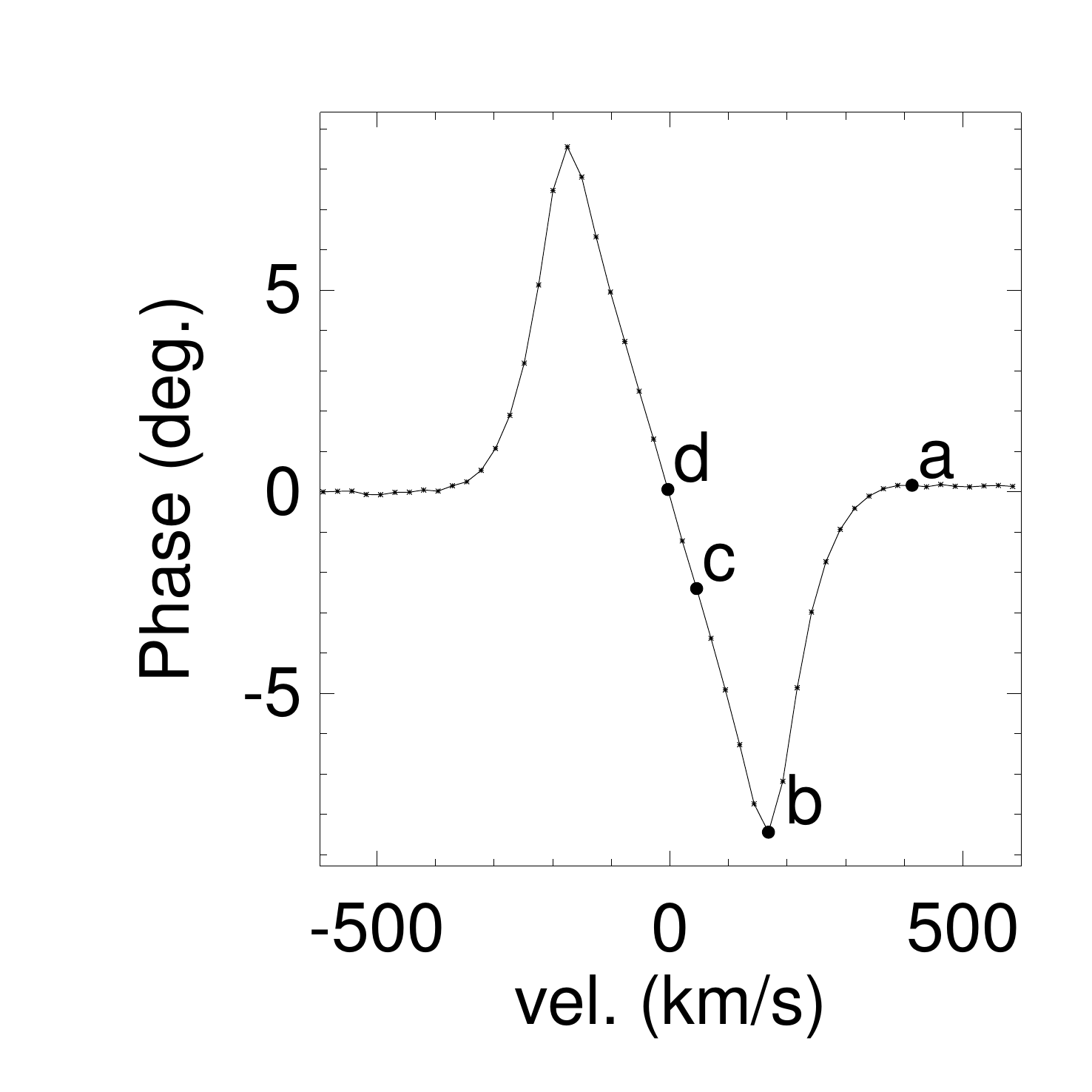}
\includegraphics[width=.62\linewidth]{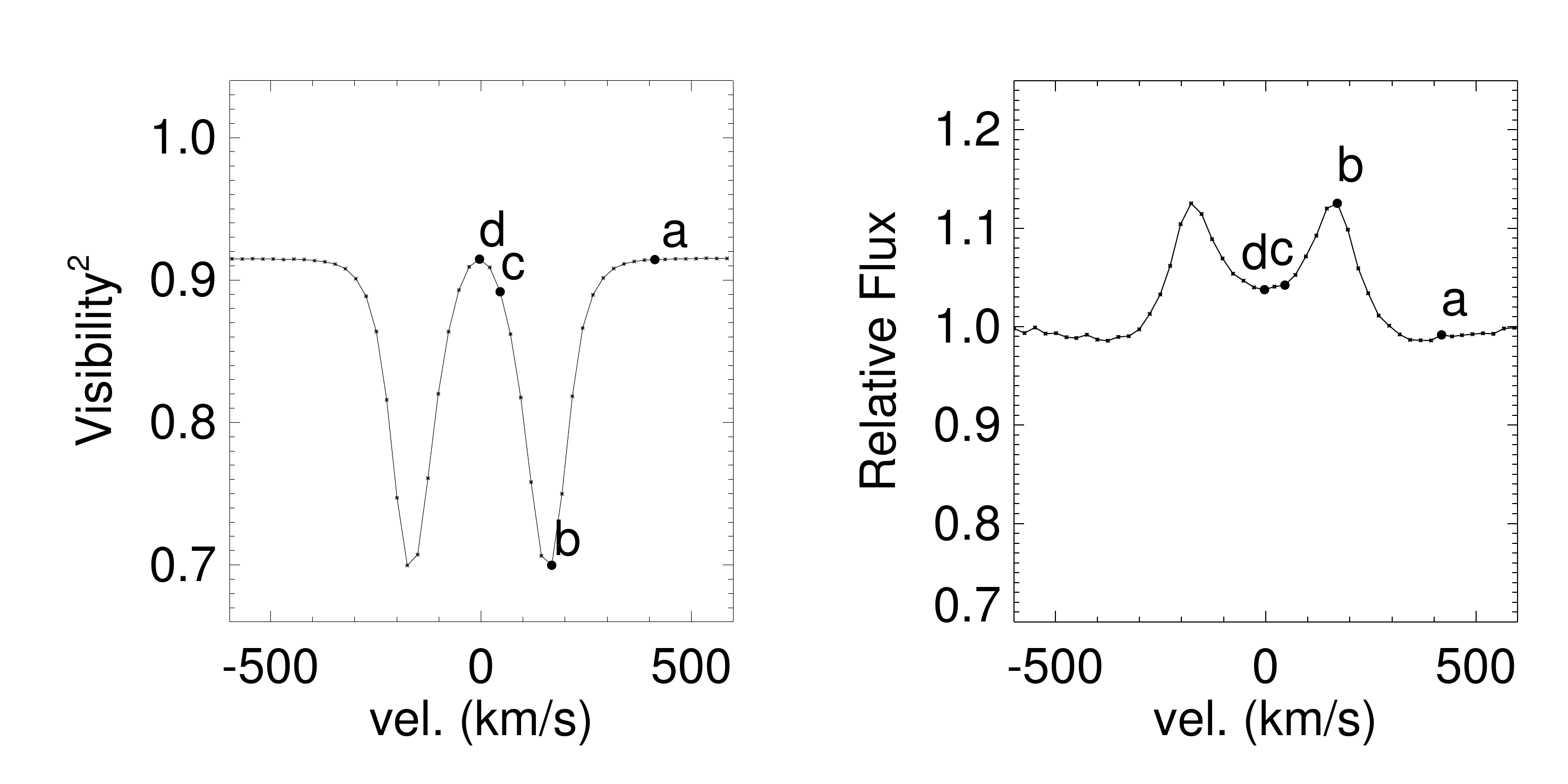} \\
\includegraphics[width=.9\linewidth]{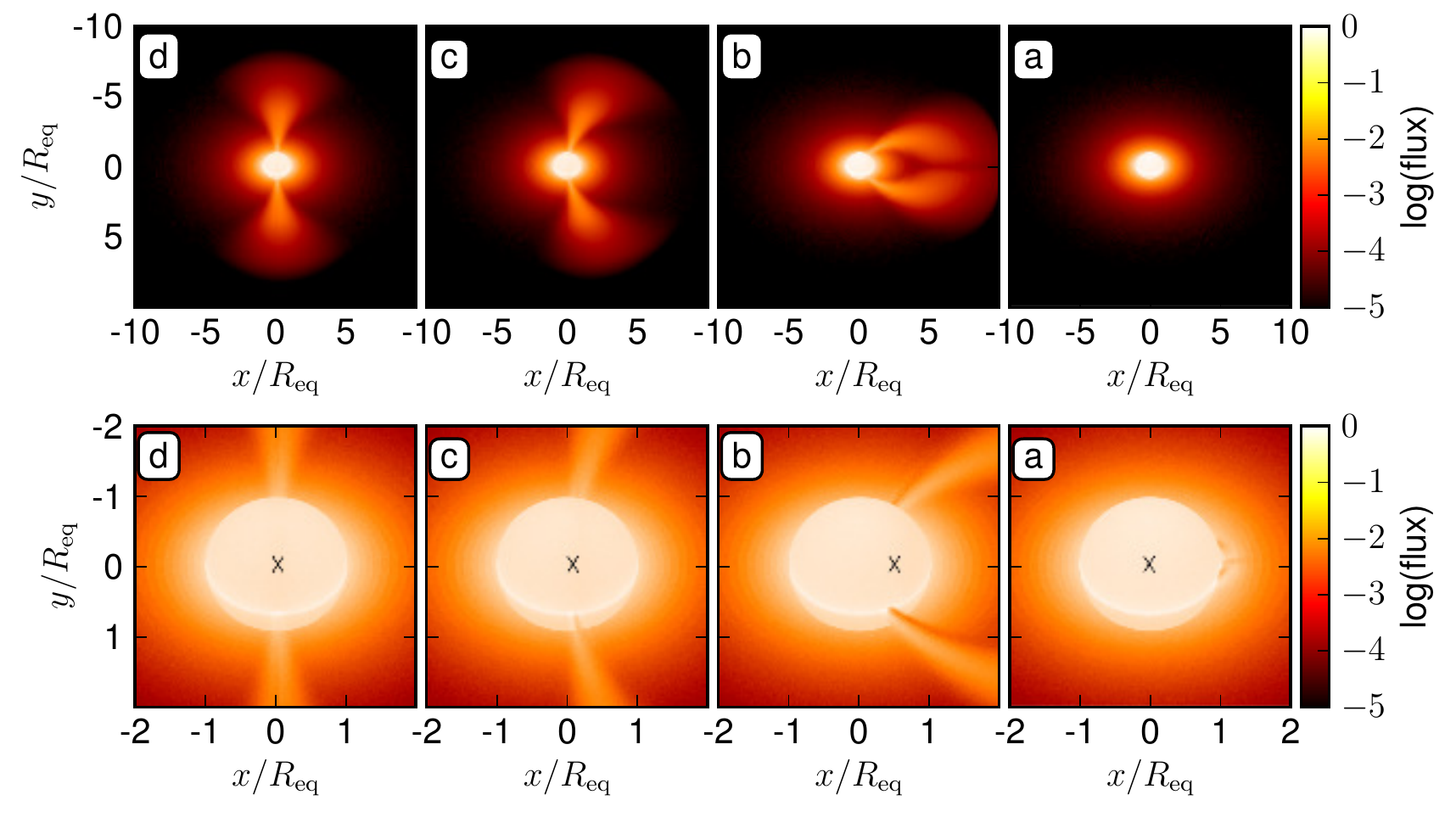} \\
\caption[Interferometric Br$\gamma$ differential phases, visibilities, and line profile for a Be star reference model]{\textit{Top 3 panels}: Interferometric Br$\gamma$ differential phases, visibilities, and line
  profile for the reference model.  Calculations
  were made for $\nu_{\rm obs}=1\,\rm m\,pc^{-1}$ {(target marginally resolved)}, and baseline orientation
  parallel to the disk equator. \textit{Bottom panels:} Model images for different spectral channels (as indicated in the upper
  panels), at different spatial scales. {The photocenter position is indicated by a black cross.} The disk inclination angle is $i=45^\circ$. Absorption bands can be identified at velocities (b) and (c) but they little alter the stellar flux in the line-of-sight. \citep{fae13a}.}
 \label{fig:cqe1}
\end{figure}

The monochromatic on-sky brightness distribution of a Be star with a circumstellar disk can be highly asymmetric for wavelengths across the emission line profile. This is illustrated in Fig.~\ref{fig:cqe1}. At continuum wavelengths (velocity \textit{a}) the images show the star surrounded by a centro-symmetric continuum emission. We defined the DP to be zero at those wavelengths. As we move from the continuum toward line center (\textit{a} to \textit{d}), the line flux at a given wavelength range initially increases as a result of progressively larger emission lobes and then decreases as the emission area decreases toward line center (\textit{d}). This creates the familiar W-shaped pattern in the visibilities and the S-shape pattern in the DPs.

\begin{figure}
\centering
\includegraphics[width=.5\linewidth]{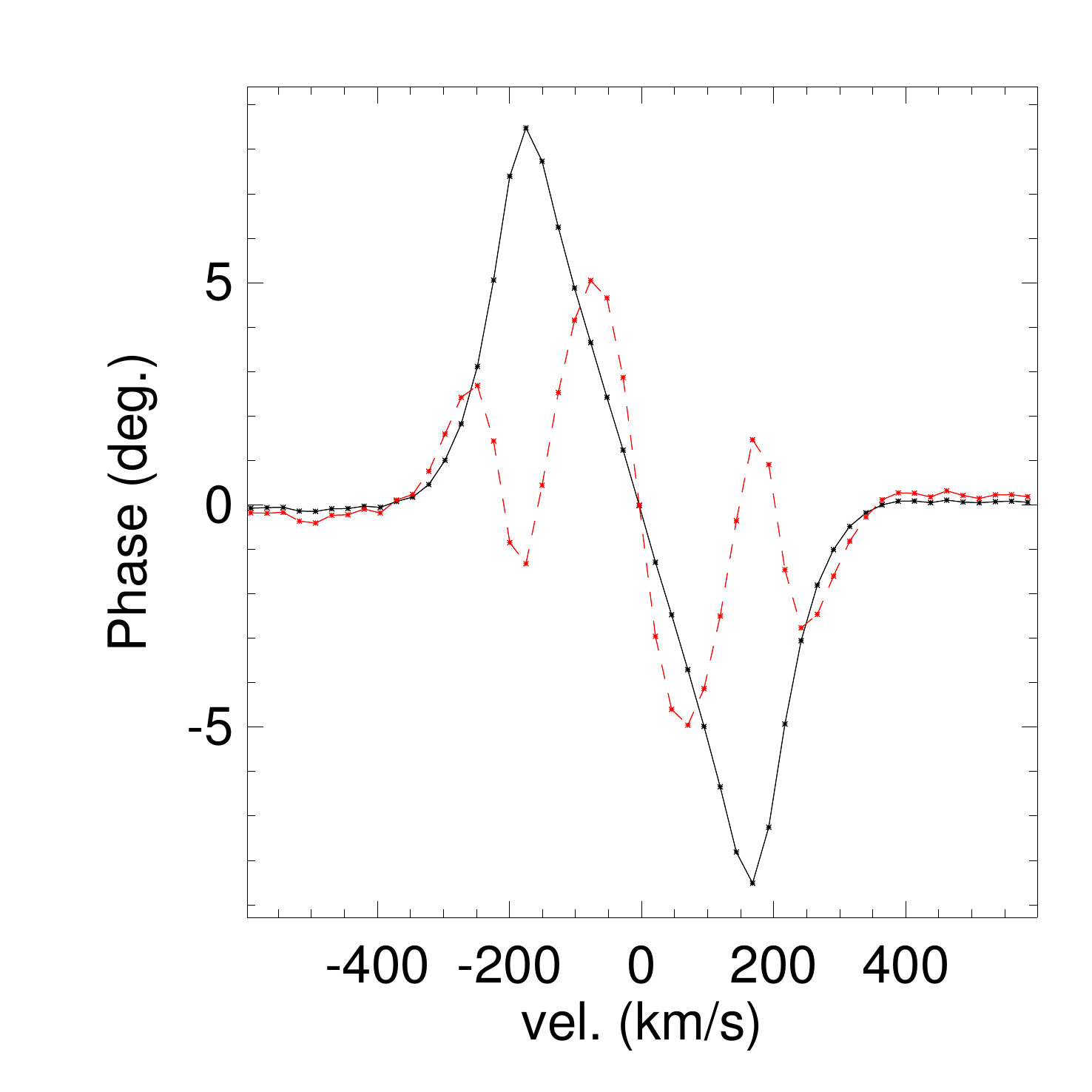}
\caption[Differential phase signal for the Be reference model in the astrometric and non-astrometric regimes]{Differential phase signal for the Be reference model. The black (full) line corresponds to $\nu_{\rm obs}= 1\,\rm m\,pc^{-1}$ (astrometric regime) and the  red (dashed) line to $\nu_{\rm obs}=2.5\,\rm m\,pc^{-1}$ (non-astrometric regime). \citep{fae13a}.}
 \label{fig:cqe2}
\end{figure}

Another important characteristic of differential phases arises when the target is \textit{resolved}, i.e., $\vec{u}\cdot\vec{r} \gtrsim 1$. The photocenter position is no longer proportional to the phase shift, but the phase value is still a function of the photocenter position. This corresponds to the so-called \textit{non-astrometric regime} and means that the phase still bears a relation to the photocenter position, but this relation is no longer simple. The astrometric and non-astrometric regimes are illustrated in Fig.~\ref{fig:cqe2}. 

\begin{figure}
\centering
\includegraphics[width=.7\linewidth]{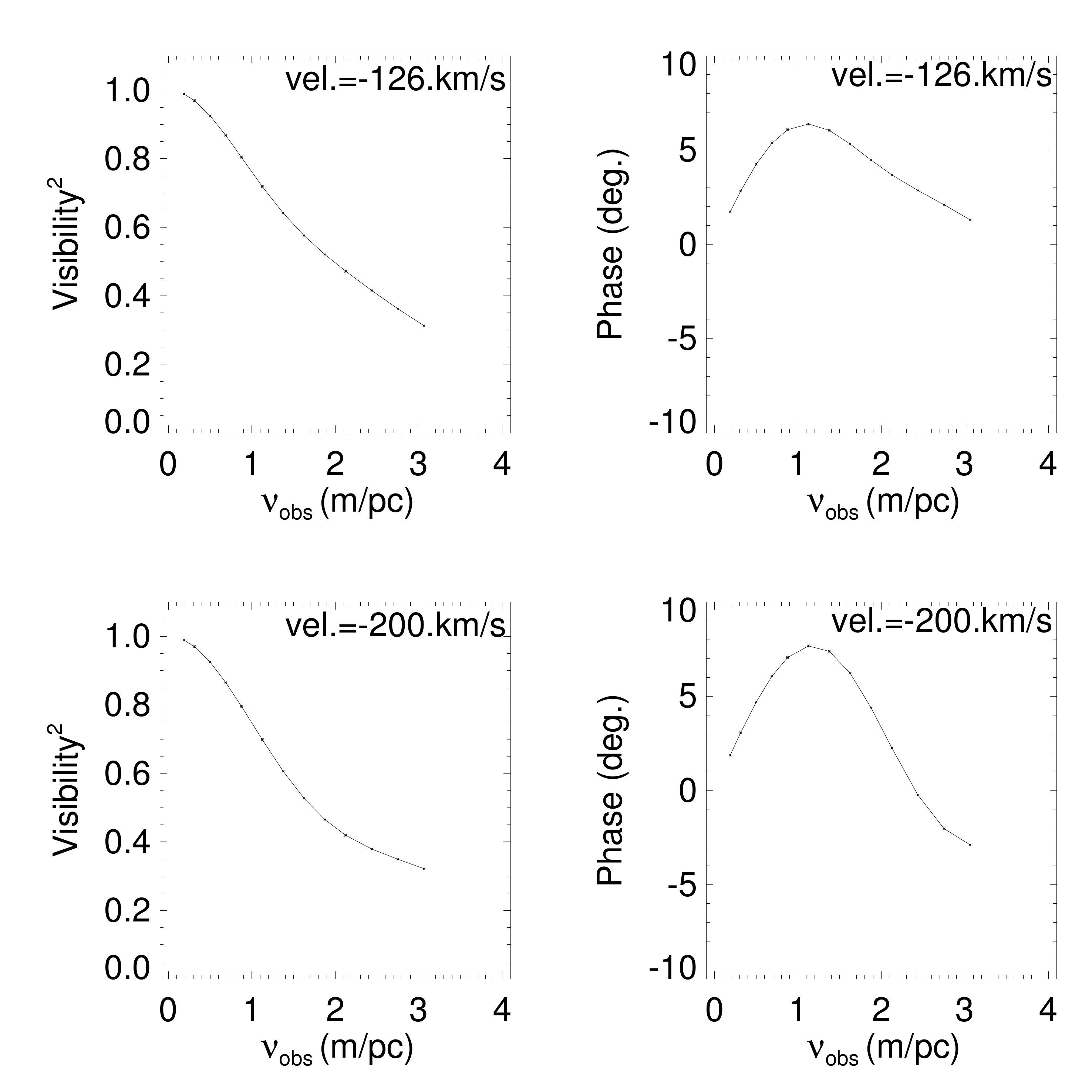}
\caption[The squared visibility $V^2$ and differential phases for the Be star reference model at specific radial velocities as function of $\nu_{\rm obs}\equiv\|\vec{B}_{\rm proj}\|/d$]{The squared visibility $V^2$ and differential phases for the Be star reference model at radial velocities of vel.$=-126\rm\,km\,s^{-1}$ and vel.$=-200\rm\,km\,s^{-1}$ {as function of $\nu_{\rm obs}$, the ratio between the length of the baseline and the distance of the target $(\|\vec{B}_{\rm proj}\|/d)$}. The linear {correspondence} of the phase with the increasing spatial resolution ($\nu_{\rm obs}\lesssim 1.5\rm\,m\,pc^{-1}$) characterizes the astrometric regime. \citep{fae13a}.}
 \label{fig:cqe3}
\end{figure}

For a typical Be star, the threshold limiting the two regimes is the ratio $\nu_{\rm obs} = \dfrac{\|\vec{B}_{\rm proj}\|}{d} \sim 1.5$~m\,pc$^{-1}$, where $\vec{B}_{\rm proj}$ is the projected baseline of the interferometer and $d$ the target distance in parsecs. For $\nu_{\rm obs} < 1.5$~m\,pc$^{-1}$ one has the astrometric regime; otherwise, the non-astrometric. This corresponds roughly to $\vec{u}\cdot\vec{r} \sim 0.37$. The transition of between the phases regimes can be clearly seen in Fig.~\ref{fig:cqe3}.  

\citet{kra12a} proposed an explanation to the phase reversal that is not valid for $\beta$\,CMi. According to them, the visibility function of the line emitting region would pass through a visibility null, transiting from the first to the second visibility lobe. This effect could reverse the direction of the photocenter vector, but is not expected observationally for Be stars. As an example, in the bottom panels of Fig.~\ref{fig:cqe3} a phase signal is reversed at $\nu_{\rm obs} \sim 2$~m\,pc$^{-1}$ with a visibility value of $V^2\sim0.4$. To reach the visibility minimum, $\nu_{\rm obs}\gg 3$~m\,pc$^{-1}$, while $\beta$\,CMi was observed with $\nu_{\rm obs} \sim 2$~m\,pc$^{-1}$. Note that the recent modeling by \citet{kle15a} points to an inclination angle $i\sim45^\circ$, very close to our reference model.

\section{The CQE-Phase Signature (CQE-PS) \label{sec:cqeps}}
Line emission from the rotating disk is the most important factor controlling the detailed shape of the DP for our reference model. However, in some circumstances the line absorption of photospheric light by the disk can have a strong impact on these observables. The absorption that generates the CQE-PS effect changes the S-shaped phases by introducing an inflection in the central part of the line or, if it is strong compared to disk emission, the absorption even transforming the phases to a double S-shape as a result of a central phase reversal. 

\begin{figure}
\centering
\includegraphics[width=.31\linewidth]{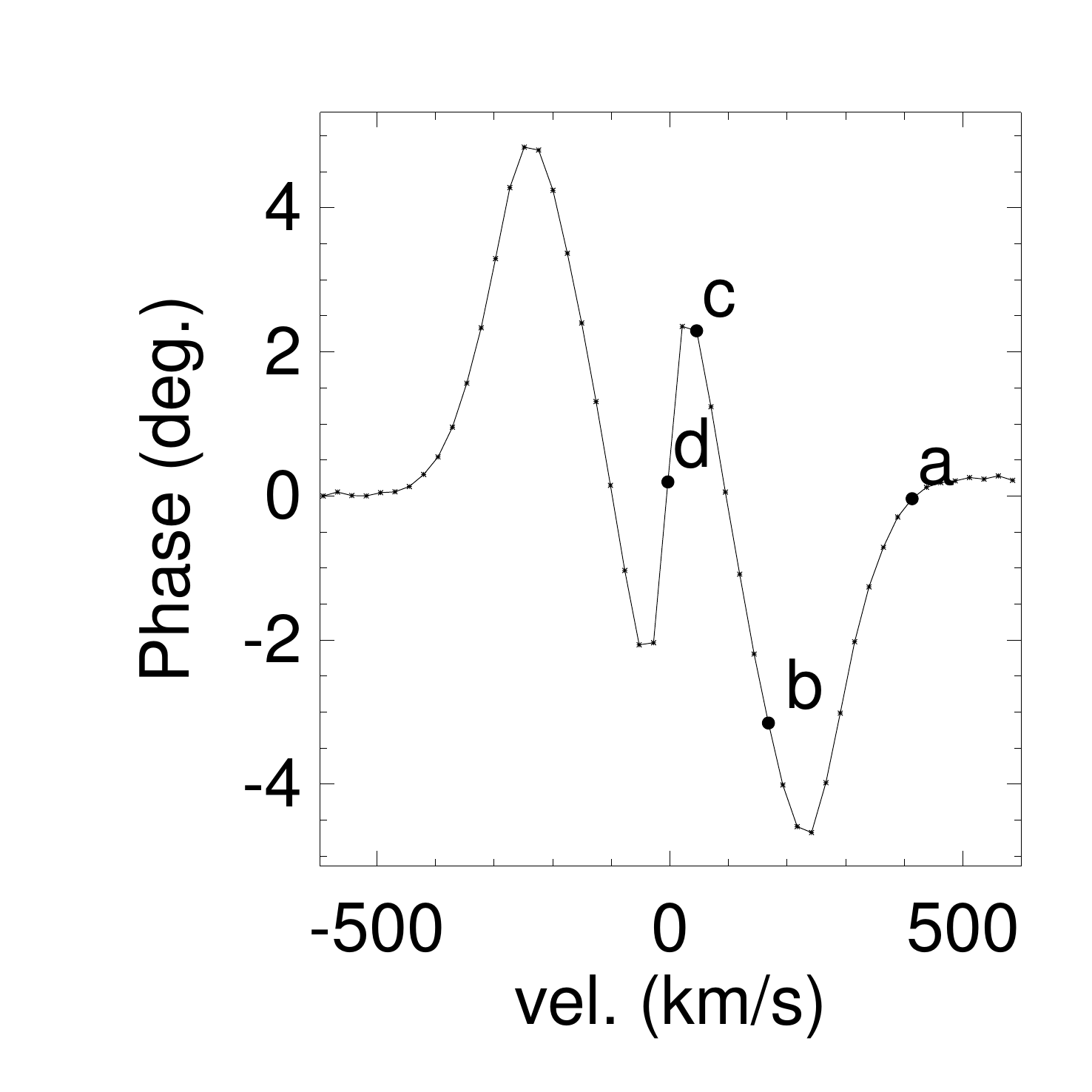}
\includegraphics[width=.62\linewidth]{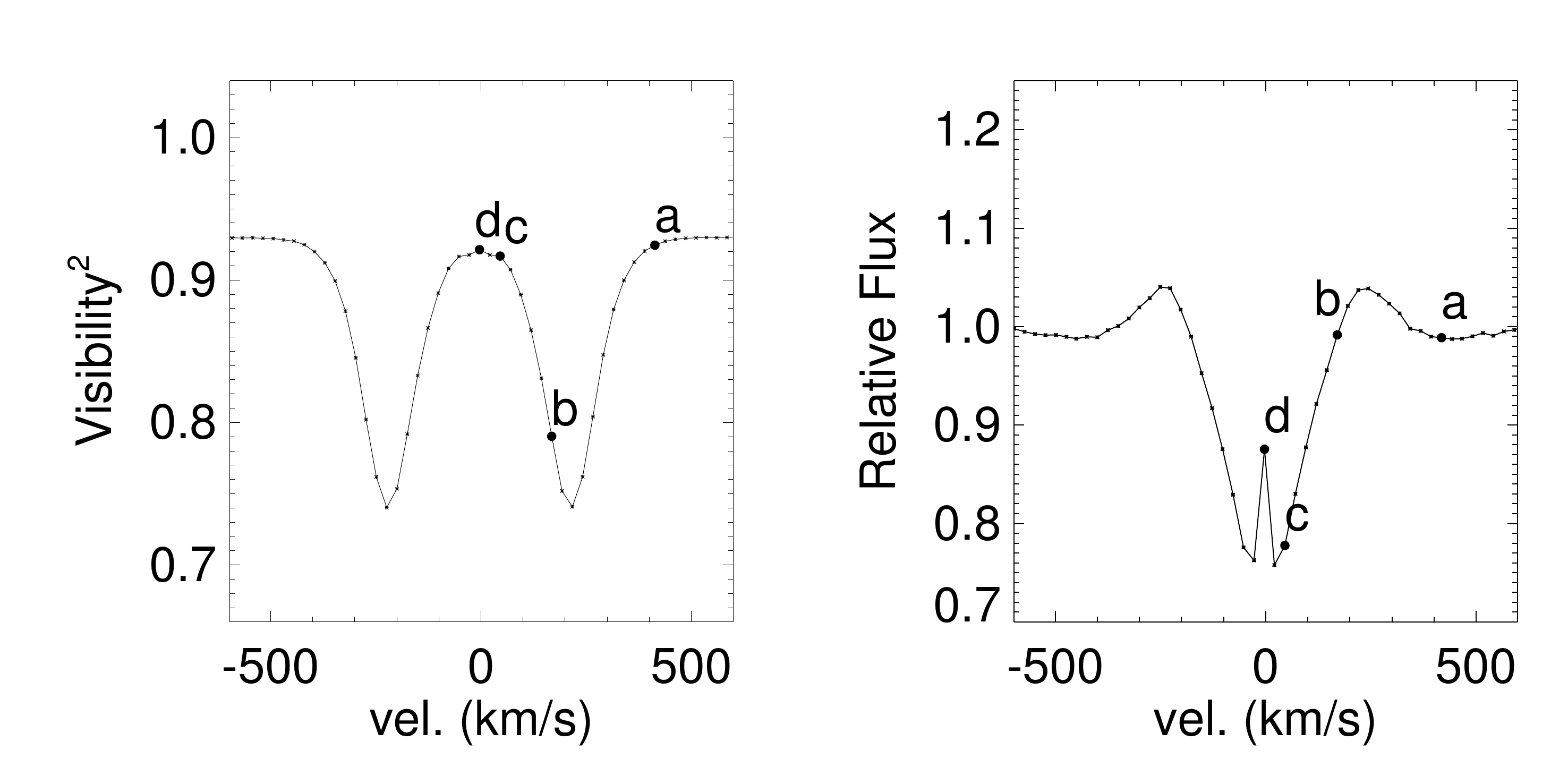} \\
\includegraphics[width=.9\linewidth]{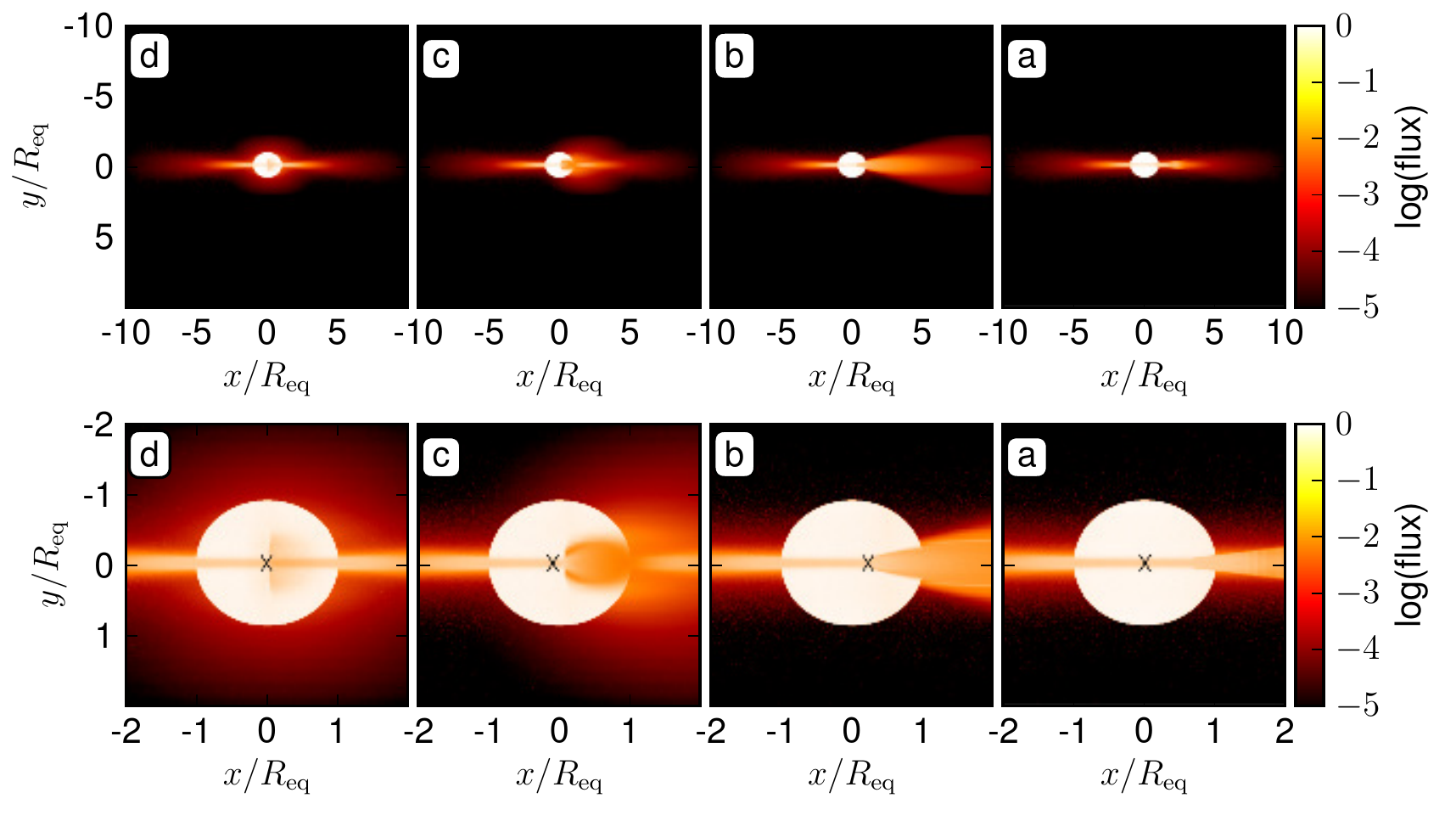} \\
\caption{Same as Fig~\ref{fig:cqe1}, but for $i=90\deg$. \citep{fae13a}.}
 \label{fig:cqe4}
\end{figure}

This process is better understood with the aid of Fig.~\ref{fig:cqe4}, where we plot the results for the reference case seen at an inclination angle $i=90^\circ$ (edge-on). At continuum wavelengths, the dark lane across the star is caused by continuum (free-free and bound-free) absorption in the CS disk (Fig.~\ref{fig:cqe4}, {velocity \textit{a}}). For the short wavelength range corresponding to the emission line profile, the continuum opacity may be regarded as constant, so we do not expect to see changes in continuum absorption in the narrow frequency range covered. In spectral channels across the Br$\gamma$ line, additional line (bound-bound) absorption of photospheric light by {H}{\sc i} atoms is seen ({velocities \textit{b} to \textit{d}}). This line absorption has a differential aspect because it depends on the line-of-sight velocity of the absorbing material. For high line-of-sight velocities, $v_{\rm proj}$, only a minor part of the star is affected by line absorption ({velocity \textit{b}}). Going towards lower velocities, i.e., towards the line center, this fraction increases, until it reaches a maximum at about 50\,km\,s$^{-1}$ ({velocity \textit{c}}). The precise value depends on the outer disk size and density. Closer to the line center, the lowest line-of-sight velocities are no longer absorbed ({velocity \textit{d}}), giving rise to the spectroscopic CQE effect. Spectroscopic CQEs are thus, in spite of their name, not related to any emission process but are an absorption phenomenon. 

\begin{figure}
\centering
\includegraphics[width=.7\linewidth]{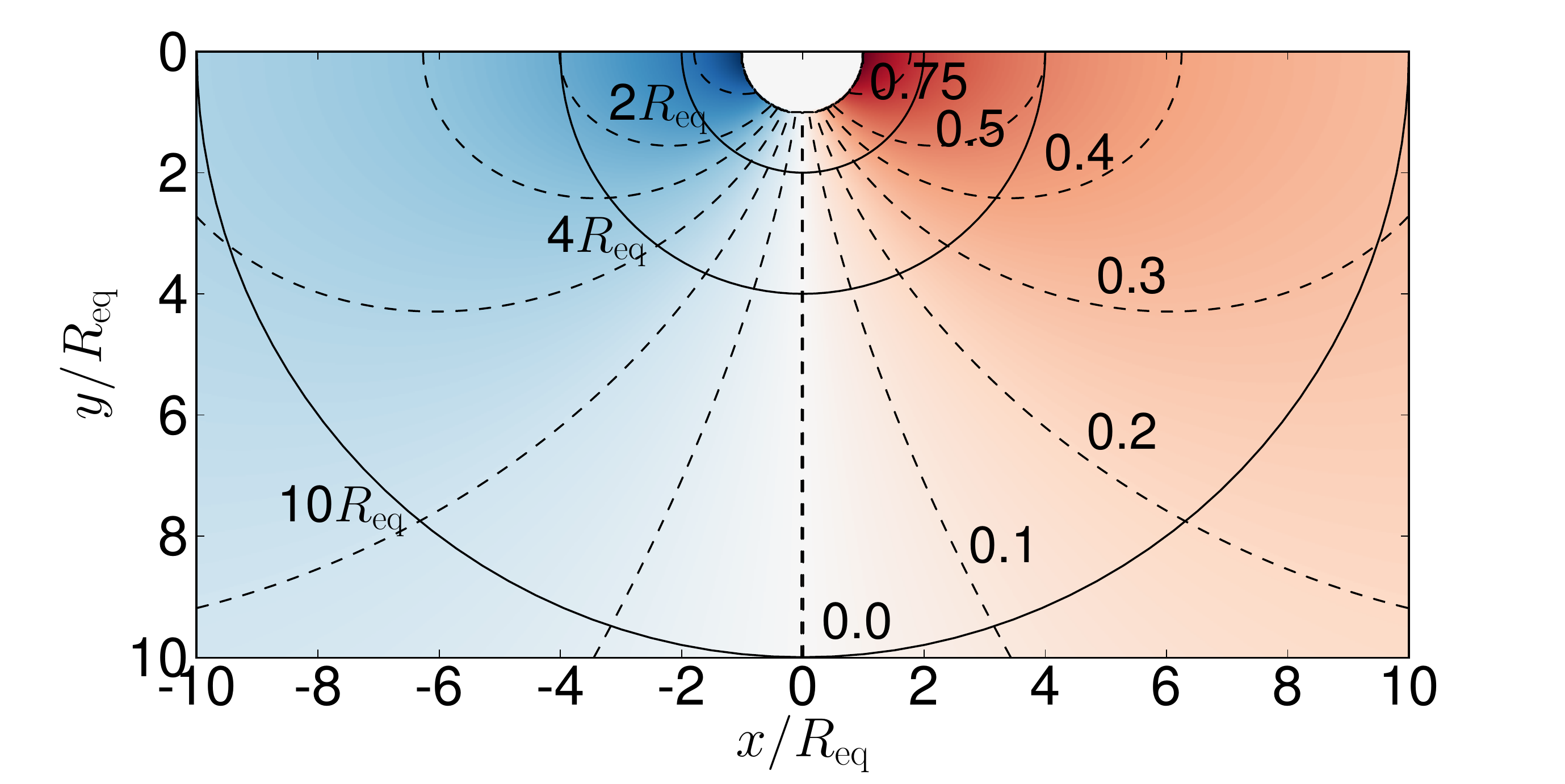} 
\caption[Isovelocity contours for a Keplerian disk]{Isovelocity contours (dashed lines) for a Keplerian disk. Each curve corresponds to a fraction of the projected orbital speed at the base of the disk ($v_{\rm orb}\sin i$), as indicated. The solid lines indicate different disk outer radii.}
 \label{fig:cqe6}
\end{figure}

As shown in Fig.~\ref{fig:cqe6}, the projected Keplerian iso-velocities over the stellar surface cover bigger disk areas as smaller velocities are considered and also the bigger the disk is. This indicates that the highest opacities occur under these conditions, i.e., large disks for small projected speeds, stating the importance of the disk size for the occurrence of the CQE-related phenomena.

The line absorption across parts of the stellar disk generates a considerable decrease in the stellar flux, which in turn will affect the photocenter position of the system, with a corresponding signal in the DP. This is clearly seen in Fig.~\ref{fig:cqe4} as the ``wiggle'' in the center of the DP profile; the highest distortion is seen where the CS absorption is strongest, i.e., at about 50\,km\,s$^{-1}$. We observe the competition between spectral line emission from the disk at line-of-sight distances from the meridian higher than $1 R_{\rm eq}$, which tends to shift the monochromatic photocenter towards one side, and disk spectral line absorption at line-of-sight distance from the meridian lower than $1 R_{\rm eq}$, which shifts the photocenter toward the opposite side. So, at wavelengths close to the line emission peak (Fig.~\ref{fig:cqe4}, velocity b), the disk emission dominates over disk absorption and the photocenter position is shifted toward the emission lobe. For wavelengths closer to the line center (Fig.~\ref{fig:cqe4}, velocity c), however, even though there is some disk emission, the absorption is more important and the photocenter position is shifted to the opposite direction of disk emission, with a corresponding change of sign of the DP. We conclude that the spectroscopic phenomenon of the CQE has an interferometrical counterpart (dubbed CQE-PS) that presents itself as a central reversal in the DP profile. We note that this central reversal is an intrinsic phenomenon and is independent of the reversals of resolved signals discussed above.

The CQE-PS occurence depends on the inclination angle on which the system is seen, since the disk must block the photospheric emission. This means that the effect will to be maximum for stars viewed equator-on, a characteristic of the so called shell stars. A list of nearby Be shell stars can be found in \citet{cat13a} and are good candidates to the look for this phenomenon observationally.

\section{ The CQE-PS diagnostic potential \label{sec:cqediag}}
\begin{figure}
\centering
\includegraphics[width=.48\linewidth]{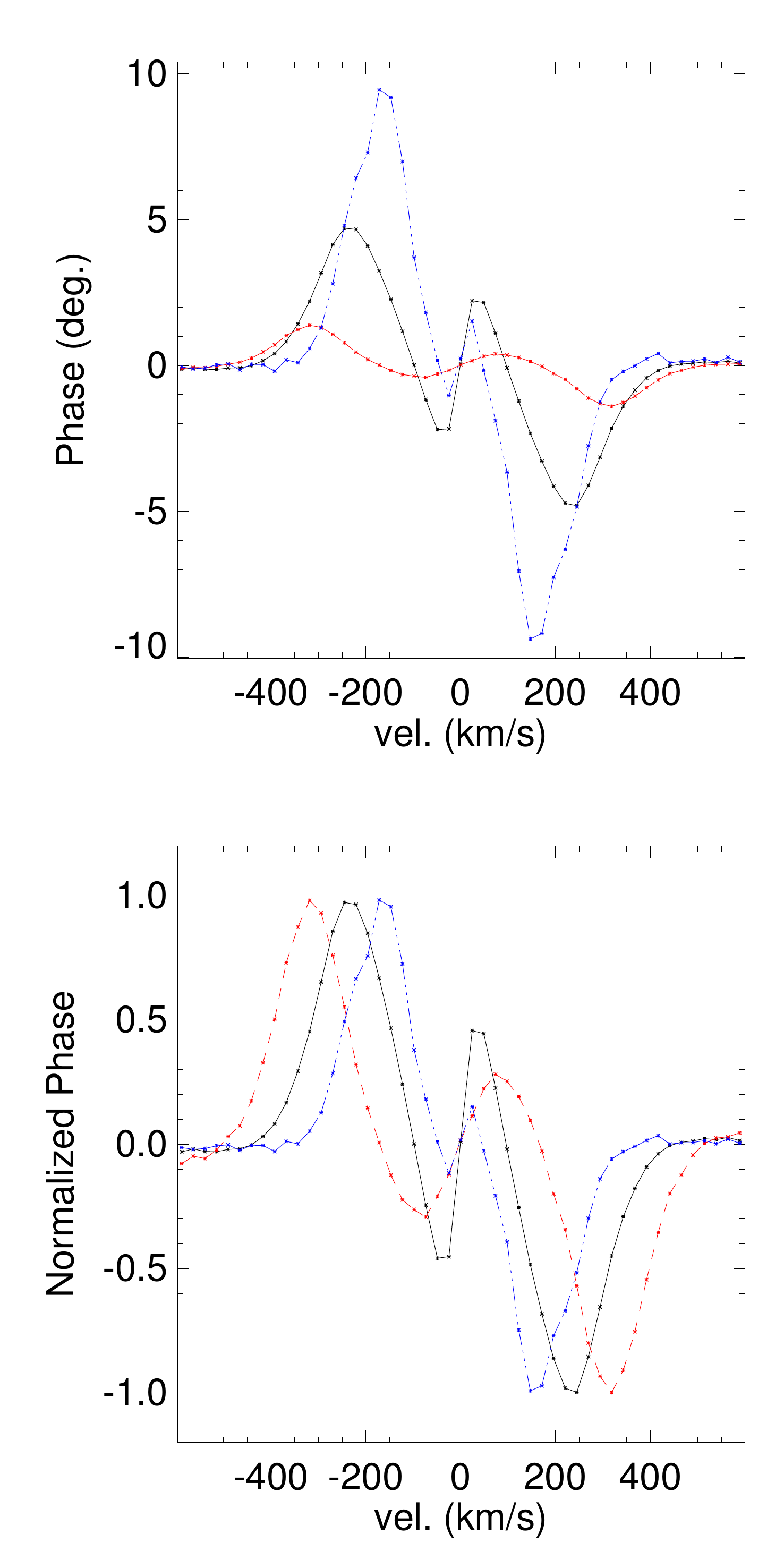}
\includegraphics[width=.48\linewidth]{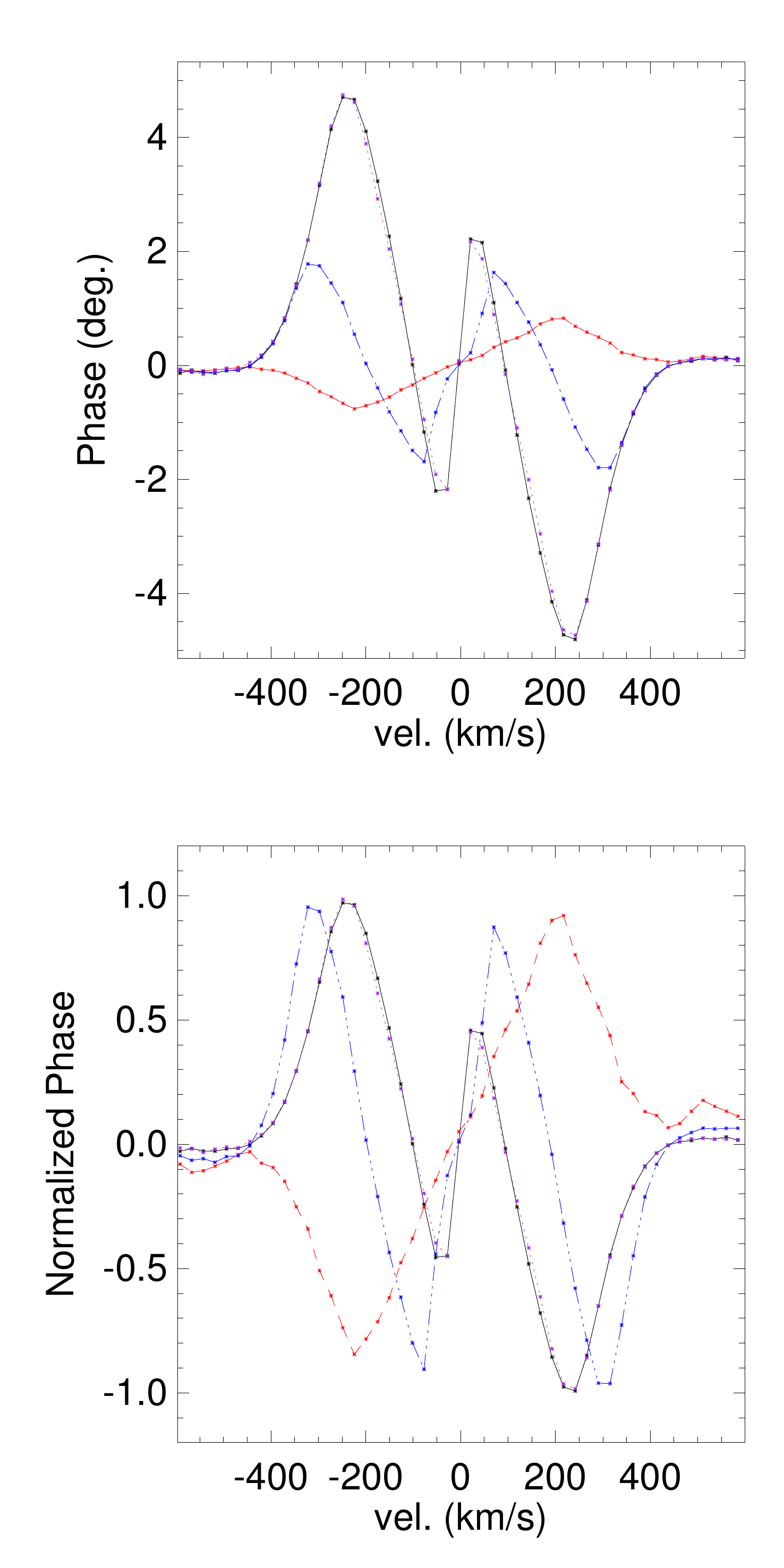}
\caption[Interferometric Br$\gamma$ differential phases, visibilities, and line profile for a Be star reference model at $i=90^\circ$]{\textit{Left:} Differential phase signal for the Be reference model at $i=90^\circ$ for low ($10^{12}\,\rm cm^{-3}$, red dashed line) and high density ($10^{14}\,\rm cm^{-3}$, blue dot-dashed line) models. The reference model is the intermediate value (black full line). \textit{Right:} The reference model at $i=90^\circ$ for different disk sizes: $R_{\rm disk}=2$ (red dashed line), 4 (blue dot-dashed line), 10 (black full line) and $20\,R_{\star}$ (purple pointed line). {The signal for $R_{\rm disk}=10$ and $20\,R_{\star}$ are almost coincident, indicating that they have the same pseudo-photospheric size (as seen in Chap.~\ref{chap:tools}). \citep{fae13a}.}}
 \label{fig:cqe1112}
\end{figure}

One example of angular displacements of photocenter within a line profile used as diagnostic of the CS envelope was done by \citet{ste96a}, where he studied the response of the photocenter as function of the envelope rotational law. So far we described the CQE-PS as dependent of the projected disk speeds, opacity and size. Thus, the actual differential phases shape defined by the CQE-PS depend on the disk
properties, having a comprehensive diagnostic potential. \citet{fae13a} demonstrated that the signal is sensitive to the CS disk density, disk size, and the radial density distribution of the CS material. Varying each of these parameters generated a characteristic shape in the differential phase diagram. This can be seen in Fig.~\ref{fig:cqe1112}, where the DP profiles of different disk base densities and sizes are compared.

Importantly, when Be stars change their brightness, this can originate in the star, the disk, or both. Although most of the observed variations are attributed to disk changes (e.g., \citealp{hau12a}), the CQE-PS can be used to assess this question because it is very sensitive to the disk part whence most of the excess flux in the visible range comes from.

The CQE-PS may even provide an estimate for the (maximum) stellar angular size. The reason is that the effect should take place over the stellar surface, since it is related to absorption of photospheric light by the disk (Fig.~\ref{fig:cqe4}). As interferometric phases are sensitive to angular displacements, this shift is proportional to the size of the stellar photosphere. More specifically, \citet{fae13a} discuss that $0< \Delta\epsilon <\dfrac{2}{5}\dfrac{R_{\rm eq}}{d}$, where is $\Delta\epsilon$ the angular displacement with the presence of photospheric absorption.

\section{Chapter summary \label{sec:cqesum}}
The interferometric DPs can directly map the target's photocenter if its marginally-resolved. For the rotating disks of Be stars, the expected DP profile is a smooth S-shape. Departures of S-shaped in (symmetric) Be disks can occur in two independent ways: (i) the target is actually resolved, and complex phase behavior is expected; (ii) additional physical processes are occurring in addition to the (pure) line emission. 

We call \textit{CQE-PS} the phase signature of the phenomenon equivalent to the spectral CQE (Central Quasi Emission) in shell stars. Its origin relies on the disk absorption of the photospheric light so that it can considerably change the photocenter position close to the line rest wavelength. The shape and amplitude of the CQE-PS depends on a number of parameters, both intrinsic to the system and related to the observational setup (baseline length and orientation, and spectral resolution). 

Using a realistic model consisting of a rotating B1\,V star with a viscous decretion disk, we made an initial study to assess to diagnostic potential of CQE-PS. Certain conditions and parameters make the effect more pronounced. Observationally, the effect is strongest for edge-on viewing (shell stars) and for baselines oriented parallel to the disk equator. 

Among the several model parameters explored, we found that the signal is sensitive to the CS disk density, disk size, and the radial density distribution of the CS material. Varying each of these parameters generated a characteristic shape in the differential phase diagram, which demonstrates the diagnostic potential of the CQE-PS. Keplerian rotation deviations could also be observed. 

A remarkable result of the CQE-PS is its ability to make a (lower) estimate of the stellar angular size even when the observed squared visibilities are close to unity. This result may extend the distance range at which such an estimate can be made with interferometry.
The full power of this diagnostic tool will be realized when applied to observations over a rotational of structural variability, as $V/R$ cycles of disk growing/dissipation process.

In Chapter~\ref{chap:aeri} we make the first report of the observation of the CQE-PS for the Be star Achernar.

\FloatBarrier
\section{Publication: Diff. phases as a sensitive physical diagnostic of circumstellar disks \label{pub:cqe}}
\href{http://adsabs.harvard.edu/abs/2013A\%26A...555A..76F}{ADS Page: http://adsabs.harvard.edu/abs/2013A\%26A...555A..76F}

%% file: chap/phots_arxiv.tex
\chapter{Resolving the photospheres of fast rotating stars \label{chap:phots}}
The possibility of resolving stellar photospheres, specially for main sequence stars, has been a dream of astronomers for centuries. To know in detail the photospheric emission of other stars is critical for verifying our understanding of the process involved in stellar astrophysics. Rotation is an ingredient that causes great impact on the stars: it prolongs stellar lifetime, modify chemical yields, and is possibly reflected to the stellar remnant at the end of their lifetime. 

Interferometric observations are sensitive to rotational effects, namely geometrical oblateness and gravity darkening. In order to understand these effects and verify a number of contemporary stellar rotational models being developed, a detailed investigation of the photosphere of rotating stars is needed. 

Bn and Be stars are main sequence B-type stars with high rotation rates. On the one hand, Be stars can be easily recognized by their episodic emission lines. On the other hand, Bn stars are characterized by their broad absorption lines with the complete absence of circumstellar emission. Therefore, they are viewed roughly equator-on. A Bn star can resemble exactly as a Be star in its quiescent phase (i.e., a phase where the circumstellar disk is absent) if it is seen edge-on. The opposite, namely that a Bn star suddenly develops strong emission line could, in principle, occur. However, there is no such observational record. So, due to their similarity, Bn and Be stars in quiescent phases can share the same techniques for investigation and reconstruction of their photospheres. An-type stars possess similar properties to Bn stars, while having smaller masses (and correspondingly, smaller surface temperatures).  

\subsubsection*{Comments on fast B-type star evolution}
Since Bn and Be stars display common properties, the question that arises is if the Be phenomenon belongs to a separate branch or a common phase in the evolution of rapidly rotating B star. 

The numbers of Be and Bn stars contained in the Bright Star Catalogue \citep{hof95a} are about equal. But global occurrence of the phenomenon is not easy to estimate: while the Be phenomenon is easily seen from any inclination angle, the Bn star are only recorded as such when seen close to equator-on viewing. Another issue is, while Bn fast rotators should remain as such throughout the main sequence due to the lack of braking mechanism for the angular momentum, the variability of the Be phenomenon makes it difficult to estimate how long the phenomenon is present during the main sequence phase.

 
In any case, occurrence of Be/Bn transitions appears to be absent. There are three main characteristics that differentiate Be and Bn stellar types and point to independent evolutionary paths: (i) no record in decades of Bn stars observations displaying any circumstellar emission; (ii) the complete absence of magnetic fields in Be stars, in contrast with a high incidence them among Bn stars; and (iii) the absence of non-radial pulsation in Bn stars, in contrast with the relatively high incidence among Be stars.

This chapter presents newly developed techniques for the photospheric characterization of  fast rotating stars applied to the Be star Achernar. Before detailing these new techniques, we reproduce part of the interferometric review of this star presented in \citet{fae15a}.

\FloatBarrier
\section{The photosphere of Achernar probed by interferometry}
Achernar ($\alpha$ Eridani, HD\,10144) is the closest and brightest Be star in the sky. The successive generations of beam combiners of ESO-VLTI have been used to study Achernar. This section summarize these early interferometric studies. All parameters derived from interferometry strongly depend on modeling (i.e., accurate brightness distribution), which is wavelength dependent. This is particularly important for stars with (potential) circumstellar emissions in addition to the photospheric one. Nevertheless, fundamental physical properties from both photosphere and CS environment could be constrained based on the high-angular resolution information available.

\subsubsection*{First oblateness determination with VLTI-VINCI}
\citet{dom03a} determined for the first time the oblateness of Achernar (and of a Be star) with $K$-band visibility data from the VINCI interferometer. It was also the first estimation of the on-sky orientation of the star. The diameter ratio between the equator and the pole $R_{\rm eq}/R_{p} = 1.56$ was found adjusting $V^2$ to each projected position angle (PA) of the observations the size of an equivalent uniform disk, an then fitting an ellipse. This resulted in an estimate of a minimum value for the radii ratio since it was a projected ratio where Achernar is seen at a viewing angle $i\neq90^\circ$. 

Although undoubtedly indicating a high rotation rate for the star, the results did not necessarily imply a rotation above the Roche limit (i.e., $W>1$): the radii ratio depends on the underlying stellar model (the fit of multiple uniform disks is just a first approximation of the stellar brightness distribution). Knowing that the star presented activity at the time of the observations, albeit small (e.g., \citealt{vin06a}), the authors have also explicitly ignored the contribution of the CS component in the data since it was a much weaker emission than the stellar one ($<5\%$).

\subsubsection*{On a polar emission in the interferometric signal}
\citet{ker06a} analyzed Achernar's VINCI data from both $K$ and $H$ bands together. They have done a single fit on them adopting a uniform ellipse brightness distribution for the star. The diameter ratio between the equator and the pole was then $R_{\rm eq}/R_{p}= 1.41$ ($W=0.906$; again a minimum value due to viewing-angle $i\neq90^\circ$).
 
In this scenario, the addition of a elongated CS envelope in the polar direction with Gaussian brightness distribution superimposed to the stellar model significantly improved the quality of the fitting. This CS component had $\approx4.5\%$ of the total flux and removed the trend that appeared in the residuals of the uniform ellipse in PAs around the polar direction. This study was made using only the VINCI data as constraint, and did not relate the polar CS emission with other observables.

\subsubsection*{The first multi-technique analyzes} 
\citet{car08b} and \citet{kan08a} have independently done a multi-technique fitting of Achernar at the time of VINCI observations. They have used SED and H$\alpha$ line profiles as constraints, and linked them with polarimetric estimates. Although taking into account the brightness distribution by the von Zeipel effect and the Roche geometry due to the high rotation of the star, they found distinct results: while \citeauthor{car08b} argue that the presence of a residual CS disk in Achernar is sufficient to explain the observed quantities with a near-critical rotating star, \citeauthor{kan08a} conclude that at the time of the VINCI observations Achernar had either a small or no CS disk, but it did have a polar, stellar wind.
\FloatBarrier
\section{Photospheres seen by Differential Phases \label{sec:dpphot}}

\begin{figure}
    \centering
    \includegraphics[width=\linewidth]{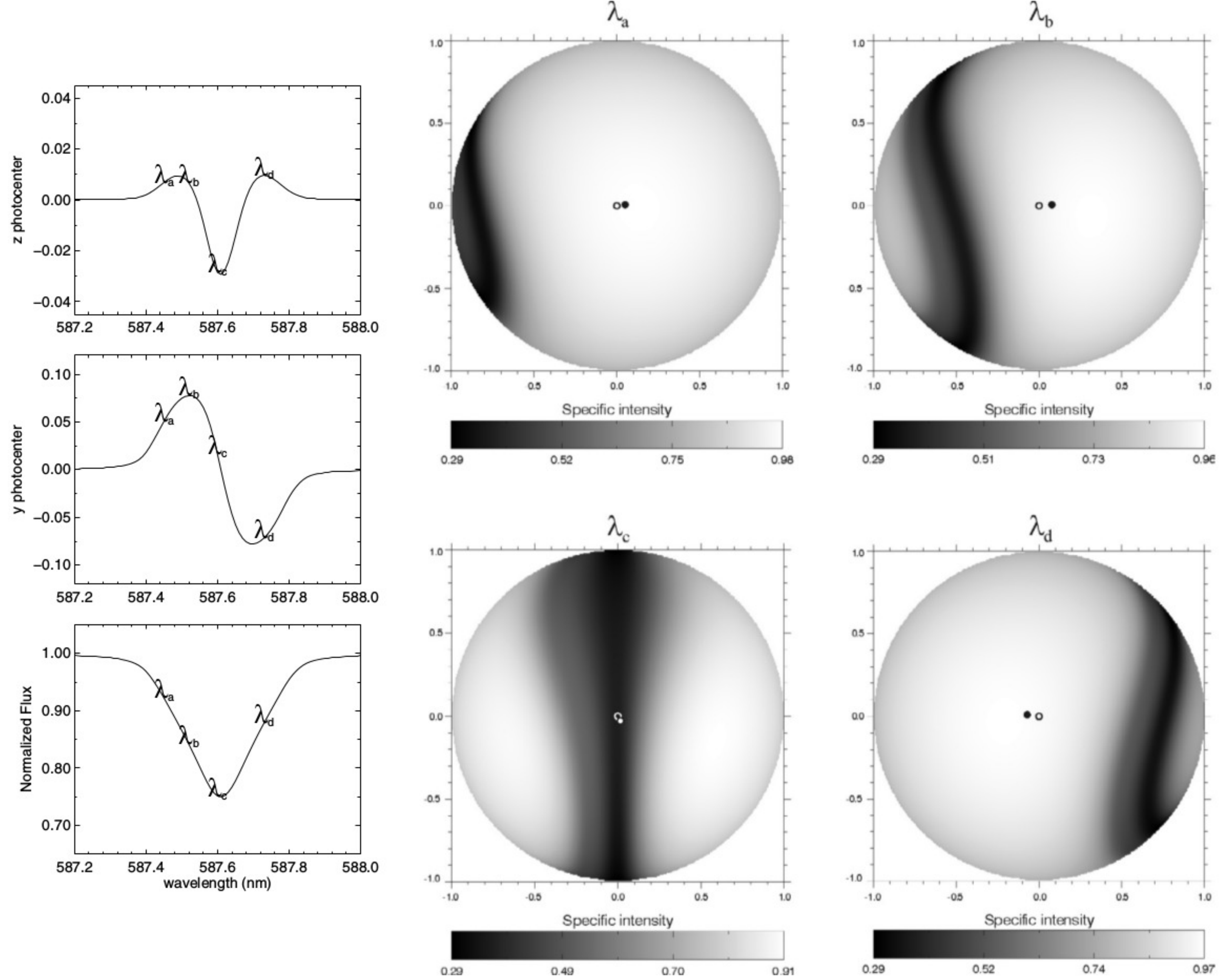}
    \caption[A differential rotating star and photocenter displacements within a spectral line (He\,{\sc i}\,5876\,\AA)]{A differential rotating star and photocenter displacements within a spectral line (He\,{\sc i}\,5876\,\AA). \textit{Left}: photocenter components, vertical (z; top) and horizontal (y; middle), and the normalized spectral flux (bottom) across the (asymmetric) line. Photocenter components are given in units of angular stellar radius. The letters indicate selected wavelengths corresponding to a region in the blue line wing ($\lambda_a$), the highest y value ($\lambda_b$), the central wavelength ($\lambda_c$), and the most positive value of z ($\lambda_d$). \textit{Right}: intensity maps associated to the spectrointerferometric observables on the left side at the four selected wavelengths. The curved dark patterns correspond to Doppler shifts of the local line profile caused by differential rotation. These non-symmetrical intensity maps result in a displacement of the stellar photometric barycenter, i.e., the photocenter (filled circles), relative to the geometrical center (opened circles). \citep{dom04a}.}
    \label{fig:photc}
\end{figure}

We have seen at that the quantity recorded in interferometry is the normalized complex visibility $\tilde{V}=\|V\|\exp(i\phi)$, where a common quantity used in high spectral resolution interferometry are the differential phases $\phi_\textrm{diff}$ (Sect.~\ref{sec:optinterf}). Also, the interferometric DPs can be associate to photocenter positions, as discussed in Chap.~\ref{chap:bes}. The relation of the phase signal and the photocenter position is useful when interpreting the differential phases signal of fast rotating stars. When looking at a rotating star in spectrointerferometry at a given spectral line, we are actually seeing at each point of the photosphere, the locally emitted flux shifted by the projected velocity in the line-of-sight. An example is Fig.~\ref{fig:photc}, where the photocenter displacements are shown for a differential rotating star. For a rigidly rotating star seen equator-on, for instance, as we go from $-v_{\rm rot}$ to $v_{\rm rot}$ spectral channels, we see a moving strip in all latitudes from one edge to another, passing over the rotation axis when $v_{\rm proj}=0$ (line rest wavelength). The width of the line is proportional to the spectral channel range, i.e., the higher the spectral resolution on which the stellar surface is seen, the narrower the strip will appear (limited by atmospheric thermal or turbulent velocities).  

When measuring the photosphere in continuum wavelengths, the interferometric phase signal points to the center of the projected disk (or ellipsoid, in case of fast-spinning stars) due to the lack of asymmetries in brightness. The presence of the strip within the line profile spectral channels will alter this. Since it is an absorption line, the photocenter is shifted towards the position opposite to the strip appearance. The interferometric phase signal for this scenario is a smooth S-shaped curve, centered at rest wavelength and with signal reaching the continuum at $\pm v\sin i$ wavelengths. This make An/Bn stars, with high $v\sin i$ values, excellent targets for applying spectrointerferometry.

The amplitude of the phase signal is proportional to the distance of the strip relatively to the stellar center, to the line depth in relation to continuous emission and to the length of the strip. The interferometric observations do not need to be in the marginally-resolved regime to apply this technique. What happens is that the phase signal loses the correspondence with the photocenter position but still depends on it. It is still possible to obtain information about the photocenter position, but this requires assumptions about the brightness distribution of the target (this was already discussed in Chap.~\ref{chap:bes}).

The detailed DPs values, in any regime, will depend on the stellar parameters and can be used to recover the brightness distribution seen by the interferometer. \citet{dom12a} show the expected DPs signal and its dependence with different stellar parameters.

I participated on the first determination of stellar parameters based on spectrointerferometry done by \citet{dom12a} for the Be star Achernar based on VLTI-AMBER data, taken between October to November 2009, and the \textsc{charron} code. Since it is a Be star, it was important to the parameters assertion that at the time of the interferometric observations no circumstellar emission was present. My main contribution to this study focused on this problem, determining if spectroscopic or polarimetric data indicated any circumstellar activity at the times of the AMBER observations. 
During all the considered period, the linear polarization levels were $P\lesssim0.01(1)\%$ and the difference spectra, compared from the pure photospheric reference, were $\Delta$flux~$\lesssim0.01(1)$\%. This allowed us to employ purely photospheric models when analyzing the AMBER data. 

The derived equatorial angular diameter to Achernar was compatible with previous values from visibilities, and the diameter ratio was below the Roche limit ($R_{\rm eq}/R_{p}=1.45; W=0.949$). The list of best-fitted and respectely derived parameters is in Table~\ref{tab:ambaeri}.
This first photospheric study with DPs was followed by \citet{had14a}, who applied the technique to the An/Bn stars Altair, $\delta$ Aquil\ae{}, and Fomalhaut using the \textsc{scirocco} code \citep{had13a}.

\begin{table}
\centering
\caption[Physical parameters of Achernar derived from AMBER differential phases at Br$\gamma$ line]{Physical parameters of Achernar derived from AMBER differential phases at Br$\gamma$ line with the \textsc{charron} code \citep{dom12a}.}
\begin{tabular}[]{cc}
\toprule
    {Model free parameters fitted} & {Values and uncertainties}\\ 
\midrule
    Equatorial radius ($R_\odot$) & 11.6(0.3) \\
    Equatorial rotation velocity (km\,s$^{-1}$) & 298(9) \\
    Rotation-axis inclination angle ($^\circ$) & 78.5(5.2) \\
    Position angle of the visible pole ($^\circ$) & 214.8(1.6) \\
\midrule
    {Fixed and Derived params.} & {Values} \\
\midrule
    Gravity darkening coefficient $\beta$ & 0.20 \\
    Rotation rate: $W$ ($v_{\rm rot}/v_{\rm orb}$) & 0.96 \\
    Radii ratio: $R_{\rm eq}/R_{p}$ & 1.45 \\
\bottomrule
\end{tabular}
\label{tab:ambaeri}
\end{table}

The DPs allows assessing information that goes beyond the diffraction limit of the optical interferometers since it is sensitive to photocenter position (sometimes referred as ``super-resolution''). 
However, this analysis is based on strong assumptions about the photospheric absorption component: (i) the continuum emission is symmetric and determined only by the photosphere (i.e., there are no CS emission features); (ii) the star must have a known pattern of rotation (e.g., rigid rotation) and stellar shape (e.g. projected Roche surface); (iii) the effect of gravitational darkening is properly considered. Deviation from these assumptions will most likely alter the derived photospheric parameters. 
Although the method reproduce the spectrointerferometric data, the results obtained by \citet{had14a} indicate that a better error determination for the derived parameters is needed (see Sect.~\ref{sec:photconc}).

\section{Photospheres seen by Visibilities amplitudes and Closure Phases \label{sec:pion}}
In Sect.~\ref{sec:optinterf} we said that closure phases add quantitative new information to the interferometric measurements. It also enables the photospheric images techniques to operate, where Fourier analysis allows to rebuild arbitrary brightness distributions. However, these tools may require fine-tunning of numerical parameters for data fitting and sometimes strong hypotheses about the brightness distribution properties of the target need to be made.

One way of avoid using these tools is, instead of analyzing the signal to rebuild the image, to create the expected images from the much fewer unknown parameters (i.e., the stellar photosphere parameters) and verify how they reproduce the observational data. With this methodology the Be star Achernar was revised applying the high precision visibilities measurements with closure phases information from the VLTI-PIONIER interferometer by \citet{dom14a}.

The parametrization of the photosphere contained in the \citeauthor[(op. cit.)]{dom14a} work is based on the Roche surface for a rigid rotating star and a generalized von Zeipel description of the gravity darkening effect making use of the \textsc{charron} code. This is here called RVZ (Roche-von Zeipel) model. 
For that work, I implemented a interface between \textsc{charron} and the MCMC ensemble sampler \textsc{emcee}. This enabled the determination of the photospheric parameters that best-fitted the PIONIER data, their uncertainties, and possible correlations between the parameters determined. 
\begin{figure}
    \centering
    \includegraphics[width=1.\linewidth]{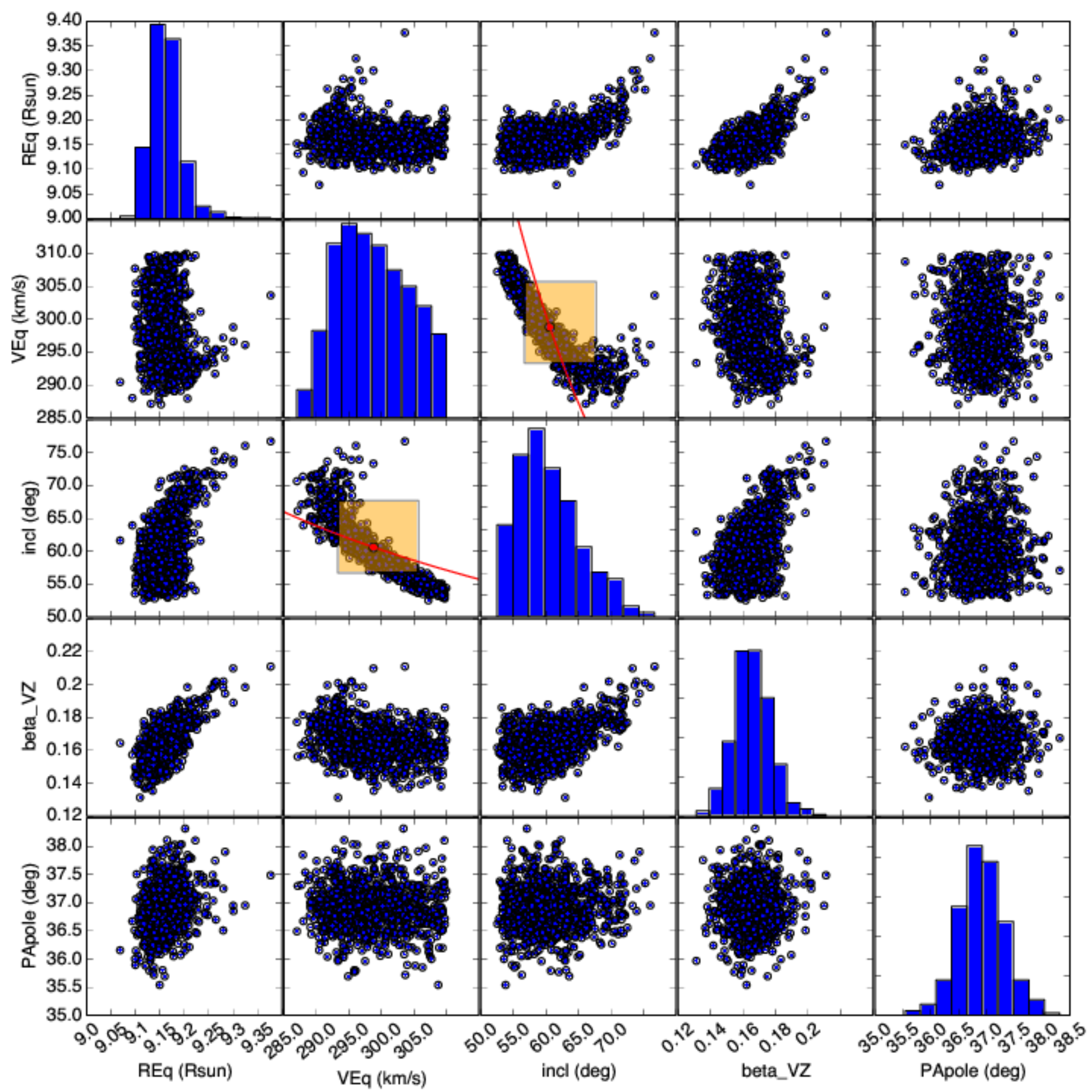}
    \caption[Probability density functions of the derived stellar parameters for Achernar and their correlation maps, corresponding to the PIONIER visibilites and closure phases minimization]{Probability density functions (PDFs) of the derived stellar parameters for Achernar and their correlation maps, corresponding to the PIONIER visibilites and closure phases minimization with \textsc{charron} code. \textit{From left to right (or top-bottom):} equatorial radius (Solar unit), equatorial velocity (km\,s$^{-1}$),  inclination angle (degrees), von Zeipel coefficient, and on-sky position angle of the rotation axis (degrees). The derived values to $\{V_{\rm eq}, i\}$ are indicated as a dot along the $V_{\rm eq}\sin i$ constant value line (in red).}
    \label{fig:aeri_correl}
\end{figure}


The assumptions for the (parametric) modeling of the photospheric emission are basically the same as in the differential rotation case. The list of RVZ model free parameters necessary to generate a synthetic photospheric model is in Table~\ref{tab:aeri14}. Fig.~\ref{fig:aeri_correl} shows the minimized paramaters PDFs and their correlations. The inclination angle $i$ is clearly correlated with the equatorial velocity $V_{\rm eq}$, so that the derived pair $\{V_{\rm eq}, i\}$ is indicated along the $V_{\rm eq}\sin i$ constant value. 

\begin{table}
\centering
\caption[Physical parameters of Achernar derived from the RVZ model fitting]{Physical parameters of Achernar derived from the RVZ model fitting (\textsc{charron} code) to VLTI-PIONIER $H$ band data using the MCMC method \citep{dom14a}.}
\begin{tabular}[]{cc}
\toprule
    {Model free parameters fitted} & {Values and uncertainties}\\ 
\midrule
    Equatorial radius ($R_\odot$) & 9.16$^{+0.23}_{-0.23}$ \\
    Equatorial rotation velocity (km\,s$^{-1}$) & 298.8$^{+6.9}_{-5.5}$ \\
    Rotation-axis inclination angle ($^\circ$) & 60.6$^{+7.1}_{-3.9}$ \\
    Gravity-darkening coefficient ($\beta$) & 0.166$^{+0.012}_{-0.010}$ \\
    Position angle of the visible pole ($^\circ$) & 216.9$^{+0.4}_{-0.4}$ \\
\midrule
    {Derived parameters} & {Values} \\
\midrule
    Rotation rate: W ($v_{\rm rot}/v_{\rm orb}$) & 0.838 \\
    Radii ratio: $R_{\rm eq}/R_{p}$ & 1.352 \\
\bottomrule
\end{tabular}
\label{tab:aeri14}
\end{table}

\subsubsection*{Residual disk characterization}
The method described can only be applied to the photosphere of the star alone. The presence of any circumstellar emission should be taken into account in determining the photospheric parameters. In \citeauthor[(op. cit.)]{dom14}, I actively contributed to the investigation of spectroscopic and polarimetric data at the times of PIONIER observations of Achernar, greatly extending the study presented in \citet{dom12a}. Within the observational uncertainties, no positive activity was detected (in polarization, $P\lesssim0.01(1)\%$ and the difference spectra to the photospheric profile, $\Delta$flux~$\lesssim0.01(1)$\%). The negative detection in these techniques do not exclude the possibility of a tenuous residual disk. Although not detectable by those techniques, a residual disk emission could change the interferometric signal and alter the determination of the photospheric parameters.

In order to characterize the densest possible residual disk around Achernar within the observational constrains available and how such disk might affect the interferometric measurements, simulations were performed with the \textsc{hdust} code. For that, we used the synthetic photosphere parameters determined by \textsc{charron} code and the VDD description for the disk. 

\begin{figure}
    \centering
    \includegraphics[width=.48\linewidth]{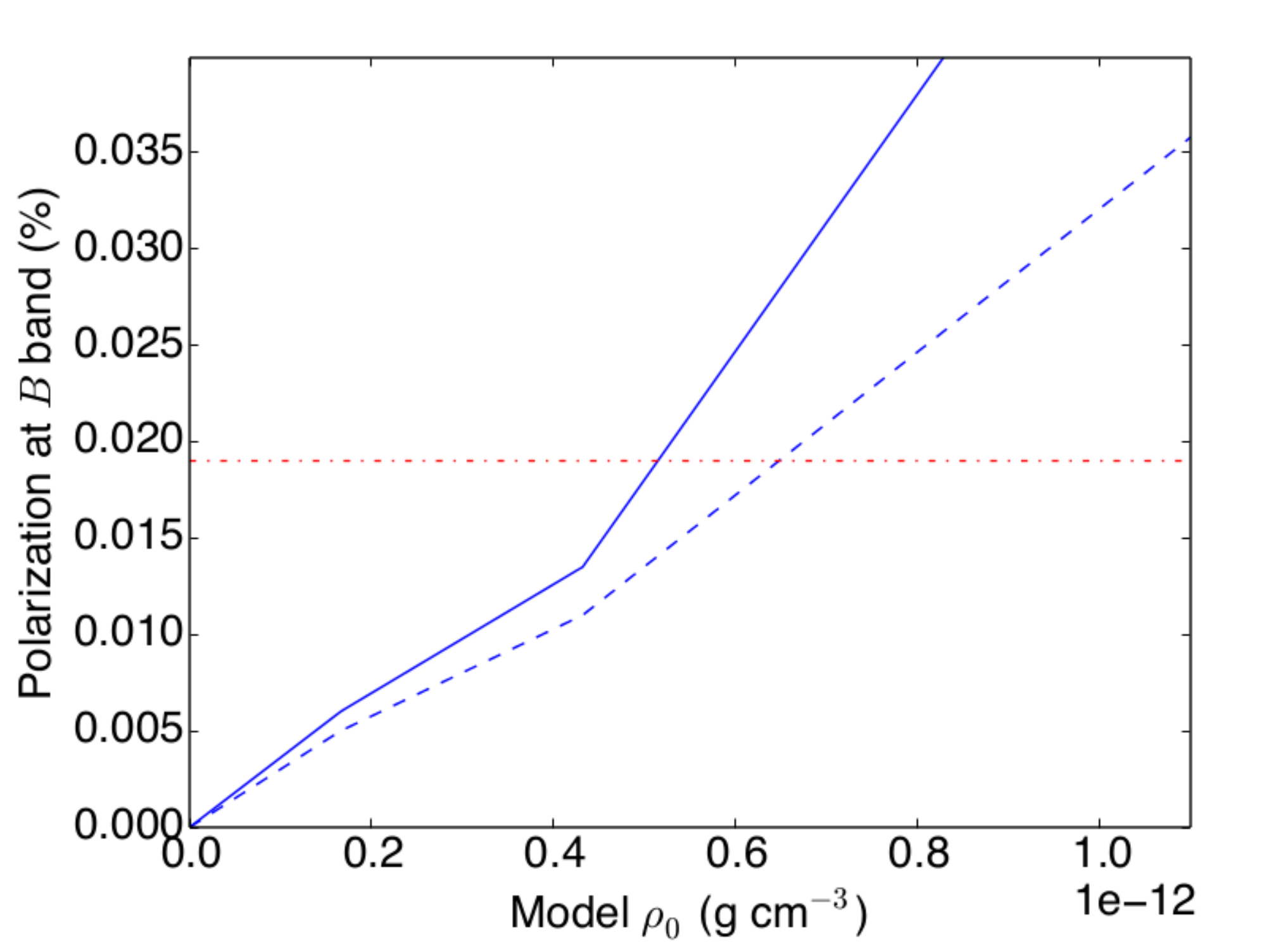}
    \includegraphics[width=.48\linewidth]{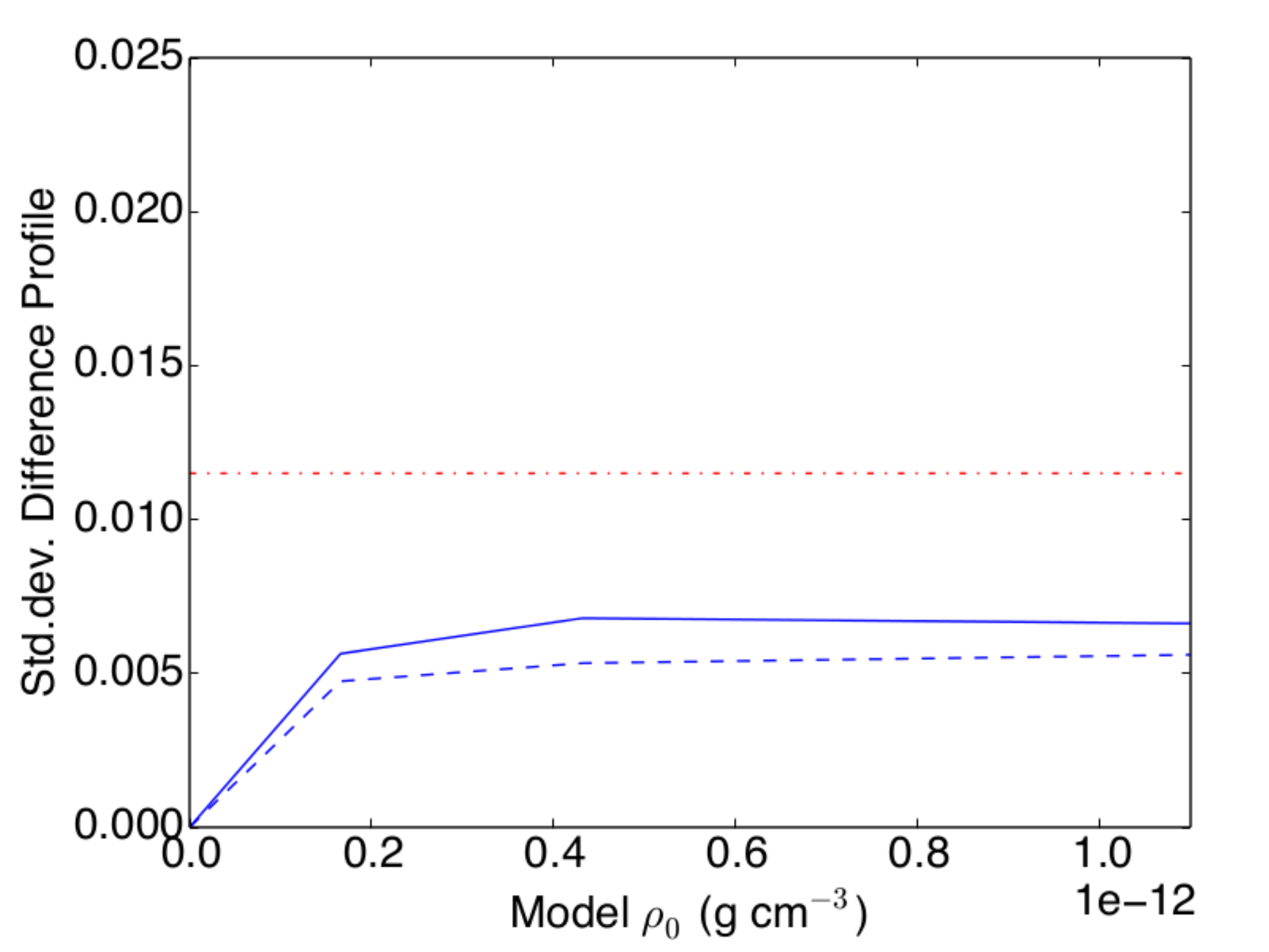}
    \caption[$B$-band polarization and standard deviation of difference H$\alpha$ profile as function of disk base density $\rho_0$ for the residual disk models based on \citet{car08b}]{(\textit{Left}) $B$-band polarization and (\textit{right}) standard deviation of difference H$\alpha$ profile as function of disk base density $\rho_0$ for the residual disk models based on \citet{car08b}. The standard deviation is the root mean square of the disk model spectrum minus the pure photospheric profile. The horizontal dot-dashed lines represent the observational limits of average polarization and difference profile determined from observations performed at the epoch of interferometric observations. While the measured H$\alpha$ difference profile level does not impose a limit to $\rho_0$, the measured average polarization sets a strict upper limit to this quantity: $\rho_0 \lesssim 0.52 \times 10^{-12}$~g\,cm$^{-3}$ \citep{dom14a}.}
    \label{fig:aeri4}
\end{figure}

Fig.~\ref{fig:aeri4}, containts the $B$-band polarization and the standard deviation of difference H$\alpha$ profile as a function of disk base density ($\rho_0$) on the residual disk models from \citet{car08a}. That set the maximum base density of $\sim0.52\times10^{-13}$~g\,cm$^{-3}$ for the disk to be undetectable. A comparison between a (pure) photometric model with the residual disk set with the upper density limit were carefully conducted to (i) H$\alpha$ line profile; the interferometric observables (ii) squared visibilities ($V^2$) in the $H$ band, and (iii) Br$\gamma$ differential phases. We concluded that the residual disk could not considerably change the PIONIER data without first alter other observables available, notably polarimetry. So, we concluded that no further component, in addition to the photosphere, were needed to analyse the PIONER data.


\subsubsection*{The von Zeipel coefficient variation with rotation}
As the photosphere of stars starting been resolved by interferometry, it became clear that the stars do not follow the von Zeipel prescription to the gravity darkening effect, namely, $\beta=0.25$ or $T_{\rm eff}\propto g_{\rm eff}^{1/4}$. The von Zeipel's $\beta$ appears to decrease the fastest the star rotates. 

\citet{esp11a} (ELR) model predicted the dependence of the $\beta$ parameter as a function of stellar rotation, assuming that the energy flux is a divergence-free vector, antiparallel to the effective gravity. So, the flux $\vec{F}$ has the dependence $\vec{F}=-f(r,\theta)\vec{g}_{\rm eff}$, where $f$ is a function of stellar surface position. In doing so, it allows the ratio $\|\vec{F}\|/\|\vec{g}_{\rm eff}\|$ to vary with latitude. The prescription was tested using fully two-dimensional models of rotating stars produced by the ESTER code.

The first Achernar parameters reconstruction \citep{dom12a} was done with von Zeipel's $\beta$ parameter fixed in 0.20. This was done, essentially, due to the weak dependence of the differential phases signal with respect to this parameter. That is not the case when looking at visibilities with closure phase information. The PIONIER information allows fitting $\beta$ simultaneously with the other photospheric parameters.

\begin{figure}
    \centering
    \includegraphics[width=.55\linewidth]{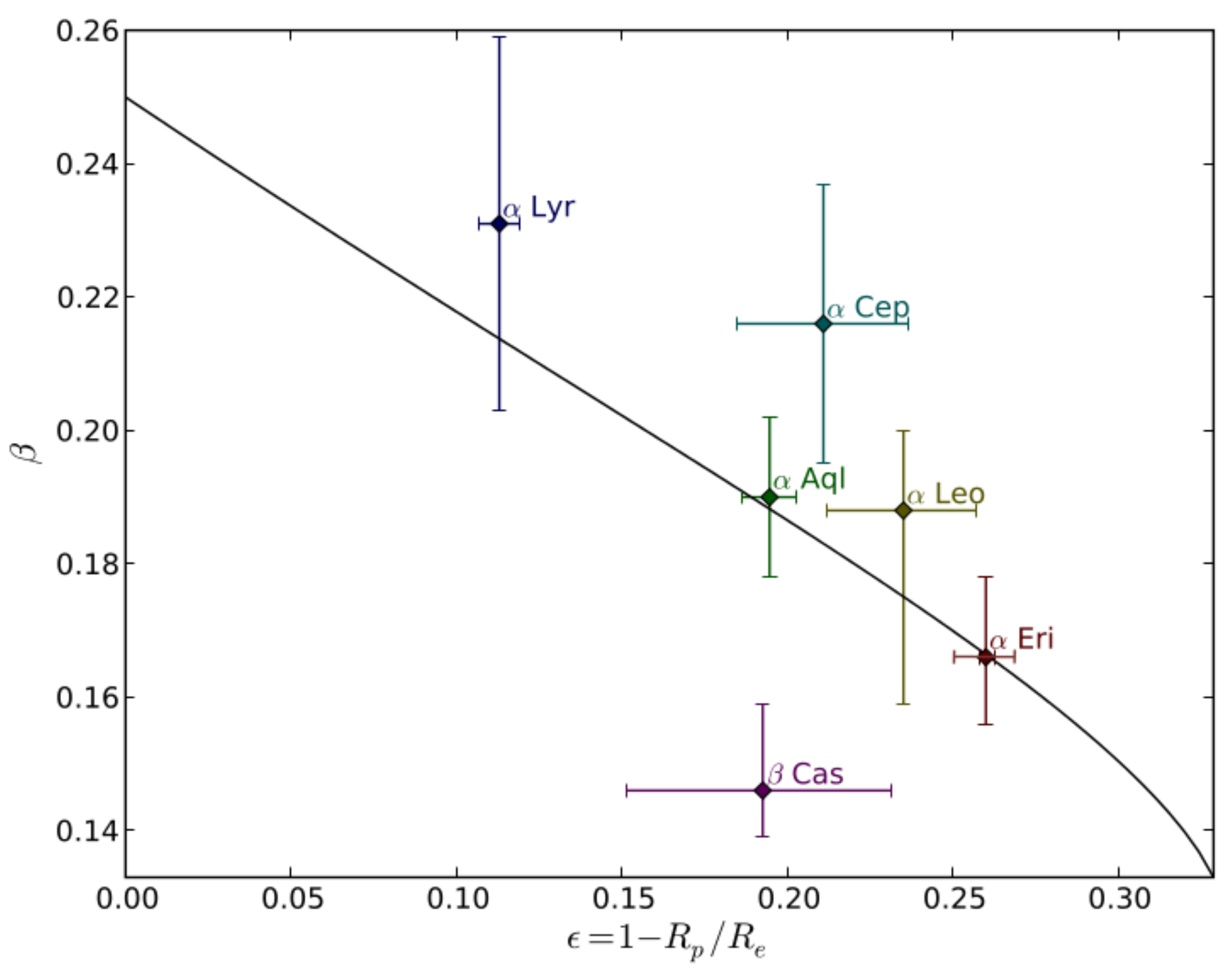}
    \caption[Gravity darkening coefficient $\beta$ from the ELR model as function of the rotation flattening compared to observations]{Gravity darkening coefficient $\beta$ estimated from the ELR model \citep{esp11a} as a function of the rotation flattening compared to values measured from interferometric observations of six rapidly rotating stars (Table~\ref{tab:introW}), Achernar being the flattest one. The estimation of $\beta$ is obtained from a fit to the $T_{\rm eff}$ vs$.$ $g_{\rm eff}$ curves directly predicted by the ELR model. The ELR model predictions and interferometric measurements have a good general agreement \citep{dom14a}.}
    \label{fig:gd}
\end{figure}

The derived $\beta$ value for Achernar, 0.166\,($^{+0.012}_{-0.010}$), matches very well the predicted value by the ELR model for its rotational rate. The derived temperature structure for the photosphere is illustrated in the Fig.~\ref{fig:gdob}. Slight differences between the RVZ and the ELR models exists, but they are much smaller than the derived uncertaints. Achernar is the fastest and more massive star with detailed photospheric information, and the results supports the dependence of the (effective) gravity darkening coefficient an the stellar rotational rate as shown in Fig.~\ref{fig:gd}.

\subsubsection*{The model-independent image reconstruction}

\begin{figure}
    \centering
    \includegraphics[width=\linewidth]{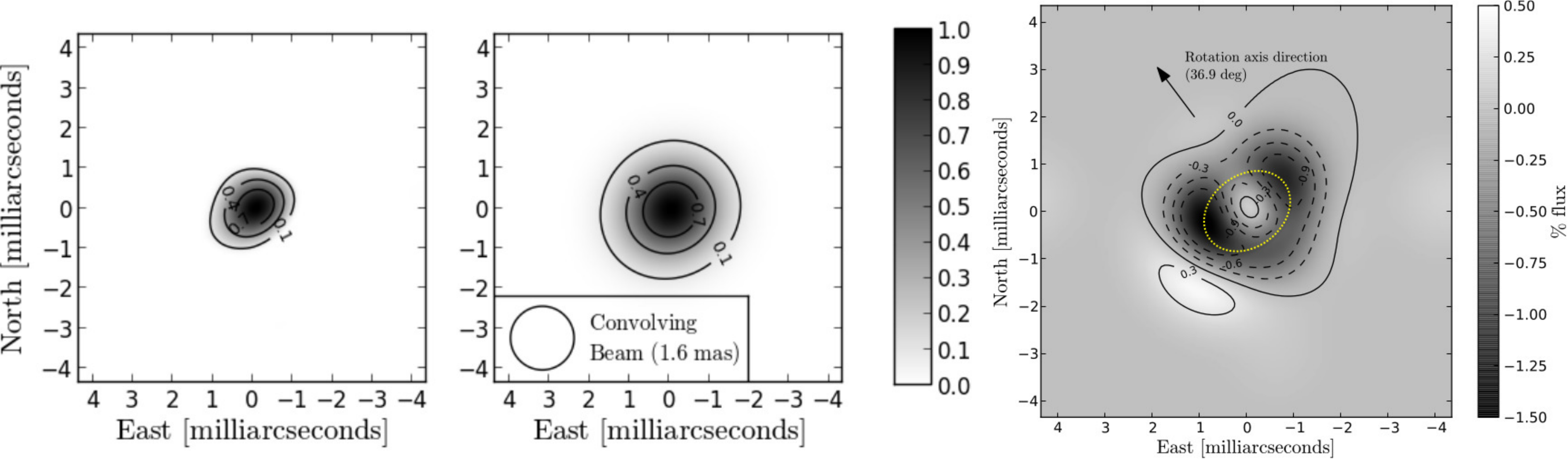}
    \caption[Image reconstruction of Achernar obtained by applying the MIRA software to the $H$ band VLTI/PIONIER observations and its comparison to the best-fit \textsc{charron} image]{\textit{Left}: Image reconstruction of Achernar obtained by applying the MIRA software to the $H$ band VLTI/PIONIER observations. \textit{Middle}: Convolution of this reconstructed MIRA image convolved by a Gaussian beam of FWHM = 1.6 mas $\left(= 0.61{\lambda}/{\|\vec{B}_{\rm proj}^{\rm max}\|}\right)$ corresponding to the diffraction limit of the PIONIER observations. \textit{Right}: Reconstructed MIRA image minus the best-fit \textsc{charron} image (convolved by the Gaussian diffraction limit beam to match the resolution of reconstructed image). The difference between the images is very small ($\lesssim1.5\%$ in modulus, relative to the total MIRA image flux), indicating that essentially only the photosphere of Achernar is contributing to the PIONIER data, without any additional circumstellar component. The dotted ellipse represents roughly the border of the apparent photosphere of Achernar given by the best-fit \textsc{charron} RVZ model.}
    \label{fig:rec}
\end{figure}

Once the closure phases information are present, it is possible to apply regular image reconstruction techniques and compare with the employed methodology of model fitting. This model-independent study was then applied to the PIONIER data of Achernar, using MIRA software (Multi-aperture Image Reconstruction Algorithm; \citealp{thi08a}) and using the parametric modeling as the prior image. 
Fig.~\ref{fig:rec}, shows the reconstruction image of Achernar and its difference to the best-fit \textsc{charron} image. The final images agreed within $\sim\pm1\%$ of the intensity level, indicating that essentially only the photosphere of Achernar is contributing to the PIONIER data, without any additional circumstellar component. It also asserts the precision of the employed methods and the corresponding derived parameters.

\FloatBarrier
\section{Chapter summary \label{sec:photconc}}
The possibility of angularly resolving main sequence stars was achieved in the beginning of this century with optical interferometry. This opens new windows to stellar astrophysics study, allowing for studying the origin, role and impact of rotation on stars. For example, the discover that Vega is a rapid rotating star \citep{auf06a} lead to significantly revisions of our view of this fundamental standard \citep{rie08a}.

Both interferometric quantities, visibilities and phases, can be used to the determination of photospheric parameters, while model-independent image reconstruction requires the combination of both. Since the absolute phase information is destroyed by the atmosphere, the combined phase information of three telescopes can be recovered with the closure phases information - enabling the interferometric reconstruction techniques.

Here we show the results to the photospheric properties determination of Achernar for three different techniques, namely parametric minimization of (i) differential phases signal; (ii) visibilites and closure phases; and (iii) model-independent image reconstruction.

\begin{figure}
\begin{center}
\includegraphics[width=0.49\columnwidth,angle=0]{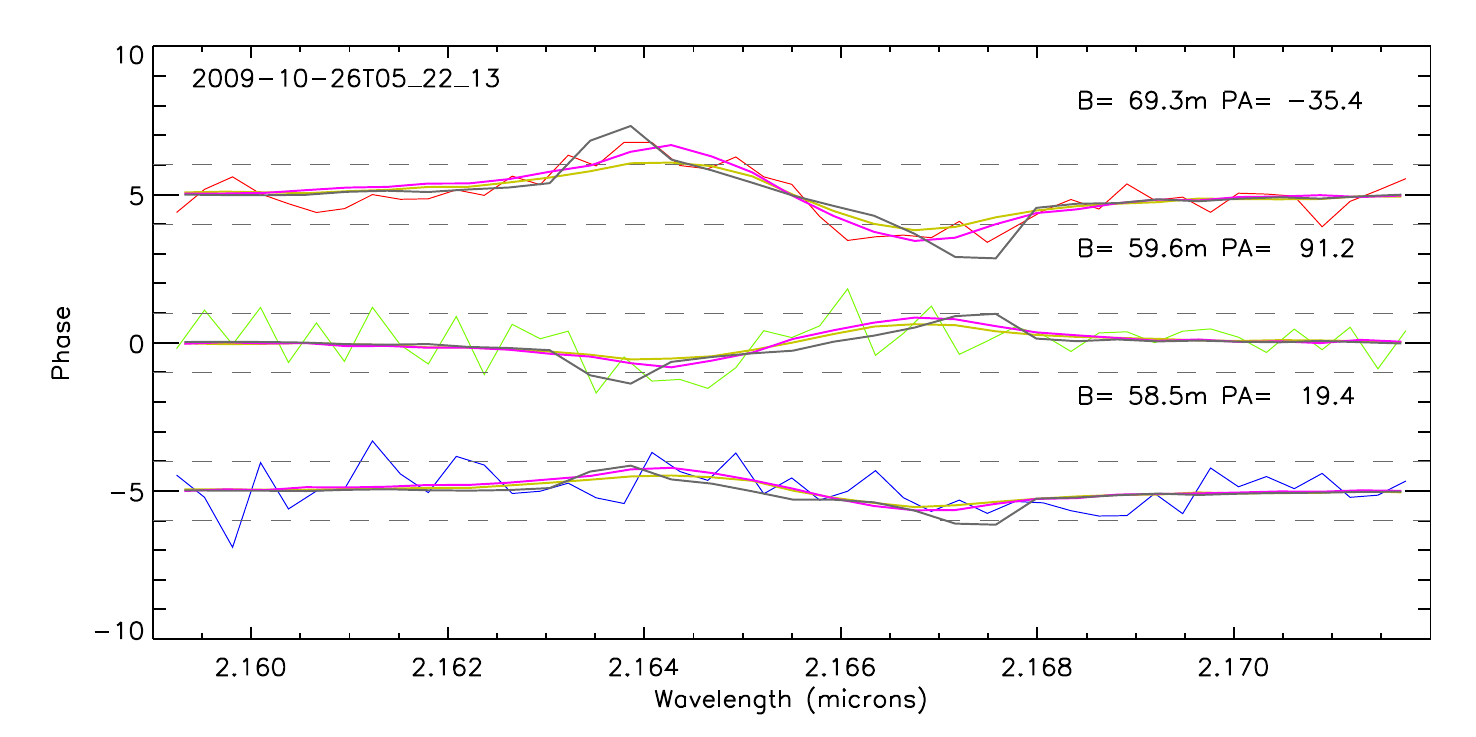}
\includegraphics[width=0.49\columnwidth,angle=0]{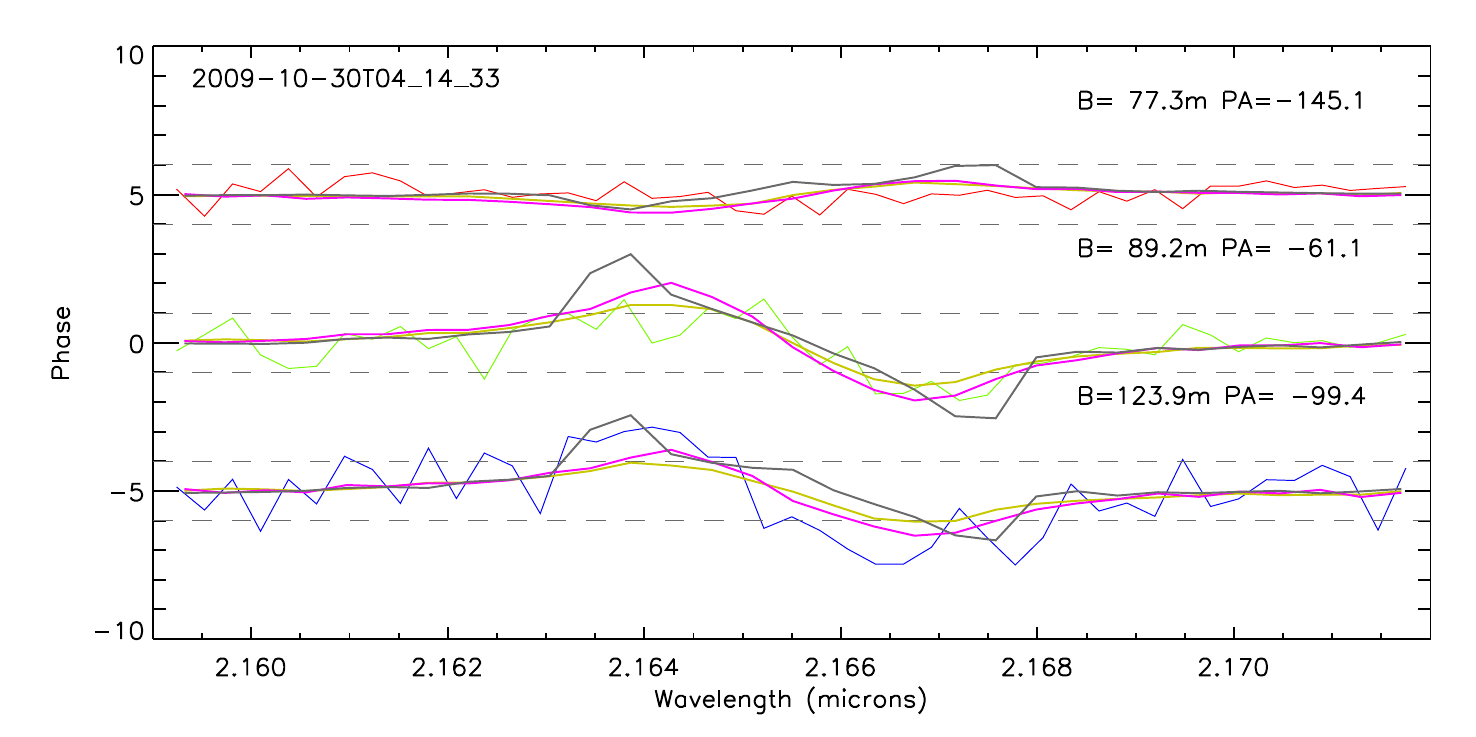}
\caption[AMBER-VLTI photospheric observations of Acherar compared to different models.]{AMBER-VLTI pure photospheric observations of Achernar compared to (i) photospheric model of \citet{dom12a} (purple line), (ii) photopheric model of \citet{dom14a} (yellow line) and (iii) the latter model with a VDD steady-state residual disk with $R=1.75R_{\rm eq}$ and volumetric particle density of $n_0=5\,10^{12}$~cm$^{-3}$ ($\rho_0\sim4.2\,10^{12}$~g\,cm$^{-3}$).} 
\label{fig:degener}
\end{center}
\end{figure}

The differential phases method was shown to be a useful quantity for determining the photospheric parameters. However, it is based on a differential quantity and strong assumptions are imposed when determining stellar parameters. In the specific case of Achernar, it resulted in a quite big star ($R_{p}\sim9~R_\odot$, by both \citealp{dom12a,had14a}) and could not constrain the gravity darkening coefficient $\beta$. Fig.~\ref{fig:degener} illustrates the degeneracy of the photometric models derived from AMBER and PIONIER data. 

The methods based on visibilities and closure phases, namely the parameters fitting and the image reconstruction, are fully compatible with each other and precisely determine the stellar photospheric properties of Achernar. The MCMC minimization tools were shown to be very useful when minimizing multiple parameters simultaneously, allowing to infer the uncertainties on their determination and the correlation between parameters. 

The photospheric characterization of Achernar is important in many aspects. As a Be star, it is the first firm characterization of a Be central star photosphere. This assures, for example, that the rigid rotation with Roche geometry is a valid assumption to this stellar type. Furthermore, the rotation rate we found ($W = 0.838$), relatively far for critical, adds constraints to the mass loss mechanism in this kind of star by allowing estimate how much extra energy is necessary for launching photospheric material into a Keplerian orbit. Finally, the data showed no signal of polar wind, contrary to previous interferometric results. These and other circumstellar features are further discussed in Chapter~\ref{chap:aeri}.

In the context of fast rotating stars, Achernar is the fastest and also the more massive star with photospheric information known with sufficient precision. This is an important point for the investigation of these stars. For example, the derived von Zeipel $\beta$ coefficient is fully compatible with the ELR model \citep{esp11a}, model which could explain the coefficient variation among of most stars with resolved photospheres. 

\FloatBarrier
\section{Publication: Beyond the diffraction limit of optical/IR interferometers; I. Angular diameter and rotation parameters of Achernar from differential phases \label{pub:amber}}
\href{http://adsabs.harvard.edu/abs/2012A\%26A...545A.130D}{ADS Page: http://adsabs.harvard.edu/abs/2012A\%26A...545A.130D}

\FloatBarrier
\section{Publication: The environment of the fast rotating star Achernar; III. Photospheric parameters revealed by the VLTI. \label{pub:aeripion}}
\href{http://adsabs.harvard.edu/abs/2014A\%26A...569A..10D}{ADS Page: http://adsabs.harvard.edu/abs/2014A\%26A...569A..10D}

%% file: chap/aeri.tex
\chapter{The Be star Achernar and the 2013 outburst \label{chap:aeri}}
Achernar is a key star to investigate the Be phenomenon. Its importance derives from its proximity, enabling the detailed investigation of the CS emission of a star whose photospheric parameters are very well known (as seen in Chap.~\ref{chap:phots}).

In the beginning of 2013 the star entered a new outburst phase (Tasso Napole\~ao, priv. comm.) after about 7 years of quiescence during which it was mostly a normal B star. Triggered by this exciting possibility, we started a broad-scope, multi-technique campaign to follow the disk evolution. Coordinating this campaign, gathering, reducing, analyzing and organizing the data took much of my time since then.

Here we report on the observational results and the initial modeling of the disk evolution. The analysis combine multi-technique data, including broadband polarimetry (OPD/LNA), spectroscopy (FEROS and others) and interferometry (VLTI-AMBER and PIONIER). The inteferometric survey, that includes 79 individual observations since 2013, corresponds to the first high-cadence interferometric survey of a Be star to date.

These data will allow us to test the VDD predictions in conditions hitherto not explored. For instance, the high cadence of the observations offers a rare opportunity to evaluate the evolution of a just formed Be disk in detail and derive the relevant physical quantities governing the system.
\section{Star overview}
Achernar displays important features that are related to the Be phenomenon: (i)~a fast rotating central star; (ii)~quasi-cyclic disk formation/dissipation, with both long- and short-term variations; (iii)~episodic mass disk injections; and (iv)~binarity. In addition, there has been suggestions in the literature about a possible polar wind emission, that still remains to be verified. Before detailing the new data, we review Achernar's characteristics known previously to the current active phase.

\subsubsection*{The photospheric determination by interferometry}
We discussed the consistent determination of multiple photospheric parameters of Achernar (radius, rotational velocity, inclination angle and on-sky orientation) by \citet{dom12a} in Section~\ref{sec:dpphot}. That work was based on AMBER spectro-interferometric differential phases along the Br$\gamma$ line, a differential quantity.

A quantitative jump was given by \citet[Sect.~\ref{sec:pion}]{dom14a}. The study was based on PIONIER 4 beam combiner observations, including high precision visibility measurements with closure phase information. The stellar diameter ratio $R_{\rm eq}/R_{p}$ was found to be 1.352, which corresponds to a not so high rotation rate ($W=0.838$). Still, it is the highest rotation rate observed in a angularly resolved star so far. 

\subsubsection*{Evolutionary stage}
Different estimations for the mass of Achernar points to the value of $\sim$6.2\,$M_\odot$ (see references in Table~\ref{tab:stpars}). The photospheric parameters obtained from interferometry show that Achernar is a considerably large star and has a higher luminosity than a main sequence star of same mass (Table~\ref{tab:stpars}).
\begin{table}
\centering
\caption[Main sequence stellar parameters compared to the ones of Achernar derived from interferometry]{Main sequence stellar parameters compared to the ones of Achernar derived from interferometry \citep{dom14a}. The main sequence data assumes $W=0.894$.}
\begin{threeparttable}
\begin{tabular}[]{cccr}
\toprule
     $X_c$ & $M (M_\odot)$ & $R_p (R_\odot)$ & $L (L_\odot$) \\
\midrule   
     0.54 & 5.50 &3.21 & 845.18 \\
     0.30 & 5.50 &4.14 & 1153.50 \\
     0.08 & 5.50 &5.38 & 1480.41 \\ \hline
     0.54 & 6.40 &3.55 & 1497.22 \\
     0.30 & 6.40 &4.45 & 1953.61 \\
     0.08 & 6.40 &5.75 & 2521.98 \\ \hline
     0.54 & 7.70 &3.71 & 2564.45 \\
     0.30 & 7.70 &4.93 & 3726.03 \\
     0.08 & 7.70 &6.37 & 4827.72 \\ \hline
     Achernar & $6.22$\tnote{a} & 6.78 & 3019.95 \\
\bottomrule
\end{tabular}
    \begin{tablenotes}
        \footnotesize
        \item[a] $M=6.1M_\odot$ \citep{har88a}; $M=6.22 M_\odot$ \citep{jer00a}.
    \end{tablenotes}
 \end{threeparttable}
\label{tab:stpars}
\end{table}

In Appendix~\ref{chap:beataeri} we apply the \textsc{BeAtlas} photospheric models of main-sequence stars to ultraviolet observations of Achernar. The results points to a star with $M\gtrsim7.2M_\odot$, but the quality of the fitting confirms that it is not a typical main sequence star. 

These elements indicate that Achernar is likely at the end of the main sequence or in the early post-main sequence phases, as already proposed in literature (e.g., \citealp{rie13a}).

\subsection{Circumstellar activity}
\subsubsection*{The variable circumstellar disk and photosphere}
\citet{vin06a} report both long-and short-term H$\alpha$ variations in Achernar. The authors argue about the existence of a 14-15 year cyclic B-Be phase transition. These variations are understood as the recurrent formation/dissipation of the equatorial CS disk, and the 2013 activity is part of one of these cycles. 
This cycle, however, should be viewed with reservations. The authors only had detailed spectra for a period of 11 years (1991 to 2002), and the cycle period was extrapolated with data from literature. The existence of activity in the years 2006-2007 and now in 2013-2015 shows that this is not a regular frequency.


\subsubsection*{The short-term variability and disk feeding}
\citet{gos11a} reported a high-precision ($\sim0.001$) magnitude observations of Achernar. Solar Mass Ejection Imager (SMEI) satellite observations were done at optical wavelengths, centered at 700\,nm and covering roughly from 350 to 1100\,nm. The observational period of Achernar covered June 2003 to November 2008, spanning 1,993 days. SMEI covered a period of great changes in disk activity, both active and quiescent phases seen through the H$\alpha$ profile \citep{riv13b}.

The authors found two main photometric frequencies, 0.775\,d$^{-1}$ (F1) and 0.725\,d$^{-1}$ (F2), where the amplitude of this last frequency changed considerably over time. The nature of F2 appears to be associated with stellar outbursts, while the main frequency F1 with $\lesssim0.04$~mag amplitude had already been reported by \citet{bal87a}, \citet{vin06a} and \citet{riv03a}, the last with amplitude of 0.02 mag. Other frequencies reported by \citet{vin06a}, namely 0.49\,d$^{-1}$, 1.27\,d$^{-1}$ and 1.72\,d$^{-1}$ were not found.
\citet{gos11a} frequencies are listed in Table~\ref{tab:gossfs}. The third low-amplitude frequency reported is 0.680\,d$^{-1}$ (F3), described as a ``pulsation or transient frequency''. 
\begin{table}
\centering
\caption[Identified photometric frequencies in Achernar by \citet{gos11a}.]{Identified photometric frequencies in Achernar by \citet{gos11a}. The starred (*) frequencies indicate with no previous published results in literature. These are frequencies for the entire observation timeseries, from 2003 up to 2008.}
\begin{tabular}[]{cccc}
\toprule
    & Frequency (d$^{-1}$) & Amplitude (mag) & SNR \\
\midrule
    F1  & 0.775177(5) & 0.0165(3) & 27.09 \\
    F2* & 0.724854(6) & 0.0129(3) & 19.05 \\
    F3* & 0.68037(3)  & 0.0027(3) & 4.11 \\
\bottomrule
\end{tabular}
\label{tab:gossfs}
\end{table}

Analyzing H$\alpha$ and high-precision photometric measurements, \citet{riv13b} argue that Achernar's photospheric lines profiles indicate a stellar spin-up by as much as $\Delta v \sin i \approx 35$~km\,s$^{-1}$ at epochs prior to stellar activity. The photospheric line widths also increase correlated with a photometric variability of $0.775$~d$^{-1}$ up to $\sim0.04$\,magnitudes in visible. 
At the end of the activity, the apparent stellar rotational speed declines to the value at quiescence after H$\alpha$ returns to its photospheric configuration, when the photometric variability is minimum. Such change at the stellar surface must be compensated by internal momentum angular changes, whose precise mechanism is unknown (see Sect.~\ref{sec:multiper}).

From polarimetry, \citet{car07a} report polarization PA changes between 26$^\circ$ and 31$^\circ$, where the polarization level changed from 0.12\% to 0.17\% within 2 hours (uncertainties $\leq2^\circ$ and $\leq0.01\%$). The linear polarization observations were taken in 2006 at the $B$ band. This variability (within a few hours) was interpreted as a signature of a large mass ejection from the photosphere, in a timescale $\sim15\times$ shorter than the the frequencies listed in Table~\ref{tab:gossfs}. 

Variations of this time scale should in principle be detected by high-precision photometry (as in \citealp{gos11a}). However, these ejections may not have a strong correspondent photometric signal given the inclination angle on which Achernar is seen. As shown by \citet{hau12a}, as the disk density grows in Be stars seen at $i\approx70^\circ$, almost no photometric changes are detected. Around this inclination angle the emitted flux at visible wavelengths by the disk is almost exactly compensated by disk absorption of photospheric light. This is the likely scenario occurring to Achernar and, if occurring, it would be an evidence that the circumstellar material is distributed until regions very close to the stellar photosphere.

\subsubsection*{The binary companion}
High angular resolution images revealed a close-in faint ($\sim30-50\times$ fainter in the infrared) A1-3V companion star orbiting the Be star \citep{ker07a}. In a following work \citep{ker08a}, the authors estimated the orbital period of $\sim15$\,years, close to the  H$\alpha$ spectroscopic variability found by \citet{vin06a}.

The orbital period of Achernar is currently been refined by Kervella et al$.$ and can contribute to define the role of the companion on Achernar's activity, and the link to the Be phenomenon. For example, during the PIONIER observations of Achernar in 2010-2011, those authors estimate that the companion star was in the apastron \citep{dom14a}. This means that the recent outburst started when the both stars were still very far apart, which suggest that there is no relation between binary and disk activity - at least for Achernar.

\subsubsection*{The polar wind emission in Achernar and Be stars}
The presence of a dense polar wind in Be stars is commonly cited in literature (e.g., \citealp{geo11a}; \citealp{mei12a}), including textbooks of stellar rotation \citep{mae09a}. Although a polar wind with the characteristics of a O-type star is expected when a B star is rotating close to its critical limit ($W\approx1$; \citealp{mae02a}), this wind would not explain the densities ($\gtrsim 10^{-13}$~g\,cm$^{-3}$) claimed by observations. The term dense here refers to densities higher than the ones expected by traditional stellar wind theory (e.g., CAK). However, these densities are low when compared to the typical densities of the circumstellar disks ($\approx10^{-11}$~g\,cm$^{-3}$).

The interferometric evidence of polar winds appear to exist for only two Be's stars, namely $\alpha$\,Ara and Achernar, as pointed out in the recent review of \citet{riv13a}. According to these authors, the evidence for $\alpha$\,Ara is very fragile: based on a single interferometric measurement in which a single calibration bias in the data would result in the alleged characteristics for such wind. Thus, it is unlikely that these features were actually present. The only case pointed out by the authors that deserves further investigation is the case of Achernar.

Although the interferometric imaging of \citet{dom14a} discard the presence of a polar wind during 2010-2011 in Achernar, its presence can be associated with disk activity. The investigation in the current active phase can provide important information about this feature.

\FloatBarrier
\section{The multiperiodicity of Achernar and other Be stars \label{sec:multiper}}

Let us now investigate whether there is any relation between the photometric frequencies reported by \citet{gos11a} and the spin-up speeds discovered by \citet{riv13a}.
The reported spin-up velocity ($\sim35$~km\,s$^{-1}$) is very roughly the difference between the stellar rotational and critical velocities 
\begin{equation}
\Delta v = (v_{\rm crit} - v_{\rm rot})\times\sin i = 34.3 \text{ km\,s}^{-1}\,,
\end{equation}
indicating that Achernar is rotating critically during its active phases (at least, in its equatorial region). Another interesting fact is that the very stable and known F1$\sim0.775$~day$^{-1}$ frequency is very closely the orbital frequency of material just above the stellar equator (F4)
\begin{equation}
F4 = \frac{v_{\rm orb}}{2\pi R_{\rm eq}} = \frac{v_{\rm orb}}{2\pi 1.352 R_{p}} \sim 0.769 \text{ d}^{-1}\,,
\end{equation}
where $R_{\rm eq}$ used was the value obtained by \citet{dom14a}. The close match between F1 and F4 suggests that F1 may be (i) a rotational frequency, perhaps associated with material freshly ejected by the star that is in an orbit very close to the equatorial radius; (ii) a non-radial pulsation frequency that naturally matches the frequency of an equatorial orbit. Note that this match could only be identified now that the photospheric parameters of Achernar are accurately known.

We also found two possible orbital frequencies with a less accurate match involving the transient frequency F3$\sim0.680$~day$^{-1}$. They are: (i) the period around the critically rotating star (F5)
\begin{equation}
F5 = \frac{v_{\rm crit}}{2\pi 1.5 R_{p}} \sim 0.657 \text{ d}^{-1}\,,
\end{equation}
and (ii) the rotational period (F6)
\begin{equation}
F6 = \frac{v_{\rm rot}}{2\pi 1.352 R_{p}} \sim 0.644 \text{ d}^{-1}.
\end{equation}

Be stars have three naturally arising frequencies. One of them is the rotation frequency (F6, above), which would be associated with purely photospheric phenomena such as stellar spots, non-radial pulsation, etc. The second is the critical frequency, when the rotation at stellar equatorial radius balances gravity (F5). The third is the orbital period at the stellar equator (F4). If the first and third periods are known, somehow, the stellar rotation rate could be easily estimated by
\begin{equation}
    W = \frac{v_{\rm rot}}{v_{\rm orb}} = \frac{\rm F6}{\rm F4}\,,
    \label{eq:WF4F6}
\end{equation} 
If the last two frequencies are known,
\begin{equation}
    \frac{v_{\rm rot}}{v_{\rm crit}} = \frac{\rm F5}{\rm F4}.
\end{equation}

If we make the associations above between the observed periods of Achernar (F1~$\equiv$~F4 and F3~$\equiv$~F6), we find
$W=0.878$, that is $\sim5\%$ higher than the one determined by \citet[$W=0.838$]{dom14a}. Assuming F3~$\equiv$~F5, the value of 0.878 is less than $1\%$ of the ratio ${v_{\rm rot}}/{v_{\rm crit}}$ determined by in the same work (${v_{\rm rot}}/{v_{\rm crit}}=0.883$).

Observationally, \citet{riv03a} state that ``the presence of these secondary [transient] periods is enhanced in the outburst phases''. The enhancement of F3 with activity was also found in \citet{gos11a} observations of Achernar. If the material ejected by the star causes an asymmetry in the stellar photosphere or it accumulates in the critical rotation radius, the frequencies related to these processes are, respectively, F6 and F5.

Applying this method to the objects in Table~4 of \citet{riv03a}, we find a (near-)constant value of ${F3}/{F1}\sim0.9$ (Table~\ref{tab:W}; at least, ${F3}/{F1}>0.8$ if we do not exclude $\mu$\,Cen for which the transient frequency was poorly determined). 
\begin{table}
\centering
\caption[NRP pulsation and transient frequencies of Achernar and other Be stars]{NRP pulsation and transient frequencies of Achernar and other Be stars listed in \citet{riv03a}.}
\begin{tabular}[]{cccc}
\toprule
    \multirow{2}{*}{Star} & F3 & F1 & F3/F1 \\
    & (d$^{-1}$) & (d$^{-1}$) & (d$^{-1}$) \\
\midrule
$\alpha$ Eri & 0.68037 & 0.775177 & 0.878 \\ \hline
$\kappa$ CMa & 1.62075 & 1.824818 & 0.888 \\
$\omega$ CMa & 0.68027 & 0.728863 & 0.933 \\
$\mu$ Cen    & $\approx$1.60772 & 1.988072 & 0.809 \\
$\eta$ Cen   & 1.55521 & 1.733102 & 0.897 \\
\bottomrule
\end{tabular}
\label{tab:W}
\end{table}

The correlation between pulsations and the ejections of matter into the circumstellar disk of Be stars was first proposed by \citet{riv01a}, firmly established by \citet{hua09a} using CoRoT observations and are confirmed by \citet{gos11a}. The theoretical foundations for these correlations are discussed by \citet{nei14a}, with the transport of angular momentum from the core to the surface due to the $\kappa$ mechanism applied for fast rotating stars. According to these authors, ``the accumulation of angular momentum just below the surface of Be stars increases the surface velocity. The surface then reaches the critical velocity so that material gets ejected from the star''. This is a suggestive interpretation of the spin-up observation in Achernar. Evidence of spin-up in other Be stars would enhance the likelihood of this model. The stars listed in in Table~\ref{tab:W} are good candidates to the presence of this phenomenon. 

A speculative resonant scenario based on the frequencies derived to Achernar is developed in Appendix~\ref{ap:reson}. Although based on strong ad-hoc assumptions, it could bring insights into the origin of the Be phenomenon.
\FloatBarrier
\section{Observations of the active phase \label{sec:aeriacti}}
After the activity report in 2013, our interferometric observations of Achernar started in August that year with a Director's Discretionary Time (DDT) proposal to AMBER 091.D-0107. They were followed by the proposal 093.D-0399, with observations from June 24th 2014 to January 19th 2015. These proposals were complemented by observations from collaborators, both from AMBER and PIONIER interferometers. The interferometric observations where complemented by high cadence polarimetric (OPD/LNA) and spectroscopic observations. Here we present the observational data obtained as well as the results of an initial modeling effort. 
\subsubsection*{Spectroscopy \label{sec:aerispec}}
Our spectroscopic campaign covers 50 nights and 163 individual spectra. Table~\ref{tab:specs} contains a summary of the observations.
The observations contains data from the following instruments: PUCHEROS (PUC-Chile; \citealp{van12b}), X-SHOOTER in experimental mode (ESO,  Martayan, priv$.$~comm$.$), FEROS (ESO; \citealp{kau99a}), ECASS and MUSICOS (OPD/LNA; both described in Sect.~\ref{sec:iraf}). The complete list of observations are in appendix (Table~\ref{ap:tabspecs}). 
\begin{table}
\centering
\caption[Summary of the spectroscopic data available for the 2013 outburst of Achernar]{Summary of the spectroscopic data available for the 2013 outburst of Achernar. For an extended list of observations see the appendix (Table~\ref{ap:tabspecs}).}
\begin{threeparttable}
\begin{tabular}[]{lrccc}
\toprule
Instrument & Approx. & Number of & First date & Last date \\
 & Resolution & spectra & & \\
\midrule
ECASS & 5000 & 1 & 2012-11-20 & 2012-11-20 \\ 
PUCHEROS & 20000 & 5 & 2013-01-18 & 2014-01-30 \\ 
Amateur & 5000 & 11 & 2013-01-22 & 2015-01-18 \\ 
XSHOOTER & 15000 & 1 & 2013-07-18 & 2013-07-18 \\ 
FEROS & 48000 & 136 & 2013-09-06 & 2014-02-25 \\ 
MUSICOS & 30000 & 7 & 2013-11-12 & 2014-11-19 \\ 
BeSS & 5000\tnote{a} & 2 & 2014-11-26 & 2014-11-28 \\ 
\bottomrule
\end{tabular}
\label{tab:specs}
    \begin{tablenotes}
        \footnotesize
        \item[a] The average value of observations used.
    \end{tablenotes}
    \end{threeparttable}
\end{table}

\begin{figure}
\begin{center}
\includegraphics[width=.8\textwidth]{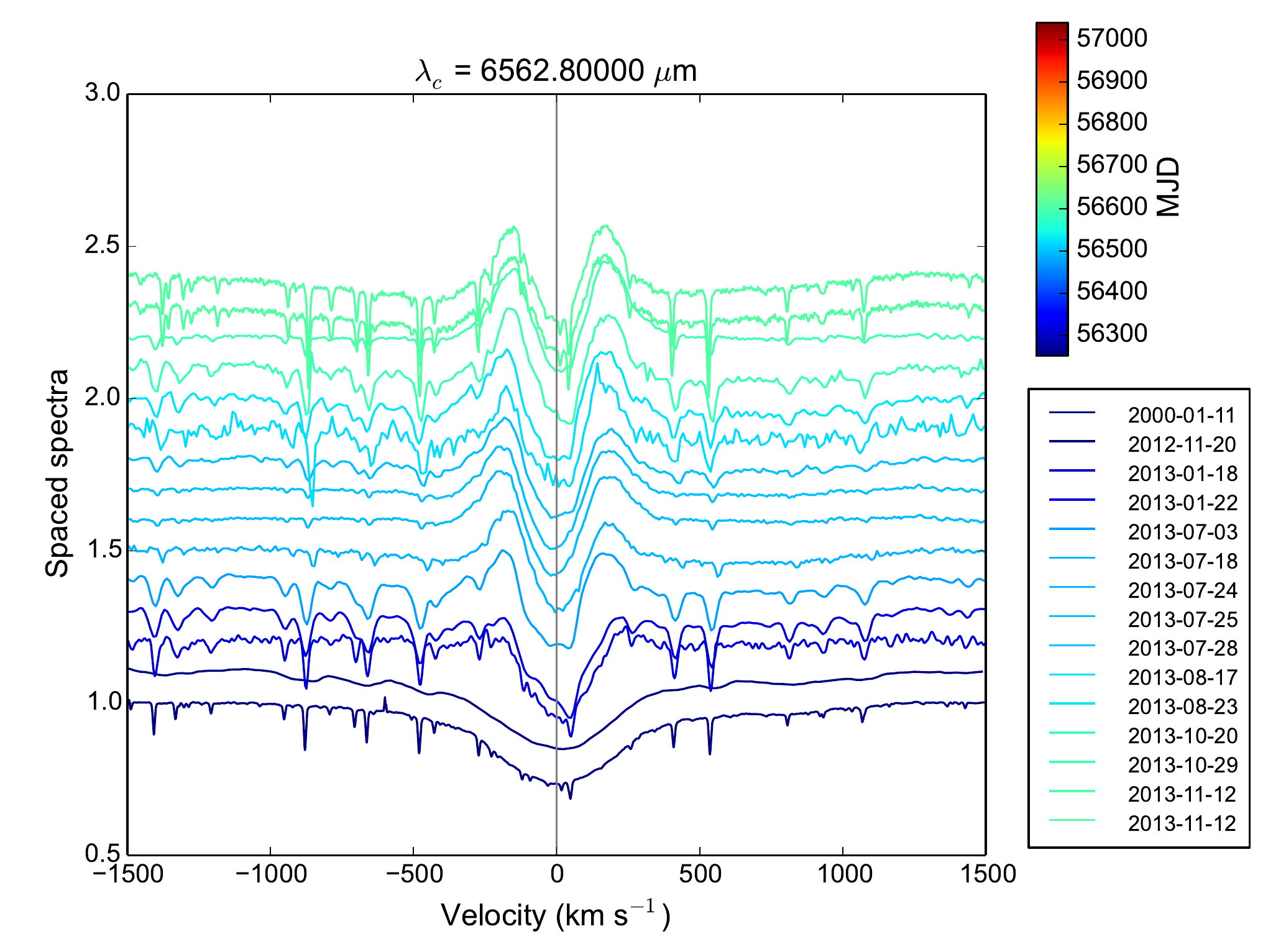} 
\caption[H$\alpha$ line profiles of Achernar between November, 2012 and November, 2013]{H$\alpha$ line profiles of Achernar between November, 2012 and November, 2013. The color corresponds to the observation date. The (purely) photospheric reference profile of \citet{riv13b} is also shown. It contains the observations from all instruments presented in Sec.~\ref{sec:aerispec}, except for FEROS/ESO, due to its high cadence.}
\label{fig:aeriincr1}
\end{center}
\end{figure}

\begin{figure}
\begin{center}
\includegraphics[width=.8\textwidth]{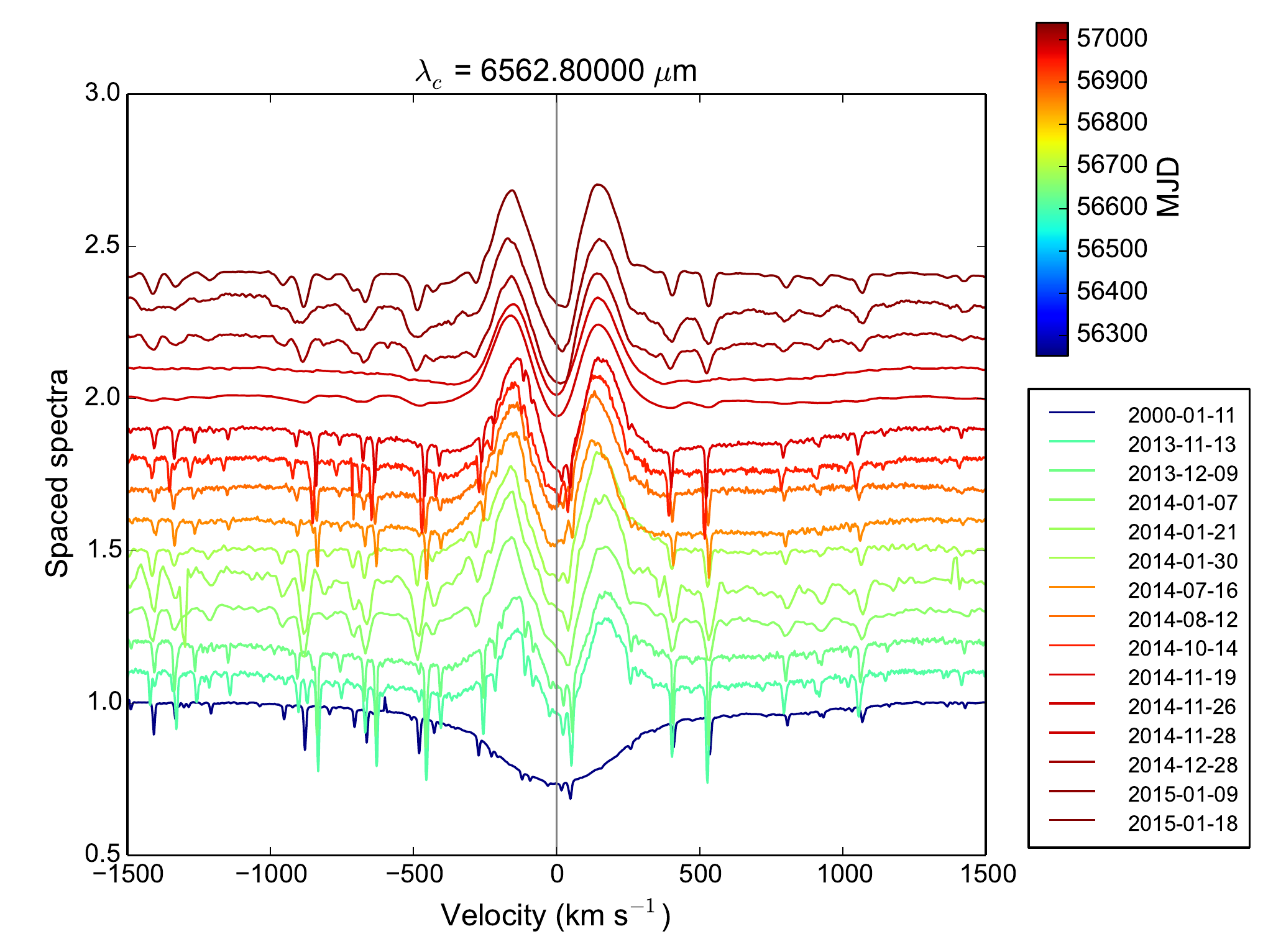} 
\caption{The same as previous figure, for observations between November 2013 to January 2015.}
\label{fig:aeriincr2}
\end{center}
\end{figure}

The observed spectra are shown in Figs.~\ref{fig:aeriincr1} and \ref{fig:aeriincr2}. They include observations from all instruments except from FEROS, due to the large number of observations and their high cadence. These spectra are also presented in the appendix overplotted and as difference to the purely photospheric reference (Figs.~\ref{fig:apaeriover1} to \ref{fig:apaeridiff2}, respectively). The reference profile was observed in January 2000 and was selected by \citet{riv13b}. FEROS spectra are presented in Figs.~\ref{fig:apfer01} to \ref{fig:apfer10}.

One problem we faced in our analysis was to combine spectra from multiple sources. Since Achernar has an extremely wide H$\alpha$ profile, with wings extending at speeds greater than $\sim1000$~km\,s$^{-1}$ from line center, in some instruments that can be mistaken as the continuous emission. This is particularly important for Echelle spectrographs that do not have a large spectral coverage at the different orders. In this case, the small difference in the level of the wings can be easily confused with the the instrumental response between the orders.

The solution was to perform the analyzes only for the center of the lines, ignoring the wings. The level on the continuum was determined by shifting the center line limits until their values correspond to the values of the purely photospheric profile. The criterion used to determine the best value for the core size was to minimize the difference of the EW values between different instruments but from a similar observational epoch, where no substantial changes were expected. The value that best described the line evolution was $\Delta v = 750$~km\,s$^{-1}$. As a consequence of the criterion adopted, the photospheric equivalent widths have systematically higher values of $\sim2$\,\AA{} since it is does not consider absorption in the line wings.

\begin{figure}
    \centering
    \includegraphics[width=\linewidth]{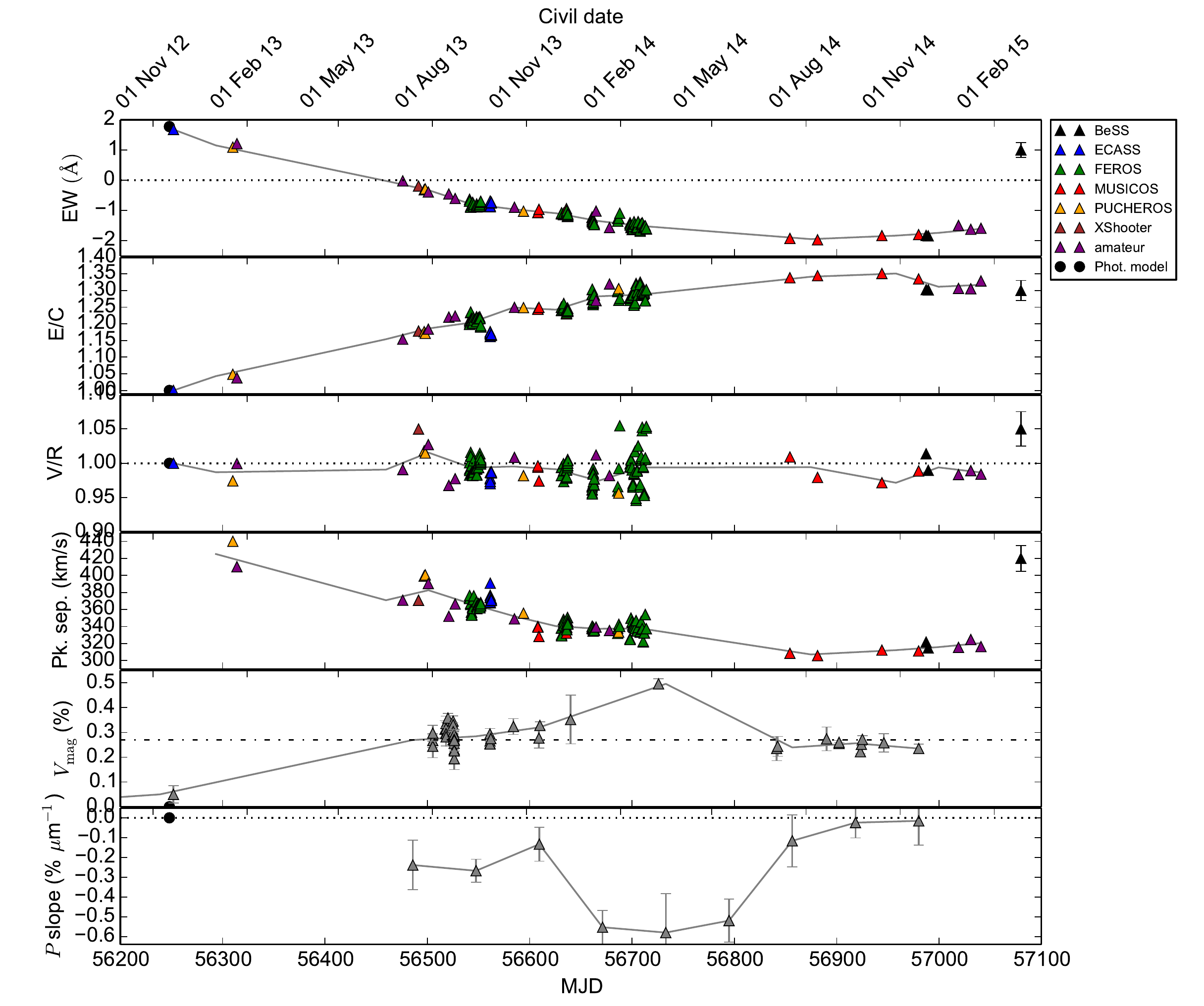}
    \caption[Secular evolution of Achernar in the 2013 outburst. ]{Secular evolution of Achernar after the 2013 outburst. Observations cover the period from November 2012 (purely photospheric observables) to February 2015. \textit{From top to bottom:} H$\alpha$ line profile equivalent width (EW), emission/continuum (E/C) ratio, $V/R$ peaks ratio, and peak separation; $V$ band polarization and $BVR$ polarization slope (or \textit{color}). The typical error of the measurements are indicated at the top right corners. The straight lines refer to the data binning in intervals of 45 days (interpolated when data is absent in the bin interval). The spectrographs used are indicated. The polarization data is from OPD/LNA (Brazil). The dotted lines indicate reference values EW\,$=0$, $V/R=1$ and flat polarized spectrum. The polarization value $P\sim0.27\%$ is also indicated by the dash-dotted line. The EW measurements consider only the core of the line, in a width of 750~km\,s$^{-1}$.}
    \label{fig:secevo1}
\end{figure}

The main spectroscopic secular characteristics are: (i) H$\alpha$ line profile is characterized by a central absorption, surrounded by emission peaks with large separation ($\Delta v> 300 $~km\,s$^{-1}$). (ii) the evolution of the profiles is slow (much slower the the polarization variations, see below) and only reach a (near-)constant regime after $\sim1.6$~years; 
The items (i) and (ii) can be seen in Fig.~\ref{fig:secevo1} and are discussed in more detail in the next section.

\begin{figure}
    \centering
    \includegraphics[width=\linewidth]{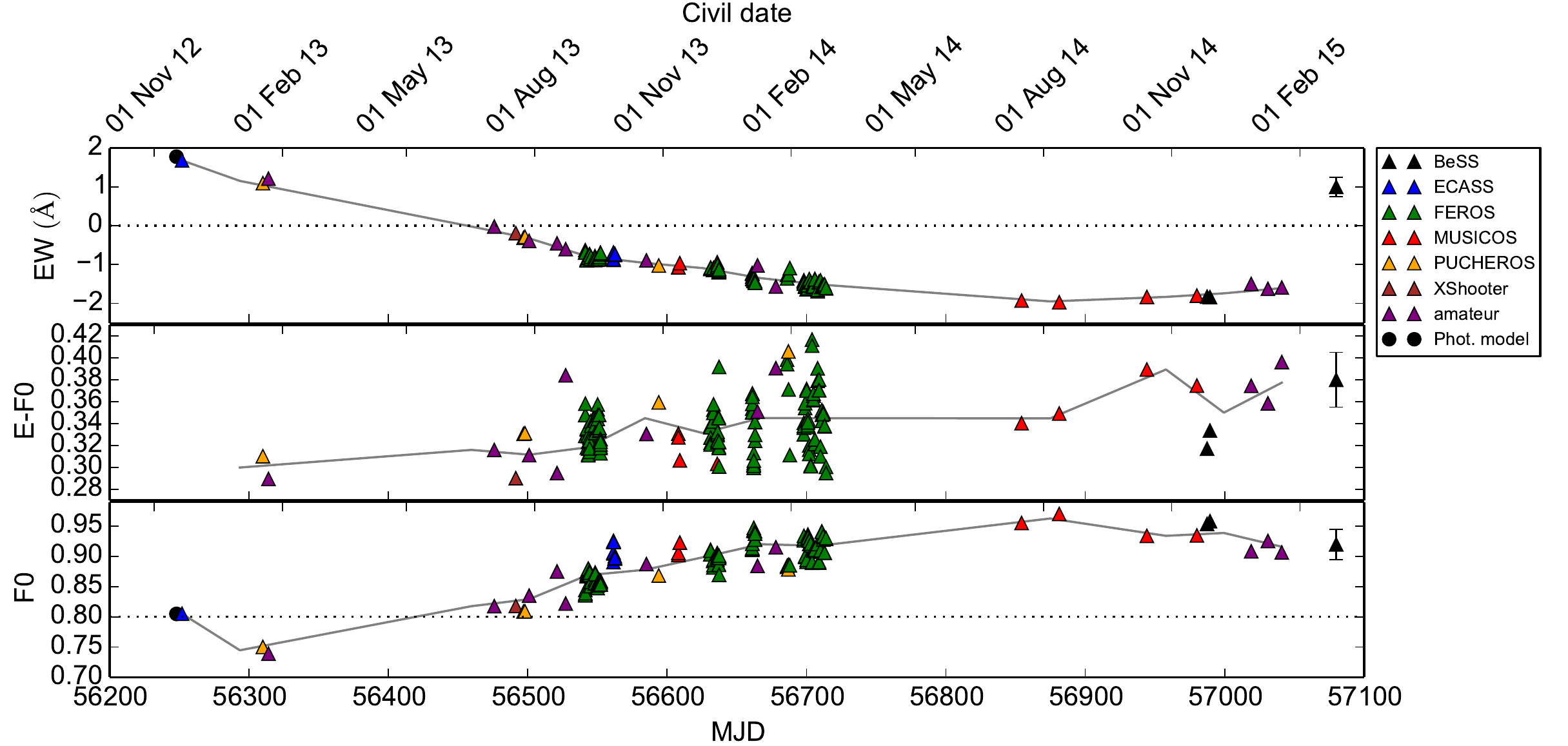}
    \caption[Secular evolution of selected spectroscopic quantities of Achernar in the 2013 outburst]{Secular evolution of selected spectroscopic quantities of Achernar in the 2013 outburst. \textit{From top to bottom:} H$\alpha$ line profile equivalent width (EW), emission-absorption depth (at rest wavelength; E-F0), and the absorption depth (F0). Observations cover the period from November 2012 (purely photospheric observables) to February 2015. The typical error of the measurements is indicated at the top right corners. The straight line refers to the interpolation of data at intervals of 45 days, interpolated when absent in the interval. The spectrographs used are indicated. The dotted lines indicate reference values EW\,$=0$ and F0 at the purely photospheric level. The EW measurements consider only the core of the line, in a width of 750~km\,s$^{-1}$.}
    \label{fig:aerisecevo2}
\end{figure}

An interesting phenomenon that was observed was the transition from the photospheric profile to a shell-star profile at the beginning of the activity (F0 quantity in Fig.~\ref{fig:aerisecevo2} and the first three profiles of Fig.~\ref{fig:aeriincr1}). This occurrence was previously reported by \citet{riv13b} in the 2006 towards the end of a disk formation phase, when the disk had reached maximum emission. This effect indicates that at the start of the activity (i) the ejected material was dense enough to absorb a considerable part of the stellar flux, (ii) the photospheric temperature changed considerably, or (iii) a combination of both effects occurred. The effect is even more intriguing since the last determination of Achernar's viewing angle $i\simeq60.6^\circ$ by \citet{dom14a} points to a angle far from the equator-on, expected for shell stars.

\FloatBarrier
\subsubsection*{Polarimetry \label{sec:aeripol}}
Our polarimetric campaign covers 35 nights and 108 individual observations. The observations consist of imaging polarimetry in the $B$, $V$, $R$ and $I$ bands obtained using the IAGPOL polarimeter attached to the 0.6~m Boller \& Chivens telescope at OPD/LNA, Brazil. The observational setup and reduction procedure are described in Sect.~\ref{sec:iraf}. The complete list of polarimetric observations are in the appendix (Table~\ref{tab:appol}). 

\begin{figure}
\begin{center}
\includegraphics[width=.8\textwidth]{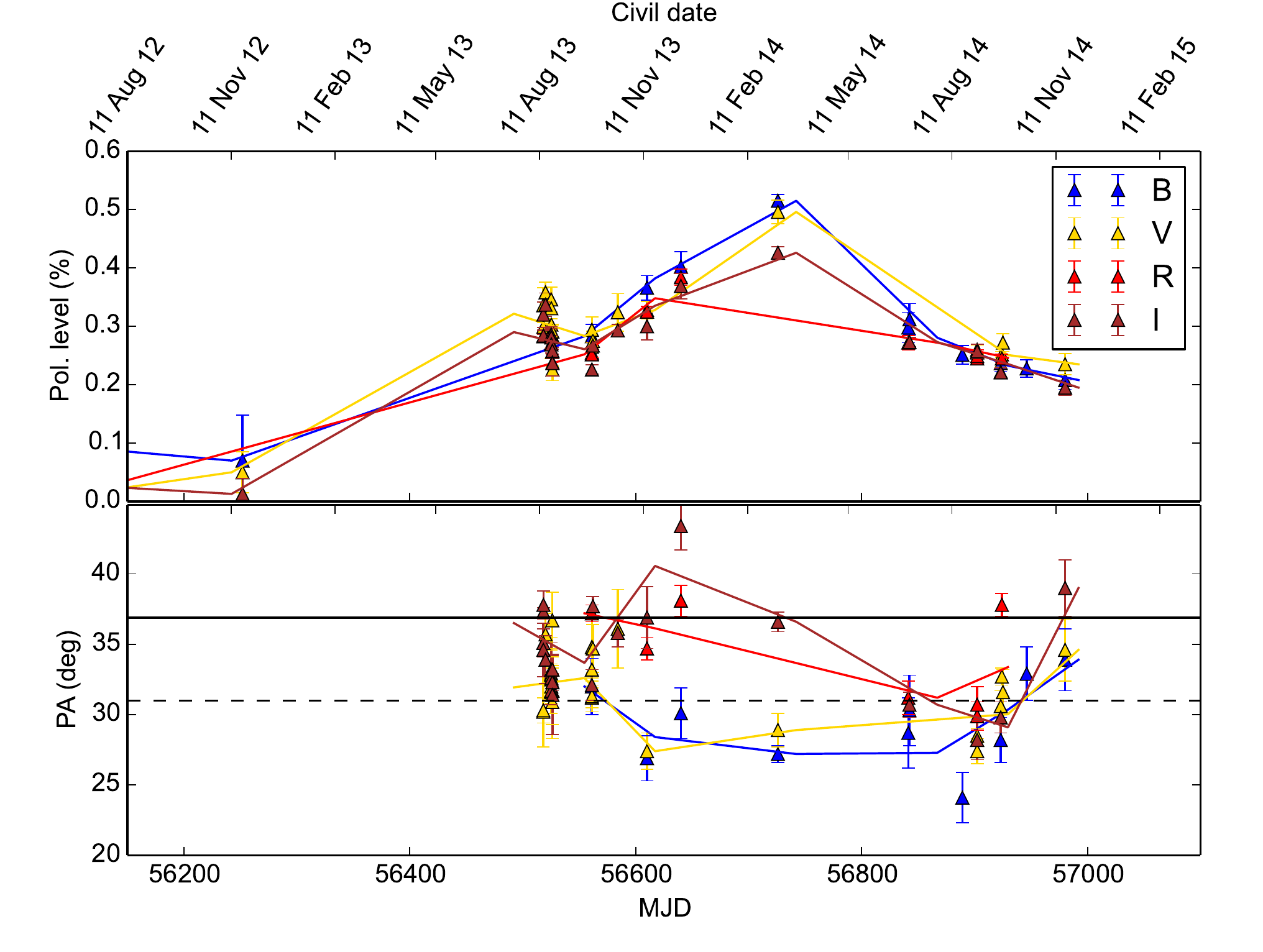} 
\caption{Broad-band linear polarization of Achernar ($BVRI$ filters). The
observational data is binned in order to increase each epoch mean precision.}
\label{fig:aerisecpol}
\end{center}
\end{figure}

The evolution of polarization at $BVRI$ bands are show in Fig.~\ref{fig:aerisecpol}. The data of $BVR$ bands were used to obtain the polarization color of Fig.~\ref{fig:secevo1}. The color is calculated as following: data intervals were selected within which no large variations were detected. For each interval the data for each filter was binned and then the resulting $BVR$ values were fitted with a linear function. $I$ band was not considered for the color fitting. As discussed in Sect.~\ref{sec:polsaw}, the $I$ band contains wavelengths beyond the Paschen series limit ($n=3$, i.e., $\lambda>8204$\,\AA). This region can have a much lower absorption opacity, since photons beyond these frequencies cannot ionized electrons at orbits $n\leq3$. Thus, this region has an abrupt increase of the polarized light once photos at these wavelengths are scattered instead of been absorbed. 

\begin{figure}
\begin{center}
\includegraphics[width=.8\textwidth]{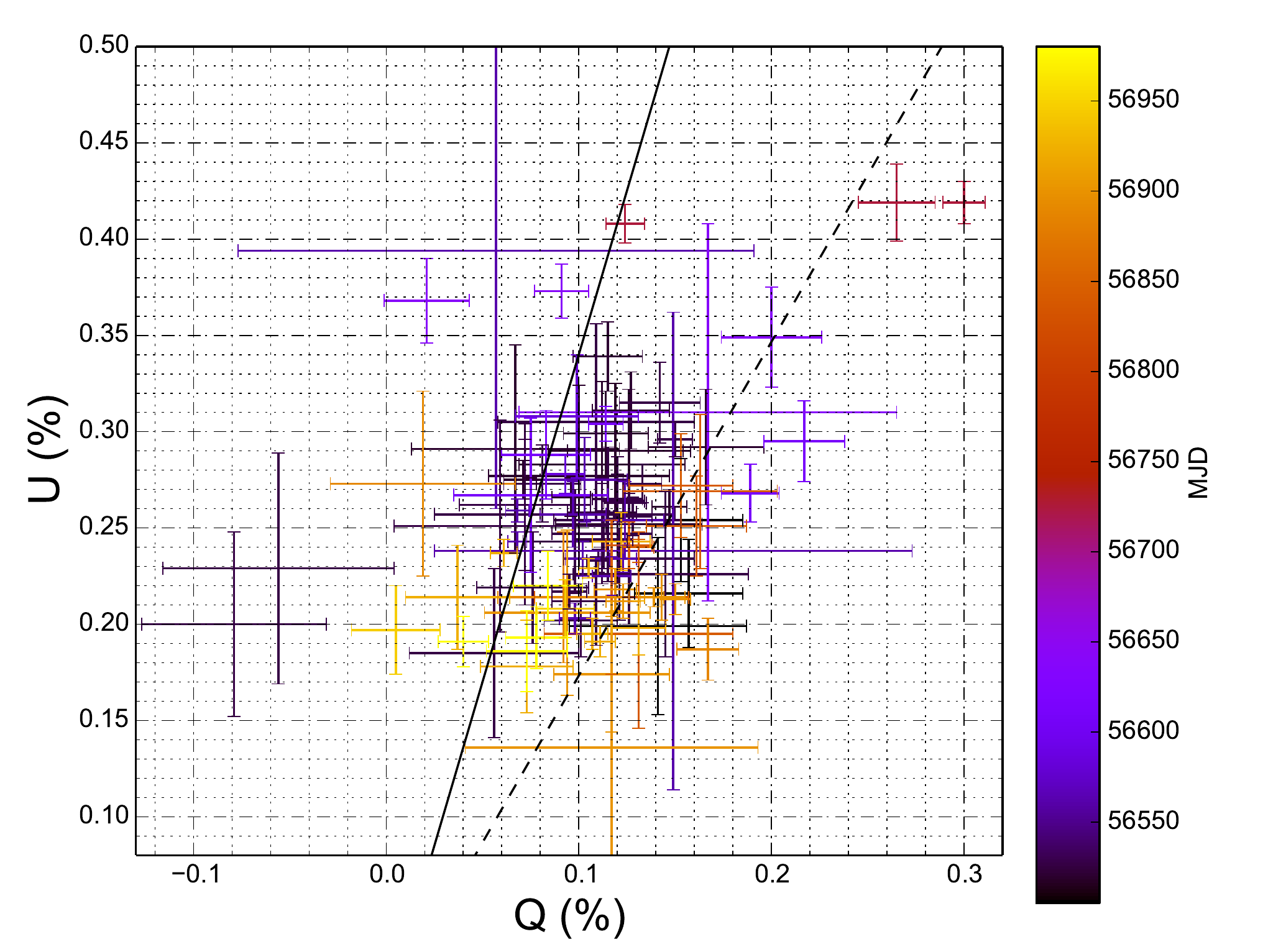} 
\caption[Polarimetric $QU$ plane of $V$ band observations of the 2013 outburst of Achernar. ]{Polarimetric $QU$ plane of $V$ band observations of the 2013 outburst of Achernar. The straight line corresponds to a (fixed) polarization orientation of 36.9$^\circ$ and the dashed line to 31.0$^\circ$.}
\label{fig:aerisecQU}
\end{center}
\end{figure}

\begin{figure}
\begin{center}
\includegraphics[width=.8\textwidth]{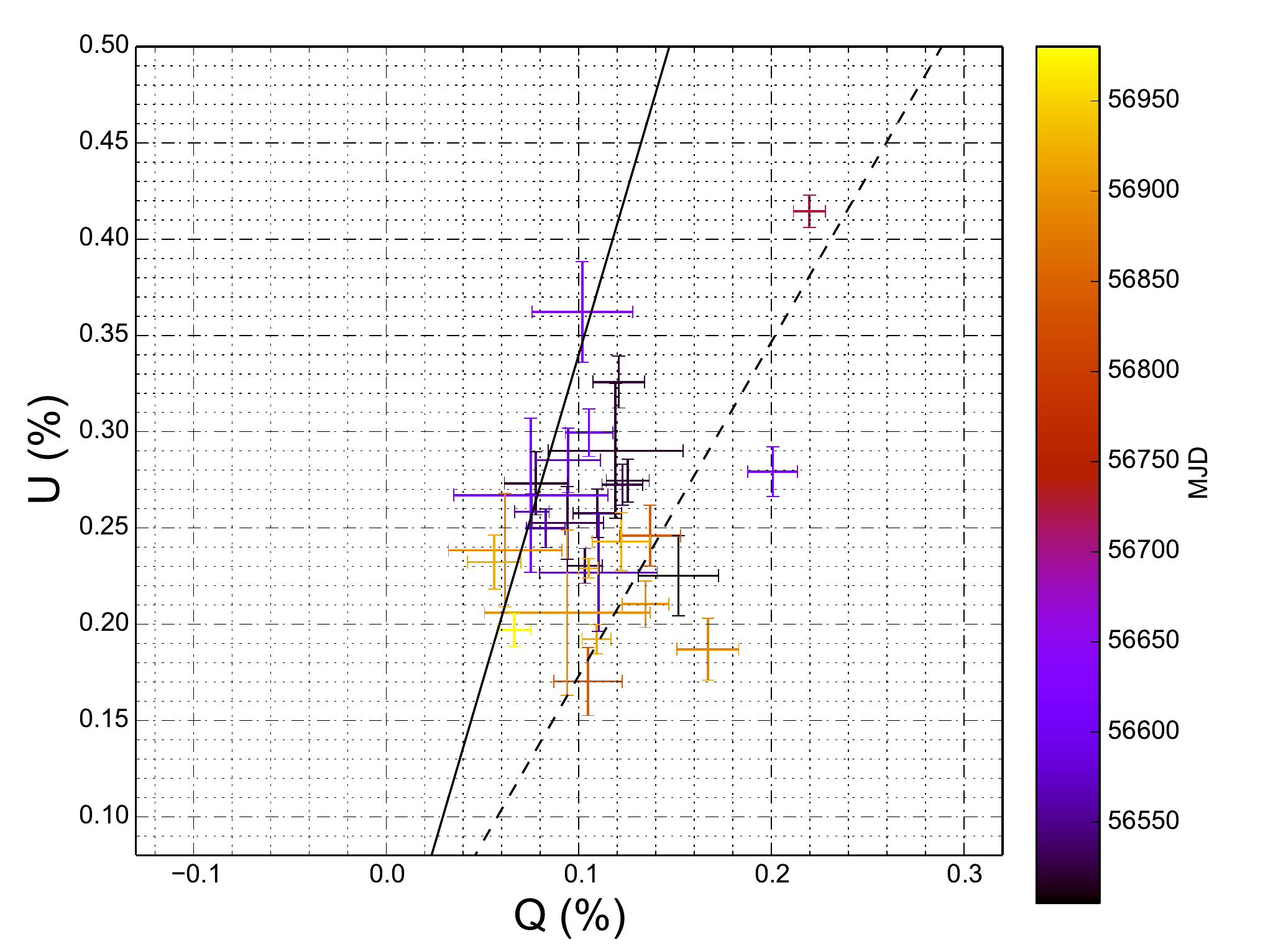} 
\caption{Same as previous figure, but data binned in time.}
\label{fig:aerisecQUbin}
\end{center}
\end{figure}

The linear polarization level at all observed filters are between $0.20-0.35\%$ for most of the considered period, starting in July 2013 up to the end of 2014. This is not true for the period around March 2014, where the linear polarization reaches a peak of $\approx0.50\%$ at $B$ and $V$ filters. Although based on observations of a single night, these do not seem to be data of bad accuracy since all observed filters indicate such increased polarization level, and the derived on-sky polarization orientation are the expected values (Fig.~\ref{fig:aerisecpol}, bottom panel). After the maximum seen in March 2014, the star reach a level similar to what was seen in the second half of 2013 ($\approx0.27\%$ in all filters), with slight downward trend.

Interestingly, this variation observed in polarization in March 2014 does not have a spectroscopic correspondence. This is indicative that the polarization originates from a smaller and closer to the star region than the spectroscopic line, which does not directly reflects short-term variations in the inner disk.

In Figs.~\ref{fig:aerisecQU} and \label{fig:aerisecQUbin} are displayed the $QU$ polarization diagrams of the $V$ band. The majority of the observations are located between the on-sky orientation angles 31$^\circ$ and 37$^\circ$. 
At the start of the outburst to polarization seems to point to $\{Q, U\} = \{0.12,0.27\} \%$. The maximum point is reaches in March 2014 at $\{0.21,0.41\} \%$, and the late period of 2014 it returns to $\{0.10,0.21\} \%$. 
The variations in the polarization angle can be interpreted as inhomogeneities in the inner disk. This allows speculating that the injection of matter in the disk does not occur with an azimuthal symmetry. Oddly, the epoch of highest polarization level ($\sim$~March 2014, i.e., approx$.$ epoch of highest mass injection) was preceded by the measurements of highest $V/R$ variations (February 2014).

\subsection{Secular evolution \label{sec:secevo}}
Our data reveals both long- (months) and short-term (weeks) variability. We start discussing the evolution of observational quantities since the start of active phase to describe the long-term variations. The precise starting date of the outburst is unknown, however spectroscopic measurements from late November 2012 (20$^{\rm th}$) show no sign of activity whereas in late January 2013 (18$^{\rm th}$) a significant activity is present. This restricts the beginning of the activity between early December 2012 and early January 2013.

The main quantities chosen to characterize the disk evolution are based on H$\alpha$ spectroscopy and broad band linear polarization and are presented in Fig.~\ref{fig:secevo1}. H$\alpha$ Equivalent Width (EW) departs from its maximum (positive) value, known to be photospheric-only, to smaller values. The decrease reveals the absorption profile been filled by disk emission, until eventually the emission overcomes the absorption (EW~$ > 0$), something that occurred around July 2013. Disk emission can also be traced by the Emission/Continuum (E/C) ratio. An E/C$>0$ means that the emission component is higher than the continuum level, a condition quickly reached (February 2013). 
The H$\alpha$ profile of Achernar exhibits a peak-separation decreasing with the increase of EW. This decrease of peak separation with time is consistent with a progressively larger disk; assuming Keplerian rotation, the fact that the peak separations decreases confirms that the emission originates in progressively larger regions.

The double peak line profile allows to measure asymmetries on the circumstellar emission through the $V/R$ quantity, or the ratio between the $V$iolet and $R$ed peak values. In the considered period, the average $V/R$ value was 0.991, with standard deviation of 0.020, i .e., completely consistent with a axisymmetric disk, for which $V/R = 1$. The maximum asymmetries records are $V/R=0.946$ and $1.054$. 

Polarization is the most sensitive quantity to detect circumstellar activity (Sect.~\ref{sec:pion}). This, and the null polarization value ($\lesssim 0.02\%$) reported by \citet{dom14a}, measured in the pre-outburst phase, indicates no considerable ISM polarization towards Achernar. For this reason no ISM correction was applied to the data. Achernar departs from its minimum (null) polarimetric value in November 2012 to a high value of $\gtrsim0.27\%$ at the first polarimetric observation in the active phase (late July 2013) until December 2013. Then, an increase in polarization occurs with peak around mid-March 2014, an in July 2014 the polarization returns to a lower value. The $V$-band polarization value of $0.27\%$, approximate value which polarimetry of Achernar is located most of the time, is twice that detected by \citet{car07a} in the 2006 activity.

The polarization of the 2013 active phase at $V$ band is presented in Fig.~\ref{fig:secevo1}, as well as the polarization color (i.e., the spectral slope of the polarization across the $B$, $V$ and $R$ bands). The polarization color is indicative of the disk density: the larger the disk, the larger in modulus (i.e., the more negative) the slope (see, e.g., \citealp[Fig.~4]{hau14a}). 
As the polarization level, that presented three clear phases (nearly constant in 2013, increase in 2014 and back to the 2013 value in August 2014) the polarization color is nearly constant in 2013, decreases sharply in the first half of 2014 and then goes back to a smaller value.
These facts alone allows us to conclude that the disk was denser between December 2013 and July 2014, and in 2013 the disk was slightly denser than in late 2014.

\FloatBarrier
\section{Interferometry}
In this section we describe the rich interferometric dataset obtained by us. In order to make an initial comparison between this data and the VDD, we developed a simple reference model. Based on the stability of the star observed between August 2014 and, at least, December 2014, we performed a modeling of Achernar based on the VDD prescription of a steady-stable disk. Our aim was to describe the spectroscopic and polarimetric quantities, and them apply it to the interferometric data. This model will set constraints on a future dynamical evolution analysis. 
As seen in Sect.~\ref{sec:secevo}, a negative slope in the polarized (visible) spectrum is indicative of high densities in the inner disk, thus suggesting a large mass injection rate. Conversely, a zero slope indicates a much smaller density, possibly related to a smaller or even zero disk feeding rate.

The steady-state (SS) assumption is likely valid for Achernar at the considered period (August 2014 - December 2014) since no big changes in the H$\alpha$ spectroscopy were recorded. Running a small grid of SS disks for Achernar, the value of $\Sigma_0=0.06$~g\,cm$^{-2}$ were found to be able to reproduce the observables, mainly $P\simeq0.27\%$ and EW~$\sim-2$~\AA{}.

To get an estimation of the disk mass injection $\dot{M}$ from this value, (and that is representative of the activity), we relate the surface density value $\Sigma_0$ with $\dot{M}$ by Eq.~\ref{eq:introMdot},
\begin{equation*}
\frac{\dot{M}R_0^{1/2}}{\alpha}=\frac{\Sigma_0 3\pi c_s^2 R_{\rm eq}^2}{(GM)^{1/2}}\,,
\end{equation*}
\begin{equation}
\frac{\dot{M}R_0^{1/2}}{\alpha}\simeq\frac{0.06\times3\pi\times1.413\cdot10^{12} \times(9.16R_\odot)^2}{(G\cdot6.2M_\odot)^{1/2}}.
\end{equation}
This results in $\dfrac{\dot{M}R_0^{1/2}}{\alpha}\simeq1.13\times10^{22}$~g\,s$^{-1}$\,cm$^{1/2}$. To have an order of magnitude estimation, we impose that $R_0\approx100R_{\rm eq}$, yielding $\dfrac{\dot{M}}{\alpha}\sim2.25\times10^{-11}$~$M_\odot$\,yr$^{-1}$.


\FloatBarrier
\subsection{Interferometric results}
Our interferometric campaign with AMBER covers 38 nights and has 79 individual observations. The observations where performed in high-resolution (HR) mode around the Br$\gamma$ line, make use of the FINITO fringe tracker. Because of its brightness, observation attempts were made in adverse weather conditions and for this reason a number have a low S/N and others were lost due to bad weather. Appendix Table~\ref{ap:aeriinterf} contains the list of observed nights, indicating for them the amount of individual observations. The data reduction was made with the software \textsc{amdlib} (e.g., \citealp{che09a}). 
During the reduction process it was attempted to perform the absolute calibration of visibilities from the standard star's observations of each night. However the visibilities were not reliable and we chose to normalize visibilities to the continuum level. 

\begin{figure}
    \centering
    \includegraphics[width=.8\linewidth]{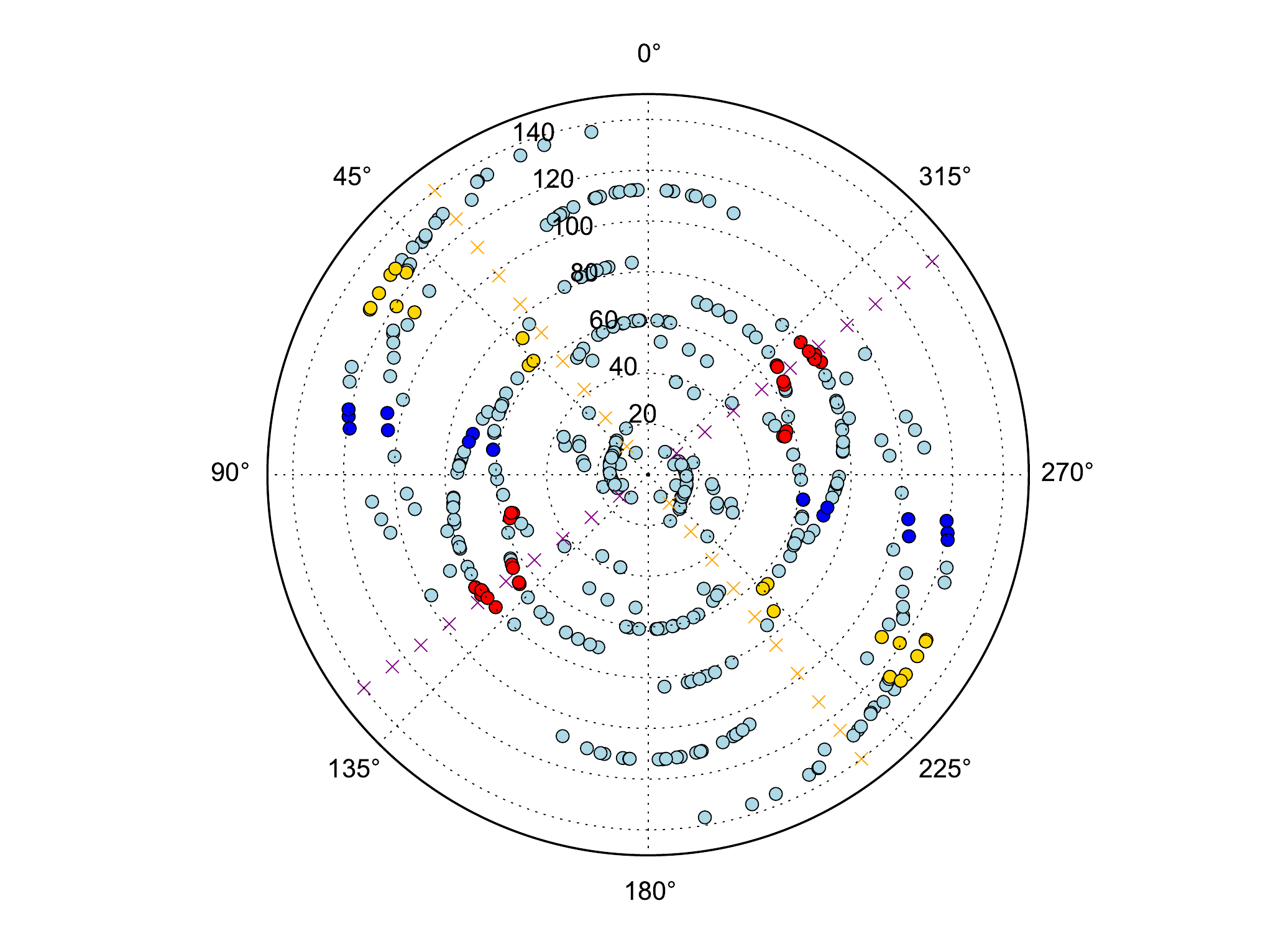}
    \caption[AMBER observations of Achernar between August 2013 and January 2015]{AMBER observations of Achernar between August 2013 and January 2015. The radial scale sets $\|\vec{B}_{\rm proj}\|$ up to 140~m, and the angle is the observed PA (deg). The selected groups in Table~\ref{tab:BPAs} are marked according to their orientation. The bottom right box indicate Achernar's polar direction (orange, PA~$=216.9^\circ$) and its equatorial direction (purple crosses, PA~$=126.9^\circ$), values from \citet{dom14a}.}
    \label{fig:polarobs}
\end{figure}

When comparing the temporal series of (spectro-)interferometric observations, differences among them can be given for two reasons: (i) changes in the configuration of the interferometer, and (ii) the disk evolution itself. In order to distinguish between the two effects, we grouped the observation on equivalent interferometric settings (i.e., similar $\vec{B}_{\rm proj}$), and analyzed the groups that had the largest number of observations. The criterion was to group observations within  $\Delta\|\vec{B}_{\rm proj}\|\leq20$~m and $\Delta{\rm PA}\leq10^\circ$. 
Table~\ref{tab:BPAs} contains the list of selected interferometers configurations. Fig.~\ref{fig:polarobs} show the data grouped by $\{\|\vec{B}_{\rm proj}\|$, PA\}.
Below, we show some of the grouped observations, guiding our discussion based on the their orientation: close to the equator, to the pole, and intermediate directions. The numbers above each observation box are, respectively, its Modified Julian Day (MJD) and the days count since January 1$^{\rm st}$ 2013, used as a reference to the activity starting date. 

\begin{table}
\centering
\caption{Selected groups of AMBER observations displaying similar $\vec{B}_{\rm proj}$.}
\begin{tabular}[]{cccccc}
\toprule
    PA & $\vec{B}_{\rm proj}$ & Number of & \multirow{2}{*}{MJD0} & \multirow{2}{*}{MJDF} & Appox. Direction \\ 
    (deg) & (m) & observations & & & (pole, interm., disk) \\
\midrule
    45 & 72 & 4 & 56518.27 & 56929.34 & pole \\
    55 & 122 & 12 & 56518.36 & 56928.25 & pole \\
    80 & 62 & 4 & 56518.39 & 56931.25 & interm. \\
    80 & 112 & 5 & 56920.30& 56924.30 & interm. \\
    105 & 52 & 4 & 56518.27 & 56920.32 & disk \\
    127 & 72 & 6 & 56558.30 & 56924.21 & disk \\
\bottomrule
\end{tabular}
\label{tab:BPAs}
\end{table}

\subsubsection*{Equatorial directions}
As in polarimetry, interferometric observations first occurred in August 2013. This is approximately 200 days after the beginning of the activity, where the disk in its inner regions was already well developed (as the constant polarization value indicates). For convenience, we employ the expression ``day $n$'' as the $n^{\rm th}$ day since January 1$^{\rm st}$ 2013 to refer to the observations dates. 

\begin{figure}
    \centering
    \includegraphics[width=.85\linewidth]{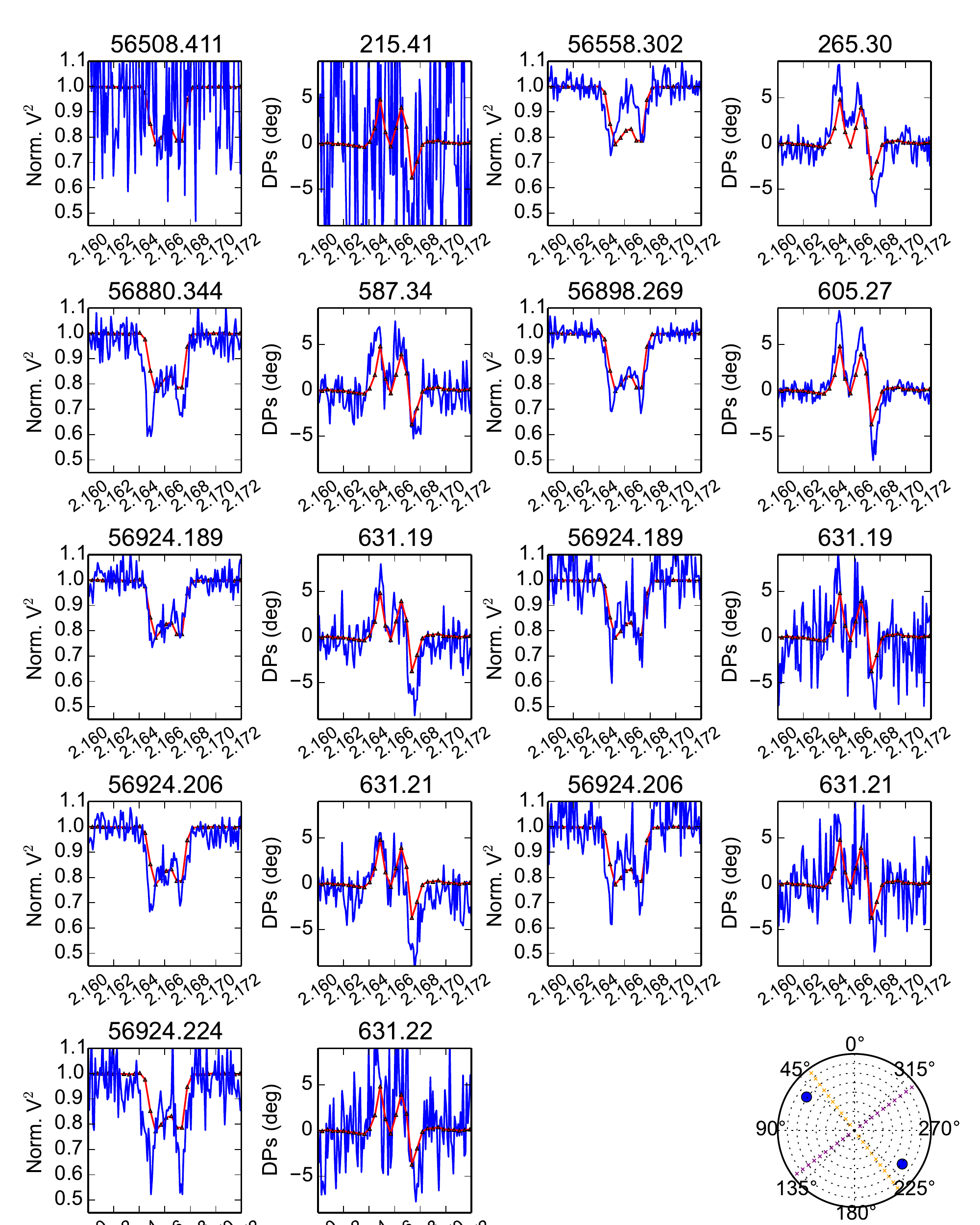}
    \caption[AMBER-VLTI Intereferometric observations for PA=111$^\circ$ and $\|\vec{B}_{\rm proj}\|=54$~m]{AMBER-VLTI Intereferometric observations. Normalized visibilities (first and third columns) and differential phases (DPs; second and fourth columns). 
    Observations are the blue curves and the steady-state VDD disk model are the red curves.
    Above the visibility boxes, is the Modified Julian Day of the observation. Above DPs, the amount of days since January 1$^{\rm st}$ 2013. 
    The bottom right box indicate Achernar's polar direction (orange) and its equatorial direction (purple crosses). They were grouped according to telescopes configuration indicated by the blue points (projected PA~$=105^\circ$ and $\|\vec{B}_{\rm proj}\|=52$~m).}
    \label{fig:ambP105B052}
\end{figure}

\begin{figure}
    \centering
    \includegraphics[width=.85\linewidth]{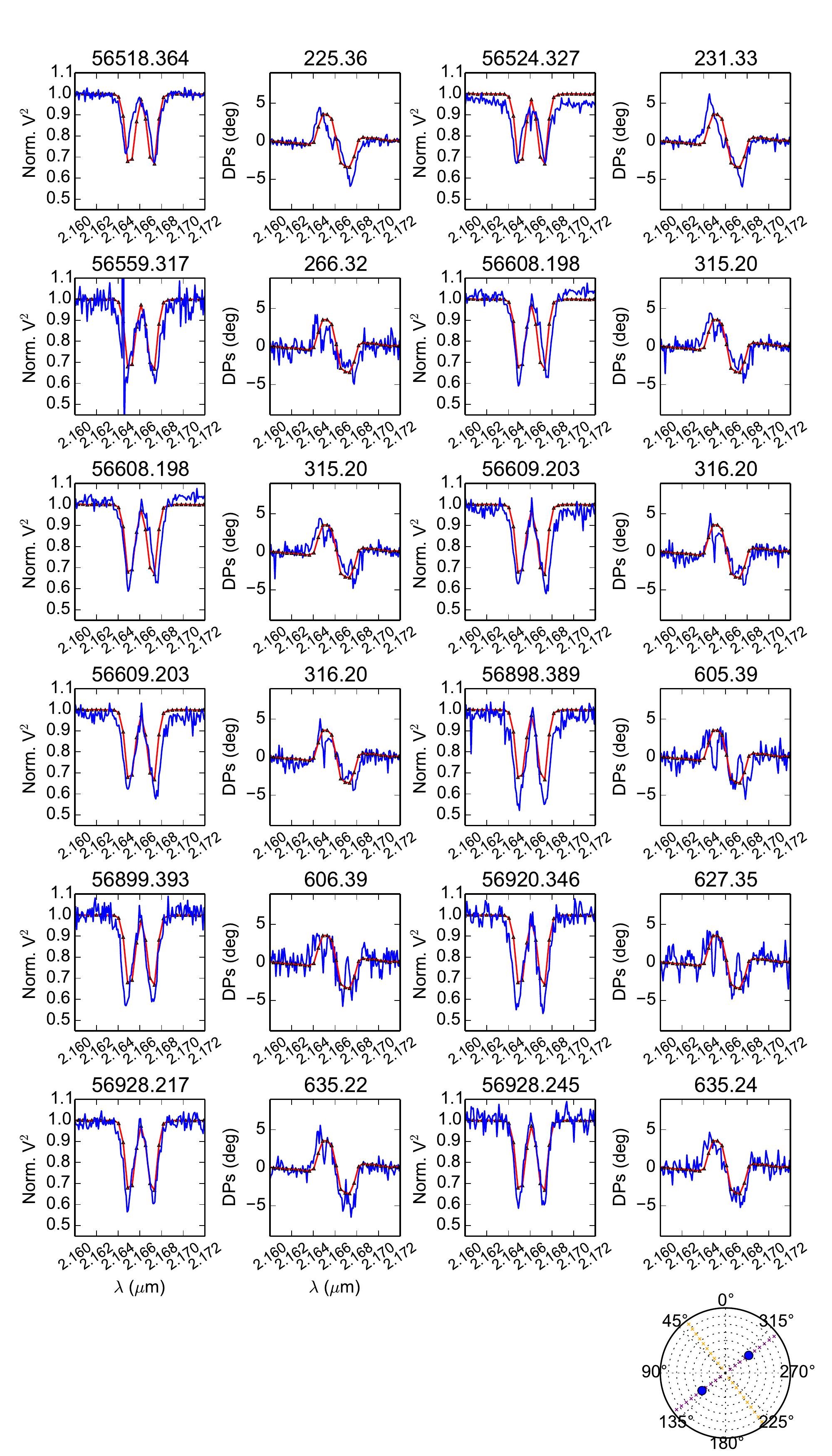}
    \caption{The same as previous figure, for PA~$=127^\circ$ and $\|\vec{B}_{\rm proj}\|=72$~m.}
    \label{fig:ambP127B072}
\end{figure}

In Fig.~\ref{fig:ambP105B052} we show the AMBER observations from days 225 to 636, where the disk growth can be observed both in visibilities and phases: the $V^2$ minimum decreases from $\sim0.8$ until $\sim0.6$, and the phase amplitudes increases from $\sim5.5^\circ$ up to $\sim8.5^\circ$. Assuming that the target brightness distribution is roughly similar in shape, a lower visibility corresponds to a angularly larger object, i.e., a larger disk, as expected from the VDD model. The same applies to the phases, if the photocenter-phase relation is valid. These observations were executed with $\|\vec{B}_{\rm proj}\|\simeq 52$~m. Although being a relatively small baseline, $\nu_{\rm obs}\simeq1.22$, indicating that the observations are in the limit of the astrometric regime. This can also be seen by the visibilities values: Fig.~\ref{fig:cqe3} shows that the astrometric regime ends when $V^2\sim0.6$, the same value registered in the latter observation. A very promissing result is the fact that our simple steady-state VDD model (red lines in Figs.~\ref{fig:ambP105B052} to \ref{fig:ambP055B122}) is already capable of reproducing most of the observed features.

Fig.~\ref{fig:ambP127B072} contains an extensive list of observations with a similar configuration, where variations of time scales of weeks to months can be studied. The $V^2$ minimum is progressively lower and the DPs show a rich morphology: from a (roughly) symmetric S-phases in at $\sim200$ days, and incursion appears on phase peaks ($\sim300$ and 635 days) to a very complex modulation in $\sim 600$ days. As the direction of this base is very close to the expected to the equator (PA~$=127^\circ$), deviations of phase symmetry are result of disk asymmetric emission. 

\subsubsection*{Intermediate directions}
\begin{figure}
    \centering
    \includegraphics[width=.85\linewidth]{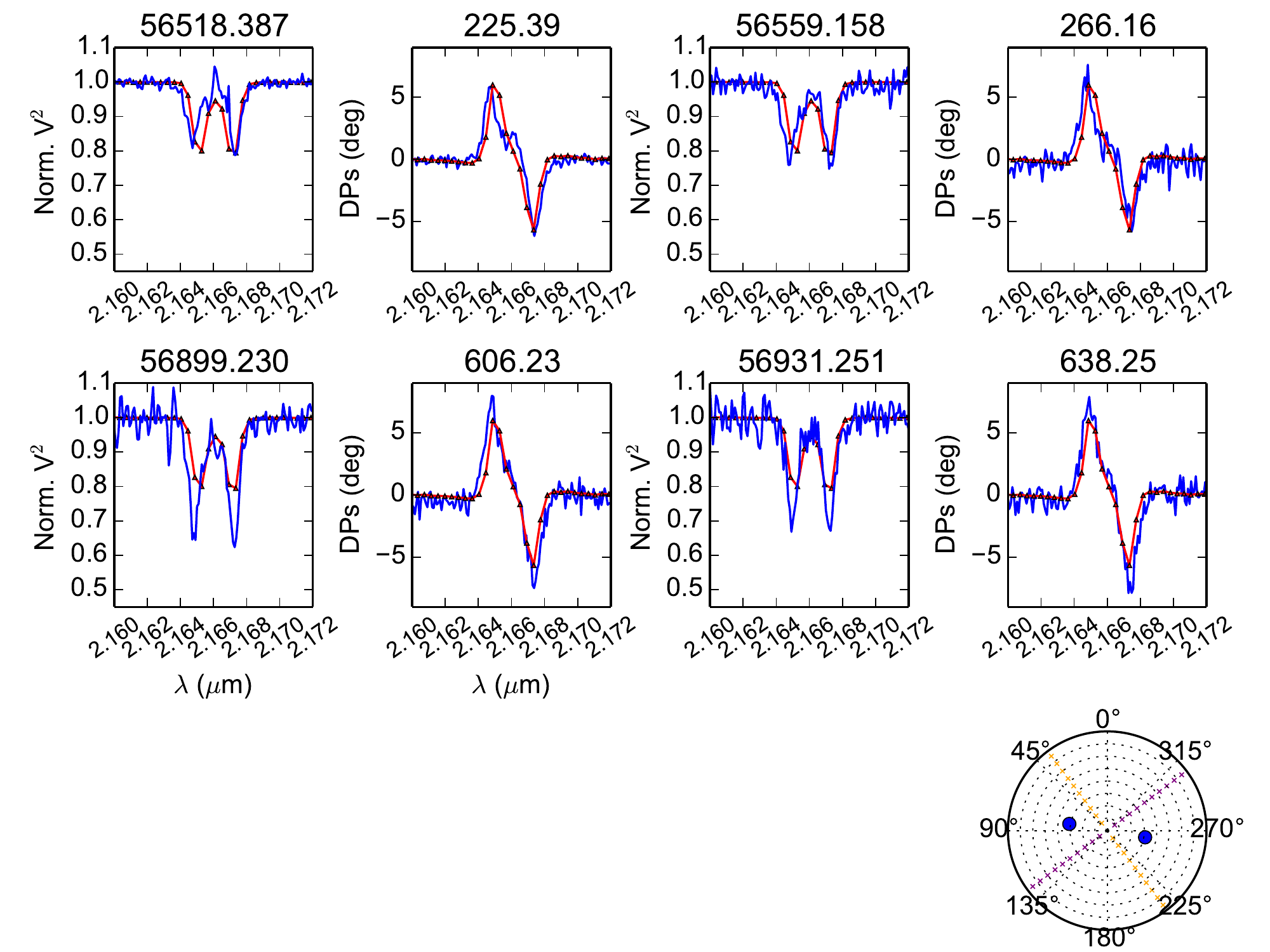}
    \caption{The same as previous figure, for PA~$=80^\circ$ and $\|\vec{B}_{\rm proj}\|=62$~m.}
    \label{fig:ambP080B062}
\end{figure}

For the rotating disk, as the Be stars, disk kinematic component is present and dominates the photocenter displacement in the astrometric regimes for most of the baseline orientations \citep[Fig.~9]{fae13a}. This only does not apply to PAs very close to the polar orientation. Fig.~\ref{fig:ambP080B062} shows the interferometric signal for a direction midway between the equator and the pole of the star at different epochs. The decrease of the visibilities and increase of the phases is readily noticed. In the first observations ($\sim200$ days) there is a substructure in the center of the phase profile.
    
\begin{figure}
    \centering
    \includegraphics[width=.85\linewidth]{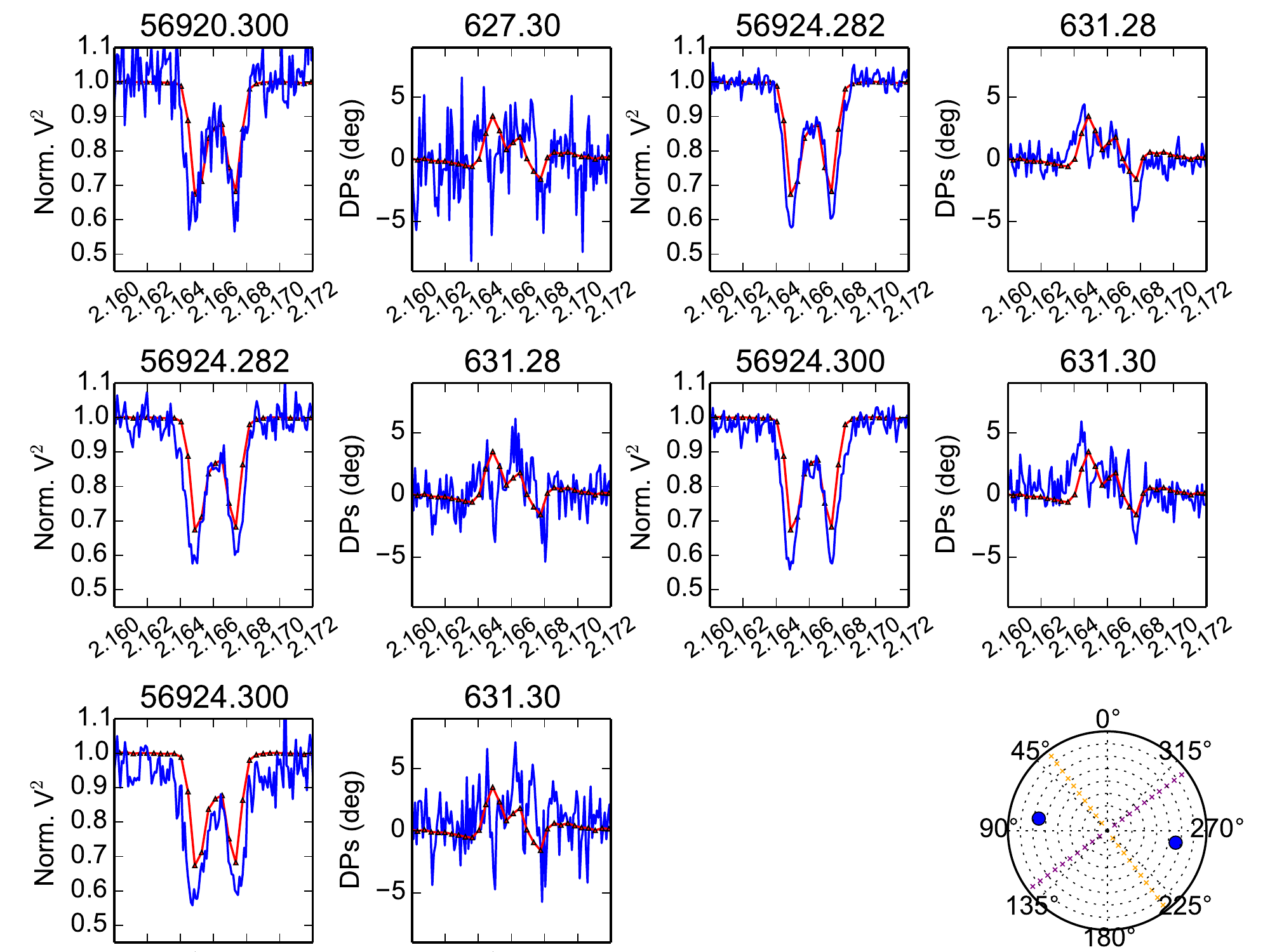}
    \caption{The same as previous figure, for PA~$=80^\circ$ and $\|\vec{B}_{\rm proj}\|=112$~m.}
    \label{fig:ambP080B112}
\end{figure}

An illustrative example of the over-resolution effects on the phase signal is at Fig.~\ref{fig:ambP080B112}. These are observations for exactly the same orientation of the previous figure, but for a longer baseline. The previous S-shaped profile almost disappears, remaining only a very irregular profile and a lower $V^2$ minimum. The displayed noise is due to bad weather conditions. Note that these complex phase profiles will like not present any challenge for the VDD model, since our initial reference model can reproduce the profiles qualitatively well.

\subsubsection*{Polar direction}
\begin{figure}
    \centering
    \includegraphics[width=.85\linewidth]{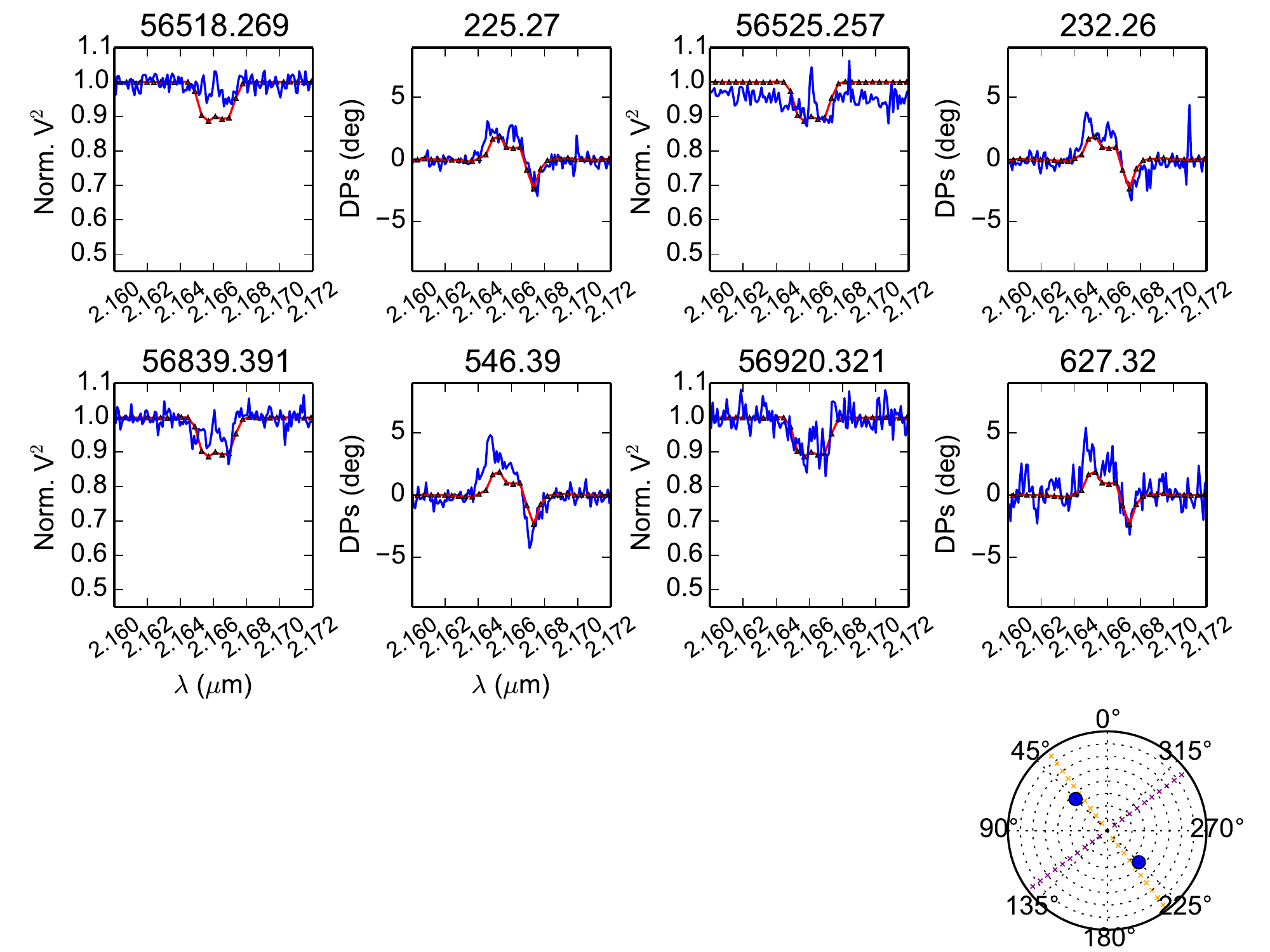}
    \caption{The same as previous figure, for PA=45$^\circ$ and $\|\vec{B}_{\rm proj}\|=72$~m.}
    \label{fig:ambP045B072}
\end{figure}

Fig.~\ref{fig:ambP045B072} contains observations in a PA very close do the polar direction. Different from observations shown until here, this configuration exhibits both low amplitude phases and visibilities (minimum $V^2\sim0.9$), but which also displays a slight evolution. 

\begin{figure}
    \centering
    \includegraphics[width=.85\linewidth]{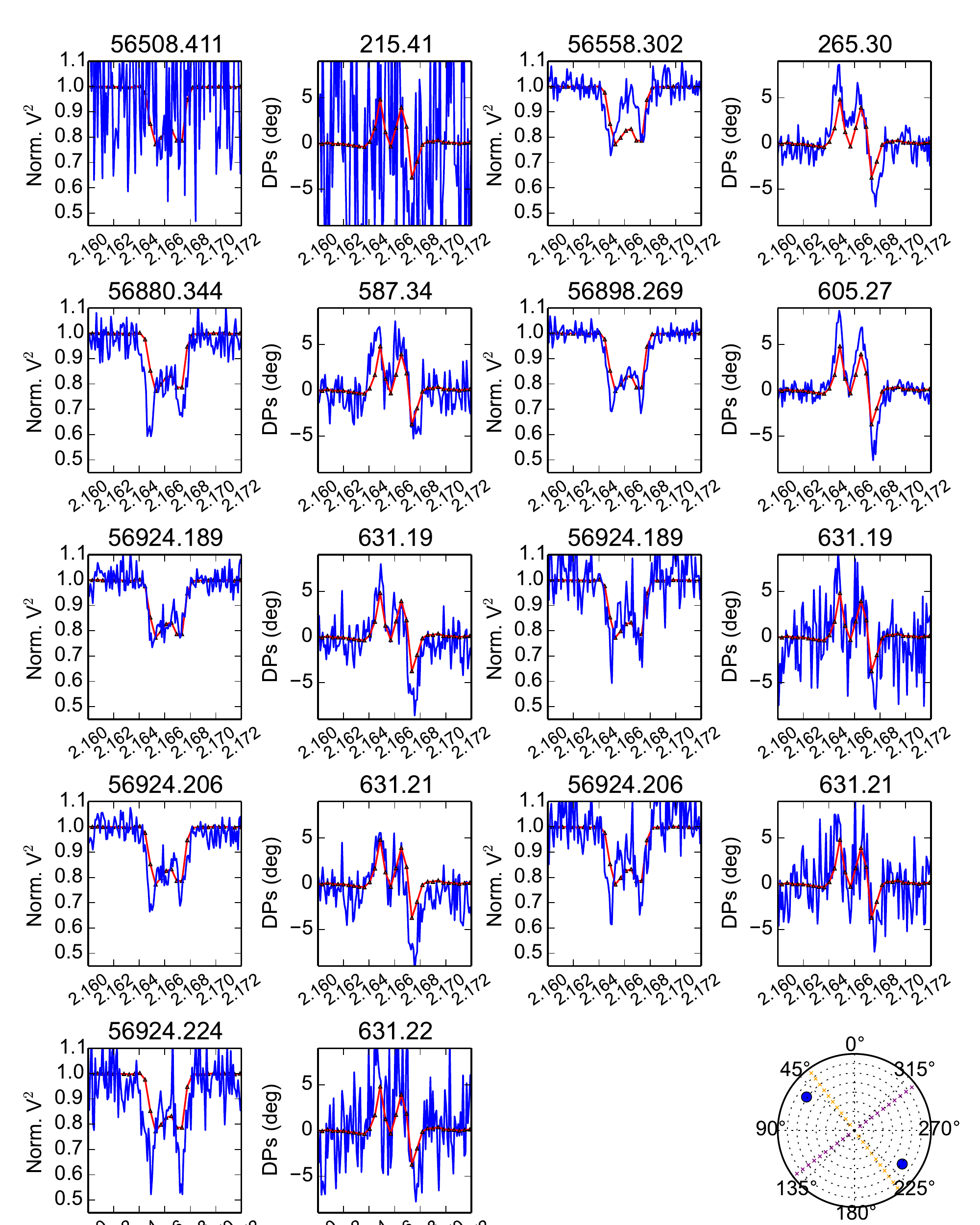}
    \caption{The same as previous figure, for PA=55$^\circ$ and $\|\vec{B}_{\rm proj}\|=122$~m.}
    \label{fig:ambP055B122}
\end{figure}

Fig.~\ref{fig:ambP055B122} contains other observations towards the polar directions, but slightly misaligned. The effects on the amplitude of the phases is clear, as pointed out in the discussion of PA's between the poles and the disk.
    
In conclusion to this inspection of the interferometric data, the observables exhibit a detectable evolution in the period (day $\sim250$ up to 700). As occurred to the polarimetric data, the emission in Br$\gamma$ appears to have evolved quickly, less than 6 months after the start of the activity. Unfortunately, no interferometric observations where performed between October 2013 and May 2014, period on which an increase of polarization was observed. 

We also highlight that almost of the data of Achernar are outside of the astrometric regime, which makes it difficult to isolate the over-resolution features of those predicted by the CQE-PS (Chapter~\ref{chap:bes}). Several phase profiles observed have a distinct shoulder in the rest wavelength (see, e.g., the profile for 266.16 days in Fig.~\ref{fig:ambP080B112}). These profiles suggest that the CQE-PS have been detected for the first time, however a more careful modeling is needed to reach this conclusion.


For the modeling, we considered the parameters from \citet{dom14a}, i.e., $d=42.75$~pc and $i=60.6^\circ$. The VDD provides a good description of the interferometric data for all considered directions. Difference between model and observations likely arise from the fact that we used a time-invariable model to fit time-variable observations. As a next step, we will study these observations with a dynamical VDD model. The main unexpected observed feature consists of the incursions near the phase peaks when the interferometer is aligned to the disk direction. Simulating different values of $\vec{B}_{\rm proj}$ and $d$ we were not able to reproduce this phenomenon. 


\FloatBarrier
\section{Chapter summary \label{sec:}}
In this chapter we presented the singular characteristics of the Be Achernar and our running observational campaign that aims at a multi-technique characterization of its current active phase. 

We started analyzing some observational phenomena, such as the $v\sin i$ variations and photometric and spectroscopic frequencies with very high precision data obtained by photospheric reconstruction. 
This led to an interesting speculation between the relationship of these frequencies observed for Be stars, traditionally attributed to non-radial pulsations, and their turnover rates.

An observational characterization of the activity was made, including H$\alpha$ spectroscopy, broad-band polarimetry, and spectro-interferometry AMBER. The observational results include: (i) variability in different time scales; (ii) variation $V/R$ and $QU$ polarimetric diagram shows that the disk is not symmetric; (iii) polarization shows periods of large-scale variation and also periods of relative stability; (iv) $H\alpha$ spectroscopy presents a slow and gradual evolution, and reaches a near stationary regime to only $\sim$1.6~years after the beginning of the activity. 

Interferometry is sensitive to the brightness distribution of a observed target. The relation of physical properties of Be stars and the interferometric data can only be directly determined if they are analyzed by realistic emission models. Both interferometry and spectrointerferometry allowed establishing the viscous decretion disk (VDD) model, solved by radiative transfer, as the most likely scenario to explain the Be star disks.

The solutions provided by radiative transfer can both determine the structure as well the emergent observables of the Be stars. This enables, for instance, the study of second order effects that may be present on these systems. Multi-technique studies, combining integrated quantities (such as photometry or spectroscopy) with ones containing spatial information (such as polarimetry or interferometry) are an important way of constraining the multiple parameters that characterized the systems in a consistent way. 

In interferometry, it is seen that the VDD model performs a good description of all baselines, ensuring that our estimates of the physical conditions of the system (such as gas density, scale high, etc.) are appropriate and that can lead to accurate estimations of parameters such as the mass loss rate. Some profile features, such as narrow cusps in visibility and phases at high speeds wavelengths, or incursions in the S-shaped driven by the disk need to be investigated. 



%% file: tex/ap-equations.tex
\chapter{Keplerian disks relations \label{ap:kepeq}}

Here we present some structure equations for Keplerian discs with isothermal scale height is steady-state regime. Steady state is defined when $\dfrac{\partial \Sigma}{\partial t}=0$.

\section{Surface density}
The integration of the mass conservation equation results
\begin{equation}
\dot{M}=2\pi r \Sigma v_r=\text{constant}.
\end{equation}
Relating the decretion rate $\dot{M}$ and the torque momentum equation
\begin{equation}
\Sigma(r)=\frac{\dot{M}R_{\rm eq}^{1/2}}{3\pi\alpha c_s^2}\frac{v_{\rm orb}}{r^{3/2}}\left[ \left(\frac{R_0}{r}\right)^{1/2}-1 \right]\,,
\label{eq:sigMdot}
\end{equation}
where $v_{\rm orb}=\sqrt{GM/R_{\rm eq}}$ and $R_0$ is the ``disk truncation'' radius obtained from momentum equation integration\footnote{$R_0$ can not be zero, otherwise it generates inconsistent quantities that and equivalent to a ``viscous disk without torque''.}. 
Under the assumption that $R_0\gg R_{d}$, Eq.~\ref{eq:sigMdot} becomes
\begin{equation}
\Sigma(r)_{R_0\gg R_{d}}=\frac{\dot{M}(GMR_0)^{1/2}}{3\pi\alpha c_s^2 R_{\rm eq}^2}\left(\frac{R_{\rm eq}}{r}\right)^2=\Sigma_0\left(\frac{R_{\rm eq}}{r}\right)^2.
\label{eq:sigSS}
\end{equation}

Once the disk is in vertically in hydrostatic equilibrium, the steady-state volumetric density $\rho(r,z)$ can be determined as
\begin{equation}
\rho(r,z)=\rho'_0(r) \exp\left(-\frac{1}{2}\frac{z^2}{H^2(r)}\right)\,,
\end{equation}
\begin{equation}
H(r)=H_0r^{1.5}=c_s\left( \frac{GM}{r} \right)^{-1/2}r.
\end{equation}
Parametrically, $m$ is the exponent of $\rho'_0(r)=\rho_0\left(\dfrac{R_{\rm eq}}{r}\right)^m$.
The surface density $\Sigma(r)$ can be written as 
\begin{equation}
\Sigma(r)=\int_{-\infty}^{\infty}\rho(r,z)dz=\int_{-\infty}^{\infty}\rho_0 \left(\frac{R_{\rm eq}}{r}\right)^{m}\exp\left(-\frac{1}{2}\frac{z^2}{H^2(r)}\right)\,;
\end{equation}
using the identity
\begin{equation}
\int_{-\infty}^{\infty}\exp-\frac{(x-x_0)^2}{a^2}dx=|a|\sqrt{\pi}\,,
\end{equation}
\begin{equation}
\Sigma(r)=\sqrt{\frac{2\pi}{GM}}c_s\rho_0R_{\rm eq}^{3/2}\left(\frac{R_{\rm eq}}{r}\right)^2\,,
\label{eq:sigrh0}
\end{equation}
where $m=7/2$ is equivalent to steady-state solution. Parametrically, $n$ is the exponent of $\Sigma(r)=\Sigma_0(R_{\rm eq}/r)^n$ and since the scale-height $H(r)\propto r^{3/2}$,
\begin{equation}
 n=m-\frac{3}{2}.
\end{equation}

\section{Disk mass injection rate $\dot{M}$}
Equaling Eq.~\ref{eq:sigSS} and Eq.~\ref{eq:sigrh0} at $r=R_{\rm eq}$,
\begin{equation}
\left(\frac{2\pi}{GM}\right)^{1/2}c_s\rho_0 R_{\rm eq}^{3/2}=\frac{\dot{M}(GMR_0)^{1/2}}{3\pi\alpha c_s^2 R_{\rm eq}^2}\,,
\end{equation}
\begin{equation}
\dot{M}_{R_0\gg R_{d}}=\left(\frac{2\pi}{R_0}\right)^{1/2}\frac{3\pi\alpha c_s^3R_{\rm eq}^{7/2}}{GM} \rho_0.
\label{ap:eqmdot}
\end{equation}

For reference, Eq.~\ref{eq:sigMdot} and Eq.~\ref{eq:sigrh0} at $r=R_{\rm eq}$,
\begin{equation}
\dot{M}=\frac{\sqrt{2\pi}3\pi\alpha c_s^3\rho_0R_{\rm eq}^3}{GM}\left[\left( \frac{R_0}{R_{\rm eq}}\right)^{1/2}-1\right]^{-1}.
\end{equation}

\section{Disk total mass}
Disk mass $M_d=M_d(c_s,M,\rho_0,R_{\rm eq},R_d)$ is
\begin{equation}
M_{d, R_0\gg R_{d}} = \int_0^{2\pi}\int_{R_{\rm eq}}^{R_d}\Sigma(r)r dr d\phi =2\pi R_{\rm eq}^2 \Sigma_0 \ln\left(\frac{R_d}{R_{\rm eq}}\right)
\end{equation}

For reference, $M_d=M_d(R_0,\alpha,c_s,M,\dot{M},R_{\rm eq},R_d)$ is
\begin{equation}
M_d =(2\pi)^{3/2}\int_{R_{\rm eq}}^{R_d}\rho(r)H(r)dr\,,
\end{equation}
\begin{equation}
M_d=\frac{2\dot{M}\sqrt{GM}}{3\alpha c_s^3}\left[\sqrt{R_0}\ln\left(\frac{R_d}{R_{\rm eq}}\right)+2(\sqrt{R_{\rm eq}}-\sqrt{R_0}) \right].
\end{equation}

\section{\textsc{BeAtlas} surface density equivalence \label{sec:}}
For each stellar mass $M$ and a given rotational rate, the base surface density $\Sigma_0$ is a function of (at least) two variables. In the case of Eq.~\ref{eq:sigrh0}, it is $\Sigma_0(c_s, \rho_0)$. To set the two parameters simultaneously, another equation needs to be satisfied. The criterion for the project was to equal the disk mass-injection rate loss equation ($\dot{M}$, Eq.~\ref{ap:eqmdot}), assuming that $R_0 =  kR_{\rm eq}$ with $k$ constant.

Thus, for two stars of different rotation rates, equality of the above mentioned equations yields, respectively,
\begin{equation}
\frac{\rho_0}{\rho_0'}=\frac{c_s'}{c_s}\frac{R_{\rm eq}'^{3/2}}{R_{\rm eq}^{3/2}}\,,
\end{equation}
\begin{equation}
\frac{\rho_0}{\rho_0'}=\frac{c_s'^3}{c_s^3} \frac{R_{\rm eq}'^{3}}{ R_{\rm eq}^{3} }.
\end{equation}

The relationship between disk temperature and equatorial radii is
\begin{equation}
{c_s'^2} = \frac{R_{\rm eq}^{3/2}}{ R_{\rm eq}'^{3/2} }{c_s^2}\,,
\end{equation}
where the reference value is $c_s=c_{s}(W=0.447) = 0.72\,T_{p}(W=0.447)$. The adopted criterion results in a decrease of disk scale height $H$ with the stellar rotation, something expected from initial studies (e.g., \citealp{mcg11a}).

\section{Relation between numerical and mass densities \label{sec:}}
The chemical composition of a gas is often defined as mass fractions of H, He and other elements, respectively $X$, $Y$ and $Z$. By definition,
\begin{equation}
X+Y+Z=1.
\end{equation}
The $X$ fraction can be written as
\begin{equation}
X=\frac{n_H m_H}{n_H m_H+n_{He} m_{He}+\sum n_i m_i}.
\end{equation}

A more convenient way of is write the mass fractions in terms of the relative abundance to the hydrogen. 
Considering $A_i$ be the mass number of an element $i$
\begin{equation}
X=\frac{1}{1+4(n_{He}/n_H)+\sum A_i(n_i/n_H)}\,,
\end{equation}
\begin{equation}
Y=\frac{4(n_{He}/n_H)}{1+4(n_{He}/n_H)+\sum A_i(n_i/n_H)}\,,
\end{equation}
\begin{equation}
Z=\frac{\sum A_i(n_i/n_H)}{1+4(n_{He}/n_H)+\sum A_i(n_i/n_H)}.
\end{equation}

\subsubsection*{Neutral atoms}
The relationship between numerical and mass density for a neutral gas with a given chemical composition is
\begin{equation}
n=\frac{\rho}{\mu m_H}\approx\frac{\rho}{m_H} \left(X+\frac{Y}{4}+\frac{Z}{\mean{A}}\right)\,,
\end{equation}
where $\mean{A}$ is the mean atomic weight of the gas, excluding H and He. Removing the dependence with $Z$
\begin{equation}
\mu=\frac{\mean{A}}{1+(\mean{A}-1)X+(\mean{A}/4-1)Y}.
\end{equation}

\subsubsection*{Full ionized atoms}
For a fully ionized gas,
\begin{equation}
n\approx\frac{\rho}{m_H} \left(2X+\frac{3}{4}Y+\frac{Z}{2}\right)\,,
\end{equation}
where $2.0\geq \mu \geq 0.5$,
\begin{equation}
\mu=\frac{2}{1+3X+Y/2}.
\end{equation}

In terms of electronic density, $n_e$,
\begin{equation}
n_e\mu_e\equiv n\mu\,,
\end{equation}
\begin{equation}
n_e=\frac{\rho}{\mu_e m_H}\approx\frac{\rho}{m_H}\left(X+\frac{Y}{2}+\frac{Z}{2}\right)\,,
\end{equation}
where $2.0\geq \mu_e \geq 1.0$,
\begin{equation}
\mu_e\approx\frac{2}{1+X}.
\end{equation}
\chapter{Useful equations for interferometry \label{ap:eqinterf}}
\section{Fourier transform}
There are several conventions for defining the Fourier transform $\hat{f}$ of an integrable function $f$. We will use the definition
\begin{equation}
\hat{f}(\xi) = \frac{1}{(2\pi)^{n/2}} \int_{-\infty}^{\infty} f(x)\exp(- i x \xi)\,dx 
\end{equation}
for every real number $\xi$.
 
When the independent variable $x$ represents a unit (let's say, \textit{time}, with SI unit of seconds), the transform variable $\xi$ represents its inverse (in this case, \textit{frequency}, in Hertz). Under typical smooth conditions, $f$ can be reconstructed from $\hat{f}$ by the inverse transform:
\begin{equation}
f(x) = \frac{1}{(2\pi)^{n/2}}\int_{-\infty}^{\infty} \hat{f}(\xi)\exp(i \xi x)\,d\xi
\end{equation}
for every real number $x$; $n$ is the dimension of the vectors $\xi$ and $x$ (in our example, $n=1$).

The Fourier transform $\hat{f}$ measures how much of an individual frequency is present in a function $f$. If we know $\hat{f}(\omega)$ for all possible frequencies $\omega$, we can perfectly approximate our function $f$. And that's what the Fourier transform does.


Fourier transform takes some function $f(t)$ of time and returns some other function$\hat{f}(\omega)=\mathscr{F}(f)$, it's Fourier transform, that describes how much of any given frequency is present in $f$. It's just another representation of $f(t)$, of equal information but with over a different domain. And we can go the other way. Given a Fourier transform, we can integrate over its all frequencies, and get the original $f(t)$ again, which we call inverse Fourier transform $\mathscr{F}(f)^{-1}$.

That is important in this context is that Fourier transform has many interesting mathematical properties (e.g., convolution is just a multiplication). Indeed, in 1807 Fourier introduced his  for the purpose of solving the heat equation in a metal plate (heat equation is a partial differential equation). 

A classical example of the utility of the Fourier transform is noise reduction on a digital image (or a sound wave). Rather than manipulating a function image(\textit{Pixel-Brightness}), we transform it and work with $\mathscr{F}$(image; \textit{Frequency-Amplitude}). Let suppose those party of high frequency that cause the noise. It can simply be cut off with $\mathscr{F}$(image; $\omega$)=0, when $\omega>w_{\rm cut}$. We transform it back, and have the image without noise. 

\section{Bessel functions}
The Bessel functions are frequently defined as solutions to the differential
equation:
\begin{equation}
x^2\frac{d^2y}{dx^2}+x\frac{dy}{dx}+(x^2-k^2)y=0. 
\end{equation}
They are generally refereed as $J_k(x)$, where $k$ is the function index:
\begin{equation}
J_k(x)=\sum_{m=0}^\infty\frac{(-1)^m}{m!\,\Gamma(m+k+1)}\left(\frac{x}{2}\right)^{2m+k}\,,
\label{eq:bessfunc}
\end{equation}
where $\Gamma(y)$ is the gamma function. $k$ is usually taken as an integer number, but Eq.~\ref{eq:bessfunc} is also valid for real numbers. There are other definitions of the Bessel functions, for examples, in form of power series.
An interesting problem on mathematics is to find the roots of the Bessel Functions. Since interferometry make use of them for simple brightness distributions that can be treated analytically, we list the initial root values in Tables~\ref{tab:BesselFuctionsRoots} and \ref{tab:BesselDerivateRoots}. Table~\ref{tab:antiseche} contains a list of useful brightness distribution models.

\begin{table}[htbp]
	\centering
		\caption{The first few roots of Bessel functions $J_k$}
		\begin{tabular}{cccc}
			$k$	&$J_0(x)$	&$J_1(x)$	&$J_2(x)$ \\ \hline
1	&2.4048	&3.8317	&5.1356 \\ 
2	&5.5201	&7.0156	&8.4172 \\
3	&8.6537	&10.1735	&11.6198 \\
4	&11.7915	&13.3237	&14.7960 \\
5	&14.9309	&16.4706	&17.9598
		\end{tabular}
	\label{tab:BesselFuctionsRoots}
\end{table}

\begin{table}[htbp]
	\centering
\caption{The first few roots of the derivative of Bessel functions $J_k'(x)$}
\begin{tabular}{cccc}
$k$	&$J_0'(x)$	&$J_1'(x)$	&$J_2'(x)$	\\ \hline
1	&3.8317	&1.8412	&3.0542	\\
2	&7.0156	&5.3314	&6.7061 \\
3	&10.1735	&8.5363	&9.9695 \\
4	&13.3237	&11.7060	&13.1704	\\
5	&16.4706	&14.8636	&16.3475	\\
		\end{tabular}
	\label{tab:BesselDerivateRoots}
\end{table}

An useful relation dealing with half integer indexes of Bessel functions are:
\begin{equation}
J_{n/2-1}(x)+J_{n/2+1}(x)=\frac{2n}{x}J_{n/2}(x)\,,
\end{equation}
\begin{equation}
J_{-n}(x)=(-1)^nJ_n(x).
\end{equation}

\section{Other valuable information \label{sec:otherinterf}}
\begin{table}
\centering
\caption{Simple interferometric reference models}
\begin{tabular}[]{p{4cm} p{4.5cm} p{4.5cm}}
\toprule
    Shape & Brightness distribution & Visibility \\    
\midrule
    Point source & $\delta(\vec{r})$ & 1 \\ \hline 
    Background & $I_0$ & 0 \\ \hline
    Binary star & $I_0[\delta(\vec{r_1})+R\delta(\vec{r_1}-\vec{r_2})]$ & $\sqrt{\frac{V_1^2+R^2V_2^2+2R\|V_1\|\|V_2\|\cos\left[2\pi\vec{u}\cdot(\vec{r_1}-\vec{r_2})\right]}{(1+R)^2}}$ \\ \hline
    Gaussian & $I_0\sqrt{\dfrac{2\sigma}{\pi}}\times\exp\left(-\dfrac{r^2}{\sigma^2}\right)$ & $\exp(-\pi\sigma|\vec{u}|)^2$ \\ \hline
    Uniform disk & $\dfrac{4}{\pi\phi^2}$ if $|\vec{r}|<\dfrac{\phi}{2}$; 0 else & $\left[\dfrac{2J_1(\pi\phi|\vec{u}|)}{\pi\phi|\vec{u}|}\right]$ \\ \hline
    Ring & $\dfrac{1}{\pi\phi}\delta\left(|\vec{r}|-\dfrac{\phi}{2}\right)$ & $J_0(\pi\phi|\vec{u}|)$ \\ \hline
    Exponential & $\exp(-k_0|\vec{r}|), k_0\ge0$ & $\left(\dfrac{k_0^2}{1+k_0^2\vec{u}^2}\right)$ \\ \hline
    Circularly symmetric & $I(r)$ & $2\pi\int_0^\infty I(r)J_0(2\pi\vec{r}\cdot\vec{u})r\,dr$ \\ \hline
    Rectangle ($xy$ coord.) & $1/lL$ if $x<l$ and $y<L$; 0 else & $\left[\dfrac{\sin(\pi xl)\sin(\pi yL)}{\pi^2xylL}\right]$ \\ \hline
    Limb-darkened (linearly) disk & $I_0[1=a_k(1-\mu)]$ if $|\vec{r}|<\dfrac{\phi}{2}$; 0 else; $\mu=\cos\left(\dfrac{2|\vec{r}|}{\phi}\right)$ & $\dfrac{\left[\alpha\dfrac{J_1(x)}{x}+\beta\sqrt{\dfrac{\pi}{2}}\dfrac{J_{3/2}(x)}{x^{3/2}}\right]^2}{\left(\dfrac{\alpha}{2}+\dfrac{\beta}{3}\right)^2}$ $\alpha=1-a_k;\ \beta=a_k\lambda$;  $x=\pi\theta_{\rm LD}|\vec{u}|$ \\
\bottomrule
\end{tabular}
\label{tab:antiseche}
\end{table}

Properties of the Fourier transform:
\begin{itemize}
    \item \textbf{linearity}: $\mathscr{F}(f+g)=\mathscr{F}(f)+\mathscr{F}(g)$
    \item \textbf{translation}: $\mathscr{F}[f(x-x_0,y-y_0)] = \mathscr{F}[f(u,v)\exp i2\pi(ux_0+vy_0)]$
    \item \textbf{similarity}: $\mathscr{F}[f(ax,by)]=\dfrac{1}{ab}\mathscr{F}\left[f\left(\dfrac{u}{a},\dfrac{v}{b}\right)\right]$
    \item \textbf{convolution}: $\mathscr{F}(f\otimes g) = \mathscr{F}(f)\times\mathscr{F}(g)$
\end{itemize}

The optical interferometry data should be stored according to the IAU standards. In this case, it is the \textsc{oifits} format defined by \citet{pau05a}. Routines to deal with this format in Python programing language are available in the \textsc{pyhdust} library.
\vspace{10pt}

The relation between angle units can be obtained as 
\begin{equation}
\frac{\text{radian}}{\text{arcsecond}}=\frac{2\pi}{360^\circ60"60'}\approx\frac{1}{206264.8}
\end{equation}

So, we have the factor of $\approx 4.848\times10^{-9}$ rad per mas.

%% file: tex/ap-polobs.tex
\chapter{Polarimetric observations of magnetospheres \label{ap:polobs}}

\begin{center}
\begin{longtable}{ccccccc}
\caption%
{$V$-band polarization measurements of \sori{}.} \label{tab:apsori} \\
\hline MJD & $P$ (\%) & $Q$ (\%) & $U$ (\%) & $\theta$ (deg) & $\sigma_P$ (\%) & $\sigma_\theta$ (deg) \\ \hline
\endfirsthead

\multicolumn{7}{c}%
{{\footnotesize  \itshape {Table \thetable{}} -- continued from previous page.}} \\
\hline MJD & $P$ (\%) & $Q$ (\%) & $U$ (\%) & $\theta$ (deg) & $\sigma_P$ (\%) & $\sigma_\theta$ (deg) \\ \hline
\endhead

\hline \multicolumn{7}{r}{\footnotesize \itshape {Continued on next page}} \\ \hline
\endfoot

\hline \hline
\endlastfoot
55808.325654 & 0.292 & -0.286 & 0.061 & 83.9 & 0.015 & 0.4 \\ 
55808.325654 & 0.292 & -0.286 & 0.061 & 83.9 & 0.015 & 0.4 \\ 
55814.297218 & 0.362 & -0.362 & -0.015 & 91.2 & 0.021 & 0.6 \\ 
55814.297218 & 0.362 & -0.362 & -0.015 & 91.2 & 0.021 & 0.6 \\ 
55808.347460 & 0.365 & -0.364 & 0.025 & 88.1 & 0.044 & 1.3 \\ 
55808.347460 & 0.365 & -0.364 & 0.025 & 88.1 & 0.044 & 1.3 \\ 
55427.338335 & 0.349 & -0.349 & -0.024 & 92.0 & 0.047 & 1.3 \\ 
55427.338335 & 0.349 & -0.349 & -0.024 & 92.0 & 0.047 & 1.3 \\ 
55813.268611 & 0.354 & -0.354 & -0.005 & 90.4 & 0.005 & 0.1 \\ 
55813.268611 & 0.354 & -0.354 & -0.005 & 90.4 & 0.005 & 0.1 \\ 
55813.309468 & 0.357 & -0.357 & 0.001 & 90.0 & 0.010 & 0.3 \\ 
55813.309468 & 0.357 & -0.357 & 0.001 & 90.0 & 0.010 & 0.3 \\ 
55813.348828 & 0.335 & -0.335 & -0.006 & 90.5 & 0.013 & 0.4 \\ 
55426.341288 & 0.346 & -0.345 & -0.033 & 92.7 & 0.036 & 1.0 \\ 
55499.133779 & 0.311 & -0.310 & -0.022 & 92.1 & 0.005 & 0.1 \\ 
55499.156024 & 0.323 & -0.322 & -0.022 & 91.9 & 0.004 & 0.1 \\ 
55499.184219 & 0.309 & -0.309 & -0.010 & 91.0 & 0.004 & 0.1 \\ 
55499.205862 & 0.310 & -0.310 & 0.000 & 90.0 & 0.006 & 0.2 \\ 
55499.227656 & 0.310 & -0.310 & 0.007 & 89.4 & 0.006 & 0.2 \\ 
55443.334203 & 0.312 & -0.312 & 0.004 & 89.6 & 0.015 & 0.4 \\ 
55443.352942 & 0.312 & -0.312 & 0.019 & 88.2 & 0.012 & 0.3 \\ 
55498.137529 & 0.333 & -0.332 & 0.015 & 88.7 & 0.006 & 0.2 \\ 
55498.160515 & 0.327 & -0.326 & 0.024 & 87.9 & 0.004 & 0.1 \\ 
55498.182436 & 0.329 & -0.328 & 0.025 & 87.8 & 0.005 & 0.2 \\ 
55498.231777 & 0.336 & -0.336 & 0.016 & 88.6 & 0.004 & 0.1 \\ 
55442.284060 & 0.326 & -0.324 & 0.034 & 87.0 & 0.013 & 0.4 \\ 
55498.253605 & 0.332 & -0.332 & 0.012 & 89.0 & 0.006 & 0.2 \\ 
55442.307880 & 0.321 & -0.321 & 0.014 & 88.8 & 0.032 & 0.9 \\ 
55498.278038 & 0.337 & -0.337 & 0.016 & 88.6 & 0.005 & 0.1 \\ 
55498.299809 & 0.328 & -0.328 & 0.007 & 89.4 & 0.007 & 0.2 \\ 
55442.333760 & 0.328 & -0.328 & 0.001 & 89.9 & 0.030 & 0.8 \\ 
55497.135596 & 0.350 & -0.350 & 0.008 & 89.4 & 0.008 & 0.2 \\ 
55497.157448 & 0.331 & -0.331 & -0.006 & 90.5 & 0.004 & 0.1 \\ 
55497.180341 & 0.331 & -0.331 & -0.004 & 90.3 & 0.009 & 0.2 \\ 
55497.201615 & 0.323 & -0.323 & -0.002 & 90.2 & 0.004 & 0.1 \\ 
55497.223744 & 0.324 & -0.323 & -0.009 & 90.8 & 0.007 & 0.2 \\ 
55497.245943 & 0.317 & -0.317 & -0.003 & 90.2 & 0.011 & 0.3 \\ 
55441.293938 & 0.287 & -0.287 & -0.020 & 92.0 & 0.016 & 0.5 \\ 
55497.268212 & 0.330 & -0.330 & -0.008 & 90.7 & 0.009 & 0.3 \\ 
55441.320951 & 0.306 & -0.299 & -0.068 & 96.4 & 0.035 & 1.0 \\ 
55441.320951 & 0.306 & -0.299 & -0.068 & 96.4 & 0.035 & 1.0 \\ 
55497.290446 & 0.328 & -0.328 & -0.007 & 90.6 & 0.015 & 0.4 \\ 
55497.290446 & 0.328 & -0.328 & -0.007 & 90.6 & 0.015 & 0.4 \\ 
55441.348127 & 0.299 & -0.298 & -0.019 & 91.8 & 0.011 & 0.3 \\ 
55441.348127 & 0.299 & -0.298 & -0.019 & 91.8 & 0.011 & 0.3 \\ 
55496.151719 & 0.327 & -0.326 & -0.018 & 91.6 & 0.008 & 0.2 \\ 
55496.151719 & 0.327 & -0.326 & -0.018 & 91.6 & 0.008 & 0.2 \\ 
55496.174196 & 0.323 & -0.322 & -0.027 & 92.4 & 0.008 & 0.2 \\ 
55496.174196 & 0.323 & -0.322 & -0.027 & 92.4 & 0.008 & 0.2 \\ 
55496.195573 & 0.337 & -0.336 & -0.027 & 92.3 & 0.010 & 0.3 \\ 
55496.195573 & 0.337 & -0.336 & -0.027 & 92.3 & 0.010 & 0.3 \\ 
55496.230110 & 0.335 & -0.334 & -0.026 & 92.2 & 0.009 & 0.3 \\ 
55496.230110 & 0.335 & -0.334 & -0.026 & 92.2 & 0.009 & 0.3 \\ 
55808.226678 & 0.328 & -0.327 & 0.013 & 88.8 & 0.019 & 0.5 \\ 
55808.226678 & 0.328 & -0.327 & 0.013 & 88.8 & 0.019 & 0.5 \\ 
55496.251962 & 0.329 & -0.328 & -0.008 & 90.7 & 0.013 & 0.4 \\ 
55496.251962 & 0.329 & -0.328 & -0.008 & 90.7 & 0.013 & 0.4 \\ 
55808.260854 & 0.335 & -0.335 & 0.010 & 89.2 & 0.013 & 0.4 \\ 
55808.260854 & 0.335 & -0.335 & 0.010 & 89.2 & 0.013 & 0.4 \\ 
55496.272980 & 0.335 & -0.335 & 0.002 & 89.8 & 0.006 & 0.2 \\ 
55496.272980 & 0.335 & -0.335 & 0.002 & 89.8 & 0.006 & 0.2 \\ 
55496.294508 & 0.337 & -0.337 & 0.005 & 89.6 & 0.009 & 0.3 \\ 
55496.294508 & 0.337 & -0.337 & 0.005 & 89.6 & 0.009 & 0.3 \\ 
55814.245312 & 0.294 & -0.291 & -0.044 & 94.3 & 0.022 & 0.6 \\ 
55814.245312 & 0.294 & -0.291 & -0.044 & 94.3 & 0.022 & 0.6 \\ 
55808.295159 & 0.363 & -0.360 & 0.045 & 86.5 & 0.030 & 0.9 \\ 
55808.295159 & 0.363 & -0.360 & 0.045 & 86.5 & 0.030 & 0.9 \\ 
55496.316047 & 0.338 & -0.338 & 0.005 & 89.6 & 0.014 & 0.4 \\ 
55496.316047 & 0.338 & -0.338 & 0.005 & 89.6 & 0.014 & 0.4 \\ 

\end{longtable}
\end{center}

\begin{center}
\begin{longtable}{ccccccc}
\caption[Feasible triples for a highly variable Grid]%
{$V$-band polarization measurements of HR\,7355.} \label{tab:aphr5907} \\
\hline MJD & $P$ (\%) & $Q$ (\%) & $U$ (\%) & $\theta$ (deg) & $\sigma_P$ (\%) & $\sigma_\theta$ (deg) \\ \hline
\endfirsthead

\multicolumn{7}{c}%
{{\footnotesize  \itshape {Table \thetable{}} -- continued from previous page.}} \\
\hline MJD & $P$ (\%) & $Q$ (\%) & $U$ (\%) & $\theta$ (deg) & $\sigma_P$ (\%) & $\sigma_\theta$ (deg) \\ \hline
\endhead

\hline \multicolumn{7}{r}{\footnotesize \itshape {Continued on next page}} \\ \hline
\endfoot

\hline \hline
\endlastfoot
55354.316111 & 0.029 & -0.018 & 0.023 & 243.5 & 0.040 & 1.1 \\ 
55355.292959 & 0.017 & 0.010 & -0.013 & 153.7 & 0.024 & 0.7 \\ 
55355.301703 & 0.025 & -0.021 & 0.014 & 253.4 & 0.024 & 0.7 \\ 
55355.307623 & 0.026 & 0.025 & 0.006 & 186.9 & 0.015 & 0.4 \\ 
55426.231794 & 0.024 & 0.003 & 0.024 & 220.9 & 0.011 & 0.3 \\ 
55428.166076 & 0.048 & -0.022 & 0.042 & 58.5 & 0.021 & 0.6 \\ 
55428.178906 & 0.021 & -0.012 & -0.017 & 116.9 & 0.020 & 0.6 \\ 
55441.163588 & 0.043 & -0.042 & 0.010 & 83.3 & 0.010 & 0.3 \\ 
55443.088255 & 0.130 & 0.102 & -0.080 & 161.0 & 0.041 & 1.2 \\ 
55443.098516 & 0.022 & 0.019 & 0.012 & 196.3 & 0.023 & 0.6 \\ 
55443.108735 & 0.006 & -0.002 & 0.006 & 56.1 & 0.027 & 0.8 \\ 
55443.118956 & 0.033 & 0.007 & -0.032 & 141.0 & 0.013 & 0.4 \\ 
55443.134418 & 0.072 & 0.035 & -0.063 & 149.5 & 0.022 & 0.6 \\ 
55443.149899 & 0.079 & -0.020 & -0.076 & 127.5 & 0.091 & 2.6 \\ 
55443.160963 & 0.115 & -0.097 & -0.061 & 106.2 & 0.069 & 2.0 \\ 
55496.978263 & 0.018 & -0.015 & 0.011 & 72.5 & 0.008 & 0.2 \\ 
55496.986972 & 0.025 & -0.024 & 0.007 & 81.7 & 0.003 & 0.1 \\ 
55727.204879 & 0.039 & 0.004 & 0.038 & 41.9 & 0.021 & 0.6 \\ 
55727.225370 & 0.011 & -0.011 & 0.004 & 80.6 & 0.019 & 0.5 \\ 
55727.246904 & 0.016 & -0.008 & 0.014 & 59.6 & 0.019 & 0.5 \\ 
55727.271788 & 0.019 & 0.001 & 0.019 & 43.7 & 0.012 & 0.4 \\ 
55741.279656 & 0.044 & -0.036 & -0.025 & 107.6 & 0.049 & 1.4 \\ 
55741.293545 & 0.040 & 0.011 & 0.038 & 36.8 & 0.031 & 0.9 \\ 
55741.306698 & 0.017 & -0.009 & 0.015 & 60.1 & 0.036 & 1.0 \\ 
55741.319470 & 0.019 & -0.001 & -0.019 & 133.5 & 0.053 & 1.5 \\ 
55741.332028 & 0.050 & 0.021 & -0.045 & 147.6 & 0.019 & 0.6 \\ 
56923.051934 & 0.017 & -0.016 & 0.004 & 83.2 & 0.012 & 0.3 \\ 
56923.055012 & 0.049 & -0.044 & -0.022 & -76.7 & 0.033 & 0.9 \\ 
56923.058097 & 0.026 & -0.026 & 0.004 & 86.1 & 0.026 & 0.7 \\ 
56923.061178 & 0.005 & -0.002 & -0.005 & -54.4 & 0.005 & 0.1 \\ 
56923.064259 & 0.019 & -0.009 & -0.017 & -59.5 & 0.012 & 0.3 \\ 
56923.067303 & 0.013 & 0.012 & -0.006 & -12.5 & 0.007 & 0.2 \\ 
56923.070385 & 0.032 & 0.008 & 0.031 & 38.1 & 0.020 & 0.6 \\ 
56923.073446 & 0.071 & 0.046 & 0.053 & 24.5 & 0.049 & 1.4 \\ 

\end{longtable}
\end{center}

\begin{center}
\begin{longtable}{ccccccc}
\caption[Feasible triples for a highly variable Grid]%
{$V$-band polarization measurements of HR\,5907.} \label{tab:aphr7355} \\
\hline MJD & $P$ (\%) & $Q$ (\%) & $U$ (\%) & $\theta$ (deg) & $\sigma_P$ (\%) & $\sigma_\theta$ (deg) \\ \hline
\endfirsthead

\multicolumn{7}{c}%
{{\footnotesize  \itshape {Table \thetable{}} -- continued from previous page.}} \\
\hline MJD & $P$ (\%) & $Q$ (\%) & $U$ (\%) & $\theta$ (deg) & $\sigma_P$ (\%) & $\sigma_\theta$ (deg) \\ \hline
\endhead

\hline \multicolumn{7}{r}{\footnotesize \itshape {Continued on next page}} \\ \hline
\endfoot

\hline \hline
\endlastfoot
55355.130869 & 0.599 & -0.262 & 0.539 & 238.0 & 0.007 & 0.2 \\ 
55355.135273 & 0.621 & -0.254 & 0.566 & 237.1 & 0.021 & 0.6 \\ 
55355.142235 & 0.601 & -0.264 & 0.540 & 238.1 & 0.011 & 0.3 \\ 
55395.960399 & 0.620 & -0.324 & 0.528 & 60.8 & 0.034 & 1.0 \\ 
55397.050492 & 0.606 & -0.333 & 0.506 & -118.3 & 0.040 & 1.1 \\ 
55397.057344 & 0.622 & -0.284 & 0.553 & -121.4 & 0.008 & 0.2 \\ 
55425.061678 & 0.588 & -0.302 & 0.504 & 60.5 & 0.042 & 1.2 \\ 
55425.073536 & 0.698 & -0.343 & 0.608 & 59.7 & 0.011 & 0.3 \\ 
55426.984575 & 0.606 & -0.265 & 0.545 & 58.0 & 0.026 & 0.7 \\ 
55426.999847 & 0.627 & -0.335 & 0.530 & 61.1 & 0.009 & 0.2 \\ 
55427.914965 & 0.613 & -0.305 & 0.531 & 60.0 & 0.029 & 0.8 \\ 
55427.923258 & 0.616 & -0.311 & 0.532 & 60.1 & 0.013 & 0.4 \\ 
55441.915871 & 0.620 & -0.282 & 0.552 & 58.5 & 0.007 & 0.2 \\ 
55727.053359 & 0.631 & -0.309 & 0.550 & 59.7 & 0.022 & 0.6 \\ 
55727.103608 & 0.617 & -0.251 & 0.563 & 57.0 & 0.019 & 0.5 \\ 
55741.046923 & 0.613 & -0.354 & 0.501 & 62.7 & 0.011 & 0.3 \\ 
55741.083277 & 0.600 & -0.351 & 0.487 & 62.9 & 0.033 & 1.0 \\ 
55812.957300 & 0.651 & -0.310 & 0.573 & 59.2 & 0.026 & 0.7 \\ 
55812.969586 & 0.566 & -0.270 & 0.497 & 59.2 & 0.045 & 1.3 \\ 
55812.981409 & 0.625 & -0.308 & 0.544 & 59.8 & 0.030 & 0.8 \\ 
56027.133345 & 0.585 & -0.297 & 0.504 & 60.3 & 0.026 & 0.8 \\ 
56027.146435 & 0.579 & -0.305 & 0.492 & 60.9 & 0.015 & 0.4 \\ 
56027.160550 & 0.634 & -0.316 & 0.549 & 60.0 & 0.020 & 0.6 \\ 
56027.175729 & 0.626 & -0.321 & 0.538 & 60.4 & 0.012 & 0.3 \\ 
56922.938028 & 0.613 & -0.275 & 0.548 & 58.3 & 0.027 & 0.8 \\ 
56922.952285 & 0.603 & -0.308 & 0.518 & 60.4 & 0.025 & 0.7 \\ 
56922.971061 & 0.666 & -0.342 & 0.571 & 60.5 & 0.072 & 2.1 \\ 
56922.985319 & 0.672 & -0.313 & 0.595 & 58.9 & 0.021 & 0.6 \\ 

\end{longtable}
\end{center}

%% file: tex/ap-beatphot.tex
\chapter{IUE data fitting to Achernar with \textsc{BeAtlas} \label{chap:beataeri}}
As an example of the work being conducted with \textsc{BeAtlas} models, we applied our photospheric models to the ultra-violet observations of the IUE satellite on Achernar. The IUE (International Ultraviolet Explorer) satellite was launched in 1978 and operated until 1996. During this period, the satellite observed Achernar in different epochs, resulting in observations over different circumstellar activity phases. 

The IUE observations analyzed here were grouped according to the circumstellar activity at the time of the observations (Table~\ref{tab:iue}). The work of \citet{vin06a} contains detailed information of H$\alpha$ line profile that can be used to assess the activity from 1991 up to 2003. It also contains the authors qualitative estimations of the emission strength before 1990 from the literature, which allows us infer the level of Achernar activity since the first observations until 1991. 

\begin{figure}
    \centering
    \includegraphics[width=\linewidth]{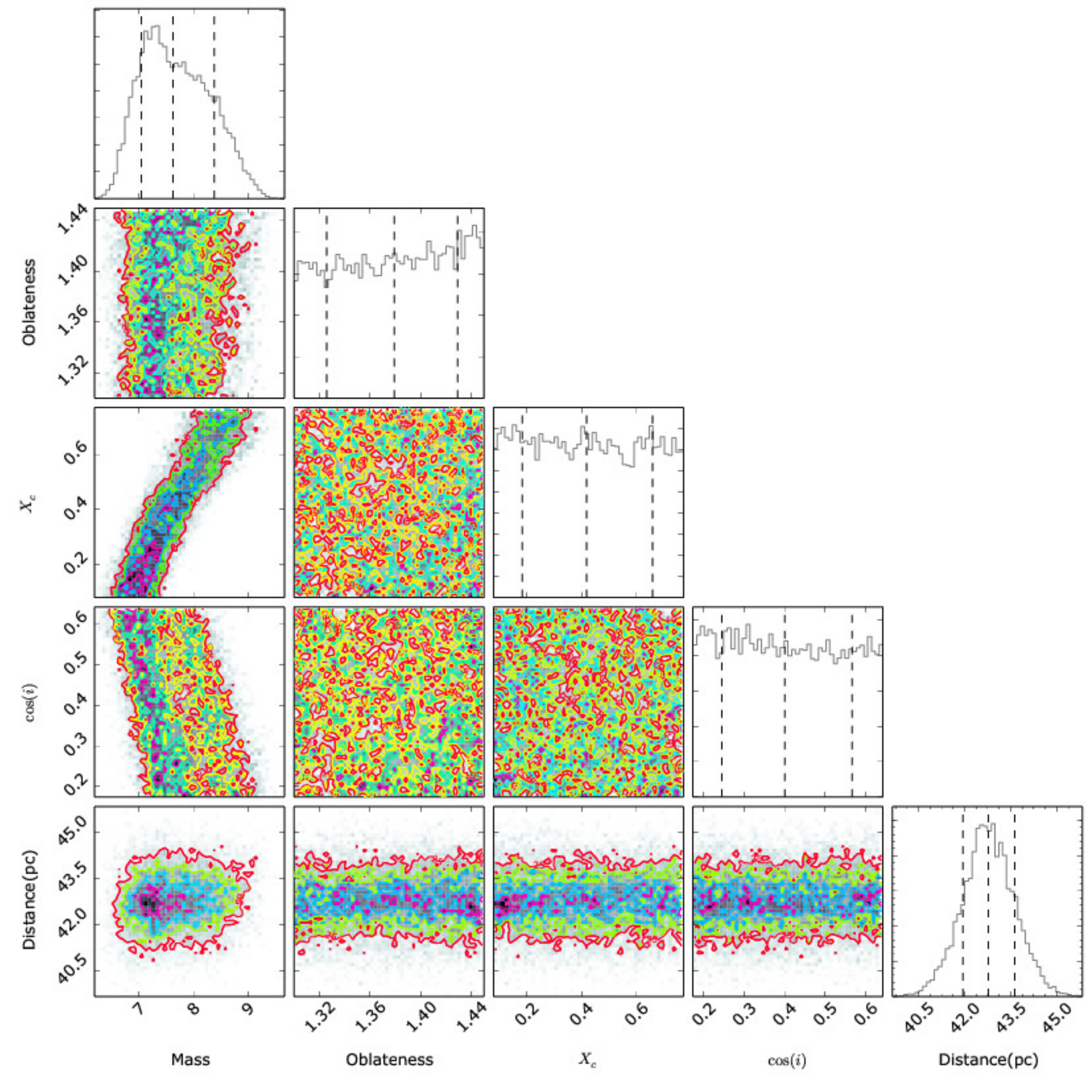} 
    \caption[Probability density functions of the derived stellar parameters for Achernar and their correlation maps. It corresponds to the set 1 of IUE observations (no disk activity).]{Probability density functions of the derived stellar parameters for Achernar and their correlation maps. It corresponds to the {set 1} of IUE observations (none or very small disk activity; Table~\ref{tab:iue}) sampled with \textsc{beatlas} models. The rainbow-like color map indicates the isocontour of highest probabilities (blue) to the lowest (red). \textit{From left to right (or top-bottom):} mass (Solar masses), oblateness, H fraction at stellar core, cosine of the inclination angle and distance (in parsec). The dashed vertical lines in the histograms indicate the percentile of 16\% ($-1\sigma$), the median and the percentile of 84\% ($1\sigma$) of each distribution.}
    \label{fig:iue1}
\end{figure}

\begin{figure}
    \centering
    \includegraphics[width=\linewidth]{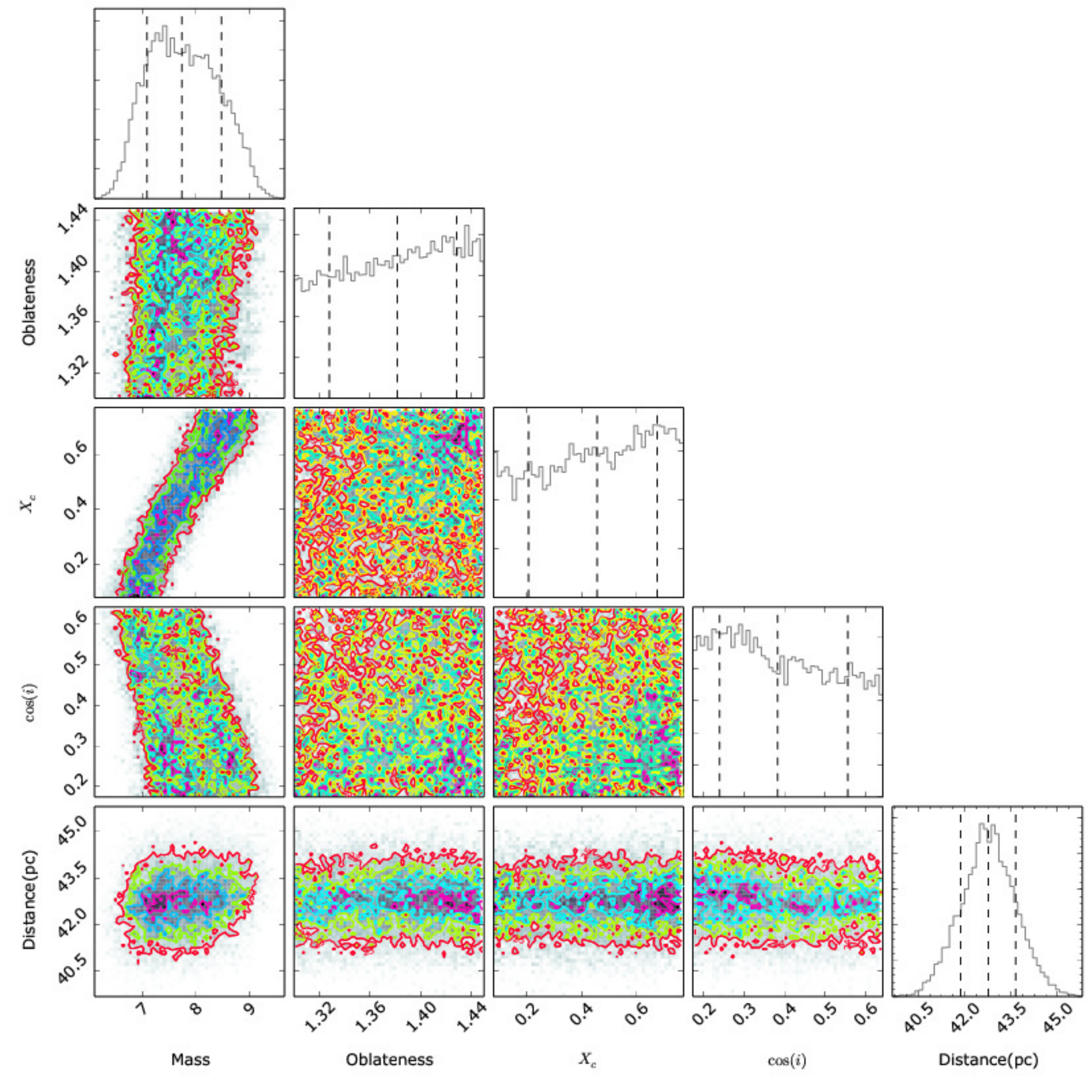} 
    \caption{Same as previous figure, but for {set 2} of IUE data (intermediate disk activity).}
    \label{fig:iue2}
\end{figure}

\begin{figure}
    \centering
    \includegraphics[width=\linewidth]{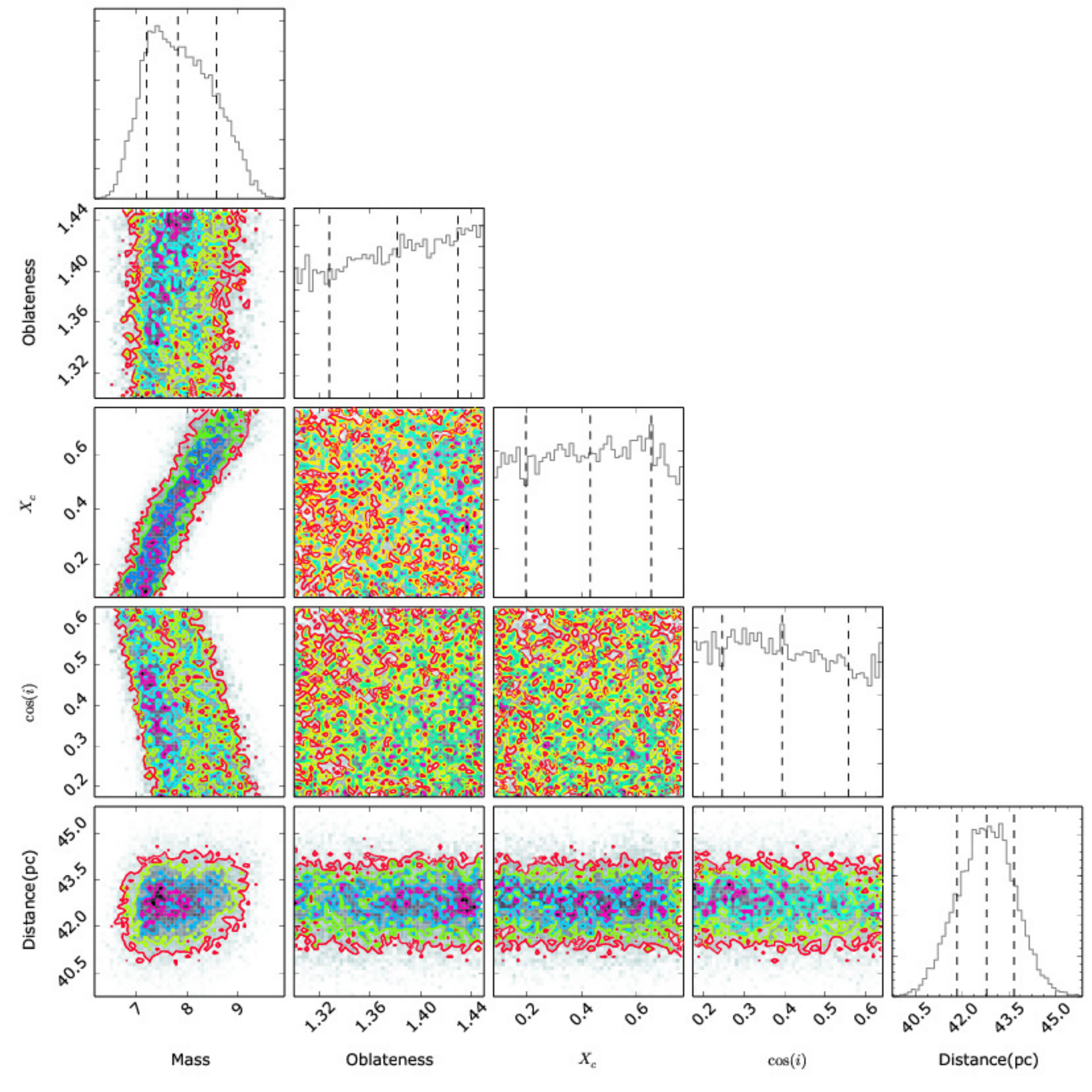}
    \caption{Same as previous figure, but for {set 3} of IUE data (full disk activity).}
    \label{fig:iue3}
\end{figure}

The \textsc{BeAtlas} MCMC minimization for the different level of circumstellar activity are shown in Figs.~\ref{fig:iue1} to \ref{fig:iue3}. The constraints, used priors in the likelyhood probability, were: (i) the oblateness $ 1.3\leq R_{\rm eq}/R_p\leq1.45$, (ii) the inclination angle $50^\circ \leq i \leq 80^\circ$, and (iii) the distance value from \citet[$d=42.8(1.0)$~pc]{van07a}. The interestellar extinction value was fixed in $E(B-V)=0$, based on the stellar unpolarized spectrum at the disk quiescent phase and its nearby position, inside the interestellar local bubble \citep{fri11a}.

As results of this minimization, we highlight: (i) the disk activity little changed the parameters determination; (ii) within the considered ranges, no clear trend exist for stellar oblateness, inclination angle, or evolutionary stage ($X_{\rm c}$) itself; (iii) the most probable stellar mass would have $\sim$7.2\,$M_\odot$, correlated with and late evolurionary age ($X_{\rm c} \lesssim 0.2$), where most of H present in the core was already consumed by the star. 

According to Mota et al$.$ (in prep$.$), the IUE spectra has the capacity of precise constrain the multiple photospheric parameters of main sequence stars, even if the interstellar extinction is present. This allows us to state that Achernar is not a typical main-sequence star. Different estimations for Achernar points to a stellar mass around the value of $\sim$6.2\,$M_\odot$ \citep{dom14a}. Comparing to the \textsc{BeAtlas} models, the photospheric parameters derived from interferometry mimics a main sequence star with a higher mass (Table~\ref{tab:stpars}). It is also known that the radius and luminosity of the star typically increase along the main sequence (e.g., \citealp{geo13a}). These elements show that Achernar likely is a star leaving the main sequence, as discussed by \citet{rie13a}.


\begin{table}
\centering
\caption{IUE observations of Achernar grouped according to the circunstellar activity at the time of observations.}
\begin{tabular}[]{cccc}
\toprule
 & Group 1 & Group 2 & Group 3 \\ 
Disk activity (H$\alpha$) & None/Small & Intermediate & Full \\   
\midrule
\multirow{6}{*}{Dates} &  1983-09-24 & 1992-08-03 & 1993-09-06 \\ 
 & 1988-05-14 & 1995-09-07 & 1993-09-24 \\ 
 & 1989-12-14 & 1995-09-08 & 1993-10-16 \\ 
 & 1990-09-30 & 1995-09-09 & 1993-11-14 \\ 
 & & 1994-06-15 & 1994-06-14 \\ 
 & & & 1994-07-26 \\ 
\bottomrule
\end{tabular}
\label{tab:iue}
\end{table}

%% file: tex/ap-other.tex
\chapter{The Be phenomenon: a resonance effect? \label{ap:reson}}
The proximity between the orbital, rotational and critical frequencies deducted for Achernar based on the parameters from \citet{dom14a} and known observational frequencies draws attention. Indeed, it points to a interesting effect on massive stars and that could be necessary for the occurrence of the Be phenomenon.

It is known to resonance that smallest is the difference between the oscillating frequency $\omega$ and the resonant frequency $\Omega$ bigger is the intensity ($I$, or squared amplitude) of the oscillations:
\begin{equation}
    I(\omega) \propto \frac{\left(\frac{\Gamma}{2}\right)^2}{(\omega - \Omega)^2 + \left( \frac{\Gamma}{2} \right)^2 }\,,
\end{equation}
where $\Gamma$ is a parameter dependent on the damping of the oscillator, and is known as the linewidth of the resonance.

For a given star at $W$ rotational rate, it is possible to define three frequencies: F4, the equatorial orbital frequency; F5, the critical frequency; and F6, the equatorial rotational frequency (if the star is a rigid rotator, F6 is the stellar rotational frequency). If $W<1$, then ${\rm F}4>{\rm F}5>{\rm F}6$. If the star rotates critically ($W=1$), ${\rm F}4={\rm F}5={\rm F}6$. 

Keeping the $W$ fixed, the difference between all these frequencies ($\Delta$F$^*$) satisfies
\begin{equation}
\Delta {\rm F}^* \propto \sqrt{\frac{M}{R_p^{3}}}\,,
\end{equation}
in the three possible cases
\begin{equation}
\Delta{\rm F} = {\rm F}4 - {\rm F}5 = \frac{v_{\rm orb}}{2\pi rR_p}-\frac{v_{\rm crit}}{2\pi 1.5R_{p}} = \frac{(GM)^{1/2}}{2\pi R_p^{3/2}}\left(\frac{1}{r^{3/2}}-\frac{1}{1.5^{3/2}}\right)\,,
\end{equation}
\begin{equation}
\Delta {\rm F}' = {\rm F}4 - {\rm F}6 = \frac{v_{\rm orb}}{2\pi rR_p}-\frac{Wv_{\rm orb}}{2\pi rR_{p}} = \frac{(GM)^{1/2}}{2\pi R_p^{3/2}}\left(\frac{1-W}{r^{3/2}}\right)\,,
\end{equation}
\begin{equation}
\Delta {\rm F}'' = {\rm F}5 - {\rm F}6 = \frac{W v_{\rm orb}}{2\pi rR_p}-\frac{v_{\rm crit}}{2\pi 1.5R_{p}} = \frac{(GM)^{1/2}}{2\pi R_p^{3/2}}\left(\frac{1}{1.5^{3/2}}-\frac{W}{r^{3/2}}\right)\,,
\end{equation}
where $1\leq r \leq1.5$ is the radius ratio $R_{\rm eq}/R_p$ for a specific $W$ value (e.g., Eq.~\ref{eq:radrat}). 

If the Be phenomenon is a resonance effect in the critical frequency then the lower the $\Delta{\rm F}^*$ differences, most likely is its occurrence. Indeed, it is know that the ratio $M/R^3$ decreases as the mass increases along the main sequence. Taking, for example, the stellar parameters from \citet{mac99a}, $\Delta $F$^*$ is smaller the more massive is the main-sequence star, with $M/R^3\lesssim0.2$ from A0V subtypes (solar units; Table~\ref{tab:deltF}).

\begin{table}
\centering
\caption[Main sequence masses and radius values.]{Main sequence masses and radius values from \citet{mac99a}.}
\begin{tabular}[]{cccc}
\toprule
    Sp & $M/M_\odot$ & $R/R_\odot$ & $M/R^3$ (Solar units) \\
    \midrule
    05V & 60 & 12 & 0.035 \\
    06V & 37 & 10 & 0.037 \\
    08V & 23 & 8.5 & 0.037 \\ \hline
    B0V & 17.5 & 7.4 & 0.043 \\
    B3V & 7.6 & 4.8 & 0.069 \\
    B5V & 5.9 & 3.9 & 0.099 \\
    B8V & 3.8 & 3.0 & 0.141 \\ \hline
    A0V & 2.9 & 2.4 & 0.210 \\
    A5V & 2.0 & 1.7 & 0.407 \\ \hline
    F0V & 1.6 & 1.5 & 0.474 \\
    F5V & 1.4 & 1.3 & 0.637 \\
\bottomrule
\end{tabular}
\label{tab:deltF}
\end{table}

This resonance could also indicates with early-type B stars have denser disks than late-type B, and rotational perturbations (inducing $\Delta$F$^*$) could easy be originate by tide effects from a binary companion star (or other process in stellar interior). However, this do not explain why the Be phenomenon is not present in O-type stars as it is in B0 ones. 
One possibility to be investigated the interaction of the ejected material with the stronger winds of these stars. Other, more likely, is the presence of NRP frequencies occurring only in B stars that are similar to near-critical equatorial orbits, working as triggers to mass-loss episodes, as the proximity between \citet{gos11a} frequency F1 and F4 for Achernar suggests. If the $\kappa$ mechanism if the origin of these pulsations and the transport of angular momentum as proposed by \citet{nei14a}, the hotter temperatures or physical dimensions of O-type stars could completely block the mechanism.

\section{Achernar frequencies}
Let us consider the following scenario: Achernar has a NRP frequency matching the orbital the equatorial orbital frequency (F4). The F4 frequency could also be associated with a tenuous residual disk \citep{car08a,dom14a}, leaving a low magnitude orbiting signature. However, its stability is hard to be explained since viscosity would dissipated the ejected material. At epochs of stellar activity, a transient period appears with the photospheric rotational frequency (F6) due to material ejection by the star, for example, associated with the presence of photospheric spots. Table~\ref{tab:freqs} contains a list of orbital frequencies calculated for Achernar, that are compared to the observed frequencies by \citet[Table~\ref{tab:gossfs}]{gos11a}.
\begin{table}
\footnotesize
\centering
\caption[Rotational and orbital frequencies for Achernar.]{Rotational and orbital frequencies for Achernar. $^c$ denote values at critical rotation.}
\begin{tabular}[]{ccccccc}
\toprule
\multirow{2}{*}{$Fn$}  & Frequency & Value & Closest observed & Diff. to closest & Vel. & Radius  \\
                       & Description  &  (d$^{-1}$) & frequency (d$^{-1}$) & obs. freq. (d$^{-1}$) & (km\,s$^{-1}$) & ($R_p$) \\
\midrule
F4 & Equatorial orbit & 0.769 & F1 & 0.0066 & 356.4 & 1.352 \\
F4$^{c}$ & Equatorial orbit & 0.657 & F3 & 0.0234 & 338.2 & 1.5 \\
F5 & Critic rotation  & 0.657 & F3 & 0.0234 & 338.2 & 1.5 \\
F6$^{c}$ & Rotation & 0.657 & F3 & 0.0234 & 338.2 & 1.5 \\
F6 & Rotation         & 0.644 & F3 & 0.0356 & 298.8 & 1.352 \\ \hline
F7 & Keplerian ($R\lesssim R_{\rm eq}$) & 0.777 & F1 & 0.0003 & 357.8 & 1.3405 \\
F8 & Keplerian ($R>R_{\rm eq}$) & 0.725 & F2 & 0.0002 & 349.5 & 1.405 \\
F9 & Keplerian ($R>R_{\rm eq}$) & 0.681 & F3 & 0.0003 & 342.2 & 1.465 \\
\bottomrule
\end{tabular}
\label{tab:freqs}
\end{table}

Having the two frequencies (F4 and F6) in the scenario described above one have a direct measurement of the stellar rotation rate (Eq.~\ref{eq:WF4F6}). The facts that (i) the transient period F3 of Achernar is roughly its critical frequency (F5; since it is a fast rotating star, F5~$\approx$~F6); (ii) the observed F1 frequency coincides with the orbital frequency F4; and (iii) the observed spin-up velocity in Achernar's line profile is exactly the one required to F4 reach F5; they together strongly suggests that the Be phenomenon can be seen as a resonant effect of these frequencies. NRP pulsations at frequency F1~$\approx$~F4 could act as a trigger to the stellar mechanical mass-loss to orbits around the equatorial radius previously its spin-up. The role of observed frequency F2 to Achernar should be determined but it could, in principle, be associate with other non-stable pulsation modes or a Keplerian orbit at radius $R>R_{\rm eq}$.

%% file: tex/ap-activelog.tex
\chapter{Achernar 2013+ active phase observational log \label{ap:obslog}}

\section{Spectroscopy \label{ap:aerispec}}
\begin{center}
{\footnotesize \begin{longtable}{cccl}
\caption
{List of spectroscopic observations covering H$\alpha$ wavelengths available after the 2013 outburst of Achernar.} \label{ap:tabspecs} \\
\hline \multicolumn{1}{c}{\textbf{Instument}} & %
\multicolumn{1}{c}{\textbf{Resolution}} & %
\multicolumn{1}{c}{\textbf{Date}} & %
\multicolumn{1}{c}{\textbf{MJD}} \\ \hline
\endfirsthead

\multicolumn{4}{c}%
{{\bfseries \tablename\ \thetable{} -- table from previous page}} \\
\hline \multicolumn{1}{c}{\textbf{Instument}} & %
\multicolumn{1}{c}{\textbf{Resolution}} & %
\multicolumn{1}{c}{\textbf{Date}} & %
\multicolumn{1}{c}{\textbf{MJD}} \\ \hline
\endhead

\hline \multicolumn{4}{r}{{Continue on next page}} \\ \hline
\endfoot

\hline \hline
\endlastfoot

ECass & 5000 & 2012-11-20 & 56251.9942 \\ 
PUCHEROS & 20000 & 2013-01-18 & 56310.0 \\ 
Amateur & 5000 & 2013-01-22 & 56314.0 \\ 
Amateur & 5000 & 2013-07-03 & 56476.0 \\ 
XSHOOTER & 15000 & 2013-07-18 & 56491.4418 \\ 
PUCHEROS & 20000 & 2013-07-24 & 56497.0 \\ 
PUCHEROS & 20000 & 2013-07-25 & 56498.0 \\ 
Amateur & 5000 & 2013-07-28 & 56501.0 \\ 
Amateur & 5000 & 2013-08-17 & 56521.0360 \\ 
Amateur & 5000 & 2013-08-23 & 56527.2199 \\ 
FEROS & 48000 & 2013-09-06 & 56541.2938 \\ 
FEROS & 48000 & 2013-09-06 & 56541.2948 \\ 
FEROS & 48000 & 2013-09-06 & 56541.2968 \\ 
FEROS & 48000 & 2013-09-06 & 56541.3519 \\ 
FEROS & 48000 & 2013-09-07 & 56542.3922 \\ 
FEROS & 48000 & 2013-09-07 & 56542.3930 \\ 
FEROS & 48000 & 2013-09-07 & 56542.3940 \\ 
FEROS & 48000 & 2013-09-07 & 56542.3952 \\ 
FEROS & 48000 & 2013-09-08 & 56543.1813 \\ 
FEROS & 48000 & 2013-09-08 & 56543.1822 \\ 
FEROS & 48000 & 2013-09-08 & 56543.3196 \\ 
FEROS & 48000 & 2013-09-08 & 56543.3218 \\ 
FEROS & 48000 & 2013-09-08 & 56543.4230 \\ 
FEROS & 48000 & 2013-09-08 & 56543.4238 \\ 
FEROS & 48000 & 2013-09-09 & 56544.2428 \\ 
FEROS & 48000 & 2013-09-09 & 56544.2436 \\ 
FEROS & 48000 & 2013-09-09 & 56544.3869 \\ 
FEROS & 48000 & 2013-09-09 & 56544.4148 \\ 
FEROS & 48000 & 2013-09-09 & 56544.4156 \\ 
FEROS & 48000 & 2013-09-10 & 56545.1311 \\ 
FEROS & 48000 & 2013-09-10 & 56545.4103 \\ 
FEROS & 48000 & 2013-09-13 & 56548.1172 \\ 
FEROS & 48000 & 2013-09-13 & 56548.1179 \\ 
FEROS & 48000 & 2013-09-13 & 56548.2305 \\ 
FEROS & 48000 & 2013-09-13 & 56548.2312 \\ 
FEROS & 48000 & 2013-09-13 & 56548.4111 \\ 
FEROS & 48000 & 2013-09-13 & 56548.4118 \\ 
FEROS & 48000 & 2013-09-15 & 56550.1151 \\ 
FEROS & 48000 & 2013-09-15 & 56550.1159 \\ 
FEROS & 48000 & 2013-09-15 & 56550.1169 \\ 
FEROS & 48000 & 2013-09-15 & 56550.1783 \\ 
FEROS & 48000 & 2013-09-15 & 56550.1793 \\ 
FEROS & 48000 & 2013-09-15 & 56550.1809 \\ 
FEROS & 48000 & 2013-09-16 & 56551.2722 \\ 
FEROS & 48000 & 2013-09-16 & 56551.2729 \\ 
FEROS & 48000 & 2013-09-16 & 56551.3209 \\ 
FEROS & 48000 & 2013-09-16 & 56551.3217 \\ 
FEROS & 48000 & 2013-09-16 & 56551.3914 \\ 
FEROS & 48000 & 2013-09-16 & 56551.3922 \\ 
FEROS & 48000 & 2013-09-17 & 56552.1087 \\ 
FEROS & 48000 & 2013-09-17 & 56552.1095 \\ 
FEROS & 48000 & 2013-09-17 & 56552.1826 \\ 
FEROS & 48000 & 2013-09-17 & 56552.1839 \\ 
FEROS & 48000 & 2013-09-17 & 56552.1846 \\ 
FEROS & 48000 & 2013-09-17 & 56552.2549 \\ 
FEROS & 48000 & 2013-09-17 & 56552.2562 \\ 
Amateur & 5000 & 2013-10-20 & 56585.1473 \\ 
PUCHEROS & 20000 & 2013-10-29 & 56594.0 \\ 
ECass & 30000 & 2013-11-12 & 56608.0633 \\ 
ECass & 30000 & 2013-11-12 & 56608.0661 \\ 
ECass & 30000 & 2013-11-13 & 56609.0978 \\ 
FEROS & 48000 & 2013-12-05 & 56631.0286 \\ 
FEROS & 48000 & 2013-12-05 & 56631.0297 \\ 
FEROS & 48000 & 2013-12-05 & 56631.1607 \\ 
FEROS & 48000 & 2013-12-05 & 56631.1616 \\ 
FEROS & 48000 & 2013-12-05 & 56631.2506 \\ 
FEROS & 48000 & 2013-12-07 & 56633.0184 \\ 
FEROS & 48000 & 2013-12-07 & 56633.1470 \\ 
FEROS & 48000 & 2013-12-07 & 56633.2091 \\ 
FEROS & 48000 & 2013-12-07 & 56633.3095 \\ 
ECass & 30000 & 2013-12-09 & 56635.9693 \\ 
FEROS & 48000 & 2013-12-10 & 56636.0266 \\ 
FEROS & 48000 & 2013-12-10 & 56636.0273 \\ 
FEROS & 48000 & 2013-12-10 & 56636.1609 \\ 
FEROS & 48000 & 2013-12-10 & 56636.1615 \\ 
FEROS & 48000 & 2013-12-10 & 56636.2349 \\ 
FEROS & 48000 & 2013-12-10 & 56636.2356 \\ 
FEROS & 48000 & 2013-12-11 & 56637.0190 \\ 
FEROS & 48000 & 2013-12-11 & 56637.0197 \\ 
FEROS & 48000 & 2013-12-11 & 56637.1407 \\ 
FEROS & 48000 & 2013-12-11 & 56637.1414 \\ 
FEROS & 48000 & 2013-12-11 & 56637.2620 \\ 
FEROS & 48000 & 2013-12-11 & 56637.2626 \\ 
FEROS & 48000 & 2013-12-11 & 56637.2633 \\ 
FEROS & 48000 & 2013-12-11 & 56637.2640 \\ 
FEROS & 48000 & 2014-01-04 & 56661.0262 \\ 
FEROS & 48000 & 2014-01-04 & 56661.0269 \\ 
FEROS & 48000 & 2014-01-04 & 56661.0995 \\ 
FEROS & 48000 & 2014-01-04 & 56661.1002 \\ 
FEROS & 48000 & 2014-01-04 & 56661.1940 \\ 
FEROS & 48000 & 2014-01-04 & 56661.1947 \\ 
FEROS & 48000 & 2014-01-05 & 56662.0422 \\ 
FEROS & 48000 & 2014-01-05 & 56662.0429 \\ 
FEROS & 48000 & 2014-01-05 & 56662.1181 \\ 
FEROS & 48000 & 2014-01-05 & 56662.1192 \\ 
FEROS & 48000 & 2014-01-05 & 56662.1200 \\ 
FEROS & 48000 & 2014-01-05 & 56662.1945 \\ 
FEROS & 48000 & 2014-01-05 & 56662.1951 \\ 
FEROS & 48000 & 2014-01-06 & 56663.0088 \\ 
FEROS & 48000 & 2014-01-06 & 56663.0096 \\ 
FEROS & 48000 & 2014-01-06 & 56663.0913 \\ 
FEROS & 48000 & 2014-01-06 & 56663.0920 \\
Amateur & 5000 & 2014-01-07 & 56664.875 \\  
Amateur & 5000 & 2014-01-21 & 56678.0027 \\ 
FEROS & 48000 & 2014-01-29 & 56686.0261 \\ 
FEROS & 48000 & 2014-01-29 & 56686.1033 \\ 
PUCHEROS & 20000 & 2014-01-30 & 56687.0 \\ 
FEROS & 48000 & 2014-01-30 & 56687.1304 \\ 
FEROS & 48000 & 2014-01-31 & 56688.0346 \\ 
FEROS & 48000 & 2014-02-10 & 56698.1137 \\ 
FEROS & 48000 & 2014-02-10 & 56698.1143 \\ 
FEROS & 48000 & 2014-02-10 & 56698.1150 \\ 
FEROS & 48000 & 2014-02-10 & 56698.1156 \\ 
FEROS & 48000 & 2014-02-10 & 56698.9999 \\ 
FEROS & 48000 & 2014-02-11 & 56699.0009 \\ 
FEROS & 48000 & 2014-02-12 & 56700.0709 \\ 
FEROS & 48000 & 2014-02-12 & 56700.0718 \\ 
FEROS & 48000 & 2014-02-12 & 56700.9995 \\ 
FEROS & 48000 & 2014-02-13 & 56701.0003 \\ 
FEROS & 48000 & 2014-02-13 & 56701.0016 \\ 
FEROS & 48000 & 2014-02-13 & 56701.0023 \\ 
FEROS & 48000 & 2014-02-13 & 56701.0029 \\ 
FEROS & 48000 & 2014-02-13 & 56701.0036 \\ 
FEROS & 48000 & 2014-02-13 & 56701.0043 \\ 
FEROS & 48000 & 2014-02-13 & 56701.9933 \\ 
FEROS & 48000 & 2014-02-13 & 56701.9941 \\ 
FEROS & 48000 & 2014-02-13 & 56701.9948 \\ 
FEROS & 48000 & 2014-02-15 & 56703.0050 \\ 
FEROS & 48000 & 2014-02-15 & 56703.0057 \\ 
FEROS & 48000 & 2014-02-16 & 56704.0037 \\ 
FEROS & 48000 & 2014-02-16 & 56704.0043 \\ 
FEROS & 48000 & 2014-02-16 & 56704.9957 \\ 
FEROS & 48000 & 2014-02-16 & 56704.9964 \\ 
FEROS & 48000 & 2014-02-16 & 56704.9973 \\ 
FEROS & 48000 & 2014-02-16 & 56704.9979 \\ 
FEROS & 48000 & 2014-02-18 & 56706.0047 \\ 
FEROS & 48000 & 2014-02-18 & 56706.0053 \\ 
FEROS & 48000 & 2014-02-19 & 56707.9940 \\ 
FEROS & 48000 & 2014-02-19 & 56707.9947 \\ 
FEROS & 48000 & 2014-02-20 & 56708.9942 \\ 
FEROS & 48000 & 2014-02-20 & 56708.9962 \\ 
FEROS & 48000 & 2014-02-20 & 56708.9969 \\ 
FEROS & 48000 & 2014-02-21 & 56709.9972 \\ 
FEROS & 48000 & 2014-02-21 & 56709.9979 \\ 
FEROS & 48000 & 2014-02-22 & 56710.9950 \\ 
FEROS & 48000 & 2014-02-22 & 56710.9957 \\ 
FEROS & 48000 & 2014-02-22 & 56710.9980 \\ 
FEROS & 48000 & 2014-02-22 & 56710.9987 \\ 
FEROS & 48000 & 2014-02-23 & 56711.9895 \\ 
FEROS & 48000 & 2014-02-23 & 56711.9911 \\ 
FEROS & 48000 & 2014-02-23 & 56711.9918 \\ 
FEROS & 48000 & 2014-02-24 & 56712.9847 \\ 
FEROS & 48000 & 2014-02-24 & 56712.9854 \\ 
FEROS & 48000 & 2014-02-25 & 56713.9918 \\ 
FEROS & 48000 & 2014-02-25 & 56713.9924 \\ 
ECass & 30000 & 2014-07-16 & 56854.2982 \\ 
ECass & 30000 & 2014-10-14 & 56944.1713 \\ 
ECass & 30000 & 2014-11-19 & 56980.0376 \\ 
BeSS & 5000 & 2014-11-26 & 56987.3834 \\ 
BeSS & 100000 & 2014-11-28 & 56989.4437 \\
Amateur & 5000 & 2014-12-28 & 57019.0116 \\ 
Amateur & 5000 & 2015-01-09 & 57031.0142 \\ 
Amateur & 5000 & 2015-01-18 & 57040.9596 \\ 

\end{longtable}}
\end{center}

\begin{figure}
\begin{center}
\includegraphics[width=.8\textwidth]{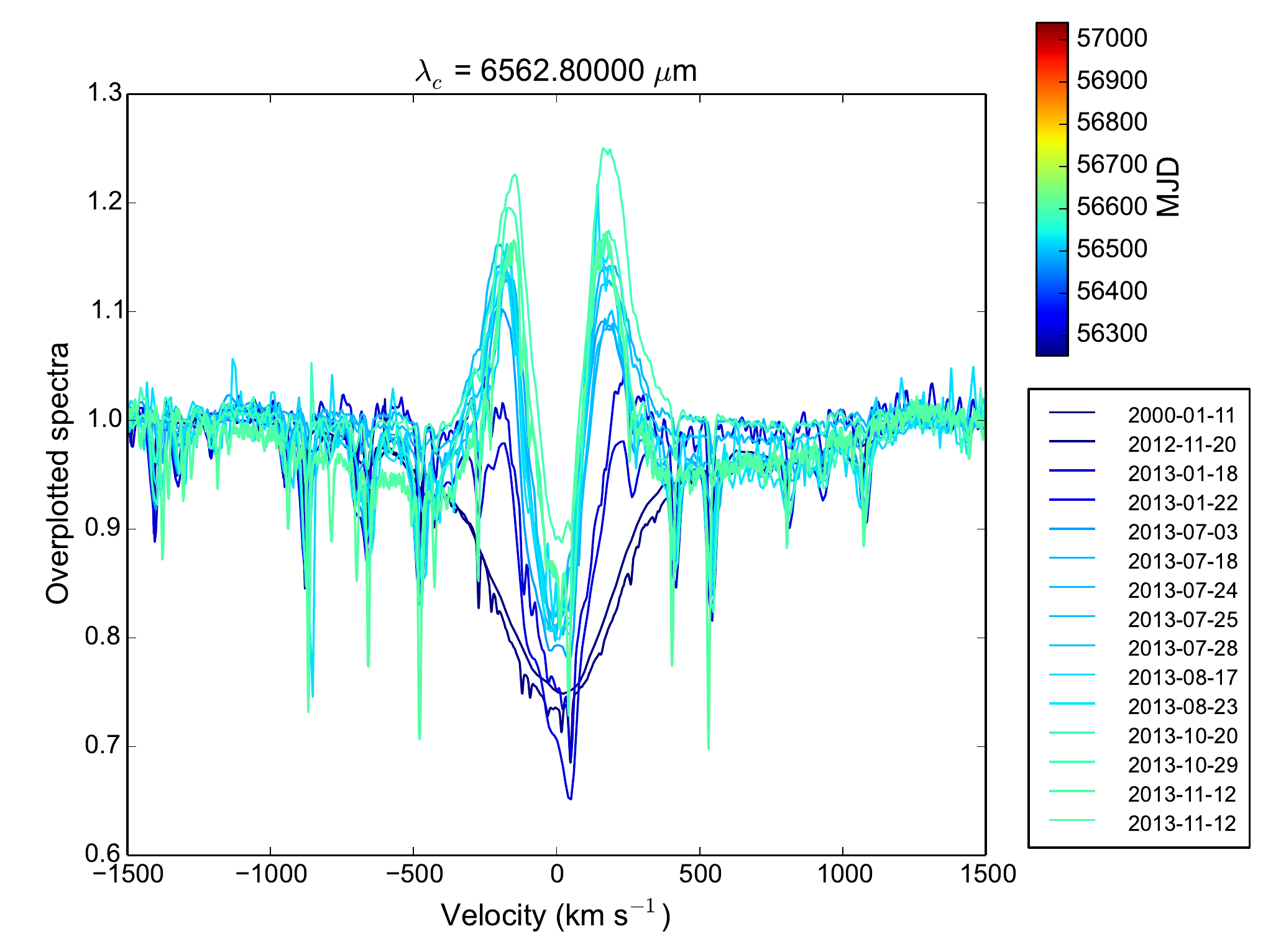} 
\caption[H$\alpha$ overplotted line profiles of Achernar during the 2013 outburst]{H$\alpha$ overplotted line profiles of Achernar during the 2013 outburst. The color corresponds to the observation date. The (pure) photospheric reference profile \citet{riv13b} is also shown. It contains the observations from all instruments presented in Sec.~\ref{sec:aerispec}, except for FEROS/ESO, due to its with frequency.}
\label{fig:apaeriover1}
\end{center}
\end{figure}

\begin{figure}
\begin{center}
\includegraphics[width=.8\textwidth]{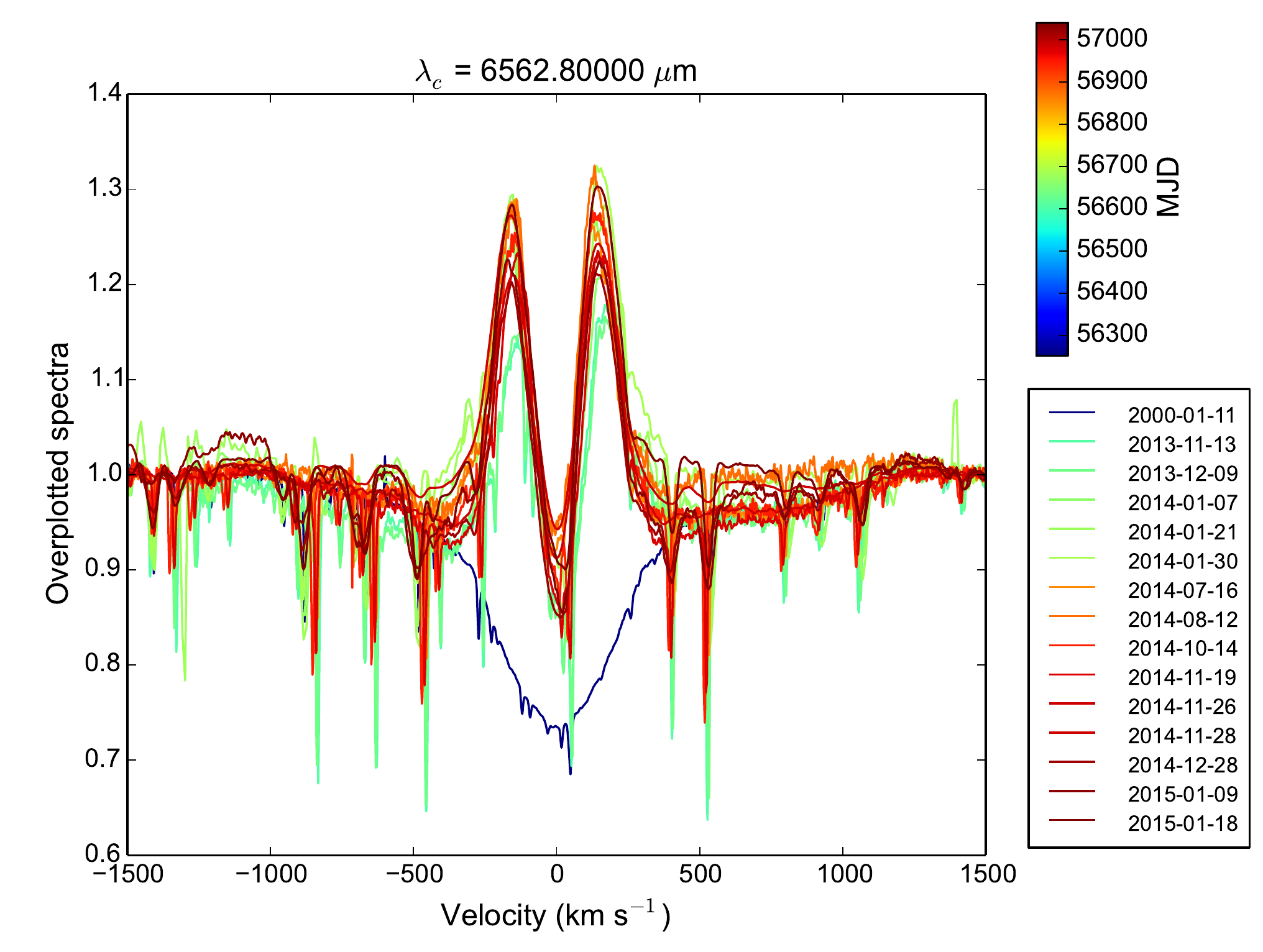} 
\caption{The same as previous figure, for posterior observations.}
\label{fig:apaeriover2}
\end{center}
\end{figure}

\begin{figure}
\begin{center}
\includegraphics[width=.8\textwidth]{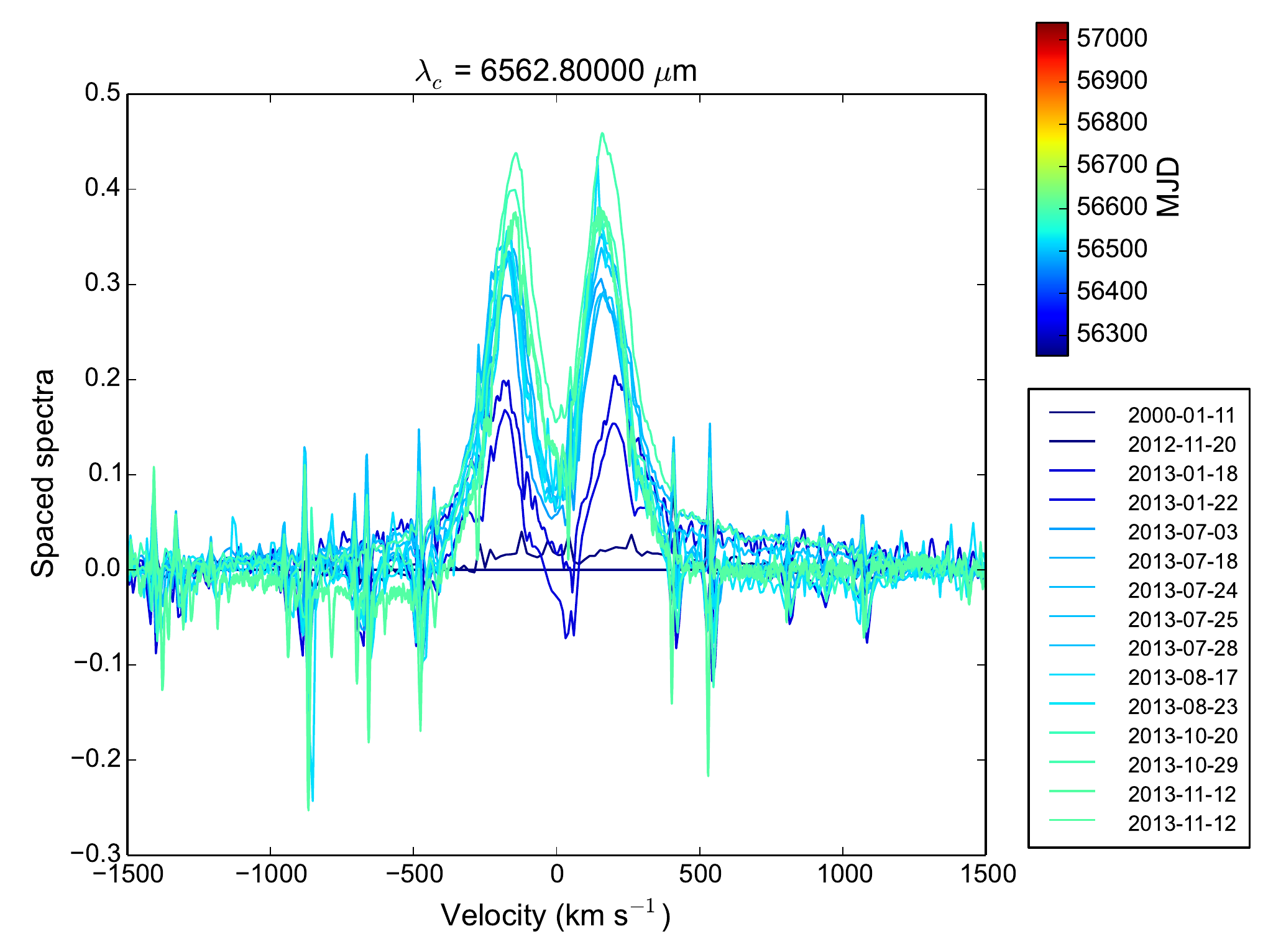} 
\caption[H$\alpha$ difference profiles of Achernar during the 2013 outburst]{H$\alpha$ difference profiles of Achernar during the 2013 outburst, in relation to the (pure) photospheric reference profile \citet{riv13b}. The color corresponds to the observation date. It contains the observations from all instruments presented in Sec.~\ref{sec:aerispec}, except for FEROS/ESO, due to its with frequency.}
\label{fig:apaeridiff1}
\end{center}
\end{figure}

\begin{figure}
\begin{center}
\includegraphics[width=.8\textwidth]{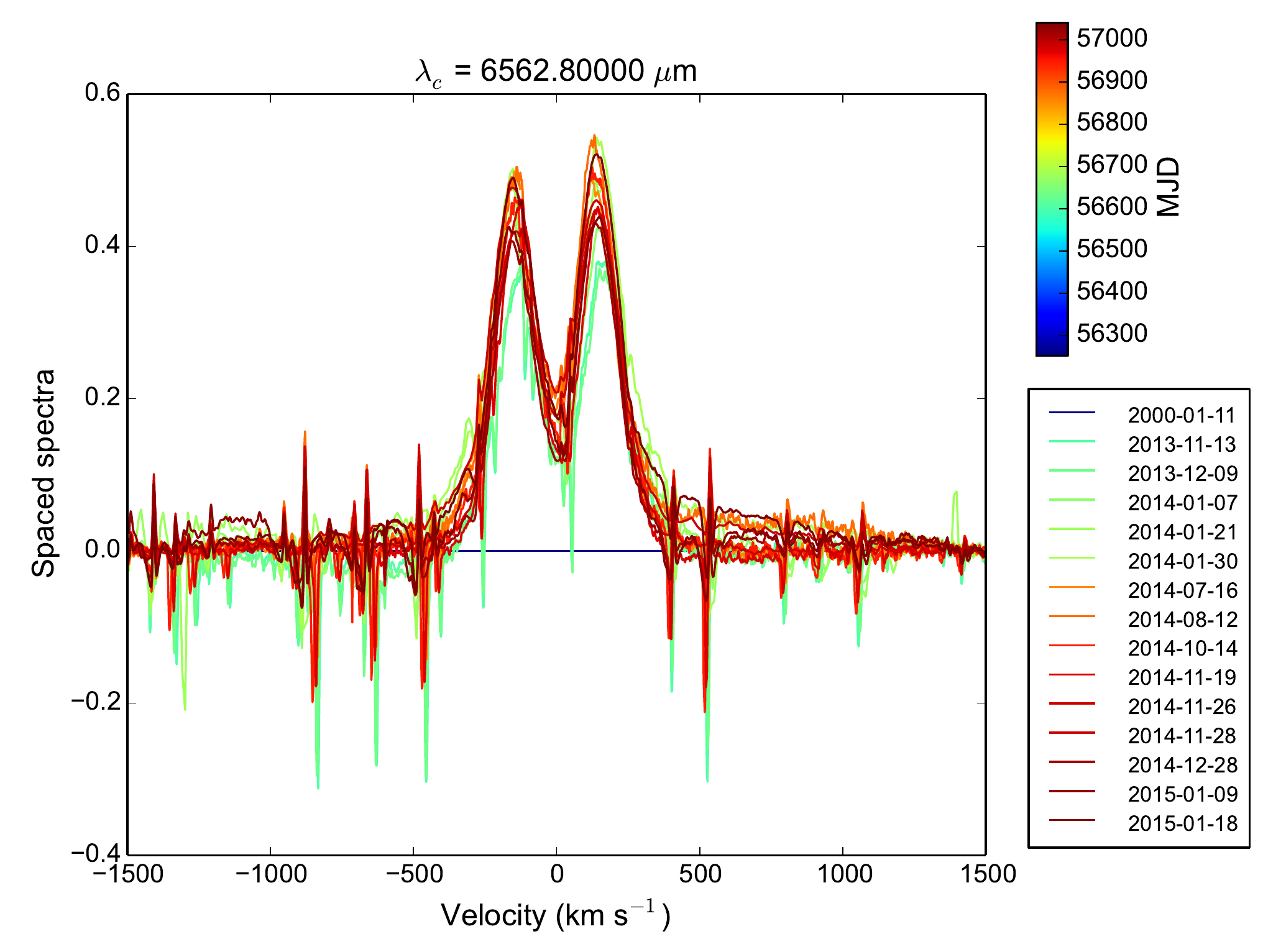} 
\caption{The same as previous figure, for posterior observations.}
\label{fig:apaeridiff2}
\end{center}
\end{figure}

\begin{figure}
    \centering
    \includegraphics[width=.8\linewidth]{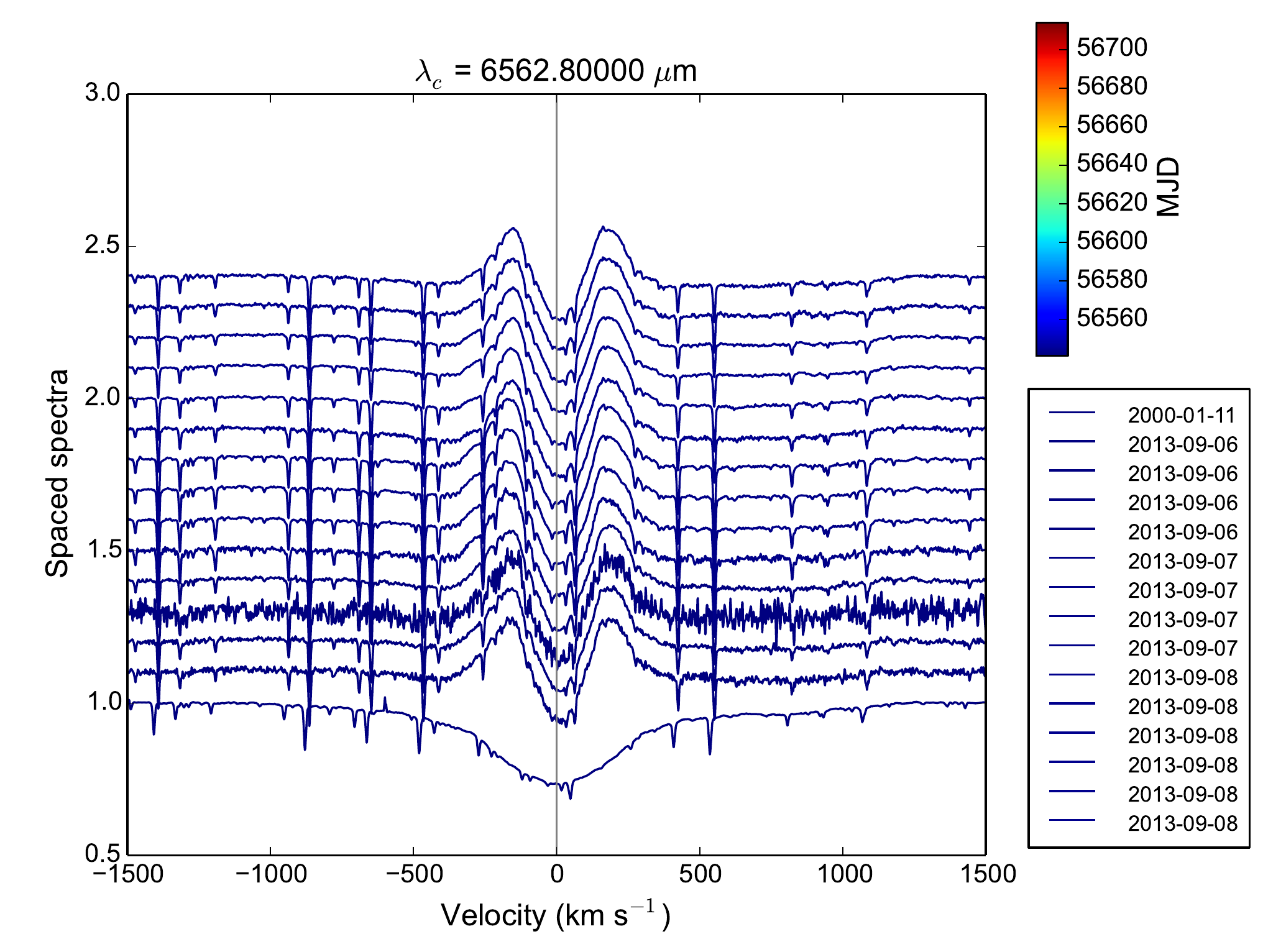}
\caption{H$\alpha$ line profiles of Achernar during the 2013 outburst. The color corresponds to the observation date. The (pure) photospheric reference profile \citet{riv13b} is also shown. It contains the observations from FEROS/ESO.}
    \label{fig:apfer01}
\end{figure}

\begin{figure}
    \centering
    \includegraphics[width=.8\linewidth]{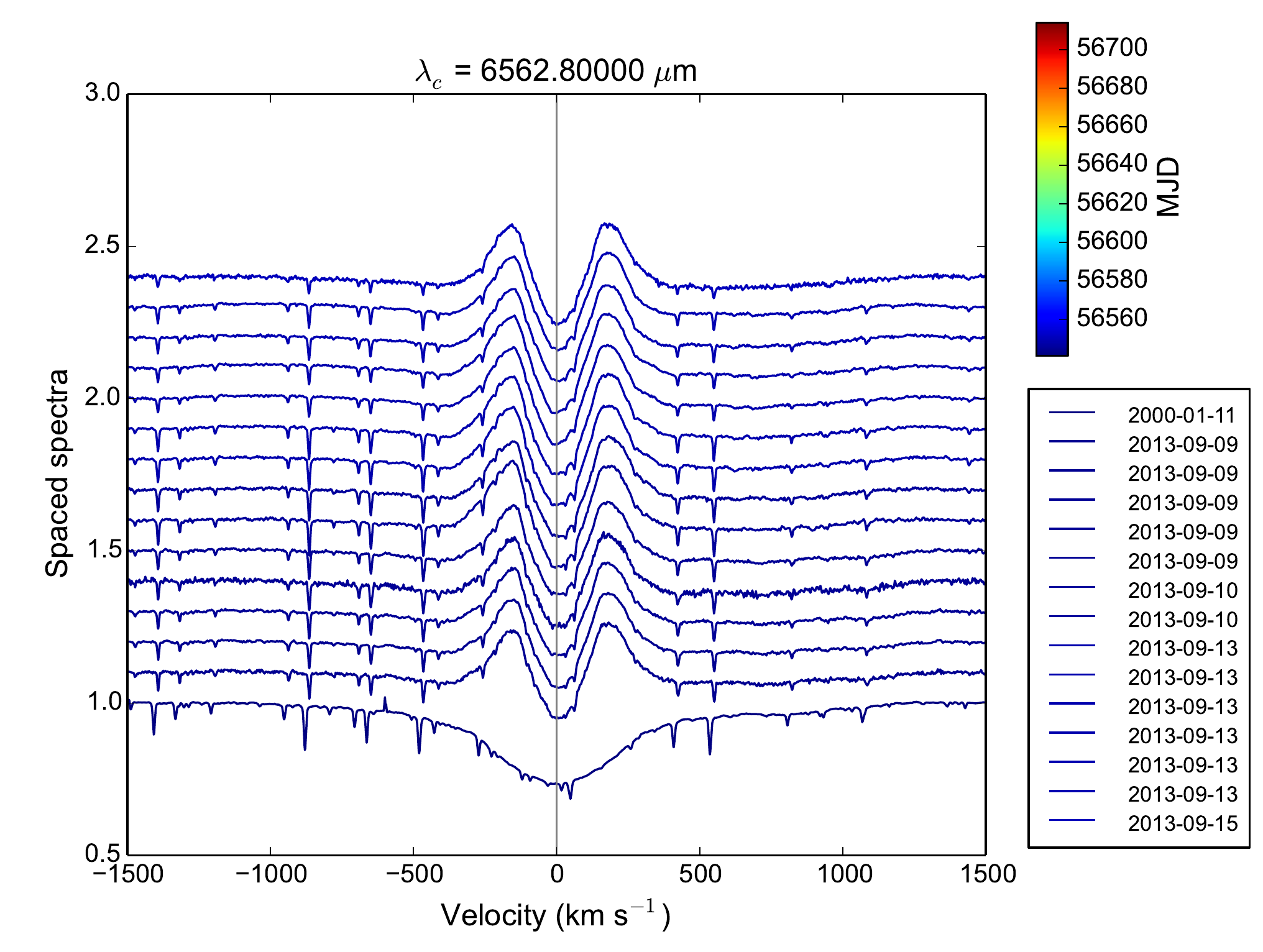}
\caption{The same as previous figure, for posterior observations.}
    \label{fig:apfer02}
\end{figure}

\begin{figure}
    \centering
    \includegraphics[width=.8\linewidth]{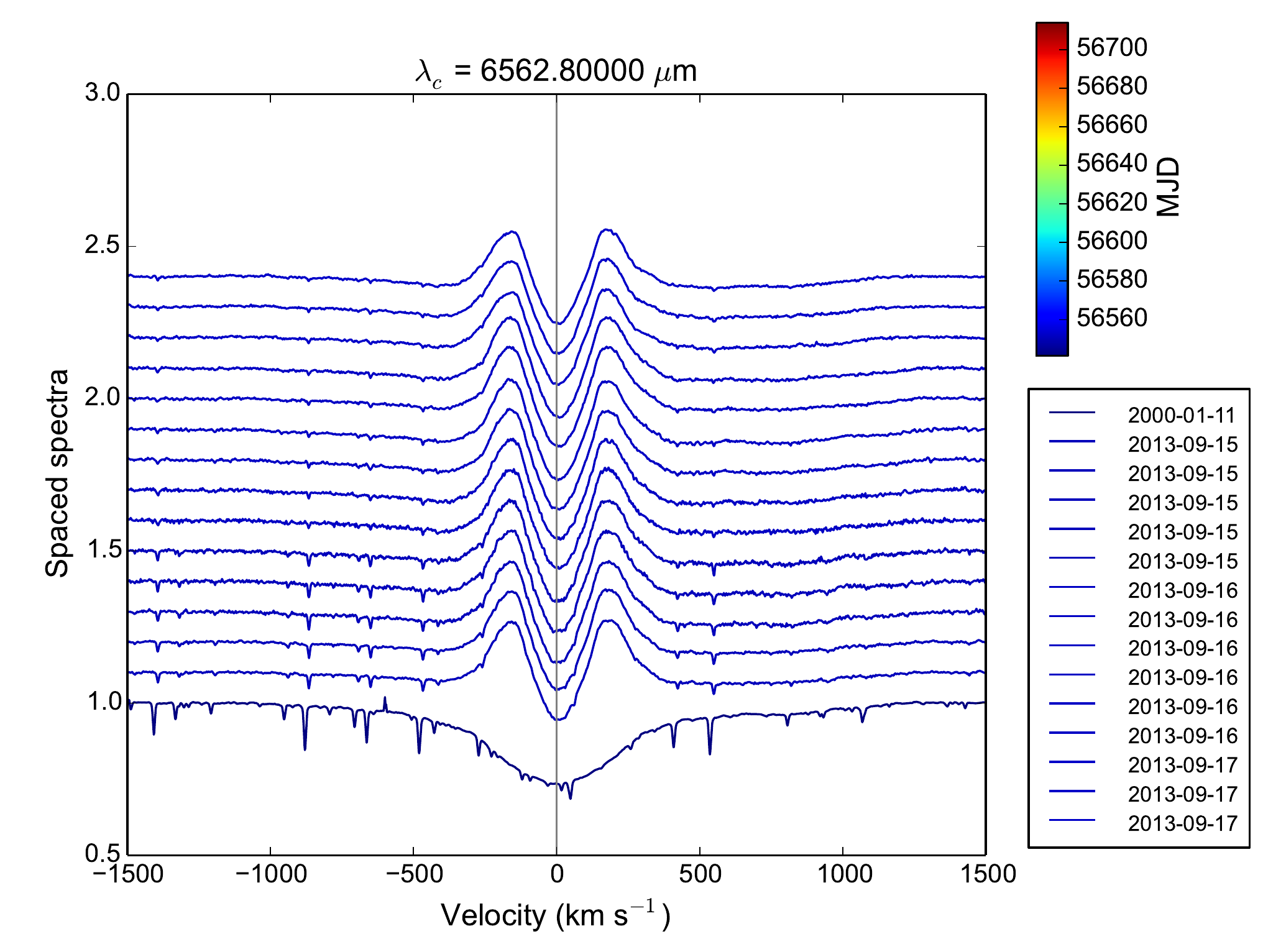}
\caption{The same as previous figure, for posterior observations.}
    \label{fig:apfer03}
\end{figure}

\begin{figure}
    \centering
    \includegraphics[width=.8\linewidth]{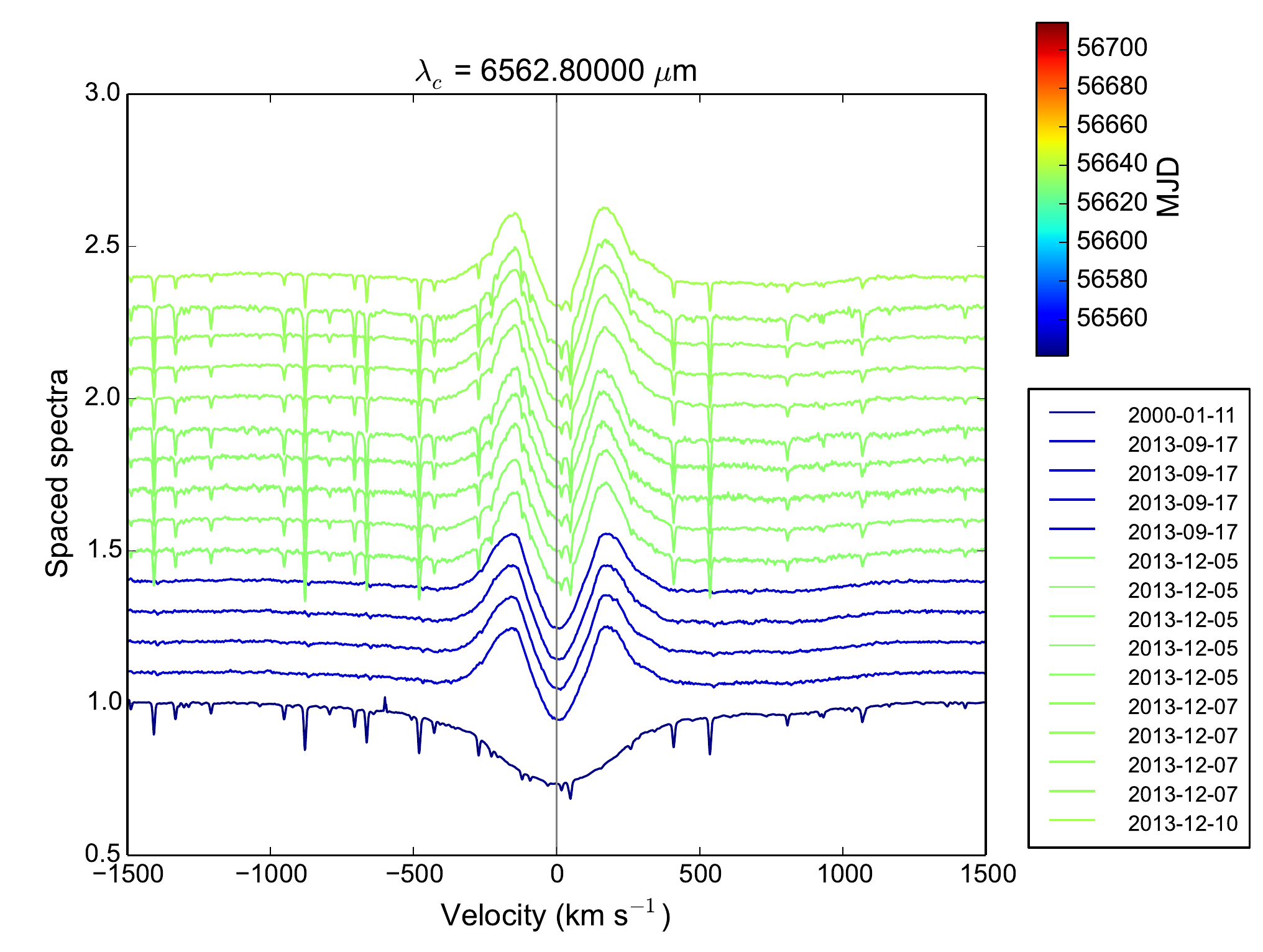}
\caption{The same as previous figure, for posterior observations.}
    \label{fig:apfer04}
\end{figure}

\begin{figure}
    \centering
    \includegraphics[width=.8\linewidth]{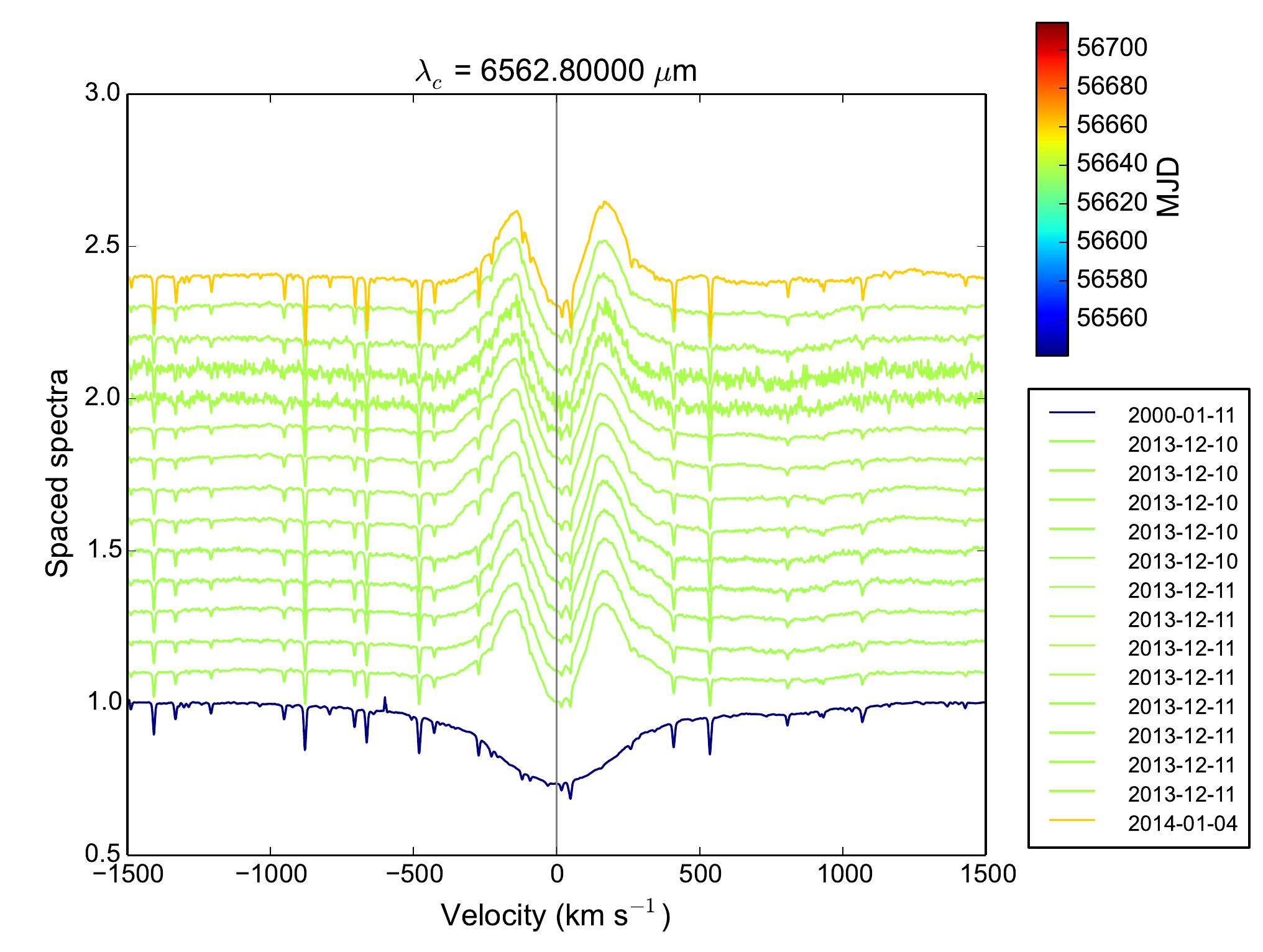}
\caption{The same as previous figure, for posterior observations.}
    \label{fig:apfer05}
\end{figure}

\begin{figure}
    \centering
    \includegraphics[width=.8\linewidth]{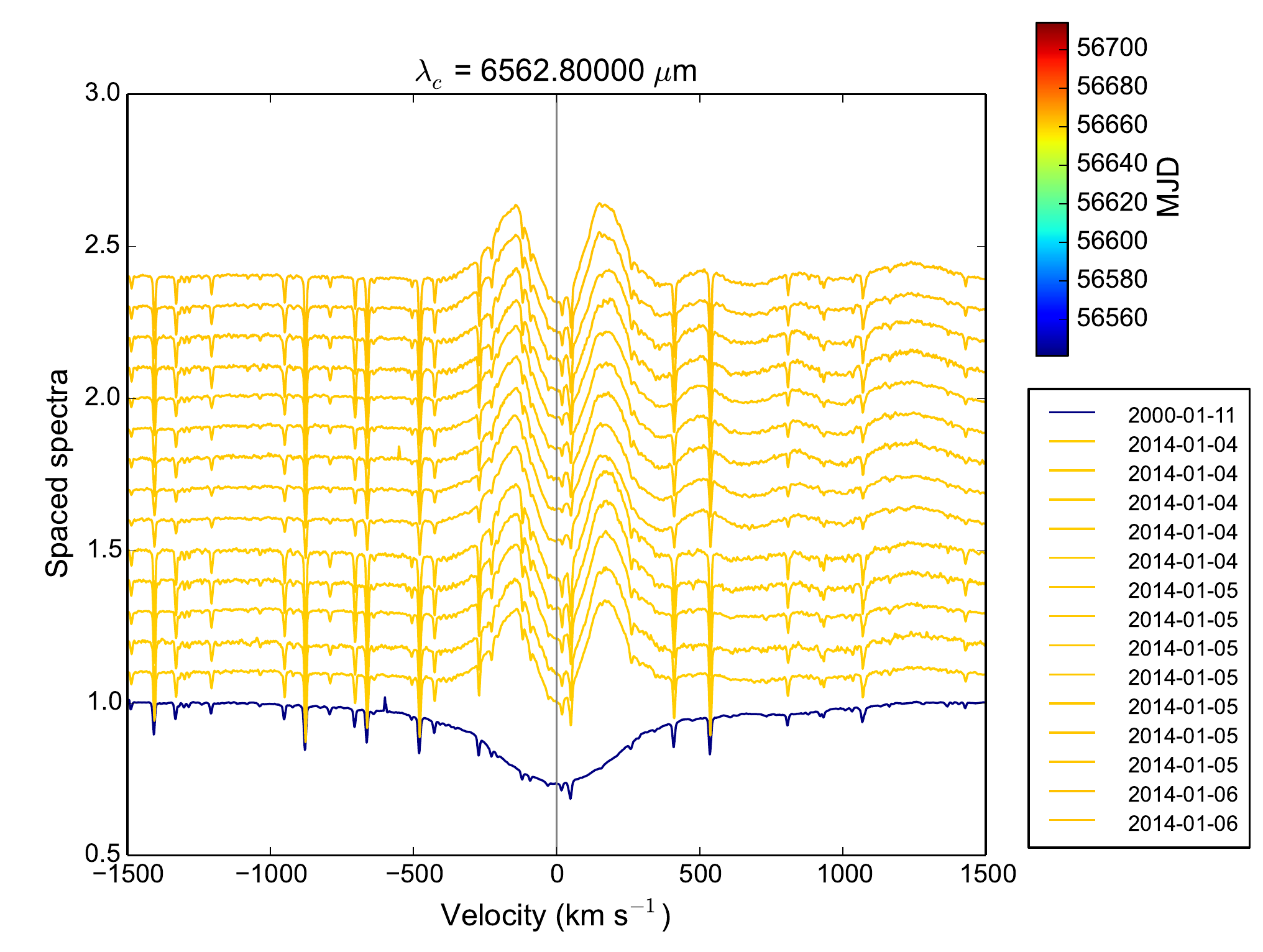}
\caption{The same as previous figure, for posterior observations.}
    \label{fig:apfer06}
\end{figure}

\begin{figure}
    \centering
    \includegraphics[width=.8\linewidth]{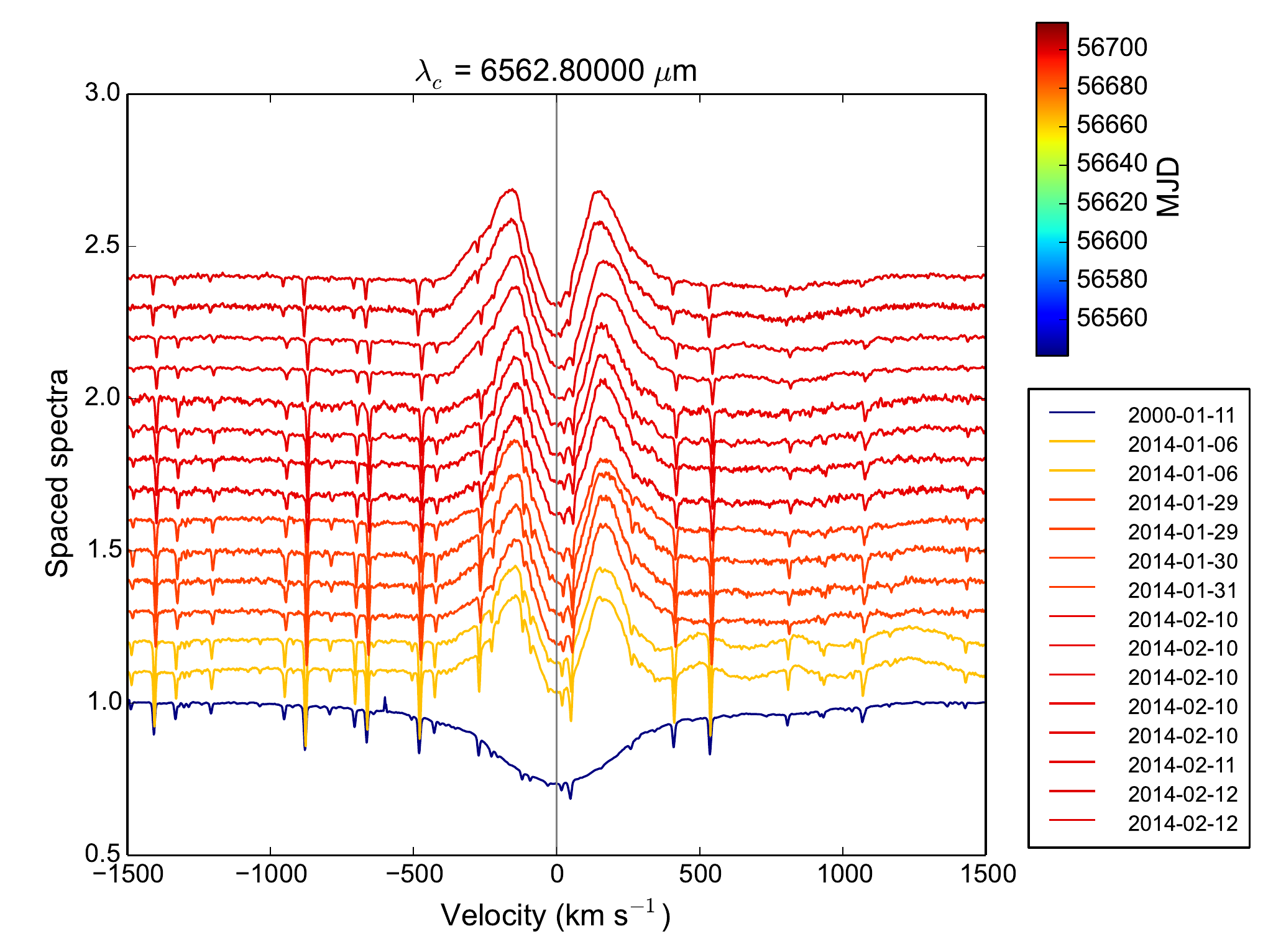}
\caption{The same as previous figure, for posterior observations.}
    \label{fig:apfer07}
\end{figure}

\begin{figure}
    \centering
    \includegraphics[width=.8\linewidth]{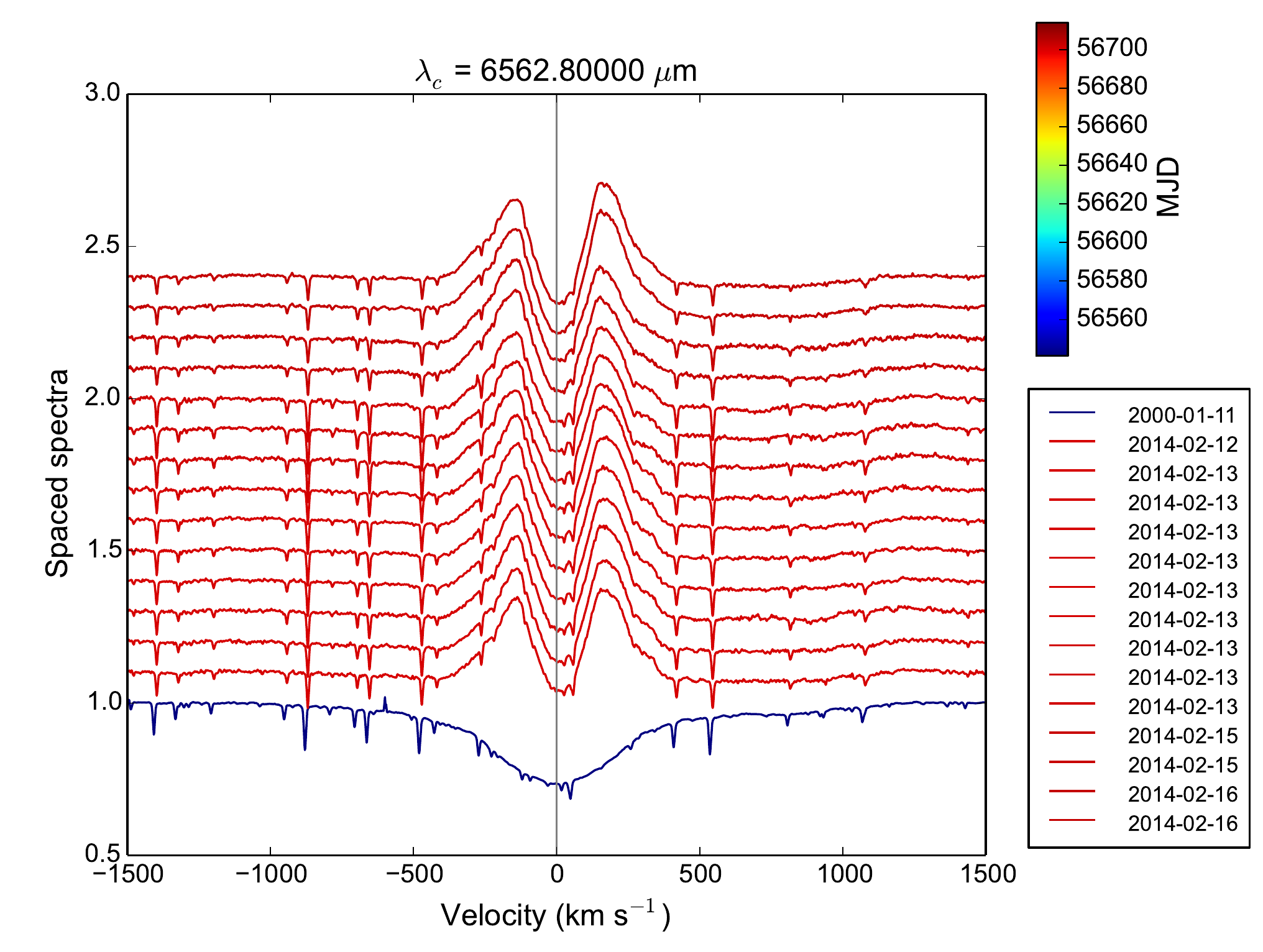}
\caption{The same as previous figure, for posterior observations.}
    \label{fig:apfer08}
\end{figure}

\begin{figure}
    \centering
    \includegraphics[width=.8\linewidth]{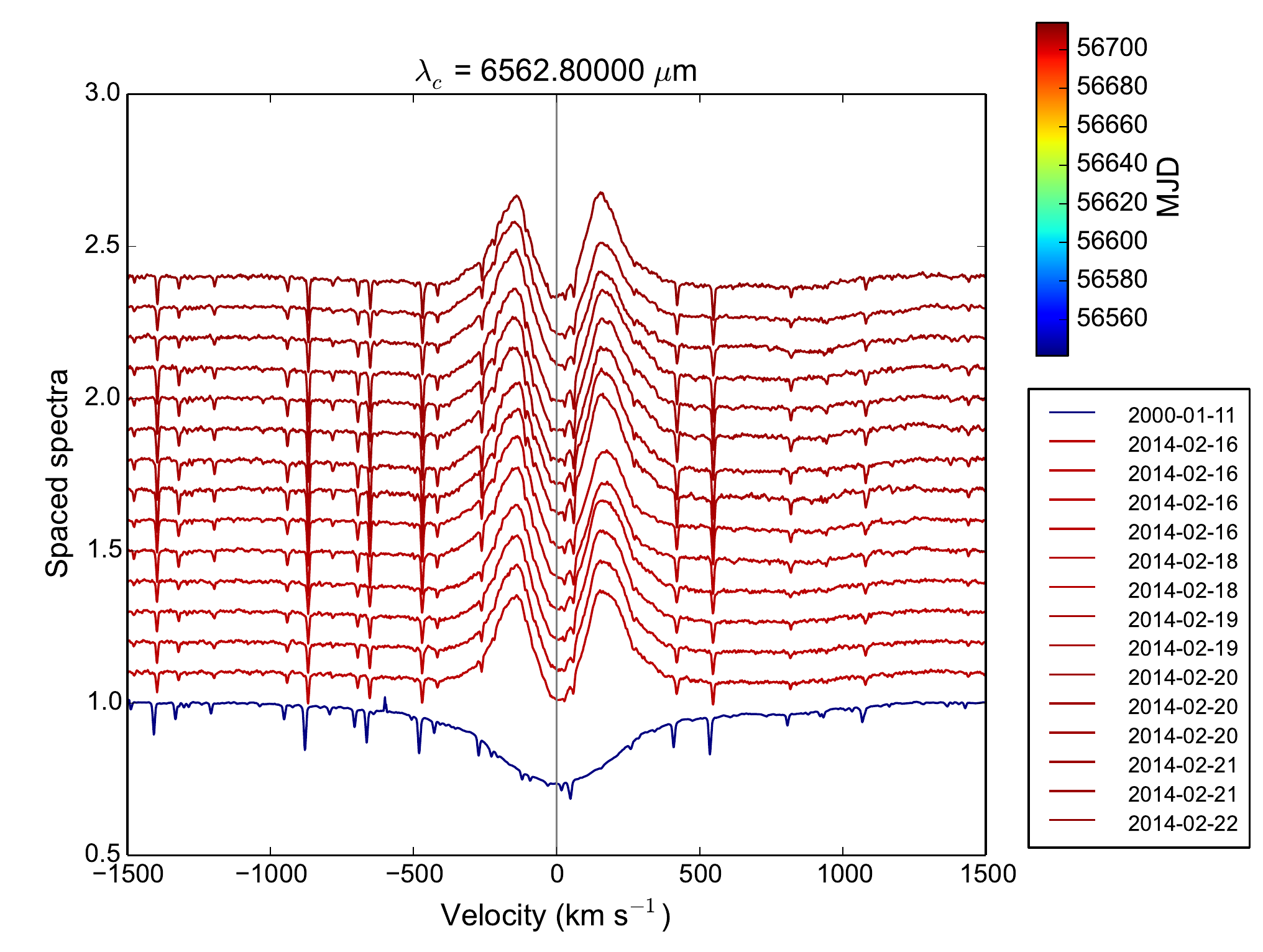}
\caption{The same as previous figure, for posterior observations.}
    \label{fig:apfer09}
\end{figure}

\begin{figure}
    \centering
    \includegraphics[width=.8\linewidth]{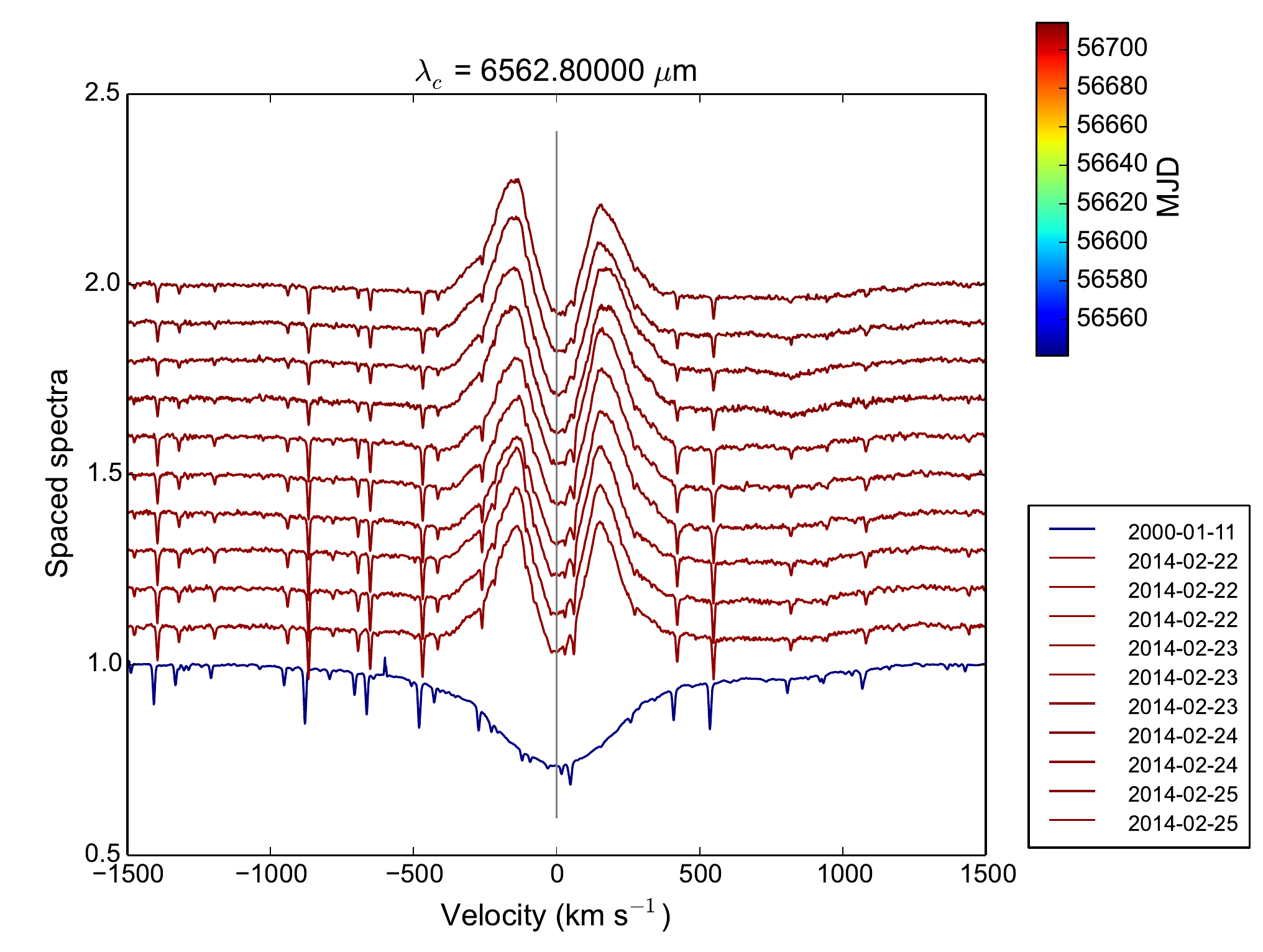}
\caption{The same as previous figure, for posterior observations.}
    \label{fig:apfer10}
\end{figure}

\section{Interferometry \label{ap:aeriinterf}}
\begin{center}
{\footnotesize \begin{longtable}{clc}
\caption
{List of VLTI-AMBER observations ($K$ band, high resolution mode) of the 2013 outburst of Achernar.} \label{ap:tabAMBER} \\
\hline \multicolumn{1}{c}{\textbf{Date}} & %
\multicolumn{1}{c}{\textbf{MJD}} & %
\multicolumn{1}{c}{\textbf{Dataset}} \\ \hline
\endfirsthead

\multicolumn{3}{c}%
{{\bfseries \tablename\ \thetable{} -- table from previous page}} \\
\hline \multicolumn{1}{c}{\textbf{Date}} & %
\multicolumn{1}{c}{\textbf{MJD}} & %
\multicolumn{1}{c}{\textbf{Dataset}} \\ \hline
\endhead

\hline \multicolumn{3}{r}{{Continue on next page}} \\ \hline
\endfoot

\hline \hline
\endlastfoot

2013-08-01 & 56506.2863503 & 2 \\ 
2013-08-02 & 56507.4056214 & 1 \\ 
2013-08-03 & 56508.4106953 & 1 \\ 
2013-08-04 & 56509.3647363 & 1 \\ 
2013-08-13 & 56518.2686651 & 5 \\ 
2013-08-19 & 56524.327139 & 1 \\ 
2013-08-20 & 56525.2566051 & 3 \\ 
2013-09-22 & 56558.3015531 & 1 \\ 
2013-09-23 & 56559.1584479 & 4 \\ 
2013-11-11 & 56608.197591 & 1 \\ 
2013-11-12 & 56609.2032108 & 1 \\ 
2013-11-15 & 56612.3060189 & 1 \\ 
2014-06-24 & 56833.4330857 & 1 \\ 
2014-06-30 & 56839.3910707 & 1 \\ 
2014-07-13 & 56852.3893149 & 2 \\ 
2014-08-04 & 56874.4086259 & 1 \\ 
2014-08-05 & 56875.4041884 & 1 \\ 
2014-08-10 & 56880.3435603 & 1 \\ 
2014-08-16 & 56886.0635461 & 1 \\ 
2014-08-28 & 56929.2058108 & 2 \\ 
2014-08-29 & 56899.3931993 & 3 \\ 
2014-09-08 & 56909.2860173 & 1 \\ 
2014-09-09 & 56910.27 & 0 \\ 
2014-09-10 & 56911.2660414 & 2 \\ 
2014-09-11 & 56912.25 & 0 \\ 
2014-09-19 & 56920.119566 & 9 \\ 
2014-09-23 & 56924.1887453 & 8 \\ 
2014-09-24 & 56925.3043003 & 4 \\ 
2014-09-27 & 56928.2170004 & 2 \\ 
2014-09-28 & 56898.3894947 & 3 \\ 
2014-09-30 & 56931.2513042 & 1 \\ 
2014-10-01 & 56932.3477905 & 1 \\ 
2014-10-18 & 56949.3 & 0 \\ 
2014-10-28 & 56958.9977269 & 5 \\ 
2014-10-29 & 56960.0185591 & 5 \\ 
2014-10-30 & 56961.2078147 & 1 \\ 
2015-01-12 & 57035.057242 & 1 \\ 
2015-01-18 & 57041.0526772 & 1 \\ 

\end{longtable}}
\end{center}

\section{Polarimetry \label{ap:aeripol}}
\begin{center}
{\footnotesize \begin{longtable}{ccccccccc}
\caption
{List of linear polarimetric observations of the 2013 outburst of Achernar. } \label{tab:appol} \\
\hline \multicolumn{1}{c}{\textbf{MJD}} & %
\multicolumn{1}{c}{\textbf{Band}} & %
\multicolumn{1}{c}{\textbf{Date}} & %
\multicolumn{1}{c}{\textbf{$P$(\%)}} & %
\multicolumn{1}{c}{\textbf{$Q$(\%)}} & %
\multicolumn{1}{c}{\textbf{$U$}} & %
\multicolumn{1}{c}{\textbf{$\theta$($^\circ$)}} & %
\multicolumn{1}{c}{\textbf{$\sigma_P$}} & %
\multicolumn{1}{c}{\textbf{$\sigma_\theta$}} \\ \hline
\endfirsthead

\multicolumn{9}{c}%
{{\bfseries \tablename\ \thetable{} -- table from previous page}} \\
\hline \multicolumn{1}{c}{\textbf{MJD}} & %
\multicolumn{1}{c}{\textbf{Band}} & %
\multicolumn{1}{c}{\textbf{Date}} & %
\multicolumn{1}{c}{\textbf{$P$(\%)}} & %
\multicolumn{1}{c}{\textbf{$Q$(\%)}} & %
\multicolumn{1}{c}{\textbf{$U$(\%)}} & %
\multicolumn{1}{c}{\textbf{$\theta$($^\circ$)}} & %
\multicolumn{1}{c}{\textbf{$\sigma_P$}} & %
\multicolumn{1}{c}{\textbf{$\sigma_\theta$}} \\ \hline
\endhead

\hline \multicolumn{9}{r}{{Continue on next page}} \\ \hline
\endfoot

\hline \hline
\endlastfoot

56500.350732 & $V$ & 2013-07-26 & 0.293 & 0.154 & 0.249 & 29.1 & 0.033 & 1.0 \\ 
56500.372017 & $V$ & 2013-07-26 & 0.251 & 0.080 & 0.238 & 35.7 & 0.038 & 1.1 \\ 
56503.228983 & $R$ & 2013-07-29 & 0.284 & 0.065 & 0.277 & 38.4 & 0.004 & 0.1 \\ 
56503.295401 & $R$ & 2013-07-29 & 0.288 & 0.062 & 0.281 & 38.8 & 0.006 & 0.2 \\ 
56504.290627 & $B$ & 2013-07-30 & 0.283 & -0.018 & 0.283 & 46.8 & 0.078 & 2.2 \\ 
56504.300175 & $V$ & 2013-07-30 & 0.227 & 0.109 & 0.199 & 30.7 & 0.029 & 0.8 \\ 
56504.317043 & $I$ & 2013-07-30 & 0.253 & 0.071 & 0.243 & 36.9 & 0.031 & 0.9 \\ 
56504.326981 & $B$ & 2013-07-30 & 0.292 & 0.188 & 0.223 & 24.9 & 0.049 & 1.4 \\ 
56504.335748 & $V$ & 2013-07-30 & 0.218 & 0.090 & 0.199 & 32.9 & 0.014 & 0.4 \\ 
56504.343907 & $R$ & 2013-07-30 & 0.247 & 0.097 & 0.227 & 33.4 & 0.011 & 0.3 \\ 
56504.350724 & $I$ & 2013-07-30 & 0.257 & 0.114 & 0.230 & 31.8 & 0.030 & 0.9 \\ 
56504.357889 & $V$ & 2013-07-30 & 0.258 & 0.134 & 0.220 & 29.4 & 0.031 & 0.9 \\ 
56504.365586 & $V$ & 2013-07-30 & 0.243 & 0.119 & 0.213 & 30.4 & 0.012 & 0.3 \\ 
56504.372831 & $V$ & 2013-07-30 & 0.229 & 0.097 & 0.208 & 32.5 & 0.024 & 0.7 \\ 
56505.349694 & $V$ & 2013-07-31 & 0.273 & 0.137 & 0.236 & 30.0 & 0.027 & 0.8 \\ 
56505.358549 & $V$ & 2013-07-31 & 0.289 & 0.134 & 0.256 & 31.2 & 0.012 & 0.3 \\ 
56505.367669 & $V$ & 2013-07-31 & 0.272 & 0.147 & 0.228 & 28.6 & 0.024 & 0.7 \\ 
56517.348524 & $V$ & 2013-08-12 & 0.266 & 0.163 & 0.210 & 26.1 & 0.071 & 2.0 \\ 
56518.184862 & $V$ & 2013-08-13 & 0.336 & 0.148 & 0.302 & 31.9 & 0.030 & 0.8 \\ 
56518.222449 & $I$ & 2013-08-13 & 0.305 & 0.112 & 0.283 & 34.2 & 0.043 & 1.2 \\ 
56518.258924 & $V$ & 2013-08-13 & 0.299 & 0.132 & 0.269 & 31.9 & 0.009 & 0.3 \\ 
56518.267888 & $I$ & 2013-08-13 & 0.293 & 0.095 & 0.277 & 35.5 & 0.024 & 0.7 \\ 
56518.278444 & $I$ & 2013-08-13 & 0.320 & 0.114 & 0.299 & 34.6 & 0.022 & 0.6 \\ 
56518.293757 & $V$ & 2013-08-13 & 0.316 & 0.109 & 0.297 & 34.9 & 0.032 & 0.9 \\ 
56518.304335 & $V$ & 2013-08-13 & 0.295 & 0.084 & 0.282 & 36.7 & 0.047 & 1.3 \\ 
56518.313461 & $I$ & 2013-08-13 & 0.299 & 0.067 & 0.291 & 38.5 & 0.054 & 1.5 \\ 
56518.333942 & $I$ & 2013-08-13 & 0.278 & 0.089 & 0.264 & 35.7 & 0.026 & 0.7 \\ 
56518.344150 & $V$ & 2013-08-13 & 0.282 & 0.111 & 0.259 & 33.4 & 0.038 & 1.1 \\ 
56518.357009 & $I$ & 2013-08-13 & 0.265 & 0.099 & 0.246 & 34.0 & 0.030 & 0.8 \\ 
56520.197235 & $V$ & 2013-08-15 & 0.358 & 0.107 & 0.342 & 36.3 & 0.018 & 0.5 \\ 
56520.208849 & $I$ & 2013-08-15 & 0.337 & 0.127 & 0.311 & 33.9 & 0.020 & 0.6 \\ 
56520.220840 & $V$ & 2013-08-15 & 0.327 & 0.099 & 0.312 & 36.2 & 0.017 & 0.5 \\ 
56520.229844 & $I$ & 2013-08-15 & 0.290 & 0.097 & 0.273 & 35.2 & 0.015 & 0.4 \\ 
56520.238658 & $V$ & 2013-08-15 & 0.303 & 0.089 & 0.290 & 36.5 & 0.021 & 0.6 \\ 
56520.247414 & $I$ & 2013-08-15 & 0.287 & 0.073 & 0.277 & 37.6 & 0.027 & 0.8 \\ 
56520.256031 & $V$ & 2013-08-15 & 0.299 & 0.089 & 0.286 & 36.4 & 0.017 & 0.5 \\ 
56520.264931 & $I$ & 2013-08-15 & 0.275 & 0.081 & 0.263 & 36.4 & 0.021 & 0.6 \\ 
56520.281870 & $I$ & 2013-08-15 & 0.297 & 0.081 & 0.286 & 37.1 & 0.012 & 0.4 \\ 
56520.290238 & $V$ & 2013-08-15 & 0.321 & 0.074 & 0.312 & 38.3 & 0.013 & 0.4 \\ 
56520.299005 & $I$ & 2013-08-15 & 0.309 & 0.023 & 0.308 & 42.9 & 0.048 & 1.4 \\ 
56520.308137 & $V$ & 2013-08-15 & 0.270 & 0.071 & 0.261 & 37.4 & 0.021 & 0.6 \\ 
56520.316621 & $I$ & 2013-08-15 & 0.295 & 0.036 & 0.293 & 41.5 & 0.036 & 1.0 \\ 
56520.325469 & $V$ & 2013-08-15 & 0.330 & 0.074 & 0.321 & 38.6 & 0.045 & 1.3 \\ 
56520.334445 & $I$ & 2013-08-15 & 0.289 & 0.049 & 0.285 & 40.1 & 0.033 & 0.9 \\ 
56520.343392 & $V$ & 2013-08-15 & 0.305 & 0.074 & 0.296 & 38.0 & 0.007 & 0.2 \\ 
56520.351650 & $I$ & 2013-08-15 & 0.305 & 0.067 & 0.298 & 38.7 & 0.009 & 0.2 \\ 
56520.359966 & $V$ & 2013-08-15 & 0.310 & 0.050 & 0.306 & 40.3 & 0.015 & 0.4 \\ 
56520.368195 & $I$ & 2013-08-15 & 0.300 & 0.045 & 0.297 & 40.7 & 0.021 & 0.6 \\ 
56525.227304 & $V$ & 2013-08-20 & 0.327 & 0.141 & 0.295 & 32.2 & 0.007 & 0.2 \\ 
56525.235956 & $I$ & 2013-08-20 & 0.289 & 0.119 & 0.264 & 32.9 & 0.038 & 1.1 \\ 
56525.247628 & $V$ & 2013-08-20 & 0.343 & 0.192 & 0.285 & 28.0 & 0.019 & 0.5 \\ 
56525.264190 & $I$ & 2013-08-20 & 0.324 & 0.109 & 0.305 & 35.1 & 0.051 & 1.5 \\ 
56525.281748 & $V$ & 2013-08-20 & 0.341 & 0.120 & 0.319 & 34.7 & 0.031 & 0.9 \\ 
56525.288762 & $I$ & 2013-08-20 & 0.326 & 0.147 & 0.292 & 31.6 & 0.061 & 1.8 \\ 
56525.298369 & $V$ & 2013-08-20 & 0.287 & 0.117 & 0.262 & 33.0 & 0.022 & 0.6 \\ 
56525.317969 & $V$ & 2013-08-20 & 0.334 & 0.140 & 0.303 & 32.6 & 0.026 & 0.7 \\ 
56525.339694 & $V$ & 2013-08-20 & 0.303 & 0.124 & 0.276 & 32.9 & 0.042 & 1.2 \\ 
56525.348207 & $I$ & 2013-08-20 & 0.281 & 0.115 & 0.257 & 32.9 & 0.035 & 1.0 \\ 
56525.355788 & $V$ & 2013-08-20 & 0.298 & 0.127 & 0.270 & 32.4 & 0.012 & 0.3 \\ 
56525.363791 & $I$ & 2013-08-20 & 0.328 & 0.178 & 0.276 & 28.6 & 0.061 & 1.8 \\ 
56526.194579 & $V$ & 2013-08-21 & 0.284 & 0.081 & 0.273 & 36.7 & 0.020 & 0.6 \\ 
56526.203473 & $I$ & 2013-08-21 & 0.267 & 0.072 & 0.257 & 37.2 & 0.047 & 1.3 \\ 
56526.212322 & $V$ & 2013-08-21 & 0.256 & 0.121 & 0.226 & 30.9 & 0.024 & 0.7 \\ 
56526.255973 & $V$ & 2013-08-21 & 0.291 & 0.124 & 0.263 & 32.3 & 0.011 & 0.3 \\ 
56526.265342 & $I$ & 2013-08-21 & 0.253 & 0.072 & 0.243 & 36.7 & 0.019 & 0.5 \\ 
56526.274266 & $V$ & 2013-08-21 & 0.272 & 0.072 & 0.262 & 37.3 & 0.034 & 1.0 \\ 
56526.284301 & $I$ & 2013-08-21 & 0.262 & 0.098 & 0.243 & 34.1 & 0.027 & 0.8 \\ 
56526.299261 & $V$ & 2013-08-21 & 0.231 & 0.076 & 0.219 & 35.4 & 0.029 & 0.8 \\ 
56526.301019 & $I$ & 2013-08-21 & 0.258 & 0.059 & 0.251 & 38.4 & 0.055 & 1.6 \\ 
56526.319659 & $I$ & 2013-08-21 & 0.238 & 0.109 & 0.212 & 31.4 & 0.023 & 0.7 \\ 
56526.321651 & $V$ & 2013-08-21 & 0.194 & 0.056 & 0.185 & 36.6 & 0.044 & 1.2 \\ 
56526.341934 & $I$ & 2013-08-21 & 0.174 & 0.023 & 0.172 & 41.2 & 0.042 & 1.2 \\ 
56526.342246 & $V$ & 2013-08-21 & 0.226 & 0.101 & 0.202 & 31.7 & 0.019 & 0.5 \\ 
56526.360707 & $I$ & 2013-08-21 & 0.266 & 0.126 & 0.234 & 30.9 & 0.034 & 1.0 \\ 
56526.360829 & $V$ & 2013-08-21 & 0.268 & 0.145 & 0.226 & 28.7 & 0.043 & 1.2 \\ 
56561.214752 & $B$ & 2013-09-25 & 0.282 & 0.140 & 0.245 & 30.1 & 0.028 & 0.8 \\ 
56561.226332 & $V$ & 2013-09-25 & 0.277 & 0.108 & 0.255 & 33.5 & 0.020 & 0.6 \\ 
56561.233259 & $R$ & 2013-09-25 & 0.252 & 0.068 & 0.243 & 37.2 & 0.005 & 0.1 \\ 
56561.242530 & $I$ & 2013-09-25 & 0.284 & 0.103 & 0.265 & 34.4 & 0.007 & 0.2 \\ 
56561.249179 & $I$ & 2013-09-25 & 0.221 & 0.065 & 0.212 & 36.5 & 0.011 & 0.3 \\ 
56561.249457 & $V$ & 2013-09-25 & 0.215 & 0.112 & 0.184 & 29.4 & 0.020 & 0.6 \\ 
56561.255805 & $I$ & 2013-09-25 & 0.224 & 0.095 & 0.203 & 32.5 & 0.007 & 0.2 \\ 
56561.262674 & $V$ & 2013-09-25 & 0.251 & 0.102 & 0.229 & 33.0 & 0.022 & 0.6 \\ 
56561.276627 & $V$ & 2013-09-25 & 0.244 & 0.116 & 0.215 & 30.9 & 0.018 & 0.5 \\ 
56561.352431 & $V$ & 2013-09-25 & 0.294 & 0.152 & 0.252 & 29.4 & 0.039 & 1.1 \\ 
56561.358791 & $I$ & 2013-09-25 & 0.284 & 0.119 & 0.258 & 32.6 & 0.051 & 1.5 \\ 
56562.090557 & $B$ & 2013-09-26 & 0.275 & 0.175 & 0.212 & 25.2 & 0.023 & 0.7 \\ 
56562.100996 & $V$ & 2013-09-26 & 0.270 & 0.105 & 0.248 & 33.5 & 0.012 & 0.3 \\ 
56562.108728 & $R$ & 2013-09-26 & 0.261 & 0.120 & 0.232 & 31.3 & 0.011 & 0.3 \\ 
56562.116656 & $I$ & 2013-09-26 & 0.245 & 0.054 & 0.239 & 38.7 & 0.025 & 0.7 \\ 
56584.163867 & $V$ & 2013-10-18 & 0.315 & 0.115 & 0.293 & 34.3 & 0.022 & 0.6 \\ 
56584.172344 & $I$ & 2013-10-18 & 0.284 & 0.098 & 0.267 & 34.9 & 0.025 & 0.7 \\ 
56609.097206 & $V$ & 2013-11-12 & 0.277 & 0.079 & 0.266 & 36.7 & 0.040 & 1.2 \\ 
56610.018897 & $V$ & 2013-11-13 & 0.328 & 0.190 & 0.268 & 27.3 & 0.015 & 0.4 \\ 
56610.034451 & $B$ & 2013-11-13 & 0.371 & 0.221 & 0.298 & 26.7 & 0.010 & 0.3 \\ 
56610.035586 & $I$ & 2013-11-13 & 0.300 & 0.087 & 0.287 & 36.6 & 0.023 & 0.7 \\ 
56610.049168 & $R$ & 2013-11-13 & 0.311 & 0.118 & 0.288 & 33.9 & 0.018 & 0.5 \\ 
56640.016320 & $V$ & 2013-12-13 & 0.372 & 0.157 & 0.337 & 32.5 & 0.027 & 0.8 \\ 
56640.040400 & $R$ & 2013-12-13 & 0.384 & 0.100 & 0.371 & 37.5 & 0.014 & 0.4 \\ 
56640.048971 & $I$ & 2013-12-13 & 0.369 & 0.021 & 0.368 & 43.4 & 0.022 & 0.6 \\ 
56640.059214 & $B$ & 2013-12-13 & 0.442 & 0.191 & 0.398 & 32.2 & 0.095 & 2.7 \\ 
56725.928754 & $B$ & 2014-03-09 & 0.566 & 0.313 & 0.472 & 28.2 & 0.047 & 1.3 \\ 
56725.939636 & $V$ & 2014-03-09 & 0.458 & 0.222 & 0.401 & 30.5 & 0.023 & 0.7 \\ 
56725.946704 & $I$ & 2014-03-09 & 0.423 & 0.089 & 0.413 & 38.9 & 0.029 & 0.8 \\ 
56841.327645 & $B$ & 2014-07-02 & 0.268 & 0.175 & 0.202 & 24.5 & 0.020 & 0.6 \\ 
56841.334865 & $V$ & 2014-07-02 & 0.232 & 0.133 & 0.189 & 27.4 & 0.043 & 1.2 \\ 
56841.339711 & $R$ & 2014-07-02 & 0.272 & 0.126 & 0.241 & 31.2 & 0.012 & 0.3 \\ 
56842.355721 & $B$ & 2014-07-03 & 0.380 & 0.197 & 0.326 & 29.4 & 0.056 & 1.6 \\ 
56842.362876 & $V$ & 2014-07-03 & 0.256 & 0.115 & 0.228 & 31.6 & 0.050 & 1.4 \\ 
56842.368099 & $R$ & 2014-07-03 & 0.315 & 0.163 & 0.269 & 29.4 & 0.040 & 1.1 \\ 
56842.375059 & $I$ & 2014-07-03 & 0.284 & 0.120 & 0.258 & 32.5 & 0.035 & 1.0 \\ 
56889.193547 & $B$ & 2014-08-19 & 0.258 & 0.158 & 0.204 & 26.1 & 0.017 & 0.5 \\ 
56889.294155 & $B$ & 2014-08-19 & 0.201 & 0.122 & 0.160 & 26.4 & 0.040 & 1.1 \\ 
56889.300386 & $B$ & 2014-08-19 & 0.329 & 0.309 & 0.113 & 10.0 & 0.113 & 3.2 \\ 
56889.325231 & $V$ & 2014-08-19 & 0.292 & 0.187 & 0.224 & 25.1 & 0.008 & 0.2 \\ 
56889.335314 & $R$ & 2014-08-19 & 0.279 & 0.172 & 0.220 & 26.0 & 0.018 & 0.5 \\ 
56889.342774 & $I$ & 2014-08-19 & 0.268 & 0.187 & 0.192 & 22.9 & 0.010 & 0.3 \\ 
56890.111175 & $I$ & 2014-08-20 & 0.233 & 0.092 & 0.214 & 33.4 & 0.034 & 1.0 \\ 
56890.120916 & $V$ & 2014-08-20 & 0.285 & 0.039 & 0.282 & 41.1 & 0.039 & 1.1 \\ 
56890.201550 & $V$ & 2014-08-20 & 0.292 & 0.168 & 0.239 & 27.5 & 0.031 & 0.9 \\ 
56890.210932 & $V$ & 2014-08-20 & 0.316 & 0.174 & 0.263 & 28.3 & 0.031 & 0.9 \\ 
56890.221297 & $V$ & 2014-08-20 & 0.262 & 0.159 & 0.209 & 26.4 & 0.027 & 0.8 \\ 
56890.231776 & $V$ & 2014-08-20 & 0.249 & 0.140 & 0.206 & 27.9 & 0.023 & 0.7 \\ 
56890.324839 & $B$ & 2014-08-20 & 0.249 & 0.178 & 0.173 & 22.1 & 0.077 & 2.2 \\ 
56890.333295 & $V$ & 2014-08-20 & 0.280 & 0.179 & 0.216 & 25.1 & 0.022 & 0.6 \\ 
56890.341413 & $R$ & 2014-08-20 & 0.276 & 0.182 & 0.207 & 24.3 & 0.007 & 0.2 \\ 
56890.353331 & $V$ & 2014-08-20 & 0.275 & 0.175 & 0.213 & 25.3 & 0.011 & 0.3 \\ 
56890.356256 & $V$ & 2014-08-20 & 0.285 & 0.198 & 0.206 & 23.1 & 0.015 & 0.4 \\ 
56890.359180 & $V$ & 2014-08-20 & 0.262 & 0.202 & 0.166 & 19.7 & 0.044 & 1.3 \\ 
56890.362105 & $V$ & 2014-08-20 & 0.276 & 0.179 & 0.211 & 24.9 & 0.007 & 0.2 \\ 
56890.365029 & $V$ & 2014-08-20 & 0.270 & 0.174 & 0.207 & 25.0 & 0.007 & 0.2 \\ 
56902.286837 & $B$ & 2014-09-01 & 0.239 & 0.015 & 0.238 & 43.2 & 0.039 & 1.1 \\ 
56902.299653 & $B$ & 2014-09-01 & 0.207 & 0.139 & 0.154 & 24.0 & 0.020 & 0.6 \\ 
56902.313673 & $V$ & 2014-09-01 & 0.267 & 0.145 & 0.224 & 28.5 & 0.011 & 0.3 \\ 
56902.318603 & $V$ & 2014-09-01 & 0.259 & 0.154 & 0.208 & 26.8 & 0.005 & 0.1 \\ 
56902.324711 & $R$ & 2014-09-01 & 0.245 & 0.123 & 0.212 & 29.9 & 0.009 & 0.3 \\ 
56902.329641 & $R$ & 2014-09-01 & 0.230 & 0.134 & 0.187 & 27.2 & 0.009 & 0.2 \\ 
56902.336581 & $I$ & 2014-09-01 & 0.241 & 0.142 & 0.195 & 27.0 & 0.015 & 0.4 \\ 
56902.342029 & $I$ & 2014-09-01 & 0.210 & 0.117 & 0.174 & 28.0 & 0.030 & 0.9 \\ 
56923.161809 & $V$ & 2014-09-22 & 0.222 & 0.066 & 0.212 & 36.4 & 0.009 & 0.3 \\ 
56923.169206 & $B$ & 2014-09-22 & 0.238 & 0.179 & 0.157 & 20.7 & 0.088 & 2.5 \\ 
56923.176906 & $R$ & 2014-09-22 & 0.192 & 0.073 & 0.178 & 33.9 & 0.024 & 0.7 \\ 
56923.184689 & $I$ & 2014-09-22 & 0.226 & 0.121 & 0.191 & 28.8 & 0.006 & 0.2 \\ 
56923.193235 & $V$ & 2014-09-22 & 0.203 & 0.067 & 0.192 & 35.4 & 0.043 & 1.2 \\ 
56923.199211 & $V$ & 2014-09-22 & 0.245 & 0.101 & 0.223 & 32.8 & 0.015 & 0.4 \\ 
56923.205194 & $V$ & 2014-09-22 & 0.252 & 0.094 & 0.233 & 34.0 & 0.014 & 0.4 \\ 
56923.211175 & $V$ & 2014-09-22 & 0.250 & 0.102 & 0.228 & 32.9 & 0.015 & 0.4 \\ 
56923.220260 & $B$ & 2014-09-22 & 0.172 & -0.004 & 0.172 & 45.7 & 0.105 & 3.0 \\ 
56923.227873 & $R$ & 2014-09-22 & 0.237 & 0.137 & 0.194 & 27.4 & 0.014 & 0.4 \\ 
56923.235107 & $I$ & 2014-09-22 & 0.238 & 0.144 & 0.189 & 26.3 & 0.027 & 0.8 \\ 
56923.243851 & $V$ & 2014-09-22 & 0.231 & 0.127 & 0.193 & 28.3 & 0.064 & 1.8 \\ 
56923.250751 & $V$ & 2014-09-22 & 0.241 & 0.099 & 0.220 & 32.8 & 0.018 & 0.5 \\ 
56923.257654 & $V$ & 2014-09-22 & 0.226 & 0.101 & 0.203 & 31.8 & 0.018 & 0.5 \\ 
56923.273918 & $B$ & 2014-09-22 & 0.279 & 0.219 & 0.173 & 19.1 & 0.085 & 2.4 \\ 
56923.282788 & $R$ & 2014-09-22 & 0.222 & 0.117 & 0.188 & 29.1 & 0.006 & 0.2 \\ 
56923.289502 & $I$ & 2014-09-22 & 0.233 & 0.125 & 0.196 & 28.7 & 0.025 & 0.7 \\ 
56923.296570 & $V$ & 2014-09-22 & 0.220 & 0.097 & 0.198 & 32.0 & 0.008 & 0.2 \\ 
56923.304646 & $V$ & 2014-09-22 & 0.230 & 0.105 & 0.204 & 31.4 & 0.014 & 0.4 \\ 
56924.151246 & $V$ & 2014-09-23 & 0.256 & 0.077 & 0.244 & 36.2 & 0.037 & 1.1 \\ 
56924.158357 & $B$ & 2014-09-23 & 0.268 & 0.190 & 0.189 & 22.4 & 0.023 & 0.7 \\ 
56924.165095 & $R$ & 2014-09-23 & 0.245 & 0.061 & 0.237 & 37.8 & 0.007 & 0.2 \\ 
56924.172355 & $I$ & 2014-09-23 & 0.254 & 0.108 & 0.229 & 32.4 & 0.037 & 1.1 \\ 
56924.178851 & $V$ & 2014-09-23 & 0.244 & 0.098 & 0.223 & 33.2 & 0.006 & 0.2 \\ 
56925.111848 & $V$ & 2014-09-24 & 0.272 & 0.122 & 0.243 & 31.6 & 0.015 & 0.4 \\ 
56946.111394 & $I$ & 2014-10-15 & 0.192 & -0.004 & 0.191 & 45.6 & 0.029 & 0.8 \\ 
56946.125236 & $B$ & 2014-10-15 & 0.225 & 0.083 & 0.209 & 34.2 & 0.014 & 0.4 \\ 
56980.150634 & $V$ & 2014-11-18 & 0.235 & 0.097 & 0.214 & 32.8 & 0.018 & 0.5 \\ 
56980.157276 & $I$ & 2014-11-18 & 0.195 & 0.040 & 0.191 & 39.0 & 0.013 & 0.4 \\ 
56980.163592 & $R$ & 2014-11-18 & 0.199 & 0.072 & 0.186 & 34.4 & 0.021 & 0.6 \\ 
56980.173516 & $B$ & 2014-11-18 & 0.208 & 0.078 & 0.193 & 33.9 & 0.016 & 0.5 \\

\end{longtable}}
\end{center}